\documentclass[prx,aps,twocolumn,10pt,superscriptaddress,notitlepage,longbibliography]{revtex4-2}

\usepackage[dvipsnames]{xcolor}
\usepackage[english]{babel} 
\usepackage{graphicx}
\usepackage{float}
\usepackage{braket}
\usepackage{epstopdf}
\usepackage{mathtools}
\usepackage{amsmath}
\usepackage{amssymb}
\usepackage{bbm,bm}
\usepackage[caption=false]{subfig}
\usepackage{pythonhighlight}
\usepackage{hyperref}
\usepackage{float}
\usepackage{bbold}
\usepackage[T1]{fontenc}
\usepackage{MnSymbol}
\usepackage{graphicx}
\usepackage{standalone}

\usepackage{tabularx}

\usepackage{amsmath,amssymb,amsfonts}
\usepackage{mathrsfs}

\usepackage{booktabs}
\usepackage{amssymb}
\usepackage{makecell}

\usepackage{tikz}

\usetikzlibrary{
arrows.meta,
positioning,
calc,
decorations.pathmorphing,
shapes.geometric,
decorations.markings
}

\usepackage{array}
\newcolumntype{P}[1]{>{\raggedright\arraybackslash}p{#1}}

\newcommand{\Tr}[1]{\mathrm{Tr}[#1]} 

\graphicspath{{./}{./Figures_thesis/}}

\usepackage[markup=underlined]{changes}
\makeatletter
\@namedef{Changes@AuthorColor}{magenta}
\colorlet{Changes@Color}{magenta}
\makeatother
\usepackage{hyperref}
 \hypersetup{
     colorlinks=true,
     linkcolor=blue,
     filecolor=blue,
     citecolor = magenta,      
     urlcolor=red,
     }

\usepackage{braket}

\usepackage{xargs}
\newcommandx{\greencom}[2][1=]
{\todo[inline, color=green!40,#1]{#2}}
\newcommandx{\bluecom}[2][1=]
{\todo[inline, color=blue!40,#1]{#2}}
\newcommandx{\bluemargin}[2][1=]
{\todo[color=blue!40,#1]{#2}}

\usepackage{letltxmacro}
\LetLtxMacro{\ORIGselectlanguage}{\selectlanguage}
\makeatletter
\DeclareRobustCommand{\selectlanguage}[1]{%
  \@ifundefined{alias@\string#1}
    {\ORIGselectlanguage{#1}}
    {\begingroup\edef\x{\endgroup
       \noexpand\ORIGselectlanguage{\@nameuse{alias@#1}}}\x}%
}
\newcommand{\definelanguagealias}[2]{%
  \@namedef{alias@#1}{#2}%
}

\makeatother

\definelanguagealias{en}{english}
\definelanguagealias{EN}{english}
\definelanguagealias{eng}{english}
\definelanguagealias{de}{ngerman}

\newcommand{\ii}{\mathrm{i}}
\newcommand{\dd}{\mathrm{d}}

\newcommand{\lp}{\left(}
\newcommand{\rp}{\right)}
\newcommand{\lsb}{\left[}
\newcommand{\rsb}{\right]}
\newcommand{\lcb}{\left\{}
\newcommand{\rcb}{\right\}}

\newcommand{\tket}[1]{\lvert #1)}

\newcommand{\Lket}[1]{\lvert #1\rrangle}
\newcommand{\Lbra}[1]{\llangle #1\rvert}
\newcommand{\Lbraket}[2]{\llangle #1\vert #2 \rrangle}
\newcommand{\Fket}[1]{\lvert #1)\!\rangle}
\newcommand{\Fbra}[1]{\langle\!( #1\rvert}
\newcommand{\Fbraket}[2]{\langle\!( #1\vert #2 )\!\rangle}
\newcommand{\FLket}[1]{\lvert #1)\!\rrangle}
\newcommand{\FLbra}[1]{\llangle\!( #1\rvert}
\newcommand{\FLbraket}[2]{\llangle\!( #1\vert #2 )\!\rrangle}

\usepackage{booktabs}
\usepackage{amsmath,amssymb}
\usepackage{graphicx}
\usepackage{mathrsfs}

\providecommand{\ii}{\mathrm{i}}
\providecommand{\dd}{\mathrm{d}}
\providecommand{\ket}[1]{\left|#1\right\rangle}
\providecommand{\Lket}[1]{\left|#1\right\rangle\!\rangle}

\begin{document}

\title{Floquet Quasienergy-Resolved Dissipation, Dynamics, and Spectroscopy
\\in Ultrastrong Cavity-QED}


\author{Kamran~Akbari}
\email[]{kamran.akbari@queensu.ca}
\affiliation{Department of Physics, Engineering Physics and Astronomy, Queen's University, Kingston ON K7L 3N6, Canada}
\author{Franco Nori}
\affiliation{Quantum Computing Center, RIKEN, Wakoshi, Saitama, 351-0198, Japan}
\affiliation{Physics Department, The University of Michigan, Ann Arbor, Michigan 48109-1040, USA}
\author{Stephen Hughes}
\email[]{shughes@queensu.ca}
\affiliation{Department of Physics, Engineering Physics and Astronomy, Queen's University, Kingston ON K7L 3N6, Canada}
\date{\today}

\begin{abstract}
Strong periodic driving of cavity-quantum electrodynamics (QED) in the ultrastrong-coupling regime creates nonequilibrium states whose dissipation is governed by Floquet quasienergies rather than by the undriven dressed resonances. 
However, modeling such a regime is a significant theoretical challenge, including a number of subtle problems such as the need to ensure gauge invariance for truncated matter-cavity systems with time-dependent driving.
To fill this theoretical gap, 
we introduce a nonsecular Floquet generalized master equation framework for strongly driven open cavity-QED systems, formulated in the dressed state basis of the quantum Rabi model and applicable to structured reservoirs without rotating-wave approximations.
Our theory can thus model Floquet-driven dynamics
in the ultrastrong coupling regime of 
open system cavity-QED, and demonstrates
a wide range of quantum state control. 
Using both strong optical pumping and strong parametric mechanical modulation, we compute long-time populations, resonance-fluorescence spectra, and the Floquet--Liouville eigenspectra, which resolves observable resonances into hybridized quasienergy transition channels and their 
corresponding decay rates. By systematically comparing with conventional time-independent dressed-basis generalized master equations, we show that static approaches only reproduce steady-state populations in restricted excitation regimes, and thus generally fail for frequency-resolved observables and break down under appropriate Floquet engineering, surprisingly, even for spectrally flat baths. Structured environments, such as Lorentzian--Ohmic reservoirs, further amplify these discrepancies through sideband-selective dissipation.
Our results demonstrate that dissipation in driven ultrastrong cavity-QED is intrinsically quasienergy resolved and we establish Floquet-dissipative theory as an accurate and powerful framework for predicting spectra, controlling decay pathways, and engineering nonequilibrium quantum states, and quantum reservoir engineering.
\end{abstract}

\maketitle
\section{Introduction}
\label{sec:Intro}
Light--matter interactions in cavity-quantum electrodynamics (QED) provide a versatile platform for controlling quantum states~\cite{Haroche_Exploring_2006}, generating nonclassical radiation~\cite{Birnbaum_Photon_2005}, and probing nonequilibrium quantum dynamics~\cite{Brune_Vacuum_1996}. In many realizations, including circuit-QED experiments~\cite{Niemczyk_Circuit_2010,Forn-Diaz_Observation_2010,Tomonaga_Spectral_2025,Wang_Strong_2025},
quantum-level coherent coupling between confined photons and atoms or artificial atoms can now reach the ultrastrong-coupling (USC) regime, where the cavity-dipole interaction strength becomes a sizable fraction of the generic bare subsystem frequencies~\cite{FriskKockum_Ultrastrong_2019,Forn-Diaz_Ultrastrong_2019}. In this USC regime, counter-rotating processes, vacuum dressing, and strong hybridization invalidate standard rotating-wave descriptions and require nonperturbative treatments based on the quantum Rabi model (QRM), which has emerged as the
paradigmatic framework for describing USC cavity-QED~\cite{FriskKockum_Ultrastrong_2019,Forn-Diaz_Ultrastrong_2019,DeLiberato_Virtual_2017}.

Strong periodic driving adds a further layer of quantum control and complexity. Time-dependent Hamiltonians are central to quantum technologies, enabling gate design, state preparation, and Hamiltonian engineering through optical, electrical, or mechanical modulation~\cite{Sinova_Spin_2015,EtxezarretaMartinez_Time-varying_2021,Hamazaki_Exceptional_2021,Baekkegaard_Realization_2019,Chen_Quantum_2021}. When strong driving is combined with USC cavity-QED in the presence of dissipation, however, the system enters a regime where coherent Floquet dressing, hybridized light--matter states, and environment-induced relaxation must be treated on equal footing. Developing a predictive theory for this driven-dissipative regime is a major technical challenge, and is a central motivation of the present work.
%
\begin{figure*}[t]
\centering
\includegraphics[width=.7\linewidth]{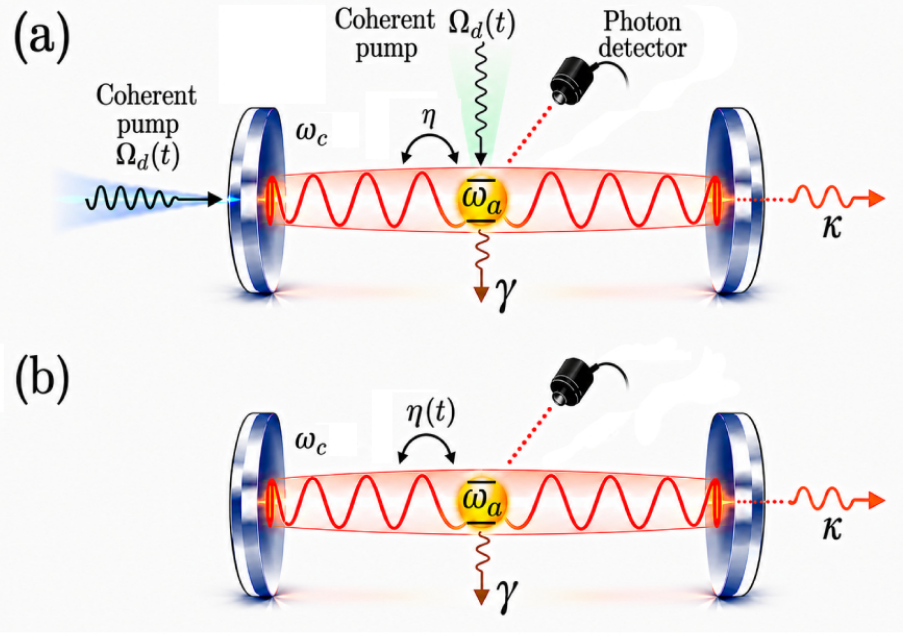}
\vspace{-0.5cm}
    \caption[Schematics]{\textbf{Time-dependent cavity-QED models.} (a) Schematic of an optically pumped and dissipative cavity-QED system via the cavity or the TLS operator. An external laser $\Omega_d(t)$ of strength $\Omega_d=\eta_d\omega_c$ (the Rabi frequency) and frequency $\omega_d$ can be used to possibly pump either the cavity or the TLS. 
    (b) An example schematic of a Floquet-engineered dissipative cavity-QED system; here, the {\it Floquet engineering} of the QRM is accomplished via the (vibration of the atom inside cavity) mechanical modulation of the cavity-TLS coupling rate, $\eta\to\eta(t)=\eta_0+\eta_M\sin(\omega_Mt)$. 
   }
    \label{fig:Schematics}
\end{figure*}

When periodic driving becomes comparable to intrinsic system energy scales, conventional
approaches formulated in the eigenbasis of the undriven Hamiltonian can
become inadequate, even state-of-the-art quantum simulation methods.
Standard time-dependent treatments, including perturbative response theories and optical Bloch-type equations, generally assume that the {\it undriven eigenstates remain the relevant basis}. Under strong driving, however, the drive itself reshapes the spectrum, since it mixes states nonperturbatively, and generates new transition channels that, under suitable driving, cannot be captured by finite-order corrections within the time-independent basis. For harmonically driven systems, Floquet theory provides the natural nonperturbative framework by replacing stationary eigenstates with quasienergy states of the full time-periodic Hamiltonian~\cite{Floquet_FloquetTheory_1883,Shirley_Solution_1965}. 
These {\it driven} states constitute the 
appropriate (and physically meaningful) basis for describing the
coherent dynamics of periodically driven quantum systems, and their
associated quasienergy spectra and Floquet sidebands have been observed
experimentally~\cite{Deng_Observation_2015}.


\begin{figure*}[htbp]
\centering
\includegraphics[width=.95\textwidth]{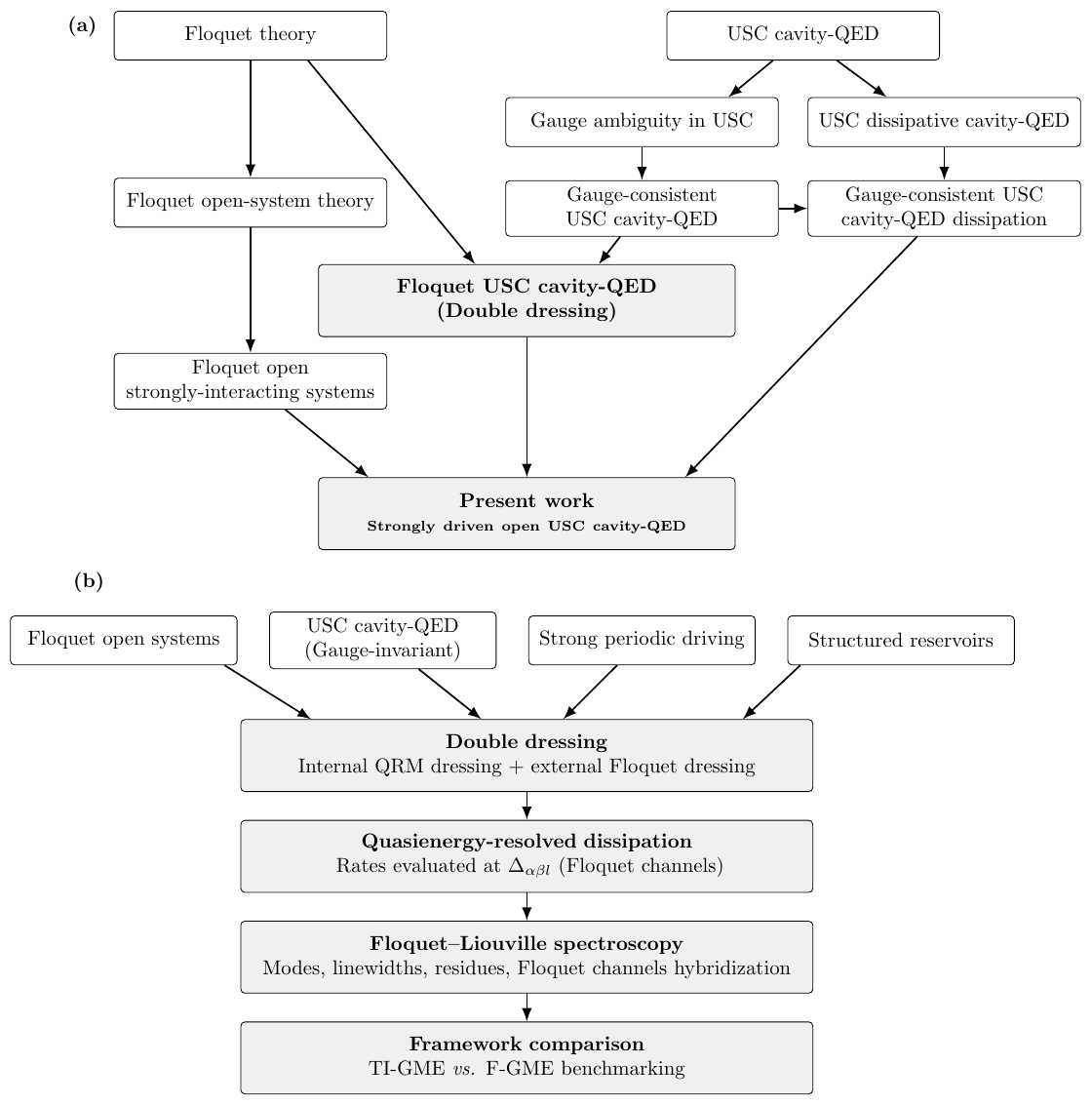}
\caption{\textbf{Historical/conceptual positioning of the presented work.}
(a) Historical development leading to the presented theoretical framework. The left branch summarizes the evolution of Floquet theory from closed periodically driven quantum systems~\cite{Shirley_Solution_1965,Sambe_Steady_1973} to open systems~\cite{Grifoni_Driven_1998,Kohler_Floquet-Markovian_1997,Kohler_Driven_2005,Hausinger_Dissipative_2010,Hausinger_Dissipative_2010PhDThesis,Hausinger_Qubit-oscillator_2010,Restrepo_Driven_2019,Restrepo_Quantum_2018}, and then to open strongly interacting systems~\cite{Akbari_Quasienergy-Resolved_2026}. The right branch summarizes the development of USC cavity-QED~\cite{Rabi_Process_1936,Rabi_Space_1937,Nataf_Ciuti_2010,Ashhab_Two-level_2007}, including the resolution of gauge ambiguities~\cite{DeBernardis_Breakdown_2018} in truncated light–matter models~\cite{DiStefano_Resolution_2019,Settineri_Gauge_2021}, and the formulation of gauge-consistent dissipation~\cite{Salmon_Gauge-independent_2022,Gustin_Gauge-invariant_2023,Gustin_Dissipation_2025} to complement the work on the open USC cavity-QED formalism without addressing the gauge-invariance~\cite{Beaudoin_Dissipation_2011,Ridolfo_Photon_2012,Ridolfo_Quantum_2012,Settineri_Dissipation_2018}. The central branch corresponds to Floquet USC cavity-QED, where internal light–matter dressing and external Floquet dressing are treated on equal footing (``double dressing''). This effort is initiated for a closed USC cavity-QED system~\cite{Akbari_Floquet_2025}, road-mapping the theoretical analysis toward dissipative USC cavity-QED. The present work combines these directions to formulate a rigorous theory of strongly driven open USC cavity-QED systems.
(b) Conceptual structure of the present theory.
The present framework combines strong periodic driving, internal light-matter dressing, structured reservoirs, and quasienergy-resolved dissipation within a unified Floquet generalized master equation framework, enabling Floquet-Liouville modal analysis, microscopic interpretation of spectral features, and direct comparisons between time-independent dressed-basis and Floquet-resolved dissipative theories.
   }
    \label{fig:positioning}
\end{figure*}

\emph{Periodic driving} can enter cavity-QED systems in two conceptually distinct ways. The first is via \emph{coherent pumping} depicted in Fig.~\ref{fig:Schematics}(a), where an external  field (for example, a laser) drives a system operator and induces transitions between light--matter states. The second is via \emph{Floquet engineering}, illustrated in Fig.~\ref{fig:Schematics}(b), where an internal system parameter---such as a resonance frequency, coupling strength, or detuning---is modulated in time, thereby reshaping the effective Hamiltonian and its accessible transitions. Both mechanisms provide powerful routes for nonequilibrium quantum control. Beyond coherent state preparation, periodic driving can also strongly modify dissipative dynamics by redistributing the spectral weight among Floquet sidebands and shifting transition frequencies relative to structured reservoirs. This enables suppression or enhancement of relaxation channels, for example in photonic bandgap or frequency-selective environments~\cite{Hughes_Phonon-mediated_2013,Wang_Dynamical_2019,Kaldewey_Demonstrating_2017,Gustin_Efficient_2020,Zhou_Accessing_2019,Ge_Accessing_2013,Vannucci_Highly_2023}. Floquet engineering has also emerged as a versatile tool for generating topological band structures, tailoring reservoir couplings, and studying nonequilibrium steady states in driven open quantum matter~\cite{Schnell_Dissipative_2024}. In this work, we consider both strong coherent pumping and parametric modulation in the USC regime.

A widely used model of cavity-QED  consists of a single quantized cavity mode of frequency $\omega_c$ coupled to a two-level system (TLS) of transition frequency $\omega_a$, with an interaction strength (rate) $g$. 
Relative to the bare frequencies $\omega_c$ and $\omega_a$, one typically distinguishes perturbative from nonperturbative light--matter interaction.
When $g \ll \{\omega_c,\omega_a\}$, the rotating-wave approximation remains valid, and the Jaynes--Cummings model (JCM)~\cite{Jaynes_Comparison_1963,Cummings_Reminiscing_2013,Larson_Jaynes-Cummings_2024} accurately describes both weak- and conventional strong-coupling regimes. Relative to the dissipation rates $\kappa$ for the cavity and $\gamma$ for the TLS, one distinguishes the weak coupling regime where $g \lesssim \{\kappa,\gamma\}$, from the conventional strong coupling regime where $g \gtrsim \{\kappa,\gamma\}$, depending on whether coherent light--matter exchange is overdamped or spectroscopically resolved. In standard Markovian dynamical regimes, the JCM is able to accurately describe both weak- and strong-coupling important phenomena, including Purcell enhancement, vacuum Rabi splitting, and related radiation-induced collective effects~\cite{Jaynes_Comparison_1963,Rybin_Purcell_2016,Kiraz_Cavity-Quantum_2001,Cummings_Reminiscing_2013,Agarwal_Control_2024,Yamamoto_Preparation_1986,Thompson_Observation_1992,Dicke_Coherence_1954,Lehmberg_Radiation_1970,Agarwal_Master-Equation_1970,Tannoudji_Atom-Photon_1991,Andreani_Strong-Coupling_1999,Lange_Superradiant_2024,Scully_Quantum_1999,Agarwal_Quantum_2012}.

However, when $g$ becomes a sizable fraction of the bare frequencies, commonly quantified by $\eta \equiv g/\omega_c \gtrsim 0.1$, the system enters the USC regime, where counter-rotating-wave and diamagnetic contributions become essential and the QRM is required.
In  USC, the eigenstates are strongly hybridized light--matter states, and even the ground state contains virtual excitations~\cite{DeLiberato_Virtual_2017,FriskKockum_Ultrastrong_2019,Forn-Diaz_Ultrastrong_2019,Qin_Quantum_2024}. In this regime, the QRM provides the minimal description, with carefully constructed gauge-consistent formulations required for truncated matter systems~\cite{DiStefano_Resolution_2019,Salmon_Gauge-independent_2022}.

Including dissipation in USC systems is itself a highly nontrivial task, since baths couple to {\it dressed} light--matter transitions rather than to bare cavity or bare TLS excitations. For open quantum systems, environmental effects are commonly incorporated through Born--Markov master equations~\cite{Carmichael_Dissipation_1999,Breuer_Theory_2002,Carmichael_Statistical_2013,Lidar_Lecture_2020}. In weak and conventional strong coupling, dissipators constructed from bare cavity and TLS operators are usually sufficient. In the USC regime, however, such treatments can become unphysical because the bath couples to hybridized dressed transitions rather than to bare excitations~\cite{Settineri_Dissipation_2018,Salmon_Gauge-independent_2022}. This has motivated the use of {\it generalized master equations} (GMEs) formulated directly in the dressed basis and without secularization, which consistently describe static USC cavity-QED and naturally incorporate structured, frequency-dependent reservoirs~\cite{Settineri_Dissipation_2018,Salmon_Gauge-independent_2022,Akbari_Generalized_2023}.

For periodically driven open quantum systems, standard Floquet--Markov approaches are developed over the bare (lab) basis and often employ full or partial secular approximations to obtain Lindblad-form generators~\cite{Grifoni_Driven_1998,Kohler_Floquet-Markovian_1997,Kohler_Driven_2005,Hausinger_Dissipative_2010,Hausinger_Dissipative_2010PhDThesis,Hausinger_Qubit-oscillator_2010,Restrepo_Driven_2019,Restrepo_Quantum_2018,Breuer_Theory_2002,Jeske_Bloch-Redfield_2015,Tscherbul_Partial_2015,Lim_Signatures_2017,NafariQaleh_Enhancing_2022,Mori_Floquet_2023,Winczewski_Intermediate-times_2024}. These approaches {\it do not} account for the internal interaction-induced dressing and the secular approximations often require quasienergy splittings to remain well separated from dissipative linewidths. 
In driven USC cavity-QED, however, quasienergy manifolds within a Brillouin zone (BZ) can become dense, exhibit near-degeneracies, strong harmonic hybridization (where Floquet sidebands
can thus strongly mix), 
and selective
coupling
 to structured reservoirs. Under such conditions, 
{\it secular truncations} can suppress interference between decay pathways and distort observable spectral weights~\cite{Restrepo_Quantum_2018,Restrepo_Driven_2019,LeBoite_Theoretical_2020}.

The central issue is therefore not whether Floquet master equations are necessary, but under what physical conditions static dressed-basis approaches appear to succeed---and 
{\it when they fundamentally fail}. We directly address this question for driven-dissipative cavity-QED {\it in the USC regime}, considering both strong coherent pumping and strong parametric modulation. Building on nonperturbative Floquet treatments of the closed QRM~\cite{Akbari_Floquet_2025} and gauge-consistent dressed-basis GME for static USC systems~\cite{Salmon_Gauge-independent_2022}, which we refer to as the time-independent-basis GME (TI-GME), we develop a fully nonsecular Floquet-Markov generalized master equation (F-GME) that spectrally resolves dissipation directly in the quasienergy basis~\cite{Akbari_Quasienergy-Resolved_2026}.

We further develop a Floquet–Liouville (FL) supermatrix formulation of the GMEs, providing a complementary modal description of strongly driven open quantum systems~\cite{Chu_Beyond_2004}. Beyond its applications to nonequilibrium critical phenomena and dissipative phase transitions~\cite{Mercurio_Floquet_2026}, the FL framework enables a direct decomposition of observable resonances into hybridized quasienergy-transition channels and their associated decay rates. This analysis yields a microscopic interpretation of fluorescence spectra, reveals the interplay between Floquet sidebands and dissipative processes, and provides a systematic basis for comparing TI-GME and F-GME descriptions of open quantum dynamics.

We stress that the present framework combines ingredients that have
previously been treated separately, namely, gauge-consistent USC
cavity-QED, strong periodic driving, structured reservoirs,
quasienergy-resolved dissipation, and Floquet--Liouville modal analysis; see the structural and developmental flow of the present work in Fig.~\ref{fig:positioning}.
To our knowledge, no existing approach simultaneously incorporates all
of these elements within a unified nonsecular framework.
With the selected examples, 
our results show that dissipation in strongly driven ultrastrong cavity-QED is intrinsically quasienergy resolved, while establishing quantitative criteria for the breakdown of TI-GMEs. More generally, we  demonstrate that Floquet-dissipative theory is essential for accurately predictive spectroscopy, engineered quantum control, and reservoir engineering. Furthermore, the theory and methods are generally and widely applicable to many quantum systems.

The major contribution of this work is not the use of Floquet theory itself, but rather its integration with a nonsecular dressed-basis open-system treatment appropriate for the USC regime of cavity-QED, to {\it adequately address} the \emph{double dressing} scheme, when also the dissipation coexists, on an equal footing. In particular, we formulate a quasienergy-resolved Floquet generalized master equation in which dissipation is evaluated at the driven transition frequencies $\Delta_{\alpha\beta l}$, rather than at the static dressed-state splittings~\cite{Grifoni_Driven_1998,Mori_Floquet_2023,Akbari_Quasienergy-Resolved_2026}. This allows us to directly benchmark the commonly used time-independent dressed-basis GME~\cite{Salmon_Gauge-independent_2022} against a fully Floquet-resolved dissipative theory and to identify the regimes in which static dissipators remain reliable or fail.

A second key element is the FL modal analysis, applied to the GMEs, which resolves each observable spectral feature into dissipative eigenmodes, linewidths, residues, and dominant quasienergy-transition channels. This provides a microscopic interpretation of the breakdown of time-independent dissipators, rather than only a numerical comparison of spectra.

The rest of the paper is organized as follows. Section~\ref{sec:BackgroundTheory} summarizes relevant driven and dissipative USC cavity-QED models. Section~\ref{sec:Theory} presents the theoretical Floquet dissipative framework and Floquet-Markov generalized master equation. Section~\ref{sec:ResultsDisscussion} compares populations and spectra for a driven QRM with static approaches, and Section~\ref{sec:Conclusions} concludes.

Additionally, in the Appendices, we also provide further details that are complementary to the arguments in the main text; specifically, we discuss the gauge consistency of the Floquet theory for the driven QRM in
App.~\ref{secS:DQRM_GaugeIndependence}, Floquet theory and the extended space configuration for the driven QRM in App.~\ref{secS:DQRM}, derivation of the Floquet-Born-Markov GME for the driven-dissipative QRM in App.~\ref{secS:FGME}, FL diagonalization and modal decomposition in the FL extended space in
App.~\ref{secS:FL}, and finally provide additional 
numerical results and discussions in
App.~\ref{secS:AdditionalResults}. 

\section{Background Theory}
\label{sec:BackgroundTheory}
In this section, we first give a 
brief 
review of the relevant USC cavity-QED models in the existing literature, including some important and recent advances. We then discuss
the QRM with no dissipation (closed quantum system), highlighting the
differences needed in the USC regime versus the JCM. Then, we 
discuss the methods to include dissipation
(open quantum system) for undriven and driven systems. 

\subsection{Nondissipative quantum light-matter interaction cavity-QED models in ultrastrong coupling}

\subsubsection{Undriven time-independent quantum Rabi model} 

Beyond the weak and strong coupling regimes of cavity-QED, higher-order, nonlinear, and nonperturbative light-matter effects are inevitable;
this is because the system-level Hamiltonian can admit non-number-conserving and squeezing processes, in the interaction between the quantum matter system and the quantized photons. These processes are due to the presence of the counter-rotating wave terms, 
i.e., terms proportional to $a\sigma^-$ and $a^\dagger\sigma^+$, or pondermotive forces, i.e., terms proportional to $(a+a^\dagger)^2$, in light-matter interaction Hamiltonian, where ${a}$ (${a}^\dagger$) is the cavity photon annihilation (creation) operator and ${\sigma}^+=\vert{e}\rangle\langle{g}\vert$ (${\sigma}^-=\vert{g}\rangle\langle{e}\vert$)  the atomic raising (lowering) operator.

The QRM Hamiltonian, in the dipole gauge
  (using units of $\hbar=1$) is
\begin{equation}
    \begin{split}
{\mathcal{H}}_\mathrm{QRM}^{\rm D}&=\omega_\mathrm{c}{a}^\dagger{a} 
+ \frac{\omega_{a}}{2}\sigma_z + \ii \eta\omega_c \left(a^\dagger-a\right)\sigma_x,
    \end{split}
    \label{eq:HD_QRM}
\end{equation}
up to a constant ($\sigma_x^2\, g^2/\omega_c$), with $\sigma_i$ ($i=x,y,z$) being the TLS Pauli operator.
In the usual Coulomb gauge, 
which we term the {\it naive Coulomb} (`NC') gauge,
since it produces the wrong result in the USC, one 
has
\begin{equation}
  {\mathcal{H}}^{\rm NC}_{\rm QRM}=\omega_\mathrm{c}{a}^\dagger{a} + (\omega_a/2)\sigma_z+ \eta\omega_a({a} + {a}^\dagger){\sigma}_y + \eta^2\omega_a({a} + {a}^\dagger)^2.
  \label{eq:HNC_QRM}
\end{equation}

The reason that
Eq.~\eqref{eq:HNC_QRM}
fails in the USC regime, 
is due to a breakdown of  
gauge-invariance~\cite{DeBernardis_Breakdown_2018,DiStefano_Resolution_2019},
which is  caused by the matter/atom truncation to a simple TLS (i.e., a reduced Hilbert space); this manifests in a nonlocal potential, which causes the minimal coupling replacement to take on a modified form that satisfies the gauge principle~\cite{DiStefano_Resolution_2019,Gustin_Gauge-invariant_2023}.
 This can be solved by a Power-Zienau-Woolley gauge transformation~\cite{Power_Coulomb_1951,Atkins_Interaction_1970,Woolley_Molecular_1971}
 in the reduced Hilbert space via~\cite{DiStefano_Resolution_2019,Gustin_Gauge-invariant_2023} 
 \begin{equation}
    \mathcal{H}_{\rm QRM}^{\rm C}=\mathcal{U}\,\mathcal{H}_{\rm QRM}^{\rm D}\,\mathcal{U}^\dagger, 
    \quad
    {\cal U}=\exp[-\ii\eta({a}+{a}^\dagger){\sigma}_x],
    \end{equation}
 yielding the 
 ({\it correct}) Coulomb gauge-corrected QRM, 
\begin{equation}
    \begin{split}
        {\mathcal{H}}^{\rm C}_\mathrm{QRM} &= \omega_\mathrm{c}{a}^\dagger{a}
\\
&\hspace{0.5cm}+ \frac{\omega_{a}}{2}\left(\sigma_z\cos[2({a}+{a}^\dagger)\eta ]+\sigma_y\sin[2({a}+{a}^\dagger)\eta]\right),
    \end{split}
    \label{eq:HC_QRM}
\end{equation}
which now yields identical eigenergies to those of
Eq.~\eqref{eq:HD_QRM}, as it must.

Thus, in the Coulomb gauge, 
the nonperturbative cavity-TLS interaction manifests in a gauge-invariant QRM that
necessarily contains photon terms to all orders in the transcendental form of Eq.~\eqref{eq:HC_QRM}.
However, with time-dependent system interactions, 
then additional terms are needed in the dipole
gauge, and it becomes far more 
convenient to solve 
the
Coulomb gauge-corrected  QRM
~\cite{Settineri_Gauge_2021,Gustin_Gauge-invariant_2023,Akbari_Floquet_2025}. 
For example, if $\eta$ becomes time depenendent, then one simply replaces 
$\eta$ by $\eta(t)$ in
Eq.~\eqref{eq:HC_QRM}. This is 
{\it not} 
equivalent to replacing
$\eta$ by $\eta(t)$ in
Eq.~\eqref{eq:HD_QRM}, which 
would fail (without the awkward consideration of a another time-dependent gauge term)~\cite{Settineri_Gauge_2021,Gustin_Gauge-invariant_2023}. 

Note that without dissipation and time-dependent interactions, the QRM is analytically solvable by the parity symmetry into two solvable disjoint subspaces~\cite{Braak_Integrability_2011}. However, often the numerical diagonalization is preferred to avoid the complexity of the analytical solutions. 
Letting the light-matter interaction coupling rate be arbitrary, leads to the nonperturbative treatment and diagonalization in the dressed picture. The time-independent atom-cavity Hamiltonian in Eqs.~\eqref{eq:HD_QRM} and \eqref{eq:HC_QRM} can be diagonalized straightforwardly~\cite{Braak_Integrability_2011,DiStefano_Resolution_2019} as a set of ${\rm Dim}(\mathbb{H}_{\rm dressed})$ (ideally, infinite number, but practically truncated with sufficiently large number of) joint (QRM-dressed) light-matter eigenstate basis~\cite{Braak_Integrability_2011,LeBoite_Theoretical_2020,DiStefano_Resolution_2019,FriskKockum_Ultrastrong_2019}, 
\begin{equation}
    \mathsf{B}_{\rm dressed}=\lcb\ket{j}\equiv\sum_{na}c^{(j)}_{na}\,\ket{n}\otimes\ket{a}\rcb,
\end{equation}
of energies, $E_j$, with the photon number $n=0,1,2,\dots$, and TLS states $a=g,e$; here, these QRM-dressed states are the joint states of the atom with truncated internal states represented as a TLS (with saturation effects possible already in vacuum), and the cavity with suitably large photon number states.

Consequently, system operators must be consistently expressed in the dressed eigenbasis of the QRM. Any bare operator 
$O^{\rm bare}$ is then represented as 
\begin{equation}
    O=\sum_{jk}O^{\rm bare}_{jk}\ket{j}\bra{k},
\end{equation}
with $O^{\rm bare}_{jk}\equiv \bra{j}O^{\rm bare}\ket{k}$.
Of particular importance are the ladder operators in the dressed picture, defined separately for each subsystem. The cavity and TLS raising operators take the form~\cite{Settineri_Dissipation_2018}, respectively:
\begin{equation}
\begin{split}
    {s}^{\rm cav+} &= \sum_{j,k>j}\braket{j|{\Pi}|k}\ket{j}\bra{k},
    \\
    {s}^{\rm TLS+} &= \sum_{j,k>j}\braket{j|{\sigma_x}|k}\ket{j}\bra{k},
    \end{split}
\label{eq:splus}
\end{equation}
with the corresponding lowering operators given by $s^{\Lambda -}=\lp s^{\Lambda +}\rp^\dagger$ for the $\Lambda$-subsystem. These dressed operators correctly encode the allowed transition matrix elements and selection rules in the USC regime.
Here, the field quadrature $\Pi$ operator is gauge-dependent as the eigenstates are, and we assume that ${\Pi}$ 
has electric field coupling, such that
${\Pi}^{\rm C}=\ii({a}^\dagger-{a})$ in the Coulomb gauge~\cite{DiStefano_Resolution_2019,Salmon_Gauge-independent_2022}.

Naturally, the QRM also predicts the correct physics in the weak and strong coupling regimes
(usually when $\eta \lesssim0.1$ as mentioned in the introduction), and the undriven closed system is identified within the joint states of the bare light and the bare matter. 
The sufficient model for these regimes is the JCM Hamiltonian~\cite{Jaynes_Comparison_1963,Cummings_Reminiscing_2013,Larson_Jaynes-Cummings_2024},
\begin{equation}
\begin{split}
{H}_{\rm JCM}&=\omega_\mathrm{c}{a}^\dagger{a} + \frac{\omega_{a}}{2}\sigma_z+ \ii g({a}^\dagger{\sigma}^--{a}{\sigma}^+),
\end{split}
\label{eq:HD_JCM}
\end{equation}
where the effects of the counter-rotating waves are neglected,
which is appropriate  
when $\lvert\delta\rvert\ll\Sigma$, with $\delta=\omega_c-\omega_a$ and $\Sigma=\omega_c+\omega_a$.

The above JCM Hamiltonian  is introduced on the joint bases of the light part (cavity photons) identified by the Hilbert space of $\mathbb{H}_c$ with the dimension of ${\rm Dim}(\mathbb{H}_c)\geq1$, and the matter part identified by the Hilbert space of $\mathbb{H}_a$ with the dimension of ${\rm Dim}(\mathbb{H}_a)=2$. Hence, the bare state representation of this space is based on the basis set composed of the joint bare eigenstates of these two components, i.e., $\mathbb{H}_{\rm bare}=\mathbb{H}_c\otimes\mathbb{H}_a$ having the dimension of ${\rm Dim}(\mathbb{H}_{\rm bare})={\rm Dim}(\mathbb{H}_c)\times {\rm Dim}(\mathbb{H}_a)$.

In reality, both ${\rm Dim}(\mathbb{H}_c)$ and ${\rm Dim}(\mathbb{H}_a)$ are infinity (for real atoms),  but can be truncated with a sufficiently large number and two, respectively. The basis set of this (complete) Hilbert space is \begin{equation}
    \mathsf{B}_{\rm bare}=\lcb\ket{j}\equiv\lvert n,a\rangle=\ket{n}\otimes\ket{a}\rcb.
\end{equation}
Similar to the QRM, the JCM is also analytically solvable, however, thanks to the U(1) symmetry, the JCM has a much simpler structure~\cite{Jaynes_Comparison_1963,Cummings_Reminiscing_2013,Scully_Quantum_1999}. The state $\lvert n=0,a=g\rangle\equiv\lvert j=0\rangle$ is the unique ground state of the JCM in the weak and strong coupling regimes, with eigenenergy $E_{n=0,a=g} = -\omega_a/2\equiv E_{j=0}$. 

With the intermixing of the light-matter states via the finite $g$, the excited eigenstates are $\lvert n>0,a\rangle\equiv\lvert j=n^\pm\rangle$,
 with the corresponding eigenenergies~\cite{Jaynes_Comparison_1963,Cummings_Reminiscing_2013,Scully_Quantum_1999}:
\begin{equation}
    E_{j=n^\pm}=\lp n-\frac{1}{2}\rp\omega_c\pm\frac{1}{2}\sqrt{4g^2n+\delta^2}.
    \label{eq:EE_JCM}
\end{equation}
Thus, the JCM can be easily diagonalized into the dressed basis of the joint light-matter states, $\{\lvert j=0,1^\pm,2^\pm,\dots\rangle\}$.
The splitting of the JCM excited states with the same photon number reads,
\begin{equation}
    E_{n^+}-E_{n^-}=\sqrt{4g^2n+\delta^2},
    \label{}
\end{equation}
which yields an initial splitting of $\lvert\delta\rvert$ for $g\to0^+$, where the light and matter are decoupled.

In contrast to the simplicity of the JCM ground state, the QRM contains nonlinear saturation effects even in the vacuum,
and the hybrid ground state is an entangled state of photonic and atomic excitations, defined from
\begin{equation}
\lvert j=0\rangle=\sum_{k=0}^\infty\langle g,2k\vert j=0\rangle\lvert j=0\rangle+\langle e,2k+1\vert j=0\rangle\lvert j=0\rangle.
\end{equation}

\subsubsection{Driven (or strongly perturbed) time-dependent\\ quantum Rabi model}
Harmonic drives play a central role in both fundamental and applied physics, particularly since the advent of lasers~\cite{Schaack_Lasers_1963,Bromberg_Birth_1988,Slusher_Laser_1999}, which are now routinely available at high powers~\cite{Strickland_Compression_1985,Maine_Amplification_1988,Mourou_Extreme_2019}. Such sources serve as efficient coherent pumps across a wide range of platforms, including nonlinear and strong-field optics~\cite{Mourou_Optics_2006,Teubner_High-order_2009,Krausz_Attosecond_2009,Fennel_Laser-driven_2010,Kruchinin_Strong-field_2018,Dombi_Strong-field_2020}, strong-field QED~\cite{Seipt_Nonlinear_2011,DiPiazza_Extremely_2012,Fedotov_Advances_2023}, cold-atom and cavity-QED systems~\cite{Ritsch_Cold_2013}, and ultrafast light–matter interactions. A harmonic drive is characterized by its oscillation frequency, coupling strength to the system, and phase. Together, these parameters determine how energy is injected, how transitions are activated, and how the system response is structured relative to the driving field. Harmonic modulation is also a key resource in quantum circuits and quantum-information processing architectures based on QED platforms.

For a single monochromatic drive, such as a coherent harmonic pump field, the absolute drive phase typically has limited dynamical significance. In contrast, in multi-tone driving or in Floquet-engineered 
settings---where intrinsic system parameters are periodically modulated---the relative phases become essential. In such cases, phase differences can generate effective dc components, induce nontrivial Hamiltonian renormalizations, and qualitatively modify the dynamics. The physical regime is controlled primarily by the drive amplitude and frequency relative to intrinsic energy scales of the system. 
While their interplay ultimately governs the response, we first discuss their roles separately to clarify the underlying mechanisms.

To describe a {\it driven} quantum Rabi model (DQRM), we consider a general time-periodic Hamiltonian $\mathcal{H}_{\rm DQRM}(t)=\mathcal{H}_{\rm DQRM}(t+T)$,
which can be expanded as a Fourier series,
\begin{equation}
    \mathcal{H}_{\rm DQRM}(t)=\sum_{m\in\mathbb{Z}} \mathcal{H}_m\,\mathrm{e}^{-\ii m\omega_dt}= \mathcal{H}_0+\mathcal{H}_d(t),
    \label{eq:HDQR_m}
\end{equation}
 where $\mathcal{H}_d(t)\equiv\sum_{m\neq0} \mathcal{H}_m\,\mathrm{e}^{-\ii m\omega_dt}$ contains the time-dependent components, and hermiticity implies $\mathcal{H}_{-m}=\mathcal{H}_m^\dagger$. 
 Two forms for the harmonic drive are widely used, which we next discuss below.


\paragraph{Floquet coherent pumping.} The first and more conventional form of periodic driving is what we call a \emph{Floquet coherent pump}, in which a system operator $S_d$ (note that in the USC this system operator must be defined in the dressed picture [cf. Eq.~\eqref{eq:splus}]) is coherently pumped with amplitude $\Omega_d$ (Rabi frequency), single frequency $\omega_d=2\pi/T$, and phase $\phi_d$. 
The \emph{nondissipative} coherently pumped quantum Rabi model (CPQR) reads~\cite{Salmon_Gauge-independent_2022}
\begin{equation}
\begin{split}
    {\mathcal{H}}_\mathrm{CPQR} (t)&= {\mathcal{H}}_\mathrm{QRM}+\mathcal{H}_d(t),
\end{split}
\label{eq:H_CPQR}
\end{equation}
where we work in the Coulomb gauge, so that ${\mathcal{H}}_\mathrm{QRM}\equiv{\mathcal{H}}^{\rm C}_\mathrm{QRM}$ is the QRM (time-independent) Hamiltonian~\cite{DiStefano_Resolution_2019,Salmon_Gauge-independent_2022}, given in Eq.~\eqref{eq:HC_QRM}.

Without applying a RWA to the drive, then we have
the following harmonic drive term,
\begin{equation}
    \mathcal{H}_d(t)=\Omega_d\, S_d\,\cos(\omega_dt+\phi_d).
\end{equation} 
Thus, for a single-frequency coherent drive, the Fourier components are
$\mathcal{H}_0=\mathcal{H}_{\rm QRM}$, 
$\mathcal{H}_{\pm1}=\Omega_d\,\mathrm{e}^{\mp\ii\phi_d}\,S_d/2$, and $\mathcal{H}_m=0$ for $m\neq0,\pm1$.
Near resonance and for weak coupling, one may apply the {\it drive-RWA}~\cite{Salmon_Gauge-independent_2022}, which reads
\begin{equation}
    H_d^{\rm RWA}(t)=\Omega_d(s^-_d\mathrm{e}^{-\ii\omega_d t}+s^+_d\mathrm{e}^{\ii\omega_d t}),
\end{equation}
where $s_d^\pm$ are the positive/negative-frequency components of the dressed operator $S_d=s_d^++s_d^-$. This approximation is valid for $\Omega_d\ll\omega_d$ and small detuning.
However, the RWA fails to capture higher-order effects, such as dynamical Stark shifts, multiphoton resonances, and drive-induced renormalizations of relaxation and dephasing rates~\cite{Hausinger_Dissipative_2010}. In the USC regime, where counter-rotating contributions are intrinsic (even with no driving), the full drive Hamiltonian must be retained.

The drive Hamiltonian describes photon-field pumping of the QRM-dressed states and thus manipulates the transitions between the dressed QRM eigenstates~\cite{Salmon_Gauge-independent_2022}. In this work, we focus on driving the TLS operator, 
$S_d^{\rm bare}=\sigma_x$, so that 
\begin{equation}
  s_d^\pm=s^{{\rm TLS}\pm}.  
\end{equation}
 This choice strategically avoids the buildup of a large coherent cavity background field, which is typically filtered out experimentally when probing quantum correlations~\cite{Fischer_Pulsed_2018,Liu_Dynamic_2024}, and has little effect on the ensuing quantum correlations (apart from trivial coherent scattering).

\paragraph{Floquet engineering via parametric modulation.} The second class of periodic driving we consider is \emph{Floquet engineering}, in which a parameter of the static Hamiltonian is modulated, for example the light–matter coupling rate~\cite{Akbari_Floquet_2025}, $g\to g(t)$, 
or a system generic frequency, $\omega_{c,a}\to\omega_{c,a}(t)$. 
Unlike the coherent pump considered above—where the drive appears as an additive term to the static Hamiltonian—the  mechanical modulation reshapes the Hamiltonian parametrically, modifying its internal structure in time. Although this modulation can formally be expanded in a Fourier series and recast as a sum of additive harmonic components acting on a static Hamiltonian, such a representation necessarily introduces an infinite hierarchy of higher-frequency terms at integer multiples of the fundamental drive frequency, able to inject time-dependence via a combination of operators. Consequently, the effective description involves a richer set of harmonic contributions than in the simple coherent pump case.

With regards to related and emerging experiments,
harmonic parameter modulation is experimentally accessible in several 
quantum systems, such as those exhibiting the dynamical Casimir effect~\cite{Johansson_Dynamical_2009,Macri_Nonperturbative_2018,Nation_Ultrastrong_2016},
surface acoustic waves~\cite{Lin_Strong_2015,Srinivasan_Cavity_2019,2019_Delsing_2019,Iikawa_Optical_2016,Manenti_Circuit_2017,Dumur_Quantum_2021,Weiss_Optomechanical_2021,DeCrescent_Large_2022,Wang_Gated_2024,Iyer_Coherent_2024,Patel_Surface_2024,Chen_Hybrid_2026},
optomechanical platforms~\cite{Cirio_Amplified_2017}, and molecular optomechanics~\cite{Roelli_Molecular_2015,Schmidt_Quantum_2016,Dezfouli_Molecular_2019,PhysRevB.104.045431}. These \emph{system-level modulations} enable controlled engineering of effective light–matter interactions, and are directly relevant to superconducting circuits, optomechanical platforms, and hybrid systems where mechanical or parametric modulations are used to engineer effective light–matter interactions.

We will consider an amplitude-modulated TLS-cavity system in the USC regime, that is the nonperturbative mechanical (or vibrational) perturbation
of a TLS ultrastrongly coupled to 
a single quantized cavity, described by the Floquet-engineered quantum Rabi model (FQRM)~\cite{Akbari_Floquet_2025}
\begin{equation}
\begin{split}
 {\mathcal{H}}_\mathrm{FQRM}(t)&=\omega_ca^\dagger a+\frac{\omega_a}{2}\lcb\sigma_z\cos\lsb2({a}+{a}^\dagger)\eta(t)\rsb\right.
\\
&\hspace{3cm}\left.+\sigma_y\sin\lsb2({a}+{a}^\dagger)\eta(t)\rsb\rcb,
\end{split}
\label{eq:HC_FQR}
\end{equation}
with the time-dependent cavity-TLS coupling 
rate (normalized), 
\begin{equation}
    \eta(t)=\eta_0+\eta_M\,\sin\omega_M t,
\end{equation}
where $\eta_0$ is the static and $\eta_M$ is the dynamic (normalized) coupling rates, and the mechanical oscillation frequency is $\omega_M$ (here, the drive frequency $\omega_d$ is equal to the mechanical oscillation frequency $\omega_M$).

Since $\eta(t)=\eta(t+T)$, the Hamiltonian is periodic and admits a Fourier expansion of the form~\eqref{eq:HDQR_m}. Using the Jacobi–Anger expansion~\cite{Abramowitz_Handbook_1965}, one obtains~\cite{Akbari_Floquet_2025}, with the negative sign convention in the Fourier series,
\begin{equation}
\begin{split}\displaystyle
 \mathcal{H}_m
&=\omega_\mathrm{c}{a}^\dagger{a} \,\delta_{m0}+\dfrac{\omega_{a}}{2}
\left \{
\dfrac{\sigma_z+\ii\sigma_y}{2}\,\mathrm{e}^{-\ii2({a}+{a}^\dagger)\eta_0} \right.
\\
& 
\hspace{-0.5cm}\left.
+(-1)^m\dfrac{\sigma_z-\ii\sigma_y}{2}\,\mathrm{e}^{\ii2({a}+{a}^\dagger)\eta_0} 
\right \}
J_m[2({a}+{a}^\dagger)\eta_{M}],
\end{split}
\label{eq:HFQR_m}
\end{equation}
where $J_m$ is the Bessel function of the first kind of order $m$.
When
$\eta_{M} \rightarrow 0$, only
$J_0(0)=1$ survives, and the time-independent QRM Hamiltonian is recovered. 

In practice, the Fourier expansion must be truncated to $\lvert m\rvert\leq m_{\rm max}$. The $m=0$ component contains the static Hamiltonian that is a renormalization of the QRM arising from the modulation~\cite{Akbari_Floquet_2025}, while $m\neq0$ terms account for genuine time-dependent interactions. Thus, Eq.~\eqref{eq:HC_FQR} accounts for static dressing, whereas Eq.~\eqref{eq:HFQR_m} represents a full periodic dressing of the cavity-QED system.

\paragraph{Weak versus strong driving regimes.}
The drive amplitude distinguishes perturbative and nonperturbative regimes. For coherent pumping, a weak-drive condition may be expressed as~\cite{Salmon_Gauge-independent_2022} 
\begin{equation}
\eta_d\equiv\Omega_d/\omega_c\ll\eta \quad \mathrm{(or,} \ \Omega_d<0.1 g),
\end{equation}
though more generally this can be even smaller in a broadband coupling regime \cite{Gustin_Dissipation_2025},
and dissipation rate can also play a role to unblock or suppress higher-order drive-assisted processes~\cite{Akbari_Quasienergy-Resolved_2026}. In this weak drive regime,  perturbative methods in the drive can suffice. As the drive strength increases, higher-order processes become relevant, leading to Stark shifts, multiphoton resonances, and modified avoided crossings in the spectrum.

However, while higher-order perturbative expansions can capture specific corrections~\cite{Hausinger_Dissipative_2010,Hausinger_Qubit-oscillator_2010,Kockum_Deterministic_2017}, a fully nonperturbative treatment is required to consistently account for the interplay of strong driving, detuning, and ultrastrong light–matter coupling. For harmonic drives, Floquet theory provides such a framework by transforming the time-dependent problem into an equivalent time-independent quasienergy problem in an enlarged space. The original eigenenergy basis is replaced by a (formally infinite) set of quasienergy states, restoring the applicability of many tools developed for time-independent systems while retaining the full nonlinear drive physics.

\subsection{Dissipative quantum light-matter interaction cavity-QED models in ultrastrong coupling}
In the USC regime, the cavity and matter degrees of freedom hybridize nonperturbatively. Thus, dissipation and detection cannot be formulated in terms of bare subsystem operators. Instead, the environment couples to transitions of the fully dressed eigenstates of the QRM Hamiltonian, and the hybrid cavity–TLS system must be treated as a single quantum entity diagonalized in its dressed basis~\cite{Settineri_Dissipation_2018,Salmon_Gauge-independent_2022,Hughes_Reconciling_2024}.

Therefore, the nonsecular dressed-basis GME (TI-GME) is adopted, which,
for certain regimes studied, has been shown to correctly reproduce USC physics and gauge-consistent open-system dynamics~\cite{Settineri_Dissipation_2018,Salmon_Gauge-independent_2022}. In this approach, no rotating-wave (secular) approximation is imposed in the dissipator, and all transition frequencies of the dressed spectrum are retained.

Depending on how the system is perturbed, the form of the TI-GME can slightly differ. When the system is  time-independent, and we are looking for the generic transitions in the system, the pump term can also be identified time-independently and be added to the dissipator. One such term is known as an {\it incoherent} pump term which can also be used to model thermal fluctuations (for example). In this case, the TI-GME for the incoherent pump reads
\begin{equation}
    \begin{split}
        \frac{\dd\rho(t)}{\dd t}&=\mathcal{L}\rho(t)
    \end{split},
    \label{eq:TIGME_inc}
\end{equation}
where the total Liouvillian 
\begin{equation}
    \mathcal{L}=\mathcal{L}_{\rm S}+\mathcal{L}_{\rm diss}+\mathcal{L}_{\rm pump},
\end{equation}
is time-independent as it contains three time-independent ingredients of the system, dissipation and incoherent pump Liouvillians. 
In this model, the system Liouvillian only includes the ({\it time-independent}) QRM via 
\begin{equation}
    \mathcal{L}_\mathrm{S}\rho(t)=-\ii\lsb\mathcal{H}_{\rm QRM},\rho(t)\rsb.
\end{equation}

An incoherent pump, when included to model thermal or noise-induced excitation, is easily included, and is written as
\begin{equation}
    \mathcal{L}_{\rm pump}\rho(t)=
     P^\Lambda_{\rm inc}\,\mathcal{D}[s^{\Lambda-}]\,{\rho}(t),
    \label{eq:L_incPump}
\end{equation} 
with  
\begin{equation}
{\cal D}[{O}]{\rho} = \frac{1}{2}(2{O}{\rho}{O}^\dagger - {\rho}{O}^\dagger{O} - {O}^\dagger{O}{\rho}).
\end{equation}

\subsubsection{Dissipator in the dressed basis}
The nonsecular dissipator reads
\begin{widetext}
\begin{equation}
    \begin{split}
      \mathcal{L}_{\rm diss}\rho(t)&=\sum_{\Lambda}\sum_{\omega,\omega'>0}
      \lcb\Gamma^{\Lambda}(\omega)\lsb{S}_{\rm B}^{\Lambda+}(\omega) {\rho}(t){S}_{\rm B}^{\Lambda-}(\omega')- {S}_{\rm B}^{\Lambda-}(\omega'){S}_{\rm B}^{\Lambda+}(\omega){\rho}(t)\rsb\right.
        \\
        &\hspace{3cm}\left.+
\Gamma^{\Lambda*}(\omega')\lsb {S}_{\rm B}^{\Lambda+}(\omega){\rho}(t){S}_{\rm B}^{\Lambda-}(\omega')-{\rho}(t) {S}_{\rm B}^{\Lambda-}(\omega') {S}_{\rm B}^{\Lambda+}(\omega) \rsb \rcb,
    \end{split}
    \label{eq:Ldiss_TIGME}
\end{equation}
\end{widetext}
where 
${S}_{\rm B}^{\Lambda+}(\omega)=\langle j\lvert {S}_{\rm B}^{\Lambda,{\rm bare}}\rvert k\rangle\,\vert j\rangle\langle k\vert$ with $k> j$, $\omega=\omega_{jk}\equiv E_k-E_j$ and ${S}_{\rm B}^{\Lambda-}(\omega)=\lsb{S}_{\rm B}^{\Lambda+}(\omega)\rsb^\dagger$, and $S^{\Lambda,{\rm bare}}_{\rm B}$ is the system operator to be coupled to the bath and observed for the $\Lambda$ subsystem.

Importantly, the operator coupled to the bath,
is defined as 
\begin{equation}
    S_{\rm B}=s^+_{\rm B}+s^-_{\rm B}=\sum_{\omega>0}[{S}_{\rm B}^{\Lambda+}(\omega)+{S}_{\rm B}^{\Lambda-}(\omega)],
\end{equation}
which need not coincide with the operator coupled to the coherent drive~\cite{Hughes_Reconciling_2024}. 
For example, in the coherent Floquet pump case, the driven operators are the same as the system (raising/lowering) operators, $s_d^\pm\equiv s^\pm$; however, $s_\mathrm{B}^\pm$ can be potentially different~\cite{Hughes_Reconciling_2024,Salmon_Gauge-independent_2022}.
To reinforce this argument, recently~\cite{Gustin_Dissipation_2025}, using a first-principles quantized QNM theory, the correct cavity-bath coupling operator appears to be just $\Pi =a$ (in the Coulomb gauge), 
which coincides very well 
with the heuristic
model in Ref.~\onlinecite{Hughes_Reconciling_2024}, 
recovering the equivalence between classical and quantum models for the open system quantum Hopfield model, in any 
arbitrary coupling regime. For convenience, we choose the same dressed-basis ladder operators, $s_{\rm B}^{\Lambda\pm}=s^{\Lambda\pm}$ given in Eq.~\eqref{eq:splus}, but our  model can clearly support
any general bath operator.

\subsubsection{Structured versus flat baths}
The bath function enters 
through the frequency-dependent rates $\Gamma^\Lambda(\omega)$, which are the Fourier transforms of the bath correlation functions.
Popular and widely used forms of the bath function include the flat, Ohmic family, and Lorentzian models~\cite{Hughes_Phonon-mediated_2013,Wang_Dynamical_2019,Salmon_Gauge-independent_2022,Akbari_Generalized_2023}. 
For the cavity, one can examine the flat bath model with a constant rate $\kappa$ via
\begin{equation}
    \begin{split}
\Gamma^{\rm cav}_{\rm flat}(\omega)&=\frac{\kappa}{2},
    \end{split}
    \label{eq:CavityBathModel_flat}
\end{equation}
and the Ohmic-family bath model with 
\begin{equation}
    \begin{split}
\Gamma^{\rm cav}_{\rm Ohm}(\omega)&=\frac{\kappa}{2}\lp\frac{\omega}{\omega_c}\rp^k,
    \end{split}
    \label{eq:CavityBathModel_OhmicFamily}
\end{equation}
where $k=1$ identifies the Ohmic bath, $0<k<1$ identifies the sub-Ohmic bath, and $k>1$ identifies the super-Ohmic bath.

The Lorentzian bath model, on the other hand, is given by
\begin{equation}
    \begin{split}
\Gamma^{\rm cav}_{\rm Lor}(\omega)=\Gamma^{\rm cav}_{\rm max}\frac{(\kappa/2)^2}{(\omega-\omega_0)^2+(\kappa/2)^2},
    \end{split}
    \label{eq:CavityBathModel_Lorentzian}
\end{equation}
where $\omega_0$ is the bath's central frequency (i.e., at the dissipation peak).

In practice, all these bath functions may be  spectrally windowed or have a cut-off, which is often the case also for real systems (such as with photonic bandgaps, electron-phonon coupling, etc.). Therefore, an interesting variety of the families of the above functions may also be used in the theoretical modeling.
Although the Lorentzian bath model may be considered unusual for a cavity bath channel, we use it here more as a proof of principle.
Specifically, 
to showcase the effect of a structured cavity bath, we use a Lorentzian bath with the Ohmic cut-off,
defined from
\begin{equation}
    \begin{split}
\Gamma^{\rm cav}_{\rm Lor-Ohm}(\omega)&=\Gamma^{\rm cav}_{\rm max}\frac{(\kappa/2)^2}{(\omega-\omega_0)^2+(\kappa/2)^2}\times\frac{\omega
}{\omega_c}.
    \end{split}
    \label{eq:CavityBathModel_LorOhmic}
\end{equation}
To relate this to other common bath models with the same rate, we also choose $\Gamma^{\rm cav}_{\rm max}=\kappa$ to have the FWHM of $\kappa/2$ at $\omega=\omega_0\pm\kappa/2$, i.e., 
\begin{equation}
    \Gamma^{\rm cav}_{\rm Lor}(\omega_0\pm\frac{\kappa}{2})=\frac{\kappa}{2}.
\end{equation}

For the TLS bath, however, we only consider the flat bath model with a constant rate $\gamma$ via 
\begin{equation}
    \begin{split}
\Gamma^{\rm TLS}_{\rm flat}(\omega)&=\frac{\gamma}{2},
    \end{split}
    \label{eq:TLSBathModel_flat}
\end{equation}
because the TLS decay is much weaker than the cavity decay, 
so that for simplicity we only assume that the TLS bath to be constant and negligible compared to that of the cavity~\cite{Salmon_Gauge-independent_2022}. 

Since no secular approximation is imposed, frequency-dependent baths are incorporated exactly at the level of dressed (self-consistent and problem-dependent) transition frequencies.

Finally, we highlight that the bath function of a dissipatve cavity is actually gauge variant, but the observables are of course gauge invariant, when formulated correctly~\cite{PhysRevLett.134.123601,nqys-jz1w,Gustin_Dissipation_2025}.

\subsubsection{Including time-dependent drive in the TI-GME}
For a driven cavity-QED system described by the time-dependent Hamiltonian 
$\mathcal{H}_{\rm DQRM}(t)$, the density matrix obeys the TI-GME via
\begin{equation}
    \begin{split}
        \frac{\dd\rho(t)}{\dd t}&=\mathcal{L}(t)\rho(t),
    \end{split}
    \label{eq:TIGME_driven}
\end{equation}
where the total Liouvillian can now admit time-dependence and the coherent dynamics are governed by the full time-dependent Hamiltonian (either coherently pumped or Floquet-engineered), so that 
\begin{equation}
    \mathcal{L}(t)=\mathcal{L}_{\rm S}(t)+\mathcal{L}_{\rm diss}.
\end{equation}

Using the DQRM Hamiltonian, 
the time-dependent system Liouvillian is defined through \begin{equation}
    \mathcal{L}_\mathrm{S}(t)\rho(t)=-\ii\lsb\mathcal{H}_{\rm DQRM}(t),\rho(t)\rsb,
\end{equation}
while the dissipative part retains the dressed transition structure of the undriven QRM, as in Eq.~\eqref{eq:Ldiss_TIGME}. Note the dissipator term  here is the same as before in the incoherent pump case and still time-independent.

\subsubsection{Limitation of the time-independent dressed dissipator}
In the TI-GME, the coherent drive is included nonperturbatively in the Hamiltonian part, but the dissipator is constructed from the static dressed transition frequencies $\omega_{jk}$ of the undriven QRM. As a result, decay channels are resolved only with respect to the static dressed spectrum.

When the drive is sufficiently weak, one expects that this separation is accurate: the drive modifies populations and coherences, but the dominant decay pathways remain those of the static dressed transitions. However, when the drive becomes comparable to the intrinsic energy scales---particularly in the USC regime---periodic driving redistributes transition weights among Floquet sidebands and can open new quasienergy-resolved decay channels. These processes are not explicitly resolved in the TI-GME dissipator.

One of the central questions addressed in this work is therefore as follows:
when and how does quasienergy-resolved dissipation become essential for correctly predicting observables in strongly driven USC cavity-QED?

To answer this quantitatively, we next formulate a F-GME in which dissipation is resolved directly in the quasienergy basis of the dressed picture, enabling a sideband-resolved treatment of structured baths.

\section{Floquet theory for arbitrary cavity-QED systems}
\label{sec:Theory}

In this section, we now turn our attention to a specific open cavity system of interest.
We consider a Floquet driven cavity-QED system as depicted in Fig.~\ref{fig:Schematics}, which can be a  coherent-pump Floquet problem [optical coherently pumped cavity-QED system, Fig.\ref{fig:Schematics}(a)] or a Floquet engineering problem, that is a time-modulation of the internal parameters in the closed time-independent cavity-QED system---for example, via mechanical vibration [an example is shown earlier in Fig.\ref{fig:Schematics}(b)].

In such systems, there can be significant {\it dressing} from the atom-field coupling (strongly driven cavity-QED in the USC regime), capable of modifying the energy basis of the system 
(cf.~Figs.~\ref{fig:EnergyBasis_Optical} and \ref{fig:EnergyBasis_Mechanical}).  
Generally, an external drive field such as a laser is used to coherently pump the system (Floquet coherent pump) or modulate the TLS frequency (Floquet engineering via frequency modulation), but mechanical drives are often used to modulate the cavity frequency or cavity-TLS coupling (Floquet engineering via coupling modulation).
We assume that the pumping field produces a coherent state with many photons, or the mechanical force is applied nonperturbatively, so that both types of drives can be well described by a classical field, according to the strong-field approximation. 

 Once the drive is sufficiently strong, or the basis size of the system is quite large or the energy/time resolution of the system is narrow, particularly when the drive is detuned, it can considerably influence the dynamics of the system and its internal transitions. This can also be viewed as a change in the Hamiltonian energy basis from a time-independent case to a time-dependent one (the latter are formally termed 
 {\it quasienergies}).
 Even though the TI-GME can predict the correct physics for many systems and regimes, the explanation of the physics is sometimes difficult to appreciate within the time-independent-basis theory,
 as it does not account for additional 
 dressing at the initial energy basis.
So another advantage of including the drive in the dressing, is that is connects to dynamical dressing in a much more intuitive and direct way 
(very simple examples of such dressing include Mollow triplets). 
 
Furthermore, the effective splitting and near-degenerate consequences of the quasienergy lines, even in a low-field drive case, can create asymmetries and Fano-like spectral structures that might be undetectable in the TI-GME theory. Even if the TI-GME theory may predict the correct results, the explanation of the underlying transitions and physics can be ill-defined in the time-independent basis picture and one must employ a dynamical (Floquet) picture. Such scenarios were previously studied in many less complex systems, such as a Floquet-driven TLS, and the formation of Mollow multiplets, even in more complex physical contexts~\cite{Wang_Observation_2021,Uchida_Diabatic_2022}.
 
Below, we introduce a theory to solve this more general problem, and then look at its consequences
 on the energy basis eigenvalues, populations and cavity-emitted spectra.
 It is also seen that when the QRM is pumped coherently (Floquet single-frequency optical drive), the time-independent Hamiltonian is unchanged~\cite{Salmon_Gauge-independent_2022}; however when the QRM is Floquet engineered (e.g., the coupling is time-modulated via a mechanical drive), the TI Hamiltonian is renormalized so that the dc component of the Hamiltonian is different from the original QRM~\cite{Akbari_Floquet_2025,Chang_Direct_2025}. This is because some of the time dependence is introduced via additional nonlinear terms to the original time-independent Hamiltonian~\cite{Akbari_Floquet_2025,Chang_Direct_2025}.
 
Next, after introducing the general Floquet theory for a closed cavity-QED problem in an arbitrary coupling regime, we explore the different cases of the Floquet coherent pump and Floquet engineering, separately, and then we discuss the effects of the environment in an open cavity-QED problem.

\subsection{Floquet model of a time-dependent \\ quantum Rabi model}
 \label{Sec:DQRM}
To study the dynamics of a driven cavity-QED system, we seek to solve the time-dependent Schrodinger equation:
\begin{equation}
    \ii \frac{\dd\lvert\psi(t)\rangle}{\dd t}=\mathcal{H}_{\rm DQRM}(t)\lvert\psi(t)\rangle,
    \label{eq:TDSE}
\end{equation} 
and we let the light-matter interaction coupling rate be arbitrary; thus, we cover all quantum light-matter interaction regimes and in particular the USC regime, which leads to a  nonperturbative treatment and diagonalization in the dressed basis of the time-independent Hamiltonian. 
 
 Note that the static Hamiltonian, $\mathcal{H}_0$, introduced in Eq.~\eqref{eq:HDQR_m} is not necessarily always the same as the QRM Hamiltonian (before driving), because the time-dependence can alter the
 dc component of the DQRM~\cite{Akbari_Floquet_2025,Chang_Direct_2025}. 
 Nevertheless, we always diagonalize the dc component of the DQRM, $\mathcal{H}_0$, and form the dressed eigenenergy basis.
 Hence, when $\mathcal{H}_0=\mathcal{H}_{\rm QRM}$, which happens in the coherent optical pump case, then the time-independent basis set is the same as the QRM eigenenergies; we term the diagonalization and the later treatment in such diagonalized framework as the {\it QRM-dressed picture}. Whereas, when $\mathcal{H}_0\neq\mathcal{H}_{\rm QRM}$, which happens in the Floquet engineering cases, then the TI basis set is \emph{not} the same as the QRM eigenenergies; we shall call the diagonalization and the later treatment in such  framework as the 
 {\it renormalized-QRM picture}~\cite{Akbari_Floquet_2025}.

The static atom--cavity Hamiltonian, $\mathcal{H}_0$, introduced in Eq.~\eqref{eq:HDQR_m}, can be diagonalized straightforwardly~\cite{Braak_Integrability_2011,DiStefano_Resolution_2019}. This yields a set of ${\rm Dim}(\mathbb{H}_{\rm dressed})$ joint light--matter eigenstates (in principle infinite dimensional), denoted by $\{\ket{j}\}$, with corresponding eigenenergies $\{E_j\}$. Explicitly, they satisfy  
\begin{equation}
\mathcal{H}_0 \ket{j} = E_j \ket{j}, 
\qquad
\braket{j|k}=\delta_{jk},
\qquad
\sum_j \ket{j}\bra{j} = \mathbf{1},
\end{equation}
where $\{\ket{j}\}$ forms a complete orthonormal basis for the dressed Hilbert space $\mathbb{H}_{\rm dressed}$, i.e.,
$\mathsf{B}_{\rm dressed}=\{\ket{j}\}$.
In numerical implementations, this space is truncated to a finite dimension ${\rm Dim}(\mathbb{H}_{\rm dressed})$, while the underlying physical Hilbert space remains formally infinite.

Importantly, even in the absence of external driving, the interacting light--matter system is already nonperturbatively dressed by the atom--cavity coupling, as described by the QRM (or its renormalized counterpart). Therefore, system operators associated with excitation and de-excitation processes must be expressed in this dressed representation rather than in the bare tensor-product basis. This becomes essential in the ultrastrong-coupling regime, where bare operators no longer correctly describe physical transitions or emission channels. In the weak-coupling limit, the formalism continuously recovers the standard perturbative description.

For arbitrary normalized coupling strengths $\eta$, the dressed eigenstates $\{\ket{j}\}$ are thus required to construct physically consistent transition operators and to evaluate observables correctly~\cite{Settineri_Dissipation_2018,DiStefano_Resolution_2019,Salmon_Master_2021,Salmon_Gauge-independent_2022}. Consequently, the effective raising and lowering operators of both the cavity and TLS subsystems must be redefined in the unified QRM-dressed (or renormalized QRM-dressed) basis. The corresponding cavity and TLS positive-frequency operators are given in Eq.~\eqref{eq:splus}.

 Considering a periodic CW field, the externally dressed QRM-type model (QRM or renormalized QRM) eigenstates are governed by the Floquet picture. 
Therefore, we wish to diagonalize the DQRM Hamiltonian into the Floquet states~\cite{Grifoni_Driven_1998,LeBoite_Theoretical_2020}, i.e., $\lvert\psi(t)\rangle=\sum_\alpha c_\alpha\lvert\psi_\alpha(t)\rangle$, where 
\begin{equation}
\begin{split}
\lvert\psi_\alpha(t)\rangle
&=\mathrm{e}^{-\ii\varepsilon_\alpha t}\lvert\alpha(t)\rangle,
\end{split}
\label{eq:FloquetState}
\end{equation} 
with $c_\alpha=\langle\alpha\vert\psi(0)\rangle$ ($\lvert\alpha\rangle\equiv\lvert\alpha(0)\rangle$) being the TI complex coefficients, and $\vert \alpha(t)\rangle$ the Floquet modes corresponding to the Floquet quasienergy $\varepsilon_\alpha$~\cite{Nikishov_Quantum_1964}, defined from
\begin{equation}
    \begin{split}
        \mathcal{H}^\mathrm{F}(t)\vert \alpha(t)\rangle=\varepsilon_\alpha\vert \alpha(t)\rangle, 
    \end{split}
    \label{eq:FloquetTDSE}
\end{equation}
with the Floquet Hamiltonian~\cite{Floquet_FloquetTheory_1883,Shirley_Solution_1965,Grifoni_Driven_1998} ${\mathcal{H}}^\mathrm{F}(t) \equiv{\mathcal{H}}_\mathrm{DQRM} (t)-\ii\partial_t$.
  Hence, the new {\it doubly dressed} (with external drive) joint quasienergies $\{\vert \psi_\alpha(t)\rangle\}$ are required to construct the correct dressed operators utilized in the GME; 
these are the eigenstates of the full light-matter system Hamiltonian \emph{including} the interaction term and also \emph{with} the influence of the external drive field. 

Since the Floquet modes are periodic, i.e., $\ket{\alpha(t)}=\ket{\alpha(t+T)}$, then they can also be expanded in a Fourier series:
\begin{equation}
\begin{split}
\lvert \alpha(t)\rangle&=\sum_{l\in\mathbb{Z}}\mathrm{e}^{-\ii l\omega_d t}\vert {\alpha_l}\rangle,
\end{split}
\label{eq:FloquetMode_FourierExpansion}
\end{equation}
where $\lvert \alpha_l\rangle=(1/T)\int_Tdt\,\mathrm{e}^{\ii l\omega_dt}\,\ket{\alpha(t)}$ are the Fourier coefficient states called  Floquet sidebands. 
Moreover, in a general quantum  
basis set $\{\lvert j\rangle\}$, obtained for the time-independent portion of the Hamiltonian, $\mathcal{H}_0$,  one can always expand the Floquet sidebands as a superposition 
$\lvert \alpha_l\rangle=\sum_{j}f_{\alpha lj}\lvert j\rangle$, and hence expand the Floquet modes in terms of the time-independent basis.
Indeed, it is clear that all the time-dependent variables in Eq.~\eqref{eq:FloquetTDSE}
possess oscillatory exponential time dependence. 

Hence, one can consider the time-functionality as a separate (temporal) basis, i.e., $\mathrm{e}^{-\ii l\omega_d t}\to \lvert l)$, and expand the Hilbert space to a tensorially larger one: 
\begin{equation}
    \mathbb{H}_{\rm ex}\equiv\mathbb{H}_{\rm temp}\otimes\mathbb{H}_{\rm dressed},
\end{equation}
known as the extended (or, Sambe) space~\cite{Sambe_Steady_1973}. Here,  the temporal Hilbert space is the space of periodic square-integrable functions, i.e., $\mathbb{H}_{\rm temp}\equiv L_2[0, T]$, spanned by the basis set of 
\begin{equation}
    \mathsf{B}_{\rm temp}=\{\lvert l)\}_{l=-l_\mathrm{max}}^{l_{\rm max}},
\end{equation}
where ${\rm Dim}(\mathbb{H}_{\rm temp})=2l_\mathrm{max}+1$ (see 
App.~\ref{Sec:DQRM}). In the extended space, now the modes are time-independent, and can be represented in terms of the extended space basis $\Fket{l,j}\equiv\lvert l)\otimes\lvert j\rangle$ as $\lvert\alpha(t)\rangle\to\Fket{\alpha}$, where
\begin{equation}
\begin{split}
\Fket{\alpha}&=\sum_{l\in\mathbb{Z}}\lvert l)\otimes\vert {\alpha_l}\rangle
\\
&=\sum_{l\in\mathbb{Z}}\sum_j f_{\alpha lj}\,\Fket{l,j},
\end{split}
\label{eq:FloquetMode_ex}
\end{equation}
with $\lvert l)$ being the (number or Fock) basis of the temporal space $\mathbb{H}_{\rm temp}$.

Similarly, the Floquet Hamiltonian can be represented into the extended basis, $\mathcal{H}^\mathrm{F}(t)\to\mathcal{H}^\mathrm{F}_{\rm ex}$ (given explicitly in  App.~\ref{Sec:DQRM}).
Inserting, now, Eqs.~\eqref{eq:FloquetMode_FourierExpansion} and \eqref{eq:FloquetMode_ex} in Eq.~\eqref{eq:FloquetTDSE} yields a quasienergy eigenvalue equation, similar to the TI Schr\"{o}dinger equation, in an extended (temporal and spatial) Hilbert space: 
\begin{equation}
\begin{split}
\mathcal{H}^{\rm F}_{\rm ex}\Fket{l\alpha}=\varepsilon_{l\alpha}\Fket{l\alpha},
\end{split}
\label{eq:FloquetTISE_ex}
\end{equation}
where $\mathcal{H}^{\rm F}_{\rm ex}$ is the representation of the Floquet Hamiltonian and $\Fket{l\alpha}$ are the eigenvectors of the Floquet Hamiltonian, in the extended space (see full details in App.~\ref{Sec:DQRM}). 

By solving the TI eigenvalue problem in Eq.~\eqref{eq:FloquetTISE_ex}, one finds quasienergies and Floquet sidebands, and, accordingly, can construct the Floquet modes and states~\cite{Eckardt_High-frequency_2015,Restrepo_Quantum_2018,Restrepo_Driven_2019}.
Also note that the dimension of the quasienergy/state eigenvalue problem is now larger than the original Hilbert space, 
\begin{equation}
    {\rm Dim}(\mathbb{H}_{\rm ex})={\rm Dim}(\mathbb{H}_{\rm temp})\times {\rm Dim}(\mathbb{H}_{\rm dressed}).
\end{equation}

Therefore, one expects a larger number of quasienergies/states, correspondingly. 
Nevertheless, only ${\rm Dim}(\mathbb{H}_{\rm dressed})$ of them are required to produce the linearly-independent Floquet states. 
In other words, after solving the (time-independent) eigenvalue problem in the extended space, one finds that $\varepsilon_{l\alpha}\equiv\varepsilon_{\alpha}+l\omega_d$ with $\varepsilon_{\alpha}\equiv\varepsilon_{0\alpha}$, and hence, the corresponding Floquet mode follows $\lvert\alpha^{[l]}(t)\rangle\equiv \mathrm{e}^{-\ii l\omega_dt}\lvert \alpha(t)\rangle$, with $\lvert\alpha(t)\rangle\equiv \lvert \alpha^{[0]}(t)\rangle$.

This property of repeating and shifting the quasienergies,  builds up Brillouin zones (BZs) for each $l$, throughout the energy spectrum.
Thus, it will be sufficient just to examine the set of eigenvalues/vectors $\{\varepsilon_{l\alpha},\Fket{l\alpha}\}$ within one BZ, e.g., $-\omega_d/2\leq\varepsilon_{l\alpha}<\omega_d/2$,
i.e., translation to the $l$th BZ, that connects Floquet modes in different BZs, yields a solution of \eqref{eq:TDSE} physically identical to \eqref{eq:FloquetState} but with shifted quasienergy (see later examples, in Figs.~\ref{fig:EnergyBasis_Optical} and \ref{fig:EnergyBasis_Mechanical}).

\subsubsection*{Symmetries, parity and selection rules from the transitions in the Floquet extended space}
The transitions in the {\it driven} USC system now occur among the Floquet quasienergy states in the extended space. Thus, it is not only instructive but also crucial to understand the selection rules for the transitions within the original Hilbert space and its extension to the extended space, a concept that is related to identifying the symmetry properties of the system. 
For example, for the original Hamiltonian in the bare Hilbert space, we define the parity operator as 
\begin{equation}
   \mathcal{P}_{\rm bare}=\exp\{\ii\pi \mathcal{N}_\mathrm{bare}\}, 
\end{equation}
where 
\begin{equation}
    \mathcal{N}_\mathrm{bare}=\sigma^+\sigma^-+a^\dagger a, 
\end{equation}
is the total number of excitations operator, $\mathcal{N}_\mathrm{bare}\lvert j\rangle=j\lvert j\rangle$.
To identify the parity of the QRM, or the renormalized QRM-dressed states $j$, we must calculate the matrix element $\langle j\vert \mathcal{P}_{\rm bare}\vert j\rangle$. If this matrix element is negative/positive, then the parity of the state $j$ is odd/even. 
However, the external harmonic drive can modify the symmetry of the system. 

Thus, in the Floquet picture, since the transitions are among the Floquet sidebands, one must understand the parity of the sidebands in the extended space, identified by the generalized parity transformation:
\begin{equation}
    \begin{split}
        \mathcal{P}^{\rm F}&:\lcb\begin{array}{l}
            x\to-x, \\
            t\to t+T/2.
        \end{array}\right.
    \end{split}
\end{equation}

To do this, we first 
diagonalize the parity operator in the dressed picture via 
\begin{equation}
    \mathcal{P}_{\rm dressed}=\sum_{jk}\langle j\vert \mathcal{P}_{\rm bare}\vert k\rangle\,\ket{j}\bra{k},
\end{equation}
and then 
find the parity operator in the temporal Hilbert space via 
\begin{equation}
    \mathcal{P}_{\rm temp}=\exp\{\ii\pi \mathcal{N}_{\rm temp}\},
\end{equation}
where $\mathcal{N}_{\rm temp}$ is a number operator, 
\begin{equation}
    \mathcal{N}_{\rm temp}\lvert l)=l\lvert l),
    \label{eq:Ntemp}
\end{equation}
with $\lvert l)$ being the temporal number (eigen)states, in the temporal Hilbert space $\mathbb{H}_{\rm temp}$.

One can now identify the generalized parity of the Floquet modes in the Floquet extended picture by 
\begin{equation}
    \Fbra{l\alpha} \mathcal{P}^{\rm F}_{\rm ex}\Fket{l\alpha}\gtrless0
\end{equation}
for even/odd parity, where  
\begin{equation}
   \mathcal{P}^{\rm F}\to\mathcal{P}^{\rm F}_{\rm ex}=\mathcal{P}_{\rm temp}\otimes\mathcal{P}_{\rm dressed} 
\end{equation}
is the generalized parity operator in the Floquet extended space.
Proper understanding of the generalized parity helps one to identify the selection rules in the field-driven QRM, and recognize how many externally assisting quanta from the external classical field is needed for a specific transition to occur.

\subsection{Driven dissipative quantum Rabi model: Floquet-Markov generalized master equation}
\label{Sec:DDQRM}

To account for dissipation, we next develop a GME formalism for the QRM, with no secular approximations, and extend it to the case of the Floquet DQRM, in order to understand the dynamics of the driven-dissipative cavity-QED systems in the USC regime and 
accurately 
compute observables.

With non-pertubative
driving, 
the periodic time dependence of ${\mathcal{H}}_\mathrm{DQRM}(t)$
and the high (infinite) number of
potential new resonances manifests in a potentially large number of near-degeneracies
in the chosen BZ, thus making the secular approximation not suitable for
our analysis~\cite{Hone_Statistical_2009,NafariQaleh_Enhancing_2022,Farina_Open-quantum-system_2019}. In this scenario, the generalized density matrix evolves with the F-GME, in the Schrödinger picture, as~\cite{Grifoni_Driven_1998,Restrepo_Driven_2019,Mori_Floquet_2023,LeBoite_Theoretical_2020} 
$\mathrm{d}{\rho}(t)/\mathrm{d}t=\mathcal{L}(t){\rho}(t)$, 
where the total fully time-dependent Liuvillian is $\mathcal{L}(t)=\mathcal{L}_{\rm S}(t)+\mathcal{L}_{\rm diss}(t)$, with 
 the now-time-dependent dissipative Liouvillian term as:
 \vspace{1.5cm}
\begin{widetext}
\begin{equation}
    \begin{split}
      {\mathcal{L}}_\mathrm{diss}(t){\rho}(t)          &=\sum_{\Lambda}\sum_{\Delta,\Delta'>0}\lcb\mathrm{e}^{-\ii l\omega_{d}t}\Gamma^{\Lambda}(\Delta)\lsb{S}_{\rm B}^{\Lambda+}(\Delta;t) {\rho}(t){S}_{\rm B}^{\Lambda-}(\Delta')- {S}_{\rm B}^{\Lambda-}(\Delta'){S}_{\rm B}^{\Lambda+}(\Delta;t){\rho}(t)\rsb\right.
        \\
        &\hspace{3cm}\left.+\mathrm{e}^{\ii l'\omega_{d}t}
\Gamma^{\Lambda*}(\Delta')\lsb {S}_{\rm B}^{\Lambda+}(\Delta){\rho}(t){S}_{\rm B}^{\Lambda-}(\Delta';t)-{\rho}(t) {S}_{\rm B}^{\Lambda-}(\Delta';t) {S}_{\rm B}^{\Lambda+}(\Delta) \rsb \rcb,
    \end{split}
    \label{eq:Ldiss_FGME}
\end{equation}
\end{widetext}
where 
\begin{equation}
    \Delta\equiv \Delta_{\alpha\beta l}\equiv\omega_{\alpha\beta}+l\omega_d,
    \quad 
    \text{with}
    \quad
    \omega_{\alpha\beta}\equiv\varepsilon_\beta-\varepsilon_\alpha,
\end{equation}
 or, $\Delta_{\alpha\beta l}=\varepsilon_{l'\beta}-\varepsilon_{l''\alpha}$, with $l=l'-l''$, and $\Gamma^\Lambda(\Delta)$ is the same bath function as the one used in the TI-GME but with a different argument now as it is quasienergy-resolved. 

We highlight  the modification of the F-GME with respect to the TI-GME via~\cite{Restrepo_Driven_2019} $\mathcal{L}_{\rm diss}\to\mathcal{L}_{\rm diss}(t)$, since the dissipator is also time-dependent, quasienergy-resolved, and thus it contains all the sidebands.
In other words, the dissipator in the TI-GME approximates the dissipator in the F-GME as the 
first-order perturbation by only counting the primary sideband, say $l=l_0$ with $\mathcal{L}_0=\mathcal{L}_{\rm diss}$ given in Eq.~\eqref{eq:Ldiss_TIGME}. Hence, the final expression of the F-GME for the driven-dissipative QRM reads:
 \vspace{1.5cm}
\begin{widetext}
\begin{equation}
    \begin{split}
       \mathrm{d}\rho(t)/\mathrm{d}t&=-\ii\lsb{\mathcal{H}}_\mathrm{DQRM}(t),\rho(t)\rsb+\sum_{\Lambda}\sum_{\Delta,\Delta'>0}\lcb\mathrm{e}^{-\ii l\omega_{d}t}\Gamma^{\Lambda}(\Delta)\lsb{S}_{\rm B}^{\Lambda+}(\Delta;t) {\rho}(t){S}_{\rm B}^{\Lambda-}(\Delta')- {S}_{\rm B}^{\Lambda-}(\Delta'){S}_{\rm B}^{\Lambda+}(\Delta;t){\rho}(t)\rsb\right.
        \\
        &\hspace{3cm}\left.+\mathrm{e}^{\ii l'\omega_{d}t}
\Gamma^{\Lambda*}(\Delta')\lsb {S}_{\rm B}^{\Lambda+}(\Delta){\rho}(t){S}_{\rm B}^{\Lambda-}(\Delta';t)-{\rho}(t) {S}_{\rm B}^{\Lambda-}(\Delta';t) {S}_{\rm B}^{\Lambda+}(\Delta) \rsb \rcb.
    \end{split}
    \label{eq:FGME_OriginalBasis}
\end{equation}
\end{widetext}

The time-independent and time-dependent spectrally decomposed system-bath coupling operators are given by 
\begin{equation}
    {S}^\Lambda_{\rm B}(\Delta)=S^\Lambda_{{\rm B},{\alpha\beta l}}\,\lvert {\alpha}\rangle\langle {\beta}\rvert
    ,
\end{equation}
and
\begin{equation}
        S_{\rm B}^{\Lambda}(\Delta;t)=S^\Lambda_{{\rm B},{\alpha\beta l}}\,\lvert \alpha(t)\rangle\langle \beta(t)\rvert,
\end{equation}
respectively, with
\begin{equation}
    \begin{split}
        {S}^\Lambda_{{\rm B},{\alpha\beta l}}&=\frac{1}{T}\int_0^T dt\,\mathrm{e}^{-\ii l\omega_{d} t}\,\langle {\alpha}(t)\rvert S^\Lambda_{\rm B}\lvert {\beta}(t)\rangle,
    \end{split}
    \label{eq:FelementS_ex}
\end{equation}
which is the matrix element of the operator $S_{\rm B}$, in the Floquet extended space.

We  define positive-frequency operators from ${S}^{\Lambda+}_{\rm B}(\Delta)={S}^{\Lambda}_{\rm B}(\Delta)$ and ${S}^{\Lambda+}_{\rm B}(\Delta;t)={S}^{\Lambda}_{\rm B}(\Delta;t)$ for $\Delta>0$, and the corresponding negative-frequency operators as ${S}^{\Lambda-}_{\rm B}(\Delta)=[{S}^{\Lambda+}_{\rm B}(\Delta)]^\dagger$ and ${S}^{\Lambda-}_{\rm B}(\Delta;t)=[{S}^{\Lambda+}_{\rm B}(\Delta;t)]^\dagger$.

Consequently, one only needs to deal with positive net energy operators.
In Eq.~\eqref{eq:FelementS_ex}, $S^\Lambda_{\rm B}$ is a (Hermitian) system operator, to be coupled to the bath operators, which can be written in terms of the dressed transition operators as $S^\Lambda_{\rm B}=s^{\Lambda+}_{\rm B}+s^{\Lambda-}_{\rm B}$ with ${s}^{\Lambda-}_{\rm B}=[{s}^{\Lambda+}_{\rm B}]^\dagger$, in the Schr\"{o}dinger picture. 
The spectral decomposition of the dressed cavity- and TLS-bath transition operators 
is similar to previous works~\cite{Settineri_Dissipation_2018,Salmon_Gauge-independent_2022,Hughes_Reconciling_2024}, but note that
one may adopt different choices for the cavity-bath operator~\cite{Hughes_Reconciling_2024};
for simplicity and also to connect to previous work with weaker driving~\cite{Salmon_Gauge-independent_2022}, we will choose the same system transition operators as the system-bath coupling operators, i.e., $s^{\Lambda\pm}_{\rm B}=s^{\Lambda\pm}$ given in Eq.~\eqref{eq:splus}.

We highlight that due to the periodicity of the Floquet modes $\lvert \alpha(t)\rangle$, the Liouvillian superoperator also has the same periodicity: $\mathcal{L}(t)=\mathcal{L}(t+T)$.
In other words, in order to conveniently take the CW driving into account, we can express the F-GME
in the Floquet basis when deriving the master equation~\cite{Grifoni_Driven_1998,Restrepo_Driven_2019,Mori_Floquet_2023,LeBoite_Theoretical_2020,Hausinger_Dissipative_2010PhDThesis}. 

The fact that the Floquet states $\lvert\psi_\alpha(t)\rangle$ form a complete basis (as they solve the Schrödinger equation) allows for a formal simplification of the F-GME by representing the density operator in this basis, i.e., \begin{equation}
    \rho(t)=\sum_{\alpha\beta}\rho_{\alpha\beta}(t)\,\vert \alpha(t)\rangle\langle\beta(t)\vert,
\end{equation} 
with $\rho_{\alpha\beta}(t)\equiv\langle \alpha(t)\vert\rho(t)\vert\beta(t)\rangle$. Then, the F-GME reads
\begin{align}
       \frac{ \mathrm{d}{\rho}_{\alpha\beta}(t)}{\mathrm{d}t}&=-\ii (\varepsilon_\alpha-\varepsilon_\beta)\rho_{\alpha\beta}(t)
        \nonumber \\
        &+\langle\alpha(t)\vert \mathcal{L}_{\rm diss}(t)\rho(t)\vert\beta(t)\rangle,
    \label{}
\end{align}
where 
\begin{equation}
\langle\alpha(t)\vert \mathcal{L}_{\rm diss}(t)\rho(t)\vert\beta(t)\rangle\equiv \sum_{\alpha'\beta'}\mathcal{L}_{\alpha\beta,\alpha'\beta'}(t)\rho_{\alpha'\beta'}(t),
\end{equation}
with $\mathcal{L}_{\alpha\beta,\alpha'\beta'}$ being the elements of the dissipative Liuovilian super-operator in the Floquet basis (see Appendix.~\ref{secS:FGME}). 

The matrix form of the F-GME, 
in the Floquet basis reads,  
\begin{equation}
    \begin{split}
\dot{{\varrho}}(t)
&=-\ii[\mathcal{H}_{\rm F},{\varrho}(t)]
\\
&\hspace{0.5cm}
+\sum_{\Lambda}[\mathcal{S}^{\Lambda-}_{U}(t)\,\varrho(t)\,\mathcal{S}^{\Lambda+}_{\rm B}(t)-\varrho(t)\,\mathcal{S}^{\Lambda+}_{\rm B}(t)\,\mathcal{S}^{\Lambda-}_{U}(t)
\\
&\hspace{0.5cm}
+\mathcal{S}^{\Lambda-}_{\rm B}(t)\,\varrho(t)\,\mathcal{S}^{\Lambda+}_{U}(t)-\mathcal{S}^{\Lambda+}_{U}(t)\,\mathcal{S}^{\Lambda-}_{\rm B}(t)\,\varrho(t)],
    \end{split}
    \label{eq:FGME_FloquetBasis}
\end{equation}
where all the above variables are in the Floquet representation of the operators, in matrix form, i.e.,
\begin{equation}
    \begin{split}
        &{\varrho}(t)\equiv[\rho_{\alpha\beta}(t)],
        \\
        &\mathcal{H}_\mathrm{F}\equiv[\mathcal{H}^{\rm F}_{\alpha\beta}(t)]=\mathrm{diag}\{\varepsilon_\alpha\}\,\,\,\mbox{(only in the $1^\text{st}\,$BZ)},
        \\
        &\mathcal{S}^{\Lambda+}_{\rm B}(t)\equiv[\sum_{l}\mathrm{e}^{-\ii l\omega_{d}t}
\Gamma^{\Lambda*}(\Delta)S^{\Lambda}_{{\rm B},\alpha\beta l}],\,\,\,\,\,\Delta>0,
\\
&\mathcal{S}^{\Lambda+}_{U}(t)\equiv [U^{{\rm F}\dagger}_{\alpha\beta}(t)][\sum_{l}S^\Lambda_{{\rm B},\alpha\beta l}][U^{\rm F}_{\alpha\beta}(t)],\,\,\,\,\,\Delta>0,
    \end{split}
\end{equation}
with $\mathcal{H}^{\rm F}_{\alpha\beta}(t)\equiv\langle\alpha(t)\rvert\mathcal{H}^{\rm F}(t)\lvert\beta(t)\rangle$, $\mathcal{S}^{\Lambda-}_{\rm B}(t)=[\mathcal{S}^{\Lambda+}_{\rm B}(t)]^\dagger$, $\mathcal{S}^{\Lambda-}_{U}(t)=[\mathcal{S}^{\Lambda+}_{U}(t)]^\dagger$, and $U^{\rm F}(t)$ is the one-period time-evolution operator of the Floquet modes within the closed system, i.e., $\lvert\alpha(t)\rangle=U^{\rm F}(t)\lvert\alpha\rangle$ with the Floquet matrix elements $U^{\mathrm{F}}_{\alpha\beta}(t)=\langle\alpha\vert\beta(t)\rangle$. The operator $\mathcal{S}^{\Lambda\pm}_{U}(t)$ is the canonically transformed system operator in the Floquet picture.

 In general, one then needs to numerically solve Eq.~\eqref{eq:FGME_OriginalBasis} (also known as the Floquet-Redfield equation) or Eq.~\eqref{eq:FGME_FloquetBasis} (the Floquet-Redfield equation in the Floquet reference frame~\cite{Mori_Floquet_2023}), to find the dynamics of the system.
 The results of the F-GME can be compared with those obtained from the TI-GME~\cite{Settineri_Dissipation_2018,Salmon_Gauge-independent_2022,Akbari_Generalized_2023}  characterized by the dissipator in Eq.~\eqref{eq:Ldiss_TIGME} and a time-dependent DQRM Hamiltonian.
When comparison of the numerical results with those of the incoherent pump case is required, we choose the TLS operator $s^{\rm TLS-}$ for the incoherent pump Liouvillian with a small amplitude, e.g., $P_{\rm inc}=0.01\Omega_d$, in Eq.~\eqref{eq:L_incPump}, inserted back in Eq.~\eqref{eq:Ldiss_TIGME}.

We stress that the dynamics predicted by a GME are governed by three fundamental ingredients: (i) the system Hamiltonian, which determines the coherent evolution, (ii) the system--bath coupling operators, which specify how the system exchanges excitations with its environment, and (iii) the bath spectral density, which determines the frequency-dependent weighting of the dissipative processes. In principle, differences in any of these three ingredients can lead to differences in the resulting observables.

In the present work, we focus on steady-state observables, including the emission spectra---which is determined by time-dependent field correlation functions. Since all the approaches considered here are based on the same periodically driven Hamiltonian (i.e., the same system Liouvillian), the coherent part of the dynamics is treated consistently. Neglecting Lamb-shift corrections, the Hamiltonian contribution constitutes the purely imaginary part of the Liouvillian spectrum and is therefore responsible for the positions of the resonances. Consequently, one generally expects the major peak and valley locations predicted by the different approaches to be similar, as they are ultimately determined by the same underlying quasienergy structure~\cite{Akbari_Quasienergy-Resolved_2026,Akbari_Floquet_2026}.

In the lossless limit, $\Gamma\rightarrow0^+$, the dissipative contribution vanishes and both GME approaches reduce to the same unitary dynamics. This limit provides a useful reference framework for understanding the role of dissipation once environmental effects are incorporated. In this regime, Floquet theory alone yields a simplified description of the driven dynamics and forms the basis of the approach introduced in Ref.~\onlinecite{Akbari_Floquet_2025}, which we refer to as the {\it Floquet phenomenological theory}. There, dissipation is incorporated phenomenologically through a finite linewidth, and the unitary limit is subsequently recovered by taking the linewidth to zero.

The essential distinction between the three approaches lies in their treatment of dissipation. The phenomenological theory does not provide a microscopic description of decoherence processes. The TI-GME incorporates dissipation microscopically, but effectively associates all dissipative processes with the only time-independent primary ($l=0$) Floquet channel. In contrast, the F-GME distributes dissipation among all Floquet sidebands and therefore resolves the quasienergy-dependent decay channels individually. Therefore, it includes the full drive-assisted sideband spectrum in the dynamics, whether it influences the coherence or the decoherence, in complement to the first-order perturbative Floquet theory of driven-dissipative USC cavity-QED systems~\cite{Macri_Spontaneous_2022}.

This distinction originates from the Floquet decomposition of the system--bath interaction, which allows different sidebands to couple to the environment with different strengths and, in general, at different transition frequencies. Moreover, the bath spectral density can further modify the relative importance of these channels by weighting each sideband differently~\cite{Grifoni_Driven_1998,Hausinger_Dissipative_2010,Restrepo_Quantum_2018,Akbari_Quasienergy-Resolved_2026,Akbari_Floquet_2026}.
Thus, while the different approaches may predict similar resonance locations, only the F-GME consistently accounts for the microscopic distribution of dissipation among the quasienergy sidebands. Consequently, differences between the approaches are expected to appear primarily in the relative spectral weights, linewidths, and steady-state populations, rather than in the locations of the dominant resonances themselves.

The F-GME corresponds to a fully nonsecular (retaining all $\Delta\neq\Delta'$ terms) Born–Markov treatment in the Floquet basis, i.e., a Floquet–Redfield generator. While this approach captures interference effects between distinct Floquet decay channels and does not rely on a quasienergy gap condition, it is not guaranteed to be completely positive. 

In contrast, the universal Lindblad equation of Ref.~\onlinecite{Nathan_Universal_2020} constructs a completely positive Markovian generator via an alternative Markov approximation that is independent of system-level spacings. 
Similarly, the time-dependent coarse-grained master equation of Ref.~\onlinecite{Mozgunov_Completely_2020} restores complete positivity through controlled temporal coarse-graining, even for arbitrary driving and arbitrarily small level spacing. 

The unified Gorini-Kossakowski-Lindblad-Sudarshan master equation method of Ref.~\onlinecite{Trushechkin_Unified_2021} provides a rigorous clustered (partial-secular) construction that retains nonsecular couplings within near-degenerate transition manifolds, while preserving complete positivity and thermodynamic consistency in the weak-coupling limit. 

Finally, from a computational perspective, Floquet-Lindblad solvers such as {\it FLiMESolve}~\cite{Clawson_Floquet-Lindblad_2025} exploit periodicity to accelerate integration and relax strict secular truncations, though typically within phenomenological Lindblad-rate models, which can fail, and sometimes drastically.

We stress that our F-GME is 
far more general than the commonly adopted approaches discussed above, since we
 retain \emph{the full nonsecular} Floquet–Redfield tensor in the QRM-dressed basis, thereby treating the internal and external (double) dressing on an equal footing, \emph{without temporal coarse-graining, clustering (i.e., retaining only specific correlations), or modification of the Markov step}, thereby providing \emph{a direct microscopic treatment of interference between distinct Floquet decay pathways}.

More generally,  our F-GME unifies and complements the nonsecular standard Floquet-Markov (and the conventional Floquet-Redfield) master equations (in the bare states)~\cite{Grifoni_Driven_1998,Mori_Floquet_2023} that account for the drive-induced (external) dressing and the nonsecular time-independent-basis GMEs~\cite{Settineri_Dissipation_2018,Salmon_Gauge-independent_2022} that account for the interaction-induced (internal) dressing. Hence, our F-GME properly incorporates \emph{double dressing} on equal footing, and 
is a significant 
advance for modeling doubly dressed 
cavity-QED systems~\cite{Akbari_Floquet_2025},
such as a strongly driven 
cavity QED systems in ultrastrong coupling
(previously studied for closed systems).
A summary of these distinctions with the conventional Floquet- or Redfield-type master equation formalism and the TI-GME is listed in Tab.~\ref{tab:Comparison}. Those formalisms only account for a single dressing, and only TI-GME is utilized for the USC cavity-QED~\cite{Salmon_Gauge-independent_2022}.

We also stress that naive treatments of
standard Floquet solver may not even
satisfy gauge invariance, which has to be addressed with great care in the ultrastrong coupling regime, especially in the presence of dissipation~\cite{Salmon_Gauge-independent_2022,PhysRevLett.134.123601,Gustin_Dissipation_2025}
and time-dependent interactions~\cite{Settineri_Gauge_2021,Gustin_Gauge-invariant_2023}.

\begin{table*}[!htbp]
\caption{
Comparison with representative existing theoretical frameworks.
}
\label{tab:Comparison}
\centering

\vspace{1.5mm}
\setlength{\tabcolsep}{6.8pt}

\begin{tabular*}{\textwidth}{lcccccc}
\toprule
Framework &
\makecell{Floquet\\Dissipation} &
USC &
\makecell{Double\\Dressing} &
\makecell{Structured\\Baths} &
\makecell{FL\\Spectroscopy} &
\makecell{TI-GME/F-GME\\Benchmark}
\\
\midrule

Floquet-Markov /
Floquet-Redfield
&
$\checkmark$
&
$\times$
&
$\times$
&
Limited
&
$\times$
&
$\times$
\\[2mm]

Nonsecular Floquet
Theories
&
$\checkmark$
&
$\times$
&
$\times$
&
Limited
&
$\times$
&
$\times$
\\[2mm]

USC Dressed-Basis
GMEs
&
$\times$
&
$\checkmark$
&
Internal only
&
$\checkmark$
&
$\times$
&
$\times$
\\[2mm]

Present Work
&
$\checkmark$
&
$\checkmark$
&
$\checkmark$
&
$\checkmark$
&
$\checkmark$
&
$\checkmark$
\\

\bottomrule
\end{tabular*}

\end{table*}

\subsection{Example observables} 
In this subsection, we introduce important  
example observables that we will explore, which also connect to common experimental measurements (populations and emission spectra).

\subsubsection{Number of excitations}
The excitation number for each subsystem reads
\begin{equation}
N_\Lambda(t)=\langle s_{\rm B}^{\Lambda-}(t)s_{\rm B}^{\Lambda+}(t)\rangle.
\label{eq:N_t}
\end{equation}
However, with periodic driving, the quantity of practical interest is the initial- 
and long-time averaged after the system has reached the steady state:
\begin{equation}
\overline{N}_\Lambda=\overline{\langle s_{\rm B}^{\Lambda-} s_{\rm B}^{\Lambda+} \rangle}=\frac{1}{T'}\int_{T'} dt\,\lim\limits_{t\to\infty}N_\Lambda(t).
\label{eq:AveN}
\end{equation}

We note that $t\to\infty$ in practice means $t>t_{\rm ss}$, when the system reaches an equilibrium, and in general, $T'\geqslant T$ is the period of the periodic steady-state result. 
In the driven-dissipative cavity-QED case studied here, the first-order observables are $T$-periodic, while some second-order observables may exhibit subharmonic structure such as $T/2$ periodicity. Thus, it is safe to choose $T=T'$.

A simplified analytical expression can be obtained within the \emph{Floquet phenomenological theory}, in which the dynamics are described by the Floquet states of the driven system while dissipation is accounted for only through a phenomenological linewidth~\cite{Akbari_Floquet_2025} (see also the derivation in App.~\ref{Sec:DQRM}). In this approach, the time-averaged excitation number is given by
\begin{equation}
\overline{N}_\Lambda
=
\sum_{\alpha l}
|c_\alpha|^2
\langle \alpha_l |
s_{\rm B}^{\Lambda-}
s_{\rm B}^{\Lambda+}
|\alpha_l\rangle,
\label{eq:AveN_phen}
\end{equation}
where $\ket{\alpha_l}$ denotes the Floquet state in the $l$th sideband and $c_\alpha$ are the corresponding Floquet-state amplitudes.

In contrast, throughout the remainder of this work, the excitation number is computed more rigorously, and self-consistently for the baths of interest,  from the steady-state density matrix obtained from the GMEs. This allows the dissipative dynamics to be treated explicitly and enables a direct comparison between the phenomenological Floquet description, the TI-GME, and the F-GME formulations for different bath models.

Note that the expression in Eq.~(\ref{eq:AveN_phen}) contains only the coherent Floquet dynamics and therefore does not resolve the microscopic distribution of dissipation among the Floquet sidebands. In contrast, the TI-GME and F-GME approaches compute $\overline{N}_\Lambda$ from the steady-state density matrix and therefore incorporate environmental effects explicitly. The comparison between these approaches allows one to isolate the role of quasienergy-resolved dissipation in periodically driven quantum systems.

\subsubsection{First-order quantum correlation function and emission spectra}
We next define the (non-normalized) two-time first-order correlation function via
\begin{equation}
G^{(1)}_{s_{\rm B}^{\Lambda-},s_{\rm B}^{\Lambda+}}(t,t')={\langle s_{\rm B}^{\Lambda-}(t)s_{\rm B}^{\Lambda+}(t')\rangle},
\end{equation}
where the correlation function can be split into incoherent and coherent parts, $G^{(1)}_{s^{\Lambda-}_{\rm B},s^{\Lambda+}_{\rm B}}(t,t')=G^{(1){\rm inc}}_{s_{\rm B}^{\Lambda-},s_{\rm B}^{\Lambda+}}(t,t')+G^{(1){\rm coh}}_{s_{\rm B}^{\Lambda-},s_{\rm B}^{\Lambda+}}(t,t')$, with
\begin{equation}
\begin{split}
    &G^{(1){\rm inc}}_{s_{\rm B}^{\Lambda-},s_{\rm B}^{\Lambda+}}(t,t')\equiv G^{(1)}_{\delta s_{\rm B}^{\Lambda-},\delta s_{\rm B}^{\Lambda+}}(t,t')=\langle \delta s_{\rm B}^{\Lambda-}(t)\,\delta s_{\rm B}^{\Lambda+}(t')\rangle,
    \\
    &G^{(1){\rm coh}}_{s_{\rm B}^{\Lambda-},s_{\rm B}^{\Lambda+}}(t,t')=G^{(1)}_{s_{\rm B}^{\Lambda-},s_{\rm B}^{\Lambda+}}(t,t')- G^{(1){\rm inc}}_{s_{\rm B}^{\Lambda-},s_{\rm B}^{\Lambda+}}(t,t')
    ,
\end{split}
\end{equation}
where $\delta s_{\rm B}^{\Lambda\pm}(t)=s_{\rm B}^{\Lambda\pm}(t)-\langle s_{\rm B}^{\Lambda\pm}(t)\rangle$.

In practical numerical studies, one uses a finite time window, but the coherent portion manifests in a Dirac delta function if using a CW drive, so it is not interesting, from a spectral viewpoint. The important point here is that we can obtain the incoherent part using the quantum regression theorem with appropriate care (see App.~D of Ref.~\onlinecite{Salmon_Gauge-independent_2022}, where one obtains a Sinc function generally, from a finite time window).

In general, the emission spectrum can be obtained from the Fourier transform of the appropriate (two-time) field correlation function. Here, we are only interested in the (long-time) incoherent emission spectrum. To calculate the incoherent emission spectra, we define the long-time average of the incoherent correlation function as 
\begin{equation}
\overline{G}^{(1)}_{\delta s_{\rm B}^{\Lambda-},\delta s_{\rm B}^{\Lambda+}}(\tau)=\frac{1}{T'}\int_{T'} dt\,\lim\limits_{t\to\infty}G^{(1)}_{\delta s_{\rm B}^{\Lambda-},\delta s_{\rm B}^{\Lambda+}}(t,t+\tau),
\label{eq:Aveg1}
\end{equation}
and then take the Fourier transform. Thus, the (incoherent) emission spectrum is obtained from
\begin{align}
\label{eq:S}
    &\mathsf{S}_\Lambda(\omega)\propto \text{Re} \left[\int_{0}^{\infty} d\tau \, e^{\ii\omega \tau}\,\overline{G}^{(1)}_{\delta s_{\rm B}^{\Lambda-},\delta s_{\rm B}^{\Lambda+}}(\tau)\right].
\end{align}
We calculate this spectrum by using the quantum regression theorem as outlined in Ref.~\cite{Salmon_Gauge-independent_2022}, where we conduct the theorem for every $t$ in the last period (in the simulation) of the pseudo-steady-state ($t>t_{\rm ss}$).

Note that the spectrum is frequency-resolved using the long-time average of the two-time correlation function, after the system has reached the steady state. Hence, using the inverse Fourier transform of the incoherent spectrum and setting the time to the initial steady state time, we obtain
\begin{align}
    & \int \frac{d\omega}{2\pi}\,\mathsf{S}(\omega)\propto\overline{\langle\delta s^-\delta s^+\rangle}=\overline{\langle s^- s^+\rangle}-\overline{\langle s^-\rangle\langle s^+\rangle}.
    \label{eq:S-N_connection}
\end{align}
This is essentially the equal-time integral of that same correlator (up to the 
particular normalization conventions). It also implies that while $\overline{N}$ is connected to the area under the spectrum (or a closely related integrated quantity), $\mathsf{S}$ confirms how that area is distributed in frequency. This is an important fact that points out even if the average number of excitations are equal from two different approaches, the corresponding spectra can disagree as the excitations can be distributed or weighted differently in a frequency range of interest.
Moreover, another difference is that the incoherent emission spectrum excludes the coherent population contributions constructing the long-time averaged excitation number.


As a consequence of the variance relation, or equivalently, the
Cauchy-Schwarz inequality, one has
\begin{equation*}
\overline{\langle s^-s^+\rangle}
-
\overline{|\langle s^+\rangle|^2}
=
\overline{\langle\delta s^- \delta s^+\rangle}
\geq 0 .
\end{equation*}
This relation provides a useful consistency check for the physical
validity of the incoherent spectrum. In the very weak coherent-pumping
regime, however, the excited-state population and the incoherent
fluctuation contribution can become extremely small. In this limit, the
subtraction of the coherent component may become numerically delicate,
especially in the F-GME implementation, and residual machine-precision
errors can noticeably affect the normalized spectrum. Therefore, spectra
obtained in regimes with vanishingly small populations should be
interpreted with care.

\subsubsection{Numerical methods for solving the GMEs\\ and evaluating observables}

To compute steady-state populations, time-averaged observables, two-time correlation functions, and emission spectra, we employ two complementary numerical approaches.

The first is direct time-domain integration of the generalized master equations using the \texttt{mesolve} routines in QuTiP~\cite{Johansson_Qutip_2012,Johansson_Qutip_2013,Lambert_Qutip5_2026}. The density matrix is propagated under the relevant Liouvillian for either of the GMEs until the asymptotic periodic steady state is reached. Time-dependent observables are then evaluated directly from the resulting density operator. Two-time correlators are computed using the quantum regression theorem, generalized to the periodically driven case by propagating operator-conditioned density matrices under the same Liouvillian dynamics (see App.~D.2 of Ref.~\onlinecite{Salmon_Gauge-independent_2022}).

The second approach, that is computationally more time-efficient, is the Floquet--Liouville (FL) formulation in which the periodic Liouvillian is represented in a FL extended Sambe space span, by 
\begin{equation}
    \mathsf{B}_{\rm FL}=\{\FLket{l,jk}\equiv\tket{l}\otimes
\ket{j}\bra{k}\}.
\end{equation} 
This is the temporal extension of the Liouville space span by the basis set, 
\begin{equation}
    \mathsf{B}_{\rm L}=\{\Lket{jk}\equiv\ket{j}\bra{k}\},
\end{equation} 
with the dimension $\dim(\mathbb{H}_{\rm L})=\dim(\mathbb{H}_{\rm dressed})\times\dim(\mathbb{H}_{\rm dressed})$, and converted into a time-independent supermatrix eigenvalue problem. Diagonalization yields the FL eigenmodes and complex eigenvalues, whose real and imaginary parts determine decay rates and oscillation frequencies, respectively. This framework provides direct access to periodic steady states, modal decompositions of the dynamics, and analytical pole expansions of correlation spectra.

In this latter representation, the spectrum can be written as
\begin{equation}
\mathsf{S}(\omega)\propto
\mathrm{Re}\lsb\sum_{\mu\notin\mathcal P}
\frac{\mathcal{W}_\mu}{-\lambda_\mu-\ii\omega}\rsb,
\label{eq:FL_spectrum_main}
\end{equation}
where 
\begin{equation}
    \lambda_\mu=-\gamma_\mu-\ii\Delta_\mu
    \label{eq:FL_eigenvalue}
\end{equation} 
are FL eigenvalues and $\mathcal{W}_\mu$ are the corresponding residues determined by the left and right FL eigenmodes.
Note the sum excludes the population modes to represent the incoherent emission spectrum~\cite{Akbari_Quasienergy-Resolved_2026}. Each pole contributes a Lorentzian-like feature with effective peak position and linewidth
\begin{equation}
\Delta_\mu \equiv -\mathrm{Im}[\lambda_\mu],
\qquad
\gamma_\mu \equiv -\mathrm{Re}[\lambda_\mu].
\label{eq:omega_gamma_mu}
\end{equation}

In the absence of dissipation, the FL basis states are built from coherences between Floquet states and evolve at frequencies set by quasienergy differences. Once dissipation is included, these basis modes acquire finite decay rates and can hybridize through nonsecular couplings. The exact `right' eigenmodes $\FLket{R_\mu}$ of the FL supermatrix may therefore be expanded as superpositions of quasienergy-resolved channels,
\begin{equation}
\FLket{R_\mu}
=
\sum_{\alpha,\beta,l}
R_\mu^{\alpha\beta l}\,
|l)\otimes|\alpha\rangle\langle\beta|,
\label{eq:Rmu_expand_main}
\end{equation}
with coefficients $R_\mu^{\alpha\beta l}$ that quantify the contribution of each channel. This addresses the question of {\it how much does a mode align with a channel, including phase}.  To answer the question: {\it out of all identified channels, how much of a mode belongs to a particular channel}, we define
\begin{equation}
    r^{\alpha\beta l}_\mu=\frac{\lvert R^{\alpha\beta l}_\mu\rvert^2}{\sum_{\alpha'\beta'l'}\lvert R^{\alpha'\beta' l'}_\mu\rvert^2},
    \qquad
    \sum_{\alpha\beta l}r^{\alpha\beta l}_\mu=1
    .
\end{equation}

Note, the label $\mu$ does not, in general, denote a single transition $(\alpha,\beta,l)$, but rather an effective FL mode that may contain several overlapping quasienergy-resolved contributions. Only in the weak-dissipation or well-separated-channel limit does one recover an approximate one-to-one correspondence,
\begin{equation}
\mu \;\longleftrightarrow\; (\alpha,\beta,l),
\end{equation}
with
\begin{equation}
\lambda_\mu \approx -\gamma_{\alpha\beta l}-\ii\Delta_{\alpha\beta l}.
\label{eq:lambda_mu_ab_l}
\end{equation}
In this limit, the pole expansion in Eq.~\eqref{eq:FL_spectrum_main} reproduces the simpler Lorentzian channel picture. Away from that limit, however, nearby quasienergy gaps and nonsecular couplings lead to mode mixing, and the FL decomposition becomes essential for interpreting linewidths, spectral weights, and steady-state observables.

This distinction is central to the analysis in the Results section, where each individual spectral peak can now be identified via the metric amplitude (\emph{mode prominence})
\begin{equation}
\mathcal{R}_\mu\equiv\mathrm{Re}[\mathcal{W}_\mu]/\gamma_\mu,  
\label{eq:SpectralProminenceMetric}
\end{equation}
with linewidth $\gamma_\mu$. Therefore, we will use the quasienergy-resolved channel picture to identify the relevant dissipative processes, and the FL modal decomposition to determine when a spectral feature is dominated by a single transition and when it instead arises from a mixed dynamical mode.

Additionally, the same FL analysis can also be applied to the TI-GME. The only formal difference is that, in that case, the Fourier expansion of the dissipative Liouvillian contains only its zeroth harmonic,
\begin{equation}
\mathcal{L}_0=\mathcal{L}_{\rm diss},\qquad \mathcal{L}_{n\neq0}=0,
\end{equation}
with $\mathcal{L}_{\rm diss}$ given in Eq.~\eqref{eq:Ldiss_TIGME}. The resulting modal structure can therefore be analyzed in parallel with the Floquet-consistent case.

The direct integration method serves as a numerically exact benchmark within the chosen truncation, while the FL approach provides physical interpretation and computational efficiency for frequency-resolved observables. Full implementation details are provided in Appendix.~\ref{secS:FL}, and also with the application in less complicated systems~\cite{Akbari_Quasienergy-Resolved_2026}.

Finally, we note that a similar FL modal decomposition can also give the modal sum for the time-dependent population in Eq.~\eqref{eq:N_t}. However, the sum only includes the population modes in contrast to the incoherent correlators and spectra, see App.~\ref{secS:FL} for details. Averaging over a period though picks only the zero-Fourier component,   
\begin{equation}
\overline{N}_\Lambda
=
\mathcal{W}_{\mu_0}^{N_\Lambda},
\label{eq:AveN_FL}
\end{equation}
where $\mu_0$ indexes the zero FL mode, with $\lambda_{\mu_0}\simeq0$ and $\mathcal{W}_{\mu_0}^{N_\Lambda}$ is determined by the corresponding right and left FL modes and how they merge the number operator into the steady state and cycle average (see full detail in App.~\ref{secS:FL}). Since these modes have zero real parts, the prominence metric used in Eq.~\eqref{eq:SpectralProminenceMetric} for the incoherent spectra, does not apply and the strength of the modes is sufficiently determined by the factor $\mathcal{W}_{\mu_0}^{N_\Lambda}$.

This FL modal analysis of the incoherent emission spectra and the long-time averaged excitation numbers, once again confirms that the origin modes contributing to the population is different from those of the incoherent emission spectra.

For convenience, we summarize the structure of the theoretical formalism in Tab.~\ref{tab:TheoryRoadmap}, which lists, for each stage of the development in Sec.~\ref{sec:Theory}, the required inputs, a representative equation and the associated computational step, and the resulting outputs together with the quantities used in their subsequent analysis.


\begin{table*}[!htbp]
\caption{Roadmap of the theoretical formalism developed in Sec.~\ref{sec:Theory} as a summary of the theoretical structure developed in this work.}

\vspace{1.5mm}
\setlength{\tabcolsep}{6.8pt}

\label{tab:TheoryRoadmap}
\centering

\begin{tabular*}{\textwidth}{@{\extracolsep{\fill}} llll}
\toprule

Stage &
Representative Object &
Output &
Physical Role
\\
\midrule

Lab Basis
&
$
\mathcal{H}(t)
=
\sum_m e^{-\ii m\omega_d t}\mathcal{H}_m
$
&
$\{\ket{g,n},\ket{e,n}\}$
&
\makecell[l]{Bare cavity and TLS states including\\ the USC interaction and periodic drive.}
\\[3mm]

Dressed Basis
&
$
\mathcal{H}_0\ket{j}=E_j\ket{j}
$
&
$\{\ket{j}\}$
&
\makecell[l]{Internal light-matter dressing due to\\ ultrastrong coupling.}
\\[3mm]

Floquet Basis
&
$
[\mathcal{H}(t)-\ii \dd/\dd t]\ket{\alpha(t)}
=
\varepsilon_\alpha\ket{\alpha(t)}
$
&
$\{|\alpha(t)\rangle\}$
&
\makecell[l]{External Floquet dressing induced by\\ the periodic drive.}
\\[3mm]

Double Dressing
&
$
\ket{\alpha(t)}
=
\sum_{jl}
f_{\alpha jl}\,
e^{-\ii l\omega_dt}\,\ket{j}
$
&
Floquet USC states
&
\makecell[l]{Unified treatment of internal and\\ external dressing.}
\\[3mm]

Quasienergy Channels
&
$
\Delta_{\alpha\beta l}
=
\varepsilon_\beta
-\varepsilon_\alpha
+l\omega_d
$
&
Transition channels
&
\makecell[l]{Reservoir-induced transitions\\ between Floquet sidebands \\(quasienergy-resolved dissipation).}
\\[3mm]

F-GME
&
$
\dot\rho(t)=[\mathcal{L}_{\rm S}(t)+\mathcal{L}_{\rm diss}(t)]\rho(t)
$
&
\makecell[l]{Driven open-system\\ dynamics}
&
\makecell[l]{Quasienergy-resolved dissipation\\ and decoherence.}
\\[3mm]

\makecell[l]{Floquet-Liouville\\
Diagonalization}
&
$
[\mathscr{L}(t)-\dd/\dd t]
\Lket{R_\mu(t)}
=
\lambda_\mu
\Lket{R_\mu(t)}
$
&
\makecell[l]{FL modes, residues\\ $\lambda_\mu=-\gamma_\mu-\ii\Delta_\mu$, $\mathcal{W}_\mu$}
&
\makecell[l]{Decay rates, oscillation frequencies,\\ and modal weights.}
\\[3mm]

Periodic Steady State
&
$
[\mathscr{L}(t)-\dd/\dd t]
\Lket{\rho_{\rm ss}(t)}
=
0
$
&
$\rho_{\rm ss}(t)$
&
Long-time asymptotic state.
\\[3mm]

Population
&
$
\overline{N}=
\overline{\langle s^-(t )\, s^+(t)\rangle}
$
&
\makecell[l]{Long-time averaged\\ number of excitations}
&
\makecell[l]{Production of cavity real excitations\\ (photons) and TLS real excitations}
\\[3mm]

Correlation Functions
&
$
G^{(1){\rm inc}}_{s^-s^+}(\tau)
=
\overline{\langle\delta s^-(t+\tau)\,\delta s^+(t)\rangle}
$
&
\makecell[l]{Long-time averaged\\ two-time correlators}
&
Quantum fluctuations and coherence.
\\[3mm]

\makecell[l]{Incoherent Emission\\Spectrum}
&
$
\mathsf{S}(\omega)
\propto
\mathrm{Re}
\!\int_0^\infty
d\tau\,
e^{\ii\omega\tau}
G^{(1){\rm inc}}_{s^-s^+}(\tau)
$
&
Observable spectrum
&
\makecell[l]{Direct physical observables\\ and experimental signatures.}
\\[3mm]

Mode Prominence
&
$
\mathcal{R}_\mu=\mathrm{Re}[\mathcal{W}_\mu]/\gamma_\mu
$
&
Bright spectral modes
&
\makecell[l]{Dominant peaks in observables'\\ spectra and experimental signatures.}
\\[3mm]

\makecell[l]{Mode-Channel\\ Resolution}
&
$
r^{\alpha\beta l}_\mu
$
&
$\mu\leftrightarrow\{(\alpha,\beta,l)\}$
&
\makecell[l]{Hybridization of the Floquet channels\\ into FL modes.}
\\

\bottomrule
\end{tabular*}

\vspace{0.1cm}

\end{table*}

\section{Numerical Results, Analysis and key findings}
\label{sec:ResultsDisscussion}
In this section, we now show selected numerical results of the presented theories for different parameters of the cavity-QED system and the drive values. 
While the theory is 
general and works for various scenarios and parameters, we choose to analyze the cases of resonant ($\omega_c=\omega_a$) cavity-QED to showcase the dominance of our Floquet theory over the usual existing theories, and to demonstrate when it is deemed necessity to 
use the {\it correct} Floquet-modified observables. We will demonstrate how to Floquet-engineer a wide range of quantum state control, where dissipation and non-RWA effects (from the interactions and driving) also plays a dominant role. We stress that this general regime of high intensity driving in an open system in USC, is one of the most technically difficult regimes to model in quantum optics, as many of the standard approaches simply do not work.


Broadly, we  investigate the 
{\it strongly driven open cavity-QED system} in
the {\it USC regime}, hence, we mostly stay in the cavity-dipole coupling range  with $0.1\lesssim\eta\lesssim1$ (USC) and also with the strongly driven QRM range with $0.1\eta\lesssim\eta_d\lesssim\eta$. The weakly driven QRM range, $\eta_d\lesssim0.1\eta$ has been investigated in Ref.~\onlinecite{Salmon_Gauge-independent_2022}, though \emph{not} with  Floquet theory but with the TI-GME. Moreover, in the calculation of the spectra, we stay away from the range of an extreme-driven cavity-QED $\eta_d\gg\eta$ as it deforms the QRM states and we no longer benefit from the interesting physics associated with the QRM~\cite{Forn-Diaz_Ultrastrong_2019,FriskKockum_Ultrastrong_2019,DiStefano_Resolution_2019}, and we are not interested in only observing the large coherent background field. Other techniques are likely more efficient in such a semi-classical regime.

Figure~\ref{fig:Schematics} depicts the driven cavity-QED models considered in this work and schematically illustrates how periodic driving qualitatively modifies the energy structure underlying dissipative dynamics in the USC regime. Panels (a) and (b) show schematics of the two experimentally relevant driving protocols studied here, respectively: an optically coherently pumped cavity-QED system and a Floquet-engineered realization in which system parameters are periodically modulated. In both cases, the driven system is described by a time-periodic QRM Hamiltonian coupled weakly to external environments.

For each excitation regime, it is essential to first understand the energy basis of the undriven and driven QRM, as it provides direct insight into the structure of allowed transitions and their symmetry constraints; then, we explore its consequences and compute and discuss the observables.

\begin{figure*}[ht]
\centering
\includegraphics[width=.99\linewidth]
{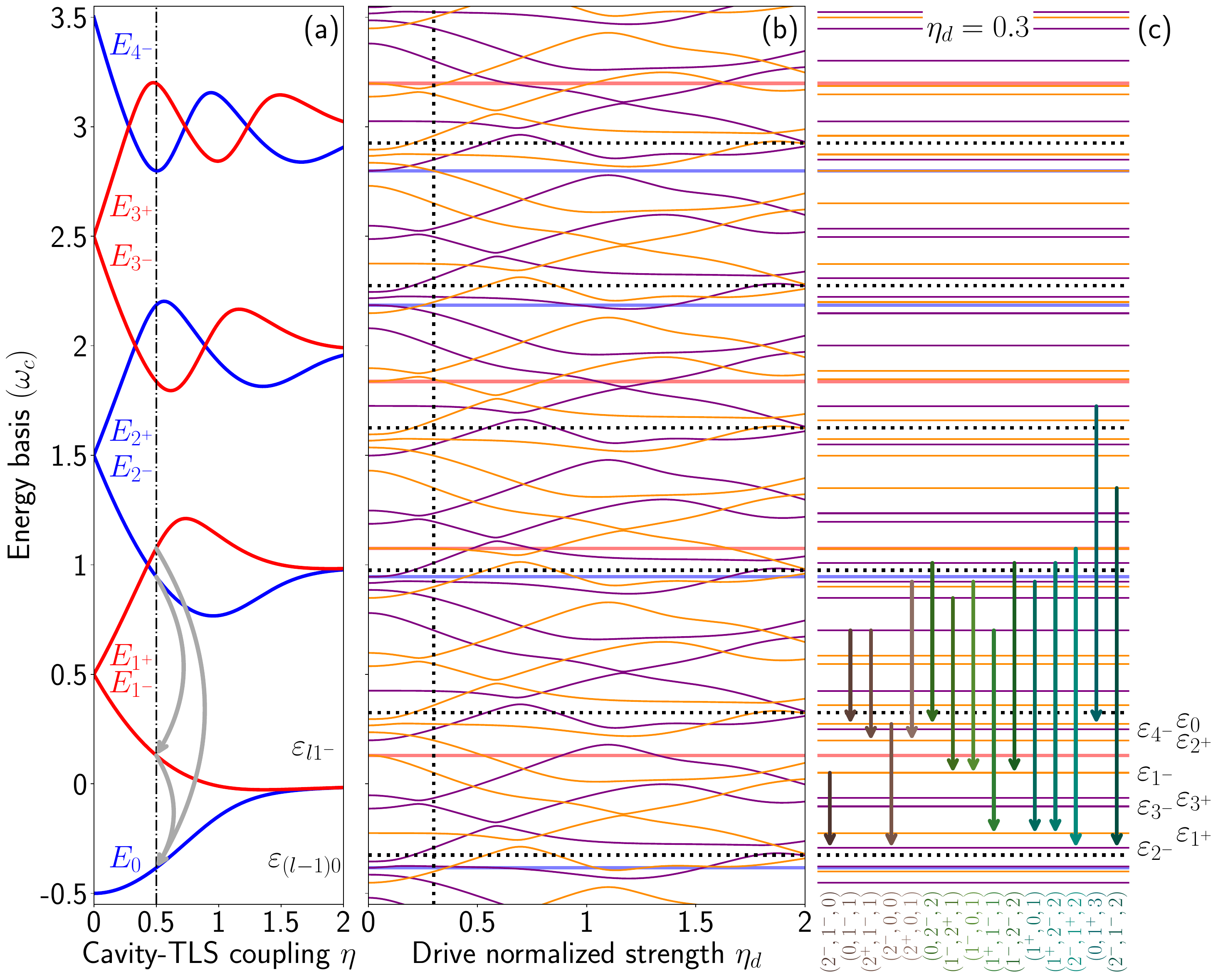}
    \caption[Schematics]{\textbf{Energy basis states for an optically driven cavity-QED system.} (a) Eigenenergies of the QRM for the first eight states, with the Hamiltonian in Eq.~\eqref{eq:HC_QRM},  plotted versus the (normalized) cavity-TLS coupling rate $\eta$ with their labels and parities (blue: even; red: odd). The vertical black dot-dashed line in panel (a) marks $\eta=0.5$, where it is used to calculate the quasienergies in panels (b), and also for future results. When the pump (here, on the TLS) is on, the energy basis of the QRM transforms into the Floquet extended-space quasienergies of the optically (coherent) pumped DQRM given in Eq.~\eqref{eq:H_CPQR} (thin solid curves), with (ideally) infinite BZs distinguished by the horizontal dotted black lines versus the pump amplitude, shown in (b) for $\omega_d=0.65\omega_c$ and with $\eta=0.5$. Note that each quasienergy curve originates from an eigenenergy (faint thick solid blue and red lines) that inherits its label and generalized parity from (purple: even; orange: odd). The dotted black lines show the drive specifications, namely, the vertical dotted black line marks $\eta_d=0.3$ which we use for future results, and the horizontal dotted black lines differentiate the consecutive BZs in the Floquet extended space of energy $\omega_d$.  Each BZ includes (nominally) eight quasienergy states (the smallest two are labeled on the left of panel (b) and the rest follow the same pattern, see text). The dominant transitions among the time-independent dressed basis in the incoherently pumped spectra are depicted by the downward gray arrows in (a).
    (c) A magnification of the drive point at $\eta_d=0.3$ shown by the black vertical dotted line in panel (b), with the downward arrows showing the dominant (top 15)  quasienergy-resolved transitions, $(\alpha,\beta,l)$, in the extended space contributing in the production of the major peaks in the emission spectra.
    For all results here, a resonant cavity-QED model is considered, i.e., $\omega_a=\omega_c$.
   }
    \label{fig:EnergyBasis_Optical}
\end{figure*}

\subsection{Floquet modulated coherent pumping the QRM}

Figure~\ref{fig:EnergyBasis_Optical}(a) displays the static dressed eigenenergies of the (undriven) 
QRM versus the cavity-TLS (static) coupling, $\eta$ (optical coherent pump case), and distinguishes their parity.
We also label the eigenenergies based on their order when nonadiabatically $\eta\to0^+$, where they match the 
analytical result of the JCM. 
It is instructive to analyze the low- and high-coupling limits of the eigenenergies.

At $\eta\to0^+$ or $g\ll\{\omega_c,\omega_a\}$, the effective splitting between the eigenenergy lines is $\lvert\delta\rvert$, and the eigenenergy lines follow the energies of the joint states of the bare state of the light and the matter as 
\begin{equation}
    E_j\to\left.E_{n^\pm}\right\vert_{g\to0^+}=(n-1/2)\omega_c\pm\lvert\delta\rvert/2,
\end{equation} 
corresponding to the states $\lvert n, a\rangle$ with $n$ cavity photons and $a=e,g$ (excited or ground) state of the TLS. 

At $\eta\to\infty$ or $g\gg\{\omega_c,\omega_a\}$, the photon states highly dress the TLS states so that the TLS internal energy can be averaged to contribute the total energy of the system which is now the joint dressed light-matter state $E_j\to n\omega_c$.
Thus, the eigenenergy spectrum represents a free displaced harmonic oscillator rather than strongly hybridized polaritons.
In this limit, where the atom sits on the symmetric and antisymmetric coherent state basis, the cavity photon cloud dressing the atom becomes tremendously large that flipping the atom would require rearranging an enormous number of virtual photons. As a result, in the time-independent regime, the photon exchange is frozen,
atomic transitions are inhibited and effective light–matter interaction vanishes, a phenomenon known as the \emph{light-matter decoupling in the deep-strong coupling regime} for $g\gtrsim\{\omega_c,\omega_a\}$~\cite{DeLiberato_Light-Matter_2014,Mercurio_Regimes_2020,Ashida_Cavity_2021}.

This becomes practically impossible.
Note the JCM's eigenenergies given in Eq.~\eqref{eq:EE_JCM} are only a good approximation for those of the QRM~\cite{DiStefano_Resolution_2019} when $g\ll\{\omega_c,\omega_a\}$.

Clearly, the number of total excitations in the QRM determines the parity of the eigenenergy states, and consequently, constrains which matrix elements of the system–bath coupling operators are symmetry-allowed. Thus, the matrix element of the parity operator follow the relation $\propto(-1)^{n}$ so that the states $j=n^\pm$ with $n$ (the number of photons in the decoupled light-matter system, $g\to0^+$) even/odd number have the even/odd parity. Therefore, in panel (a), the blue lines show the even-parity dressed states containing even number of photonic states, whereas the red lines show the odd-parity dressed states containing odd number of photonic states.

 In this {\it time-independent} picture, dissipative processes are restricted to transitions between a discrete set of dressed eigenstates, forming the basis of standard dressed-basis master-equation treatments. These static spectra, therefore, encode the transition frequencies at which the environment is sampled in TI dissipation models.

Periodic driving fundamentally alters this structure, as illustrated in Fig.~\ref{fig:EnergyBasis_Optical}(b), which also shows the corresponding Floquet quasienergy spectra in the extended Sambe space. 
The periodic modulation lifts the static dressed levels into multiple Floquet sidebands separated by integer multiples of the drive frequency.
The nomenclature for the Floquet quasienergies, is chosen such that at the limit of $\eta_d\to0^+$ (a first order perturbation), $\alpha=j$, so $\varepsilon_{\alpha=j}\pm l\omega_d=E_{j}$. 

We can distinguish the quasienergy states based on their generalized parity, and each quasienergy curve shares the same generalized parity as the parity of its originating eigenenergy line. Namely, the solid thin orange quasienergy curves of odd generalized parity stem from their originating solid thick red eigenenergy lines of odd parity, whereas the solid thin purple quasienergy curves of even generalized parity stem from their originating solid thick blue eigenenergy lines of even parity.

We also note that regarding the generalized parity of the quasienergy states, the eigenvalues of the parity operator in the extended space are determined via $\propto(-1)^{\alpha+l}$. Therefore, the variation of the generalized parity with respect to the different BZs follows $\propto(-1)^{l}$ so that the quasienergy lines swap parity (or color) in the consecutive BZs~\cite{Kohler_Driven_2005}.

We emphasize that in the optical coherent pumping case, the quasienergies inherit their originating eigenenergy parity at the limit of the vanishing drive and sustain their parity through the increase of the drive strength, as seen in Fig.~\ref{fig:EnergyBasis_Optical}(b,c); however, this is not exactly the case for the mechanical drive as we discuss later [cf. Fig.\ref{fig:EnergyBasis_Mechanical}(b)].

 In the USC regime, the Floquet modulation leads to dense quasienergy manifolds and near-degeneracies, even when the static spectrum is relatively sparse~\cite{Akbari_Floquet_2025}. As a result, dissipative transitions in the driven system are {\it no longer characterized solely by static energy differences (bare and dressed eigenenergies), $\omega_{jk}$, but instead occur through multiple channels at frequencies $\Delta_{\alpha\beta l}$}, with weights determined by the corresponding Floquet harmonics of the system–bath coupling operator~\cite{Akbari_Quasienergy-Resolved_2026,Akbari_Floquet_2026}.

According to the von Neumann-Wigner noncrossing rule~\cite{vonNeuman_Merkwurdige_1929}, eigenvalues of a Hermitian operator which belong to the same symmetry class generically do not cross. Thus, we expect to see that the curves with different colors can only form crossings, as shown in panels (a) and (b) of Fig.~\ref{fig:EnergyBasis_Optical}.
As the driving amplitude is sufficiently increased, the nonadiabaticity can mix quasienergy states of the same parity. The mixing is particularly strong if the quasienergy gap that separates the states is small.
The selection rule has a significant implication of the parity symmetry,
and help to show that the stimulated transitions between two eigenstates with the same (opposite) parity require a mediation of an even (odd) number of quanta (oscillation harmonics) from the external drive field. This is also true when the deexcitation processes are emitting quanta.

Later, we will show additional results at $(\eta,\eta_d,\omega_d)=(0.5,0.3,0.65\,\omega_c)$, and thus it is convenient to know the evolution and origin of the quasienergy states at this excitation regime. Looking at the 
drive frequency range of $-\omega_d/2\leq\varepsilon_\alpha<\omega_d/2$ in panel (b), and also the magnified version in panel (c) of Fig.~\ref{fig:EnergyBasis_Optical}, we see the quasienergy states, from bottom to the top, and the first one is $2^-$ as it has adiabatically evolved from the same eigenenergy state. The latter states are $1^+,3^-,3^+,1^-,2^+,4^-$ and $0$, which are listed in the right side of panel (c) for $\eta_d=0.3$.
The downward colored arrows represent the dominant quasienergy-resolved transitions, $(\alpha,\beta,l)$-channels, to construct the dominant FL mode $\mu$ in the correlators/spectra, listed in Tabs.~\ref{tab:FLmodes_cavity_kappa01Flat}--\ref{tab:FLmodes_cavity_kappa02LorOhmic1}.

The quasienergy-resolved structure of dissipation~\cite{Akbari_Quasienergy-Resolved_2026} forms the central physical mechanism explored in the remainder of this section. In particular, while the time-independent dressed-basis master equations sample the environment only at static dressed transition frequencies, Floquet-dressed descriptions naturally incorporate the full distribution of sideband-shifted channels generated by the drive. Whether this distinction has observable consequences depends crucially on the structure of the environmental spectral density.

\begin{figure*}[th]
\centering
\includegraphics[width=0.98\linewidth]
{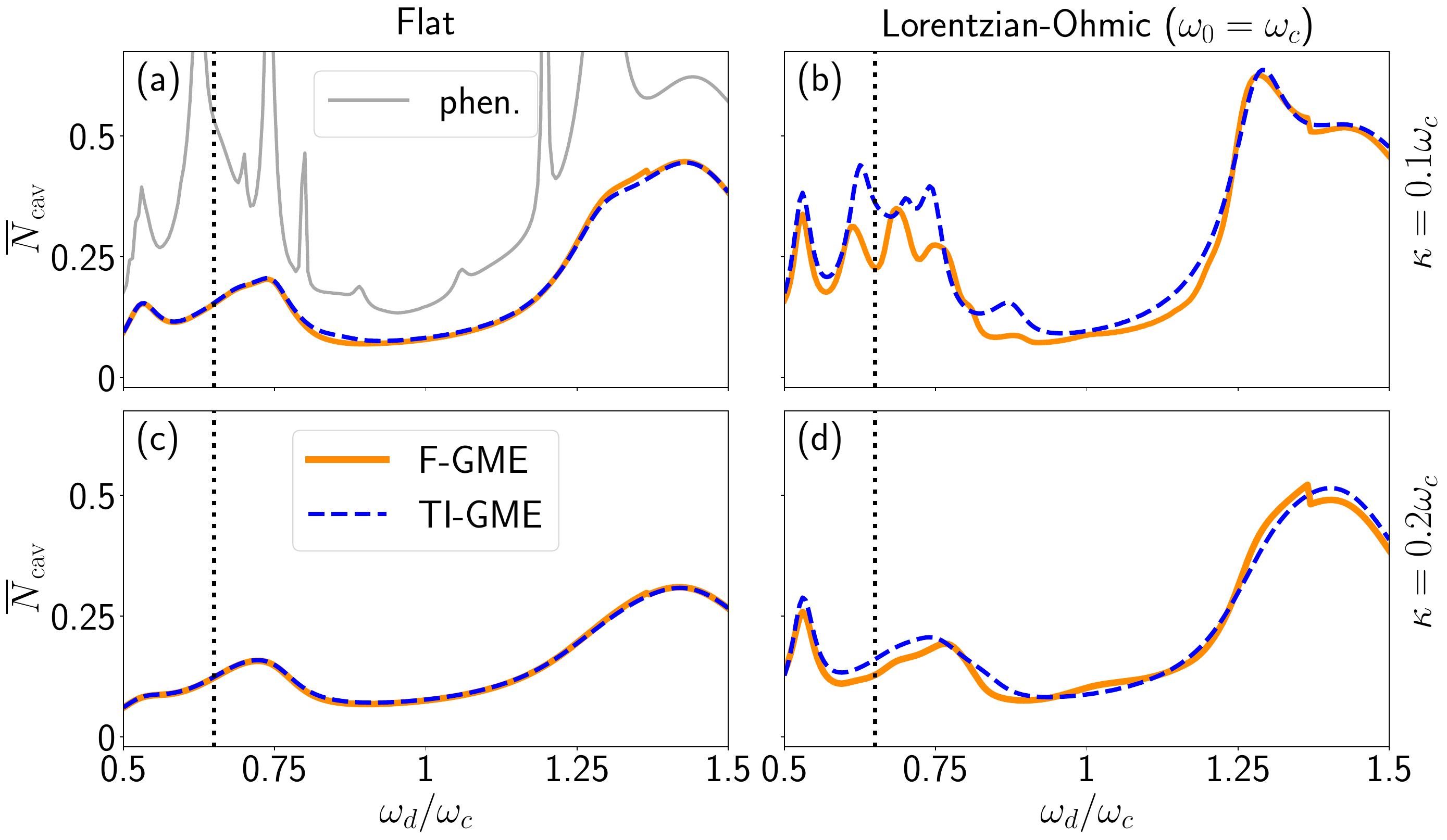} 
\caption[]{\textbf{Optically driven open cavity-QED in USC: 
Average number of cavity excitations versus the optical drive frequency $\omega_d$.} 
The other relevant parameters are: $\eta=0.5$, $\eta_d=0.3$ (drive amplitude), $\omega_a=\omega_c$, and $\gamma=0.01\omega_a$.
Temporal average of the number of cavity excitations are compared among three different approaches: (i) phenomenological Floquet theory (thin solid gray curves), from Eq.~\eqref{eq:AveN_phen}; (ii) TI-GME (blue dashed curves), obtained from Eq.~\eqref{eq:AveN} with the GME in Eq.~\eqref{eq:TIGME_driven}; and (iii) F-GME (solid orange curves), obtained from Eq.~\eqref{eq:AveN} with the GME in Eq.~\eqref{eq:FGME_OriginalBasis}. The left panels (a,c) show the results for the flat baths [Eqs.~\eqref{eq:CavityBathModel_flat} and \eqref{eq:TLSBathModel_flat}], whereas the right panels (b,d) for the structured Lorentzian cavity bath centered at $\omega_0=\omega_c$ with an Ohmic cut-off [Eqs.~\eqref{eq:CavityBathModel_Lorentzian} and \eqref{eq:TLSBathModel_flat}]. Additionally, the cavity bath dissipation rate is $\kappa=0.1\omega_c$ in the top panels (a,b) and $\kappa=0.2\omega_c$ in the bottom panels (c,d).
The phenomenological curves are included to present a general trend curve, but most importantly, as a \emph{reference map of higher-order resonances}.
The vertical dotted black line on each panel shows the representative drive frequencies at $\omega_d=0.65\omega_c$ for further investigation later.} 
\label{fig:AveNcav_wd_Optical}
\end{figure*}

\begin{figure*}[th]
\centering
\includegraphics[width=.98\linewidth]
{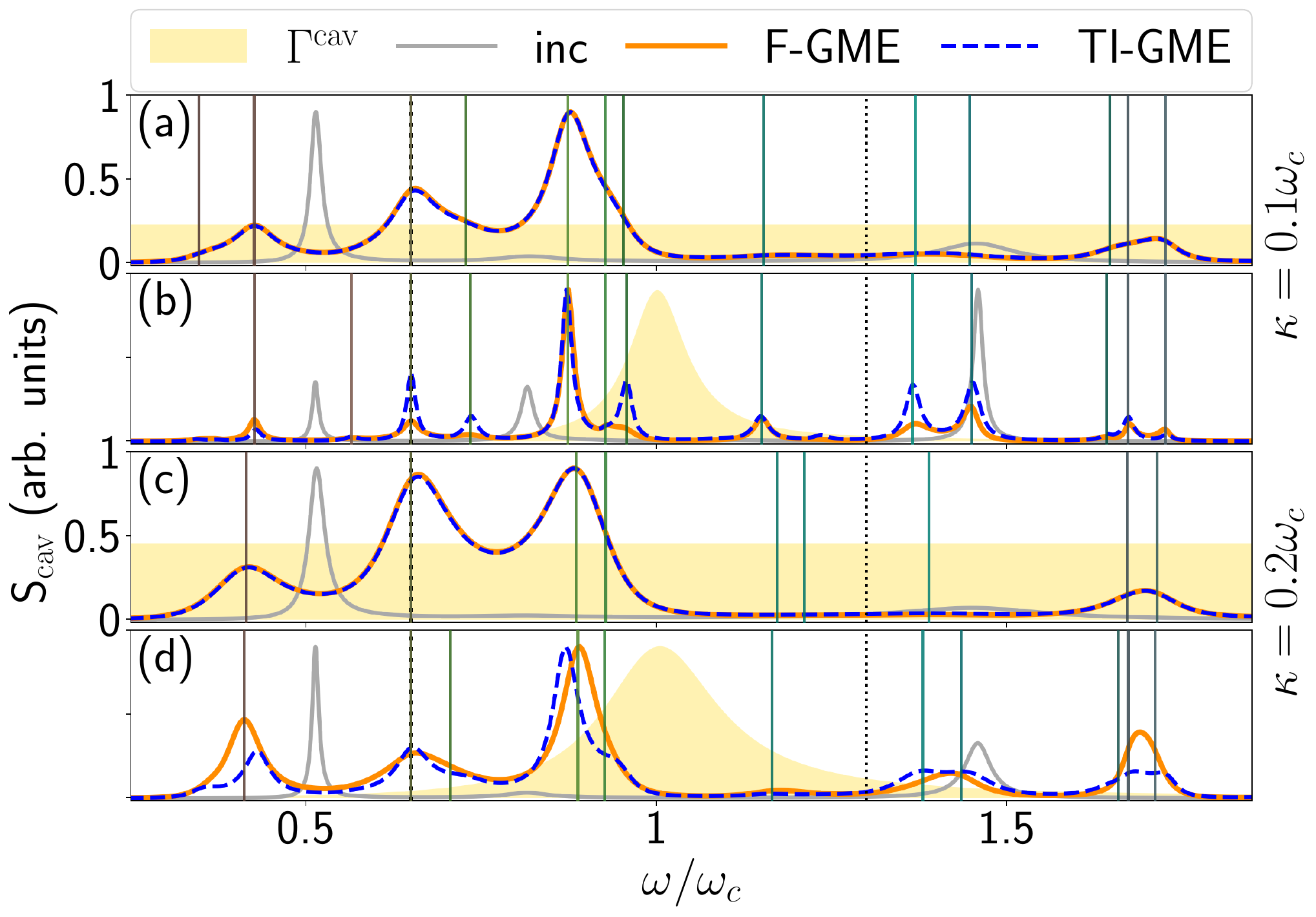} 
\caption[]{\textbf{Optically driven cavity-QED in USC: Incoherent cavity spectra.} Cavity emitted spectra, obtained from the TI-GME (dashed blue) and the F-GME (solid orange) approaches. For clarity, 
a reference spectra with weak incoherent pumping ($P_{\rm inc}=0.01\eta_d\omega_c$) 
is also plotted with the thin gray curves in each panel. Panels (a) and (c) show the spectra for the flat bath, and the right panels (b) and (d) show the spectra (only for the cavity) for the Lorentzian bath with the central frequency of $\omega_0=\omega_c$ and the Ohmic cut-off, shown by the semitransparent yellow regions. Panels (a) and (b) show the results for the dissipation rate of $\kappa=0.1\omega_c$, whereas panels (c) and (d) are for $\kappa=0.2\omega_c$. The vertical black dotted lines represent integer multiples of the drive frequency. The vertical color lines mark the FL mode gaps $\Delta_\mu$. Marker colors follow the corresponding quasienergy transition arrows ($\alpha\beta l$-channels) in Fig.~\ref{fig:EnergyBasis_Optical}; for modes composed of multiple dominant channels, colors represent a weighted average of the contributing components. The results are obtained for the minimum eight truncation number of the QRM dressed states to have good convergence, with parameters: $\omega_a=\omega_c$, $\eta=0.5$, $\omega_d=0.65\omega_c$, $\eta_d=0.3$ and $\gamma=0.01\omega_a$.
}
\label{fig:Spectracav_Optical_wd065}
\end{figure*}

Importantly, the quasienergy-resolved structure illustrated in Fig.~\ref{fig:EnergyBasis_Optical} (also, see Fig.~\ref{fig:EnergyBasis_Mechanical}) does not imply that all observables are equally sensitive to Floquet effects. While the redistribution of dissipative channels over Floquet sidebands may cancel out in quantities that depend only on the total integrated transition weight—such as time-averaged excitation proxies—frequency-resolved observables can retain clear signatures of the underlying quasienergy structure. As we demonstrate below, this distinction explains why TI- and F-GMEs may yield similar steady-state excitation numbers in certain regimes (e.g., cf. Fig.~\ref{fig:AveNcav_wd_Optical}), yet predict markedly different incoherent spectra for specific subsystems (e.g., cf. Fig.~\ref{fig:Spectracav_Optical_wd065}), even when the same dressed operators and flat bath spectra are employed.

In the following, we investigate how these differences manifest in experimentally accessible observables. 
Specifically, we concentrate on the number of cavity mode excitations and the low-energy cavity photon incoherent emission spectra for each case, where we consider the transitions among the first few QRM states, and  wish to connect the observed peak and valley structures to the specific structure in the energy basis diagrams. In addition, we discuss the effect of the structured baths in each regime.
In our analysis, we mostly concentrate on the drive's range of interest that is quantified by $0.1\eta\lesssim\eta_d\lesssim\eta$.
In this range, the QRM states are still strongly driven but have not fully changed their characteristics by the external drive. 

The discussions in Ref.~\onlinecite{Moiseyev_Conditions_2024} on the equivalence of the Floquet optical drive of a TLS (the classical Rabi model) and a quantum optically dressed TLS (the JCM) provided that the number of externally assisting photons are sufficiently large, implies that the regime of tuned and resonant coherent optical pump of the QRM can be equivalently predicted by the Floquet theory and the time-independent theory in the dressed basis, if the number of photon states as well as the number of truncated dressed states are sufficiently large. Guided by this argument, we thus begin our analysis by looking into the populations (here, the long-time averaged number of real excitations) versus the drive frequency, and hence, deduce the regimes of the detuned drive where the prediction of the Floquet theory and the time-independent-basis theory fall apart.

We first analyze the steady-state average number of the cavity excitations in Fig.~\ref{fig:AveNcav_wd_Optical} (for the corresponding TLS excitations, see Fig.~\ref{figS:AveNTLS_wd_Optical} in App.~\ref{secS:AdditionalResults}) as a function of the drive frequency, comparing predictions obtained from the F-GME with those from TI-GME as well as the phenomenological Floquet-dressed approaches.

Figure~\ref{fig:AveNcav_wd_Optical} shows the long-time averaged cavity excitation number,
$
\overline{N}_{\rm cav}
=
\overline{\langle s^{{\rm cav}-} s^{{\rm cav}+}\rangle},
$
as a function of optical drive frequency $\omega_d$ for a strongly driven cavity-QED system in the USC regime ($\eta=0.5$, $\eta_d=0.3$). Results from the F-GME and the TI-GME are compared with a phenomenological Floquet treatment of dissipation (gray curves) used as a reference. Two cavity loss rates, $\kappa=0.1\omega_c$ [panels (a,b)] and $\kappa=0.2\omega_c$ [panels (c,d)], are considered for both flat baths (left panels) and structured Lorentzian--Ohmic baths peaked at $\omega_0=\omega_c$ (right panels). The TLS decay rate is fixed at $\gamma=0.01\omega_a$.

The phenomenological Floquet treatment exhibits a sequence of peaks and valleys,
as the drive frequency changes, associated with higher-order drive-assisted resonances and avoided crossings in the quasienergy spectrum of the DQRM, discussed in Ref.~\cite{Akbari_Floquet_2025} and App.~\ref{secS:AdditionalResults}. However, including physically consistent dissipation through 
GME significantly smooths and broadens any narrow resonant features whose intrinsic linewidths are smaller than the bath broadening. Increasing the cavity loss rate from $\kappa=0.1\omega_c$ to $0.2\omega_c$ further suppresses higher-order structures, as seen by comparing panels (a,b) with (c,d).

For the flat cavity bath [panels (a,c) of Fig.~\ref{fig:AveNcav_wd_Optical}], the F-GME and TI-GME yield nearly identical long-time averaged populations over the full drive range. 
The dominant excitation pathways are shown in Fig.~\ref{figS:Fig_wd_Optical}(c), and particularly for $\omega_d=0.65\omega_c$ (marked by the dotted vertical black line), one finds the top three dominant transition probabilities are $\overline{P}^{(+2)}_{2^-\leftarrow0}$, $\overline{P}^{(+1)}_{1^-\leftarrow0}$ and $\overline{P}^{(+4)}_{2^+\leftarrow0}$, respectively.
Since the bath function is frequency independent, $\Gamma^{\rm cav}(\omega)=\mathrm{const}$, all Floquet-assisted decay channels are weighted equally. Strong driving redistributes transition amplitudes among Floquet sidebands, but for time-averaged observables the total dissipative weight entering the steady-state balance is largely preserved after summation over sidebands. Consequently, both approaches
(Floquet and TI GME)
produce nearly the same zero-Fourier steady-state component $\rho_0$, and hence similar values of
$
\overline{N}_{\rm cav}
=
\mathrm{Tr}[N_{\rm cav}\,\overline{\rho}]
$
with
$
\overline{\rho}
=
T^{-1}\int_0^T dt\,\rho_{\rm ss}(t).
$

A qualitatively different behavior emerges for the {\it structured} Lorentzian--Ohmic bath [panels (b,d)  of Fig.~\ref{fig:AveNcav_wd_Optical}], where the two theories visibly separate. Here the bath weights different transition frequencies unequally, so the redistribution of decay amplitudes among Floquet sidebands becomes physically different. The discrepancies are largest near higher-order resonances, where multiple sideband channels contribute simultaneously and where the phenomenological reference curve also shows pronounced peaks [see  Fig.~\ref{fig:AveNcav_wd_Optical}(a)].

A representative example of a clear qualitative difference between the models occurs near $\omega_d=0.65\omega_c$
[see Fig.~\ref{fig:AveNcav_wd_Optical}(b)]. From the quasienergy analysis in App.~\ref{secS:AdditionalResults}, the dominant excitation pathways involve the transitions $0\!\to\!2^{-}$ with $\Delta l=+2$, followed by $0\!\to\!1^{-}$ with $\Delta l=+1$, and weaker higher-order channels such as $0\!\to\!2^{+}$ with $\Delta l=+4$. In the time-independent dressed basis, the leading transition is associated with $\omega_{0,2^-}\approx1.329\omega_c$, whereas in the Floquet description the corresponding quasienergy-resolved channel occurs near $\Delta\approx1.215\omega_c$. Since the structured bath is larger at the latter frequency, $\Gamma^{\rm cav}(\Delta)>\Gamma^{\rm cav}(\omega_{0,2^-})$, the F-GME predicts stronger damping and therefore a lower steady-state population than the TI-GME.
This mechanism becomes even clearer in the frequency-resolved emission spectra, which we discuss later. 

Further insights into the underlying physics can be realized by studying the population modes in the FL modal decomposition given in Eq.~\eqref{eq:AveN_FL}, where a single FL population mode forms the average number of excitation at each point and the overlaps to Floquet channels can be identified. From the FL modal decomposition viewpoint, at the marked drive frequency in panel (b) of Fig.~\ref{fig:AveNcav_wd_Optical}, $\omega_d=0.65\omega_c$, the discrepancy between the
F-GME and TI-GME predictions for $\overline{N}_{\rm cav}$ can be traced directly to
the zero FL mode. Since
$\overline{N}_{\rm cav}
\propto\int_T N_{\rm cav}^{\rm ss}(t)\,dt$,
only the neutral component with $\lambda_\mu\simeq0$ survives the period average;
for the TI-GME, the corresponding zero-mode contribution is
$\mathcal{W}_{\mu}^{N_{\rm cav}}\simeq 0.363$,
whereas the F-GME gives
$\mathcal{W}_{\mu}^{N_{\rm cav}}\simeq 0.228 $.
Thus, the lower orange curve is not caused by a shift of an oscillatory mode, but by
a different steady-state population encoded in the zero FL eigenmode.

The mode decompositions further show that both descriptions are dominated by
population-sector components, as expected for an averaged occupation observable.
In the TI-GME, the zero mode contains mainly
$(3^-,3^-,0)$, with the normalized weight $R_\mu^{(3^-,3^-,0)}\simeq0.815$,
with a secondary contribution
$(2^+,2^+,0)$, with $R_\mu^{(2^+,2^+,0)}\simeq0.105 $.
In contrast, the F-GME zero mode is more concentrated in the single population
channel
$(3^-,3^-,0)$, with $R_\mu^{(3^-,3^-,0)}\simeq0.946 $.
This indicates that the quasienergy-resolved dissipator redistributes the steady
population among Floquet-dressed states differently from the time-independent
basis dissipator. Consequently, even though the relevant contribution to
$\overline{N}_{\rm cav}$ is a zero-frequency mode in both theories, the structure
and weight of the zero modes are different, leading to the observed discrepancy
between the F-GME and TI-GME curves.

There is no contradiction between the dominant excitation pathway
$|0\rangle\rightarrow|2^{-}\rangle$ obtained from the transition-probability
analysis and the diagonal components appearing in the zero-mode
FL decomposition of $\overline N$.
The former identifies how the coherent drive injects population into the
system, whereas the latter describes where population accumulates after the
combined action of pumping, relaxation, and repeated redistribution processes (from the master equation in the steady state, $\dd{\rho}_{\rm ss}(t)/\dd t=0$).

Accordingly, populations initially promoted to $|2^{-}\rangle$ (or, possibly, other less affected from dissipation and survived transitions such as to $\ket{1^-}$) need not remain
there in the steady state. Subsequent dissipative and drive-induced transfer
can populate other Floquet-dressed states, producing dominant zero-mode
population channels such as $(3^-,3^-,0)$, that is relaxing in the state $\ket{\alpha=3^-}$; and $(2^+,2^+,0)$, i.e.,  relaxing in the state $\ket{\alpha=2^+}$.
Thus, the transition analysis probes the excitation mechanism, while the
FL zero mode probes the final stationary population
distribution.

In this regard, the F-GME does not merely 
{\it remove} population uniformly; it produces a different stationary population distribution, more concentrated in one diagonal Floquet-dressed sector and with a lower measured cavity occupation, in the current example.


We next turn to the 
{\it incoherent cavity emission spectra} at the representative drive frequency $\omega_d=0.65\omega_c$, with results shown in Fig.~\ref{fig:Spectracav_Optical_wd065}. The system remains strongly driven in the USC regime ($\eta=0.5$, $\eta_d=0.3$). Flat cavity baths are shown in panels (a) and (c), while structured Lorentzian--Ohmic baths centered at $\omega_0=\omega_c$ are shown in panels (b) and (d). Two cavity dissipation rates are considered: $\kappa=0.1\omega_c$ [top rows (a,b)] and $\kappa=0.2\omega_c$ [bottom rows (c,d)]. Spectra from the F-GME (solid orange) are compared with those from the TI-GME (dashed blue), together with reference spectra obtained from weak incoherent pumping (gray). The shaded background indicates the cavity bath spectral density, and vertical dotted lines mark integer multiples of the drive frequency.

Unlike the long-time averaged populations of Fig.~\ref{fig:AveNcav_wd_Optical}, the frequency-resolved spectra reveal substantially greater sensitivity to the underlying dissipative description. Even in regimes where the two master equations give similar steady-state occupations, they can 
{\it predict visibly different peak positions, linewidths, and spectral weights}. This demonstrates that agreement in averaged observables does not imply that the underlying relaxation channels are correctly captured.

To interpret these spectra in more detail, we employ the FL modal decomposition. The colored vertical markers indicate the dominant FL spectral modes listed in Tabs.~\ref{tab:FLmodes_cavity_kappa01Flat}--\ref{tab:FLmodes_cavity_kappa01LorOhmic1}
of Appendix.~\ref{secS:AdditionalResults}. Each mode $\mu$ is characterized by a complex eigenvalue
$
\lambda_\mu=-\gamma_\mu-i\Delta_\mu,
$
whose imaginary part $\Delta_\mu$ gives the spectral peak position and whose real part $\gamma_\mu$ determines the effective linewidth. Modes are ranked by the approximate peak prominence
$
\mathcal{R}_\mu=\mathrm{Re}(\mathcal{W}_\mu)/\gamma_\mu,
$
where $\mathcal{W}_\mu$ is the corresponding spectral residue.

A central advantage of the FL approach is that each observable resonance can be decomposed into contributions from one or more quasienergy-resolved transition channels $(\alpha,\beta,l)$, with gaps $\Delta_{\alpha\beta l}$ and weights $R_\mu^{(\alpha\beta l)}$. The dominant channels retained in the tables (those exceeding $10\%$ of the largest component) reveal how individual spectral peaks arise either from single Floquet transitions or from hybridized superpositions of multiple sideband-assisted decay pathways. Colors indicate the dominant quasienergy-channel composition of each FL mode. The corresponding quasienergy transitions are indicated by the colored downward arrows in Fig.~\ref{fig:EnergyBasis_Optical}(c).

This modal decomposition provides direct physical insight which is not possible in conventional static approaches: peaks that appear as single resonances in the spectrum may in fact originate from several interfering Floquet decay channels with distinct sideband indices and bath couplings.
This illustrates that accurate coherent quasienergies alone are insufficient; the dissipative channel structure is equally essential.

We first examine the flat-bath spectra in panels (a) and (c) of Fig.~\ref{fig:Spectracav_Optical_wd065}, for which the F-GME and TI-GME predictions are nearly indistinguishable over the full frequency range shown. 
Their persistence under both master equations further confirms that, for flat reservoirs, the dominant effect of strong driving is coherent Floquet dressing rather than frequency-selective dissipation. Increasing the cavity loss rate from panel (a) to panel (c) primarily broadens the resonances and suppresses weaker higher-order features, consistent with enhanced dissipative smearing.

The change from incoherent pumping (gray curves) to coherent periodic driving is striking. Rather than the simpler few-peak structure of the incoherently driven case, coherent pumping generates a sequence of Mollow-triplet-like sideband structures centered around harmonics of the drive frequency, together with additional nonlinear mixing features arising from higher-order Floquet transitions.

The dominant spectral pattern in panels (a) and (c)
of Fig.~\ref{fig:Spectracav_Optical_wd065} is therefore a repeated {\it Mollow-resonance-like} feature organized around the drive-induced resonances. Superimposed on this structure are weaker FL modes originating from additional quasienergy channels, which distort the ideal triplet symmetry and generate asymmetric shoulders or satellite peaks. A complete mode-by-mode identification, including the corresponding $(\alpha,\beta,l)$ channel content, is provided in Tabs.~\ref{tab:FLmodes_cavity_kappa01Flat} and \ref{tab:FLmodes_cavity_kappa02Flat}
of App.~\ref{secS:AdditionalResults}.

A qualitatively different behavior appears in panels (b) and (d) of Fig.~\ref{fig:Spectracav_Optical_wd065}, where the flat reservoir is replaced by a structured Lorentzian--Ohmic bath. A detailed mode-by-mode identification of these spectra, including the corresponding $(\alpha,\beta,l)$ channel content, is provided in Tabs.~\ref{tab:FLmodes_cavity_kappa01LorOhmic1} and \ref{tab:FLmodes_cavity_kappa02LorOhmic1}
of App.~\ref{secS:AdditionalResults}. Here, clear discrepancies emerge between the F-GME and TI-GME spectra. Since both approaches share the same coherent system Hamiltonian, the dominant resonance frequencies remain largely aligned. The differences instead appear in the spectral weights and linewidths: peaks in the TI-GME are incorrectly broadened or misweighted because the underlying dissipative pathways are collapsed onto static dressed-state transitions rather than resolved into Floquet sideband channels to construct the FL modes.

Comparing panels (a) and (b) of Fig.~\ref{fig:Spectracav_Optical_wd065}, the effect of the structured bath is immediately clear. Resonances lying near the bath maximum experience stronger damping and are substantially suppressed, whereas sharper side features located away from the bath peak survive with narrower linewidths. The resulting spectra display more pronounced shoulders and asymmetries within the generalized Mollow structures.

A particularly important mismatch occurs at the  mode near $\omega=\omega_d$, in Fig.~\ref{fig:Spectracav_Optical_wd065}(b). 
Although the total integrated incoherent spectrum is constrained by the equal-time fluctuation 
$\overline{\langle\delta s^-\delta s^+\rangle}$ according to Eq.~\eqref{eq:S-N_connection}, this relation does not imply a one-to-one correspondence between a single spectral peak and a population mode in Fig.~\ref{fig:AveNcav_wd_Optical}(b). In particular, the peak near $\omega=\omega_d$ reflects the weight of nonstationary FL fluctuation modes whose quasifrequencies lie near the drive frequency. Population modes determine long-time averaged occupations, whereas they are removed from the incoherent spectrum through the subtraction of the coherent component. Hence, the area of the $\omega_d$ peak should be interpreted as fluctuation spectral weight, not as a direct population at $\omega_d$.

Another clear example of the breakdown of the TI-GME in panel (b), occurs for the nearby modes at $\Delta_\mu \approx 0.927\omega_c$ and $\Delta_\mu \approx 0.957\omega_c$. In the F-GME, these resonances remain clearly resolved as two distinct peaks, whereas the TI-GME misweights their amplitudes and overbroadenes their linewidths, causing them to merge into a single composite structure. Such errors become especially consequential when multiple modes lie close in frequency, since incorrect dissipative broadening can qualitatively alter the observed lineshape.

A second striking example appears near the right lobe of the generalized Mollow structure in Fig.~\ref{fig:Spectracav_Optical_wd065}(d), around two modes near $\Delta_\mu \approx 0.888\omega_c$ and $\Delta_\mu \approx 0.926\omega_c$; here the incorrect peak weights and linewidths predicted by the TI-GME can create the false impression that the resonance frequencies themselves are shifted. In reality, the underlying coherent peak locations remain nearly unchanged; the apparent displacement originates from erroneous real parts of the Liouvillian eigenvalues, i.e., incorrect decay rates, which distort the superposed nearby resonances.
Therefore, the presence of the quasi-degenerate spectral modes can further amplify the noticeable discrepancies.

This demonstrates that spectroscopy can misidentify transition frequencies when dissipative broadening is modeled incorrectly. To quantify these discrepancies systematically, the mode-resolved effective linewidths $\gamma_\mu$ and peak-prominence metric $\mathcal{R}_\mu$ for the dominant (bright) cavity modes are compared for both GMEs in the bar plots of Fig.~\ref{figS:CavityBrightChannels_Optical} in App.~\ref{secS:AdditionalResults}. These data provide a mode-by-mode diagnosis of how the structured reservoir reshapes the spectra differently within the F-GME and TI-GME frameworks.
Panels (a,c) of Fig.~\ref{figS:CavityBrightChannels_Optical} 
 correspond to the structured-bath spectra of Fig.~\ref{fig:Spectracav_Optical_wd065}(b), while panels (b,d) correspond to Fig.~\ref{fig:Spectracav_Optical_wd065}(d).

The bar plots show that disagreements in the full spectra can be traced to specific bright modes whose damping rates and residues are strongly misestimated by the TI-GME.
The disagreements therefore originate at the level of individual dissipative modes before manifesting in the full spectral envelope.

These results show that frequency-resolved observables 
constitute a substantially more stringent benchmark of dissipative theory than steady-state populations, especially in strongly driven USC systems with structured reservoirs.

The FL modal decomposition also clarifies why the relevant spectral resonances are not, in general, isolated quasienergy transitions. In a secular approximation, each channel $(\alpha,\beta,l)$ would evolve independently and the spectrum could be interpreted as a sum of separate transition lines. In the present nonsecular F-GME, however, the FL superoperator contains off-diagonal couplings between different quasienergy-resolved channels:
\[
(\alpha,\beta,l)\leftrightarrow(\alpha',\beta',l'),
\]
especially when their gaps are nearly resonant,
\[
|\Delta_{\alpha\beta l}-\Delta_{\alpha'\beta'l'}|\lesssim \kappa .
\]

These {\it nonsecular} couplings hybridize the bare Floquet transition channels into collective FL eigenmodes. The hybridization is further enhanced by coherent harmonic mixing induced by the strong drive and by quasienergy near-degeneracies within the Brillouin zone. Consequently, the physically observed resonances are the eigenmodes of the full nonsecular Floquet--Liouville operator, rather than individual $(\alpha,\beta,l)$ transitions.

This helps explain why several nearby channels can combine into one broadened spectral feature, or conversely why a single apparent peak can split into multiple resolved FL modes. Such collective channel hybridization is absent or strongly suppressed in static or secularized descriptions, which highlights why they fail in clustered spectral regions.
Frequency-resolved observables therefore constitute a substantially more accurate benchmark of dissipative theory than steady-state populations, especially in strongly driven ultrastrong-coupling systems with structured baths.

\subsection{Floquet Engineering the Quantum Rabi Model}
\begin{figure*}[ht]
\centering
\includegraphics[width=.99\linewidth]{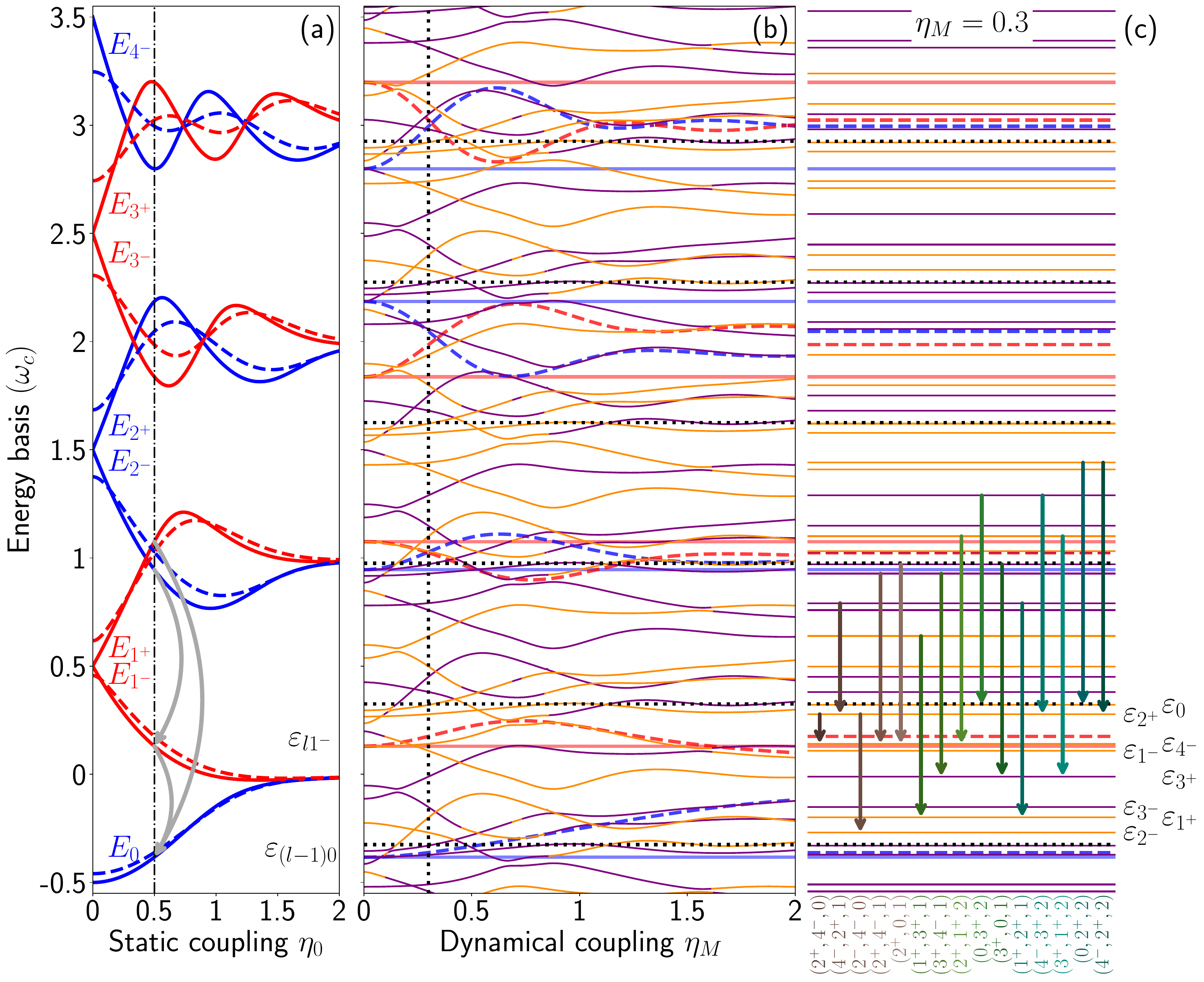}
    \caption[Schematics]{\textbf{Mechanically driven cavity-QED: Energy basis.} (a) Eigenenergies of the QRM  plotted versus the cavity-TLS static coupling rate $\eta_0$ with their labels and parities (blue: even; red: odd). When the mechanical drive (here, vibration of the TLS) is switched on with $\omega_d=0.65\omega_c$, the energy basis of the QRM, are initially renormalized [dashed curved in (a) vs. $\eta_0$ (with $\eta_M=0.3$)  and (b) vs. $\eta_M$ (with $\eta_0=0.5$) due to the finite dynamical coupling $\eta_M\neq0$, and then, transforms into the Floquet extended-space quasienergies of the DQRM (thin solid curved) with (ideally) infinite BZs distinguished by the horizontal dotted black lines versus the pump amplitude, shown in (b) for with $\eta=0.5$. Note that each quasienergy curve originates from a renormalized eigenenergy which is in turn is originated from an eigenenergy (faint thick solid blue and red lines) that inherits its label and generalized parity from (purple: even; orange: odd).
    The dominant transitions among the time-independent dressed basis in the incoherently pumped spectra are depicted by the downward gray arrows in (a).
    A magnification of the drive point at $\eta_M=0.3$ shown by the black vertical dotted line in panel (b) is plotted in panel (c) with the downward arrows showing the dominant (top 15) transitions in the extended space producing the major peaks in the emission spectra.
    Here, a resonant cavity-QED model is considered, i.e., $\omega_a=\omega_c$ for all panels.
   }
    \label{fig:EnergyBasis_Mechanical}
\end{figure*}

We next consider the mechanically driven QRM, where the periodic modulation acts parametrically through the light--matter coupling rather than through direct coherent pumping. This realizes a {\it Floquet-engineered quantum Rabi model}, where the system Hamiltonian itself is modulated in time.

Figure~\ref{fig:EnergyBasis_Mechanical}(a) shows the eigenenergies of the undriven QRM as a function of the static coupling $\eta_0$, together with their generalized parities, analogous to the optical-drive case of Fig.~\ref{fig:EnergyBasis_Optical}(a). The same state labels and avoided-crossing rules apply. In the presence of a finite mechanical modulation amplitude $\eta_M$, however, the instantaneous static Hamiltonian is already renormalized even before genuine  Floquet hybridization sets in. This is reflected by the dashed curves, which represent the renormalized QRM spectrum for nonzero $\eta_M$ ($=0.3$)~\cite{Akbari_Floquet_2025,Chang_Direct_2025}.

The corresponding quasienergies of the driven Hamiltonian $\mathcal H_{\rm FQRM}(t)$ are shown in Fig.~\ref{fig:EnergyBasis_Mechanical}(b) as a function of the dynamical coupling $\eta_M$, for fixed $\eta_0=0.5$ and $\omega_M=0.65\omega_c$. Also shown are the renormalized static eigenenergies (dashed lines) and the bare QRM eigenenergies (horizontal thick lines). Unlike the optical pumping case, where quasienergy branches largely preserve the parity inherited from the zero-drive limit, the mechanically driven system exhibits a more intricate symmetry evolution as the modulation strength increases.

The apparent discontinuities in the generalized-parity trend of individual quasienergy branches in panel (b)
of \ref{fig:EnergyBasis_Mechanical}, as the modulation strength $\eta_M$ is varied, reflect the richer symmetry structure of the parametrically driven system. Since the coupling modulation generates higher harmonics of the base drive frequency, Floquet sectors with different sideband indices can hybridize strongly, particularly near avoided crossings~\cite{Kohler_Driven_2005,Poertner_Validity_2020,Poertner_Bichromatic_2020}. This can lead to branch relabeling, exchange of dominant symmetry character, and modified transition selection rules in the extended Sambe space. Although the present model is driven by a single fundamental frequency, the resulting higher-harmonic sector mixing is conceptually related to phenomena encountered in generalized multimode Floquet theories with bichromatic (e.g., as a minimal quantum system, or strong  bichromatically driven quantum dot~\cite{Gustin_High-resolution_2021}) or incommensurate drivings, where multiple frequency sectors coexist in the extended space~\cite{Chu_Beyond_2004,Chu_Recent_1985,Poertner_Validity_2020,Poertner_Bichromatic_2020}. Such behavior lies beyond any static time-independent dressed-basis description.
Mechanical driving thus provides a more stringent test of Floquet-dissipative theory than direct coherent pumping.

A magnified view of the quasienergy spectrum at $\eta_M=0.3$ is shown in Fig.~\ref{fig:EnergyBasis_Mechanical}(c). The dominant downward transitions indicate the principal quasienergy channels that generate the emission spectra by contributing in the FL modes, discussed below in Fig.~\ref{fig:Spectracav_Mechanical_wM065}. Their color coding matches the corresponding spectral peaks, similar to those of the optical drive case.


\begin{figure*}[th]
\centering
\includegraphics[width=.98\linewidth]
{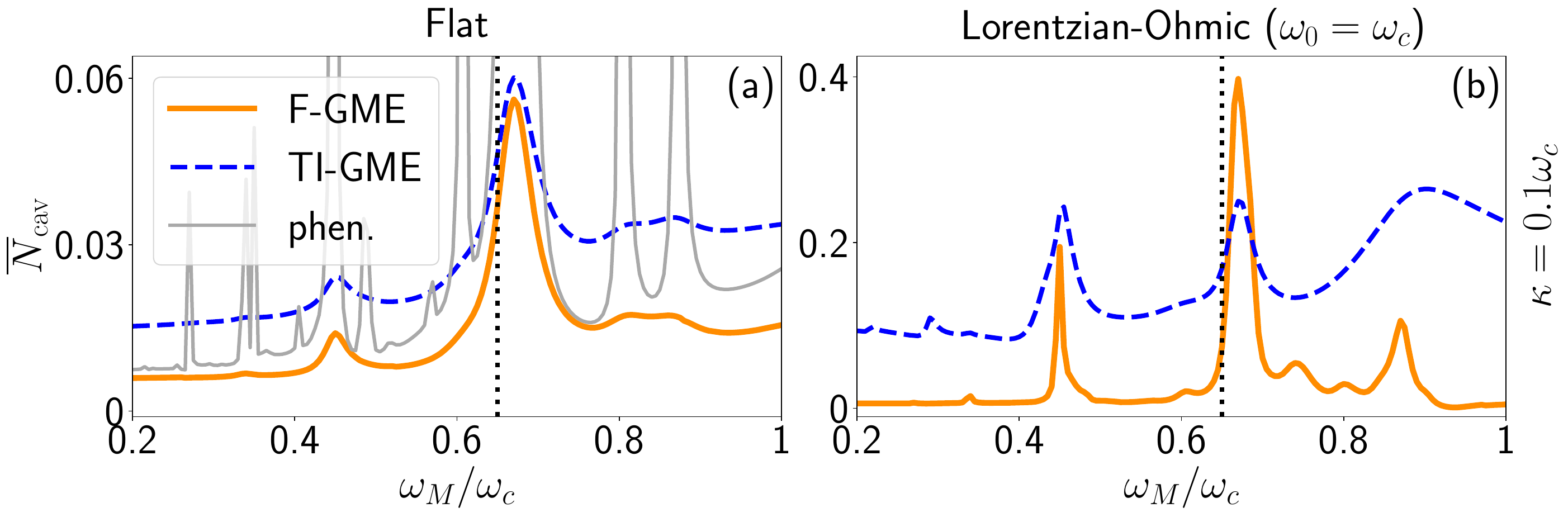} 
\caption[]{\textbf{Mechanically driven open cavity-QED in USC: Floquet-engineered QRM steady-state cavity populations.}
Temporal average of the steady-state number of cavity excitations are compared among three different approaches: (i) phenomenological Floquet theory (thin gray curves), (ii) TI-GME (blue curves) and (iii) F-GME (orange curves). Panel (a) shows the results for the flat baths [Eq.~\eqref{eq:CavityBathModel_flat}], whereas the right panels (b) for the structured Lorentzian bath centered at $\omega_0=w_c$ with an Ohmic cut-off [Eq.~\eqref{eq:CavityBathModel_LorOhmic}]. A dissipation rate of $\kappa=0.1\omega_c$ is considered. 
Other parameters are $\eta_0=0.5$, $\eta_M=0.3$, $\omega_a=\omega_c$ and $\gamma=0.01\omega_a$. 
The vertical dotted black lines show the representative drive frequencies at $\omega_d=0.65\omega_c$ for the spectra figures.
}
\label{fig:AveNcav_wd_Mechanical}
\end{figure*}

\begin{figure*}[th]
\centering
\includegraphics[width=.98\linewidth]
{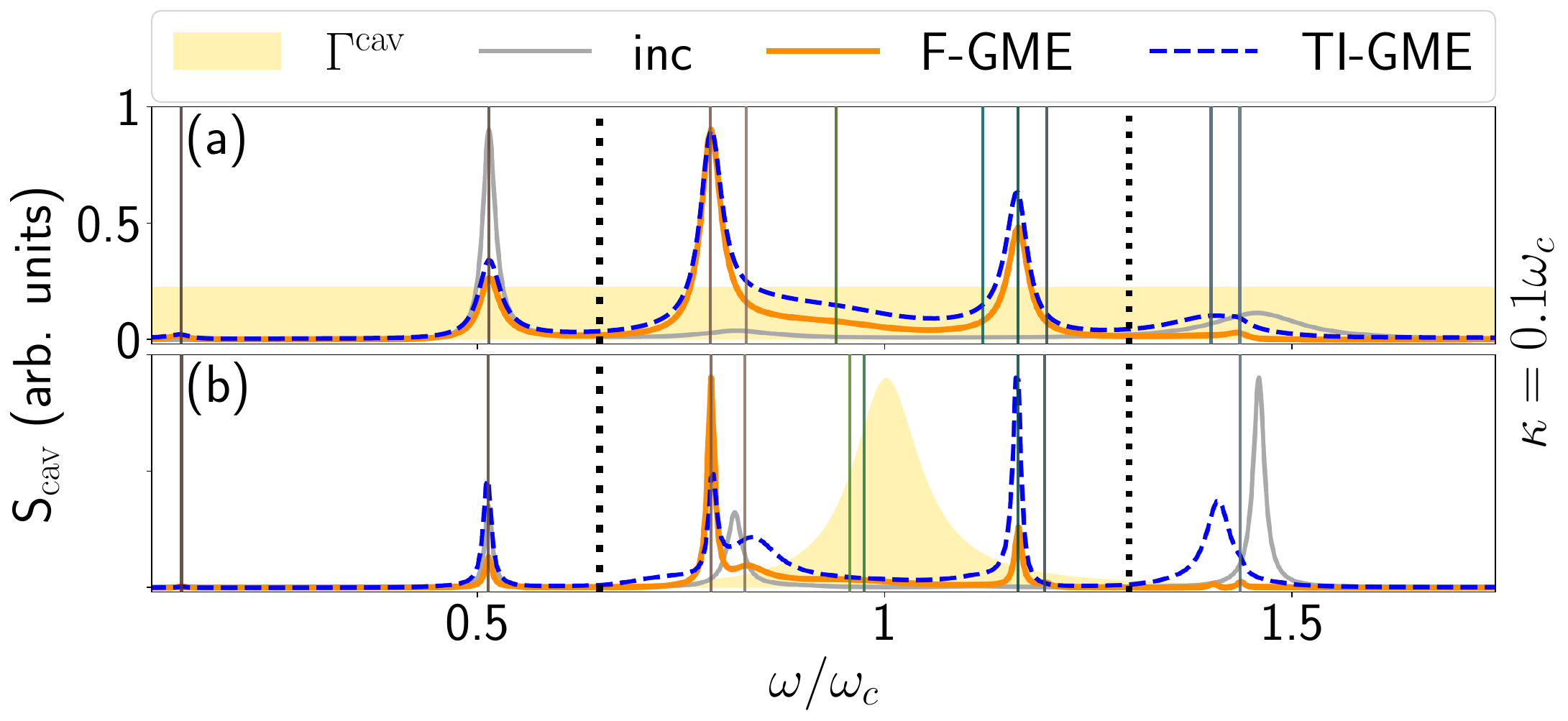} 
\caption[]{\textbf{
Mechanically driven open cavity-QED in USC (Floquet-engineering the dissipative QRM in USC): Cavity spectra.}
(Incoherent) spectra for the cavity excitations, obtained from the TI-GME (dashed blue curves) and the F-GME (thick solid orange curves) approaches. Left panel (a) shows the spectra for the flat bath, and the right panel (b) show the spectra for the Lorentzian bath with the central frequency of $\omega_0=\omega_c$ and the Ohmic cut-off, shown as the yellow shaded region. A dissipation rate of $\kappa=0.1\omega_c$ is considered. The colorful vertical solid lines  mark the dominant transitions corresponding to the same color arrows in panel (c) of Fig.~\ref{fig:EnergyBasis_Mechanical}, whereas the black vertical dotted lines represent the integer multiples of the drive frequency. 
For the reference, we also plot the result of the incoherently driven system with a small perturbation of $P_{\rm inc}=0.01\eta_M\omega_c$ applied through the TLS operator, for the cavity (black thin solid line) and the TLS (gray thin dashed lines).
The results are obtained for the minimum eight truncation of the QRM dressed states to have the convergence for $\omega_a=\omega_c$, $\eta_0=0.5$, $\omega_M=0.65\omega_c$, $\eta_M=0.3$ and $\gamma=0.01\omega_a$.
}
\label{fig:Spectracav_Mechanical_wM065}
\end{figure*}

Figure~\ref{fig:AveNcav_wd_Mechanical} shows the long-time averaged cavity excitation number $\overline{N}_{\rm cav}$ as a function of the mechanical modulation frequency $\omega_M$ for a FQRM with static coupling $\eta_0=0.5$ and modulation amplitude $\eta_M=0.3$. Results from the F-GME (orange), TI-GME (dashed blue), and phenomenological Floquet treatment (gray) are compared for both flat baths [panel (a)] and structured Lorentzian--Ohmic baths peaked at $\omega_0=\omega_c$ [panel (b)], with $\kappa=0.1\omega_c$. For an analysis of the TLS counterpart, see Fig.~\ref{figS:AveNTLS_wd_Mechanical} in 
App.~\ref{secS:AdditionalResults}.

In contrast to the optically driven case, mechanical modulation does not inject excitations through a direct coherent pump operator. Instead, excitations arise indirectly through parametric Floquet hybridization of the dressed eigenstates, followed by dissipative population redistribution. As a consequence, the steady-state occupations are intrinsically controlled by quasienergy-resolved relaxation pathways. It is important to note that these {\it real cavity photon excitations} (which will be released) are initiated by the 
mechanical drive, with no optical input, and this situation is unique to the USC regime---virtual photons are spectra are released through adiabatic perturbation of the hybrid photon-matter states. 

This concept of {\it photon creation from modified vacuum interactions} has a similar origin to the dynamical Casimir effect~\cite{Johansson_Dynamical_2009,Macri_Nonperturbative_2018,wilson_observation_2011,galego_cavity_2019}, where the rapid motion of a cavity mirror (ordinarily at the relativistic speeds) shakes virtual photons out of the vacuum fluctuations, converting them into real radiation. The emitted photons gain the required energy directly from the mechanical work done to accelerate the mirror against the radiation pressure of the vacuum.

 The dynamical Casimir effect is closely related to Hawking radiation (radiation from black holes) and the Unruh effect (an accelerated observer detecting particles in a vacuum)~\cite{Nation_Stimulating_2012}. This effect was first predicted in 1970~\cite{Moore_Quantum_1970} and was experimentally confirmed~\cite{wilson_observation_2011} in 2011 using a superconducting circuit (specifically, a SQUID-based artificial mirror) rather than a physical moving mirror~\cite{Johansson_Dynamical_2009}, as it allows for simulating relativistic movement~\cite{wilson_observation_2011}.
The photons in the dynamical Casimir mechanism are generally produced in entangled pairs to conserve momentum and energy, with frequencies that sum to the modulation frequency of the mirror. Yet, here, in the Floquet engineering mechanism, we can potentially observe peaks in various multiple drive frequencies (cf. Figs.~\ref{fig:AveNcav_wd_Mechanical} and \ref{fig:Spectracav_Mechanical_wM065}).

This distinction is immediately visible even for the flat bath in panel (a) of Fig.~\ref{fig:AveNcav_wd_Mechanical}. Unlike the optical case of Fig.~\ref{fig:AveNcav_wd_Optical}(a), the F-GME and TI-GME predictions already differ substantially despite the frequency-independent reservoir. There is therefore no analogue of the cancellation mechanism previously observed under coherent pumping; because the excitation itself is generated by Floquet manifold mixing, collapsing the dissipator onto static dressed-state transitions fundamentally changes the balance between pumping and relaxation.

The spectrally sharp structures as a function of $\omega_M$ correspond to parametric resonances, where the modulation frequency matches quasienergy splittings or Floquet-replica gaps. Near these resonances, the drive strongly hybridizes selected manifolds and dissipation determines whether population accumulates or is rapidly depleted. The F-GME captures both the locations and relative strengths of these resonances, whereas the TI-GME typically misestimates their amplitudes and relative prominence.

For the structured bath in panel (b) of Fig.~\ref{fig:AveNcav_wd_Mechanical}, the discrepancies become even more pronounced. Here the reservoir selectively enhances or suppresses decay depending on how the Floquet gaps, $\Delta_{\alpha\beta l}$, align with the bath peak near $\omega_0=\omega_c$. The F-GME naturally incorporates this quasienergy selectivity, while the TI-GME continues to sample the bath only at static dressed-state energy differences. Consequently, the TI-GME can significantly misestimate the resonance strengths and steady-state excitation numbers.

Figure~\ref{fig:AveNcav_wd_Mechanical} therefore highlights a central distinction between coherent Floquet pumping and Floquet engineering in ultrastrong cavity-QED: under parametric driving, resolving the full Floquet structure of dissipation is already essential at the level of time-averaged populations, even for flat reservoirs.

This shows that the apparent success of static master equations in some driven systems can be contingent on the drive mechanism rather than on the validity of the theory itself.

Similar to the optical drive case at the marked drive frequency, $\omega_d=0.65\omega_c$, the \emph{mechanical} drive
case is dominated by the excitation pathways (cf.~Fig.~\ref{figS:Fig_wd_Mechanical})
$P^{(+2)}_{1^+\leftarrow 0}$, then $P^{(+3)}_{2^+\leftarrow 0}$ and $P^{(+5)}_{4^-\leftarrow 0}$, depending on which of them survive the dissipation.  
To appreciate the dissipative processes in more detail, let us consider the most probable excitation transition that is $j=1^+\leftarrow0$ with $\Delta l=+2$ externally assisting quanta. From Fig.~\ref{fig:EnergyBasis_Mechanical}(b), the transition between these QRM states adiabatically correspond to $\ket{\alpha=2^-}\leftarrow\ket{\alpha=2^+}$, yielding $\Delta>\omega_{01^+}$ and $\Gamma_{\rm Lor-Ohm}^{\rm cav}(\Delta)>\Gamma_{\rm Lor-Ohm}^{\rm cav}(\omega_{01^+})$ since they both lands after the peak frequency of the bath. Thus, the discrepancy between the TI-GME and the F-GME is amplified changing the bath from flat to Lor-Ohmic.

The FL zero-mode analysis shows
how the above-mentioned excitation pathways is converted into the final cycle-averaged cavity
population.  As seen before, in the steady-state cycle averaging, 
only the zero FL mode with $\lambda_{\mu_0}\simeq0$ contributes
to the plotted scalar quantity $\overline{N}_{\rm cav}$.

For the flat bath, both dissipative descriptions give almost the same zero-mode
composition.  The F-GME zero mode has
$\mathcal{W}_{\mu_0}^{N_{\rm cav}}\simeq0.037$ and is dominated by the population channel
$(2^+,2^+,0)$ with $R_{\mu_0}^{(2^+,2^+,0)}\simeq0.957$, while the TI-GME gives
$\mathcal{W}_{\mu_0}^{N_{\rm cav}}\simeq0.075$, and an almost pure population component
$(2^+,2^+,0)$ with $R_{\mu_0}^{(2^+,2^+,0)}\simeq0.992$.  In contrast, the TI-GME gives a
much larger value,
$\mathcal{W}_{\mu_0}^{N_{\rm cav}}\simeq0.170$, and its zero mode is distributed over
several population and coherence components, including $(3^-,3^-,0)$ with $R_{\mu_0}^{(3^-,3^-,0)}\simeq0.196$, $(4^-,4^-,0)$ with $R_{\mu_0}^{(4^-,4^-,0)}\simeq0.193$,
$(2^+,2^+,0)$ with $R_{\mu_0}^{(2^+,2^+,0)}\simeq0.118$, $(2^+,4^-,0)$ with $R_{\mu_0}^{(2^+,4^-,0)}\simeq0.107$, and $(4^-,2^+,0)$ with $R_{\mu_0}^{(4^-,2^+,0)}\simeq0.107$.  This shows that the structured bath
amplifies the difference between quasienergy-resolved and time-independent
dissipation: the F-GME damps and redistributes the mechanically pumped
population into a much cleaner zero-mode population channel, whereas the
TI-GME retains a more mixed steady-state structure and therefore predicts a
larger cycle-averaged cavity population.

Similar to  the optical drive case, the excitation probability identifies the dominant injection pathway, while the
zero-mode FL decomposition identifies the stationary population
and coherence structure that remains after quasienergy-resolved dissipation.

Figure~\ref{fig:Spectracav_Mechanical_wM065} shows the 
cavity emission spectra of the \emph{mechanically} FQRM at the representative modulation frequency $\omega_M=0.65\omega_c$, for static coupling $\eta_0=0.5$ and modulation amplitude $\eta_M=0.3$. Results from the F-GME (solid orange) and TI-GME (dashed blue) are compared with weak incoherent-drive reference spectra (gray) for a flat bath [panel (a)] and a structured Lorentzian--Ohmic bath peaked at $\omega_0=\omega_c$ [panel (b)], with $\kappa=0.1\omega_c$. Colored vertical markers indicate the dominant FL modes contributed by the quasienergy transitions identified in Fig.~\ref{fig:EnergyBasis_Mechanical}(c) and listed in Tabs.~\ref{tab:FLmodes_cavity_kappa01flat_Mechanical} and \ref{tab:FLmodes_cavity_kappa01LorOhmic1_Mechanical}.
For the TLS counterpart see Fig.~\ref{figS:SpectraTLS_Mechanical_wM065} of Appendix.~\ref{secS:AdditionalResults}.

We first consider the flat-bath spectra in panel (a)
of Fig.~\ref{fig:Spectracav_Mechanical_wM065}. In sharp contrast to the optically driven case, the F-GME and TI-GME already differ substantially despite the frequency-independent reservoir. This reflects the fundamentally different role of the drive: mechanical modulation does not inject excitations through a direct pump operator, but instead periodically reshapes the Hamiltonian and generates emission through Floquet hybridization of dressed states. The resulting spectra depend not only on transition frequencies, but also on the redistribution of population among Floquet replicas and on micromotion-sensitive correlation dynamics.

The F-GME captures these effects through the explicitly time-dependent Liouvillian and quasienergy-resolved jump processes. In contrast, the TI-GME evaluates correlations using a static dressed-state Liouvillian and therefore projects emission onto stationary transition channels. Even for spectrally flat baths, where decay rates are frequency independent, this misses the operator micromotion and harmonic interference that determine the correct spectral weights. Consequently, the TI-GME can reproduce some peak locations while failing to predict their intensities.

The comparison with the weak incoherent-drive reference spectra further clarifies this distinction. Those gray curves resemble static dressed-state emission and therefore track the TI-GME more closely. The clear departure of the F-GME spectra from this reference demonstrates that mechanical Floquet engineering does not simply mimic weakly driven dressed-state decay, but instead creates genuinely new radiative channels associated with Floquet-hybridized manifolds.

For the spectrally structured bath in panel (b) of Fig.~\ref{fig:Spectracav_Mechanical_wM065}, the discrepancies become even larger. Here, the reservoir selectively enhances or suppresses decay depending on how the Floquet gaps,
$
\Delta_{\alpha\beta l}=\varepsilon_\beta-\varepsilon_\alpha+l\omega_M
$,
align with the bath maximum near $\omega_0=\omega_c$. The F-GME naturally incorporates this quasienergy selectivity, whereas the TI-GME continues to evaluate dissipation only at static dressed-state energy differences. As a result, the TI-GME strongly misestimates the relative spectral weights and line shapes, even when the dominant peak positions remain broadly similar.

The origin of these discrepancies is further quantified in Fig.~\ref{figS:CavityBrightChannels_Mechanical} of App.~\ref{secS:AdditionalResults}, 
whose bar plots compare the mode-resolved linewidths and spectral weights extracted from the Floquet--Liouville decomposition. They show that the disagreement between the F-GME and TI-GME arises at the level of individual bright resonances before propagating into the full spectral envelope.

Together with Fig.~\ref{fig:AveNcav_wd_Mechanical}, these results show that under Floquet engineering the generation of excitations and their radiative decay are inseparable aspects of the same quasienergy dynamics. Time-independent dressed-basis master equations therefore generically fail in mechanically driven ultrastrong cavity QED, whereas the Floquet generalized master equation provides a consistent description of both steady-state populations and emission spectra.

\begin{table*}[!htbp]
  \centering
  \caption{Decomposition of the TI-GME \emph{vs.} F-GME dissipator difference into
    three independent mechanisms. Summary of the distinction in the prediction between the GMEs.
    }
  \label{tab:three_factors}
  \vspace{1.5mm}
  \renewcommand{\arraystretch}{1.5}
\setlength{\tabcolsep}{6.8pt}
  \begin{tabular*}{\textwidth}{@{\extracolsep{\fill}} lllll}
    \toprule
    \# & \textbf{Difference} & \textbf{Activated by} &
    \textbf{Cancels when} & \textbf{Diagnostic}
    \\
    \midrule
    (1) &
    \makecell[l]{$S_{{\rm B},\alpha\beta l}\;\text{vs.}\;S_{{\rm B},jk}$\\ (operator content)} &
    \makecell[l]{higher-order multiphoton;\\ ac-Stark / Bloch-Siegert} &
    \makecell[l]{weak drive; flat bath\\ (populations, via Parseval)} &
    \makecell[l]{closed-system $P^{(n)}_{f\leftarrow i}$;\\ lossless limit}
      \\[3mm]
    (2) &
    \makecell[l]{$\Gamma(\Delta_{\alpha\beta l})\;\text{vs.}\;\Gamma(\omega_{jk})$\\ (bath sampling)} &
    \makecell[l]{structured bath,\\ $\Delta_{\alpha\beta l}\neq\omega_{jk}$} &
    flat bath &
    $|\partial_\omega\Gamma|\,
       |\Delta_{\alpha\beta l}-\omega_{jk}|$
    \\[3mm]
    (3) &
    \makecell[l]{$(\alpha\beta l)\;\text{vs.}\;(j,k)$\\ (nonsecularity)} &
    \makecell[l]{near-degenerate /\\ overlapping channels} &
    well-separated channels &
    $|\Delta_{\alpha\beta l}-\Delta_{\alpha'\beta'l'}|<\Gamma$ 
     \\
    \bottomrule
  \end{tabular*}
\end{table*}

We note that, in the absence of coherent driving ($\eta_d=0$), regardless of the drive type, both the TI-GME and F-GME reduce to the same static generalized master equation and therefore yield identical results, even in the presence of incoherent pumping. For finite coherent driving, the F-GME continuously approaches this static limit as $\eta_d\to0^+$. In practice, however, direct numerical verification of this convergence becomes increasingly challenging because the coherent drive induces only weak populations, causing the associated spectral signals to become vanishingly small and difficult to resolve numerically, as implied by Eq.~\eqref{eq:S-N_connection}. For this reason, when the primary objective is to probe the spectral properties of the system rather than coherent Floquet effects, an incoherent pumping scheme may provide a more practical alternative in the extremely weak-driving regime. Since the TI-GME and F-GME coincide exactly in the absence of coherent driving, such calculations may be carried out using the considerably simpler static generalized master equation.

Although the TI-GME may reproduce the F-GME under particular parameter choices, including certain flat-bath configurations, such agreement is neither universal nor predictable a priori. The apparent success of the TI-GME is generally nonmonotonic in the drive parameters and depends sensitively on the interplay between Floquet dissipative channels and static dressed-state transitions. By contrast, the F-GME remains valid throughout the driven regime and continuously reduces to the static generalized master equation in the vanishing-drive limit. Consequently, the F-GME provides a more robust and generally reliable framework for describing dissipation in periodically driven strongly interacting quantum systems. Furthermore, when combined with FL modal analysis, it offers a natural and physically transparent interpretation of the resulting dynamics and spectra.

\subsection{Origin of the differences between the TI-GME and F-GME 
results}
\label{sec:three_factors}

As stated before, the TI-GME and the F-GME share
an \emph{identical} coherent Hamiltonian and an \emph{identical} bath model; they
differ only in their dissipators. It is therefore instructive to isolate,
precisely, what makes the two dissipators differ. Comparing the static
dressed-basis dissipator [Eq.~\eqref{eq:Ldiss_TIGME}] with the
quasienergy-resolved one [Eq.~\eqref{eq:Ldiss_FGME}], the difference can be
decomposed into three \emph{independent} mechanisms, as discussed below.

\paragraph*{($i$) Operator content: $S_{{\rm B}, jk}\to S_{{\rm B}, \alpha\beta l}$.}
The system-bath transition operators differ because the Floquet matrix
elements in Eq.~\eqref{eq:FelementS_ex}
redistribute spectral weight across the sideband index $l$, whereas the
static element $\langle j|S_B|k\rangle$ carries no such structure. This
mechanism is purely \emph{coherent}: it is encoded in the closed-system
Floquet states and is already visible in the unitary transition
probabilities $P^{(n)}_{f\leftarrow i}$ and in the lossless limit. It is
activated by higher-order multiphoton processes and by large
ac-Stark/Bloch--Siegert renormalizations. Since the sideband weights obey
the Parseval relation,
\begin{equation}
  \sum_{l}\bigl|S_{{\rm B},\alpha\beta l}\bigr|^2
  = \bigl|\langle\alpha|S_{\rm B}|\beta\rangle\bigr|^2 ,
  \label{eq:parseval}
\end{equation}
then the \emph{total} transition weight is conserved. Consequently factor~(1)
cancels for flat baths in cycle-averaged \emph{populations}, but not in
\emph{frequency-resolved} observables, where the distribution over $l$ is
what matters.

\paragraph*{($ii$) Bath frequency sampling: $\Gamma(\omega_{jk})\to\Gamma(\Delta_{\alpha\beta l})$.}
The dissipative rate is evaluated at the quasienergy channel frequency
$\Delta_{\alpha\beta l}=\varepsilon_\beta-\varepsilon_\alpha+l\omega_d$
rather than at the static splitting $\omega_{jk}=E_k-E_j$. For a flat bath,
$\Gamma(\Delta)=\Gamma(\omega)$ and this factor vanishes identically; it is
activated solely by reservoir structure, scaling as
$\sim|\partial_\omega\Gamma|\,|\Delta_{\alpha\beta l}-\omega_{jk}|$. This is
the dominant mechanism for the Lorentzian--Ohmic results.

\paragraph*{($iii$) Nonsecularity: $(j,k)\to(\alpha\beta l)$ coherences.}
The nonsecular (off-diagonal Kossakowski) couplings act among the
\emph{dense} quasienergy channels $(\alpha\beta l)$ in the F-GME, rather than
among the \emph{sparse} static channels $(j,k)$ in the TI-GME. These become
significant when channels are near-degenerate or spectrally overlapping,
\begin{equation}
  \bigl|\Delta_{\alpha\beta l}-\Delta_{\alpha'\beta'l'}\bigr|\lesssim\Gamma ,
  \label{eq:neardegen}
\end{equation}
in which case, distinct channels hybridize into collective
FL modes.
This is the only factor connected to complete positivity; the same near-degenerate nonsecular couplings that hybridize the channels can drive a loss of complete positivity of the Floquet-Redfield generator, a known feature of fully nonsecular Redfield treatments~\cite{Breuer_Theory_2002,Farina_Open-quantum-system_2019}; where relevant, it can be circumvented by alternative complete-positivity-preserving constructions~\cite{Nathan_Universal_2020,Mozgunov_Completely_2020,Trushechkin_Unified_2021}.

These three mechanisms are summarized in Tab.~\ref{tab:three_factors}.
Crucially, they explain \emph{all} observed behavior self-consistently. For a
flat bath under optical driving, factor~($ii$) is absent and factor~($i$) cancels
in populations [Eq.~\eqref{eq:parseval}]; the two GMEs then agree at
$\omega_d=0.65\,\omega_c$, where the quasienergy channels are well separated
and factor~($iii$) is negligible,
yet diverge at $\omega_d=\omega_c$,
where dense BZ near-degeneracies activate factor~($iii$). 
For structured (Lorentzian--Ohmic) reservoirs,
factor~($ii$) is active at all frequencies, so the GMEs differ even when the nonsecularity among the Floquet dissipation pathways is not quite effective.
For Floquet engineering (parametric
modulation), factor~($i$) does not cancel---excitations are generated by the
operator redistribution itself rather than by additive pumping---so the GMEs
differ already for flat baths. 



\section{Conclusions and Summary Remarks}
\label{sec:Conclusions}

We have developed and applied a powerful Floquet generalized master-equation framework for driven open cavity quantum electrodynamics  in the ultrastrong-coupling regime, using the quantum Rabi model as a paradigmatic platform. Working in the dressed states basis and without secularization, the  theory self-consistently treats:
\begin{itemize}
    \item periodic driving,
    \item nonperturbative light--matter hybridization, and
    \item dissipative environments,
\end{itemize}   
on an equal footing. 
We stress that this is 
a significant advance over previous models, and is the only model we are aware of that can accurately treat a strongly driven 
open quantum system in the USC regime.
By systematically comparing the resulting Floquet generalized master equation (F-GME) with conventional time-independent dressed-basis generalized master equation (TI-GME), we established when static descriptions remain accurate and when they fundamentally break down.
Importantly, our results are also gauge invariant, which requires extra care in the construction of the Hamiltonians in a restricted Hilbert space.

Using this theory, 
one of our key findings is that dissipation in driven USC cavity-QED is generically organized by the Floquet quasienergy structure. Relaxation and emission processes occur through the full set of sideband-assisted quasienergy channels,
\begin{equation*}
\Delta_{\alpha\beta l}=\varepsilon_\beta-\varepsilon_\alpha+l\omega_d,
\end{equation*}
rather than solely through undriven dressed-state transitions. Whether this Floquet channel structure can be neglected, however, depends crucially on both the bath spectral profile and the manner in which the periodic drive enters the Hamiltonian.

Although the TI-GME and the F-GME have  the coherent Hamiltonian and bath model, they
differ in how their dissipators are constructed. We discussed three independent mechanisms that help to explain the differences, including:
($i$) Operator content:
    The system--bath transition operators differ since the Floquet matrix
elements in Eq.~\eqref{eq:FelementS_ex}
redistribute spectral weight across the sideband index $l$, whereas the
static matrix element 
$\langle j|S_{\rm B}|k\rangle$ 
carries no such structure;
($ii$)
Bath spectral sampling:
The dissipative rate is evaluated at the quasienergy channel frequency
$\Delta_{\alpha\beta l}=\varepsilon_\beta-\varepsilon_\alpha+l\omega_d$
rather than at the static splitting $\omega_{jk}=E_k-E_j$;
($iii$) Nonsecularity:
The nonsecular 
couplings act among the
dense quasienergy channels $(\alpha\beta l)$ in the F-GME, rather than
among the \emph{sparse} static channels $(j,k)$ in the TI-GME. They become
important when channels are near-degenerate or spectrally overlapping.
These three mechanisms explain 
all of the main features shown in our results.

For coherent optically pumped systems with large driving fields, we found that TI-GME can reproduce both steady-state populations and incoherent spectra when the reservoirs are flat and frequency independent. In this restricted regime, the redistribution of transition weight among Floquet sidebands does not qualitatively modify the final observables, and static dressed-basis approaches can provide an effective description. 
That is, for a
flat bath under optical driving, factor~($ii$) is absent and factor~($i$) cancels
in populations (through Parseval relation); the two GMEs then agree at
$\omega_d=0.65\,\omega_c$, where the quasienergy channels are well separated
and factor~($iii$) is negligible.
This agreement is not generic, and the apparent success of the TI-GME in doubly dressed systems is generally nonmonotonic in the drive parameters and occurs only when the dominant Floquet dissipative channels effectively map onto static dressed-state transitions.
For example, the two GMEs give predictions 
that diverge at $\omega_d=\omega_c$,
where dense BZ near-degeneracies activate factor~($iii$).

When substantial dissipation is redistributed into genuinely Floquet channels, the F-GME becomes essential, although accidental agreement can arise when $\lvert \Delta_{\alpha\beta l}-\omega_{jk}\rvert<\Gamma$, allowing interference among Floquet pathways to mimic the static dissipative dynamics.
Once structured reservoirs are introduced, such as Lorentzian--Ohmic baths, the quasienergy selectivity of dissipation becomes explicit and the TI-GME fails to predict the correct populations, linewidths, and spectral weights for both cavity and TLS observables. 
Strictly speaking, for structured (Lorentzian--Ohmic) reservoirs,
factor~($ii$) is active at all frequencies, so the GMEs differ even where the
generator is not affected by the nonsecularity among the Floquet dissipative pathways.

A qualitatively different situation arises for mechanical driving or Floquet engineering through parametric modulation of (closed-system) Hamiltonian parameters. In this case, the drive does not primarily inject excitations through a direct pump operator, but instead reshapes the spectrum and generates excitations through Floquet hybridization of dressed states. We found that under these conditions the TI-GME fails already for flat baths, for both populations and spectra of the cavity and TLS sectors, with discrepancies becoming even stronger for structured environments. Thus, for Floquet-engineered architectures, resolving dissipation in the quasienergy basis is essential even when the bath itself carries no frequency structure.

This qualitatively different behavior between the optical and mechanical driving cases is closely connected to the influence of \emph{double dressing}~\cite{Akbari_Quasienergy-Resolved_2026,Akbari_Floquet_2025}. 
The external periodic drive modifies the dissipative dynamics not only through the replacement of the bath spectral dependence from $\Gamma(\omega)$ to the quasienergy-resolved form $\Gamma(\Delta_{\alpha\beta l})$, but also through the transformation of the system-bath coupling operators from the static dressed operators $S_{\rm B}^{\pm}(\omega)$ to the Floquet-dressed operators $S_{\rm B}^{\pm}(\Delta_{\alpha\beta l})$~\cite{Akbari_Quasienergy-Resolved_2026}. 
Importantly, both GMEs developed in this work are already formulated in the dressed basis of the static QRM Hamiltonian, so that the internal light-matter dressing is incorporated in both approaches, and in a gauge invariant way.
In other words, for Floquet engineering (parametric
modulation), factor~($i$) does not cancel---excitations are generated by the
operator redistribution itself rather than by additive pumping---so the GMEs
differ already for flat baths. 

For the optically driven case we considered, the Floquet quasienergy states mainly remains {\it perturbatively} connected to the static dressed QRM states, provided that sufficiently large cavity and dressed-state truncations are used. In this regime, the static dressed basis already captures most of the relevant transition structure, consistent with the argument of Ref.~\onlinecite{Moiseyev_Conditions_2024}. Consequently, for flat baths, the additional Floquet redistribution of the operator matrix elements among the sidebands remains comparatively weak, and the primary effect of the external drive enters through the quasienergy-resolved bath sampling. As a result, the TI-GME and F-GME remain quantitatively close for flat baths, while substantial differences emerge for structured environments where the bath-frequency dependence becomes important.

For the mechanically driven case, however, the external periodic modulation induces significantly stronger Floquet hybridization and sideband redistribution of the transition operators. In this regime, the external dressing cannot be effectively absorbed into the static dressed basis alone, and the dissipative transitions become genuinely quasienergy resolved both through the bath functions and through the Floquet-dressed operator structure. Consequently, the double dressing strongly reorganizes the dissipative channels, leading to substantial deviations between the TI-GME and F-GME even for flat baths~\cite{Akbari_Quasienergy-Resolved_2026}.

A second key finding is that observable resonances are naturally dictated by the Floquet--Liouville eigenspectrum. The corresponding dissipative modes provide a transparent decomposition of spectra into peak frequencies, effective linewidths, and residues, while also revealing their composition in terms of hybridized quasienergy transition channels. This shows that experimentally observed peaks need not necessarily  correspond to isolated transitions, but can emerge from several interfering Floquet sideband processes mixed through the nonsecular Liouvillian dynamics.

Taken together, our results establish a clear hierarchy for the validity of time-independent dissipative theories in driven USC cavity-QED. They can remain accurate for optical pumping in flat reservoirs, but fail generically whenever structured baths or Floquet-engineered driving protocols make the quasienergy channel structure operationally relevant. Floquet-resolved dissipation is therefore not merely a technical refinement, but a necessary ingredient for predictive spectroscopy, nonequilibrium state control, and quantitative modeling of strongly driven light--matter systems.

Broadly, our framework 
is a significant advance for modeling strongly driven open cavity-QED systems, and opens a route toward reservoir engineering, dissipative Floquet state preparation, and the controlled design of driven superconducting circuits, polaritonic devices, and hybrid quantum platforms operating deep in the nonperturbative light--matter regime. More broadly, our results show that not all periodic drivings are equivalent from the perspective of dissipation---in USC cavity-QED, the correct basis for relaxation is determined by the Floquet quasienergy manifold itself.

\acknowledgements
 This work was supported by the Natural Sciences and Engineering Research Council of Canada (NSERC) [through a Discovery Grant and a Quantum Alliance grant],
 the Canadian Foundation for Innovation (CFI), and Queen's University, Canada.
F.N. is supported in part by the Japan Science and Technology Agency (JST)
[via the CREST Quantum Frontiers program Grant No. JPMJCR24I2,
the Quantum Leap Flagship Program (Q-LEAP), the Moonshot R{\&}D Grant Number JPMJMS256E,
and the ASPIRE program (Grant Number JPMJAP2513)].
and the Office of Naval Research (ONR) Global (via Grant No. N62909-23-1-2074).
We thank Alberto Mercurio and Daniele Lamberto for their comments on the manuscript.

\newpage

\appendix

\section{Driven quantum Rabi model and gauge consistency check}
\label{secS:DQRM_GaugeIndependence}

To solve the time-dependent Schr\"odinger equation of Eq.~\eqref{eq:TDSE}
for strongly driven USC cavity-QED systems, it is essential to employ a gauge-consistent form of the DQRM Hamiltonian~\cite{DiStefano_Resolution_2019,Salmon_Gauge-independent_2022,Akbari_Generalized_2023,Akbari_Floquet_2025}. This issue becomes particularly important once periodic driving strongly dresses the eigenstates of the static QRM.

For coherent optical driving, the Hamiltonian is written as Eq.~\eqref{eq:H_CPQR}
where $\mathcal{H}_{\rm QRM}$ denotes the static QRM Hamiltonian in a chosen gauge, and the general periodic drive term is written as
\begin{equation}
\mathcal{H}_d(t)=\sum_\Lambda \Omega_d^\Lambda \cos(\omega_d t+\phi_d^\Lambda)\,S_d^\Lambda,
\end{equation}
which satisfies $\mathcal{H}_d(t+T)=\mathcal{H}_d(t)$;  $\Lambda=\{{\rm cav},{\rm TLS}\}$ labels the subsystem being driven, $\Omega_d^\Lambda=\eta_d\omega_c$ is the drive amplitude, $\omega_d$ is the drive frequency, and $\phi_d^\Lambda$ is the drive phase.
The drive operator $S_d^\Lambda$ is Hermitian and has the form
\begin{equation}
S_d^\Lambda=s_d^{\Lambda +}+s_d^{\Lambda -},
\qquad
s_d^{\Lambda -}=(s_d^{\Lambda +})^\dagger .
\end{equation}
In the calculations of the main text, we focus on
an effective TLS driving.

Gauge consistency (invariance) is crucial because, although physical observables must be  gauge invariant, truncated light--matter models can exhibit artificial gauge dependence if the Hamiltonian and system operators are not transformed consistently~\cite{Salmon_Gauge-independent_2022}. In particular, cavity-field observables couple through the electric-field bath operator $\Pi$, whose form may depend on gauge. 
For example, in the corrected dipole gauge one must transform the cavity operator (for use in input-output and detection) according to
\begin{equation}
a \rightarrow a' = a + i\eta \sigma_x ,
\end{equation}
which modifies the corresponding field operator and derived observables. In the system Hamiltonian, however, one can indeed 
use $a$, since the interaction Hamiltonian uses the displacement field (in the dipole gauge), which is not simply a linear 
combination of bosonic operators.

\begin{figure}[!htbp]
    \centering
    \includegraphics[width=.98\linewidth]{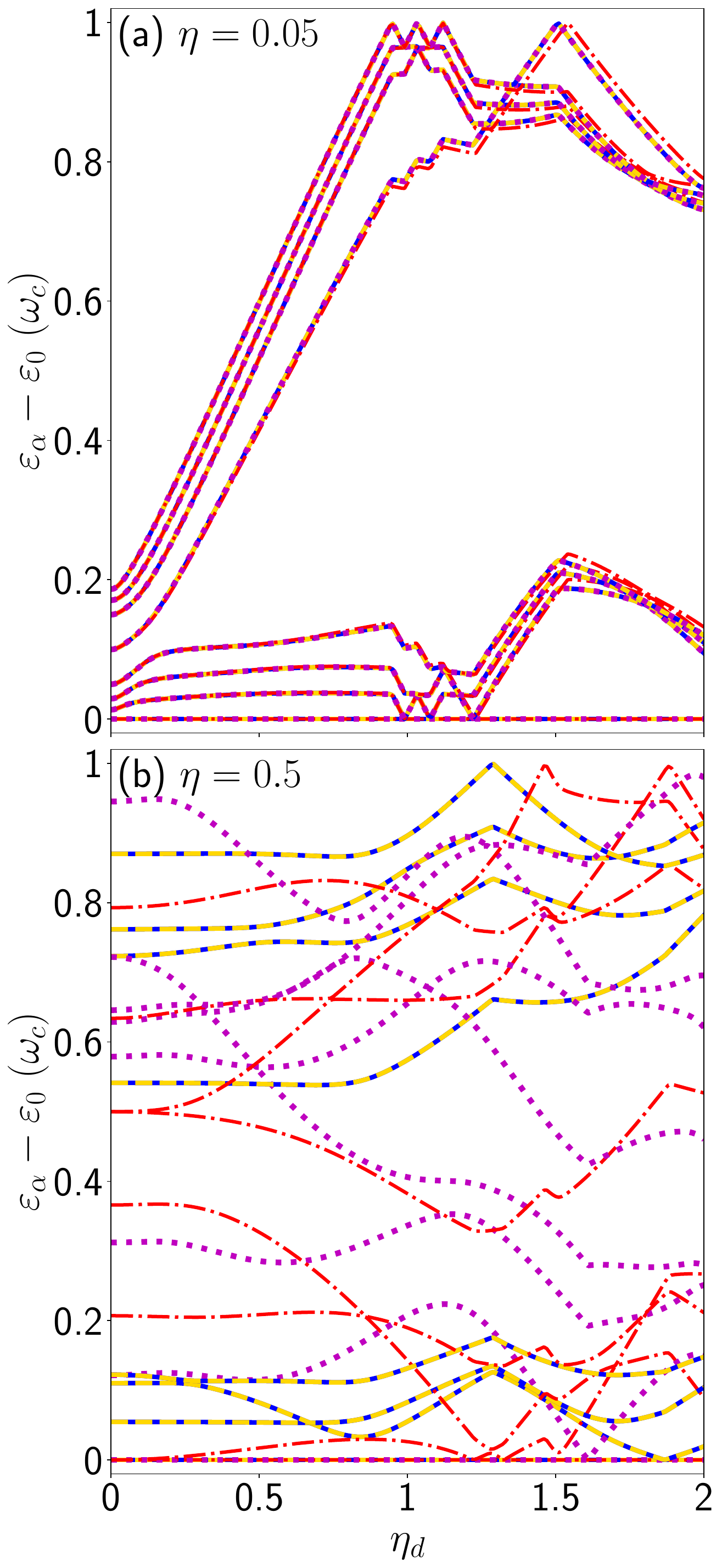}
    \caption[]{\textbf{Optically driven cavity-QED: gauge-consistency check.} We show the Floquet quasienergy curves of the DQRM as a function of the optical coherent pump drive, using two different values of the cavity-TLS coupling $\eta=0.05$ (a) and $\eta=0.5$ (b).  The comparison is made for different gauges of the DQRM, i.e., solid blue: `C' (correct Coulomb), dashed gold: `D' (correct dipole), dotted magenta: 'NC', and dot-dashed red: `JCM'.
    The latter JCM of course is not expected to work in the USC regime, but we include it for clarity, to better highlight the essential role of the counter rotating wave effects as well.
}
    \label{figS:FQE_GaugeDependence}
\end{figure}

For clarity, we compare four common prescriptions/models:
\begin{enumerate}
\item 
\textbf{C}: corrected Coulomb gauge Hamiltonian with consistent operators, and entirely bosonic field operators;

\item 
\textbf{NC}: naive Coulomb gauge, where gauge corrections are neglected;

\item 
\textbf{D}: corrected dipole gauge Hamiltonian with properly transformed operators (field operators in the interaction Hamiltonian contain TLS operators);

\item 
 \textbf{ND}: naive dipole gauge, where output/bath photon operator corrections are
 not corrected, namely they are assumed to be the same as the one in the interaction Hamiltonian (which is proportional to a displacement field). 
\end{enumerate}

Figure~\ref{figS:FQE_GaugeDependence} compares the Floquet quasienergy spectra obtained from these various model choices, for weak coupling ($\eta=0.05$) and ultrastrong coupling ($\eta=0.5$). For weak driving and weak coupling, differences among gauge model choices remain small. Under strong driving in the USC regime, however, naive gauge choices lead to visible deviations in the quasienergy spectrum, while the corrected Coulomb and corrected dipole gauges remain consistent. These discrepancies become more pronounced when the dressed-state truncation is reduced.

The results in Fig.~\ref{figS:FQE_GaugeDependence} therefore illustrates an important practical point: in strongly driven USC cavity-QED, gauge consistency is required not only for static spectra and observables, but also for the Floquet quasienergy structure that underlies dissipative dynamics. Further discussion of gauge-consistent open-system observables can be found in Ref.~\onlinecite{Salmon_Gauge-independent_2022}.

However, for the mechanical drive (Floquet engineering) case, to ensure gauge invariance
for a truncated matter system,
while
and include time-dependent interactions
in the light-matter interaction,
it is preferable and easier to use
the Coulomb 
gauge picture for the Hamiltonian~\cite{Settineri_Gauge_2021,Gustin_Gauge-invariant_2023,Akbari_Floquet_2025}.
Conveniently, in  the Coulomb gauge, the time-dependence of the cavity-matter coupling is simply included
through the time-dependent
coupling rate, with no other change. 
 Without time-modulation, then $\eta_0= g/\omega_c$ is the usual
 form for the
 truncated-matter QRM,
 ${\cal H}_{\rm QRM} = {\cal H}_{\rm FQR}(0)$; calligraphic notation is used
 to clarify that we model the truncated quantum system~\cite{DiStefano_Resolution_2019,Salmon_Gauge-independent_2022,Gustin_Gauge-invariant_2023}. 
Hence, we obtain the FQR Hamiltonian as
given in 
 Eq.~\eqref{eq:HC_FQR}.


\section{Driven quantum Rabi model and Floquet theory specifications}
\label{secS:DQRM}
Due to the periodicity of the drive Hamiltonian with a period $T=2\pi/\omega_d$, the system Hamiltonian is also $T$-periodic, i.e. $\mathcal{H}_{\rm DQRM}(t)=\mathcal{H}_{\rm DQRM}(t+T)$. Thus, the
Hamiltonian can be expressed in terms of a  Fourier expansion 
of Eq.~\eqref{eq:HDQR_m},
where, for the  optically pumped driven 
Floquet QRM, one only has  nonzero terms for the positive and negative primary frequency, but in the Floquet-engineered QRM, there are additional Fourier terms~\cite{Akbari_Floquet_2025}.

We conveyed in the main text that, according to the Floquet theorem~\cite{Floquet_FloquetTheory_1883,Chicone_Ordinary_2006}, we can write
a time-dependent solution of 
the time-dependent Schr\"{o}dinger equation, yielding a Floquet state, in the form of Eq.~\eqref{eq:FloquetState}. By using these  Floquet states in 
the time-dependent Schr\"{o}dinger equation,  Eq.~\eqref{eq:TDSE}, it turns out that the Floquet modes function as the eigenstates of the Floquet Hamiltonian $\mathcal{H}^{\rm F}(t)$, seen in Eq.~\eqref{eq:FloquetTDSE}.
One needs, in principle, to solve the eigenvalue problem of Eq.~\eqref{eq:FloquetTDSE} to find the Floquet quasienergies and the Floquet modes to construct the Floquet states. 

Moreover, because the Floquet modes are periodic, they can also be expanded in the Fourier series represented in Eq.~\eqref{eq:FloquetMode_FourierExpansion} in terms of the Floquet sidebands with the normalization condition $\sum_l\langle\alpha_l\vert\alpha_l\rangle=1$, and $\langle\alpha_l\vert\beta_k\rangle=0$ for $\alpha\neq\beta$ and $l\neq k$. 
Also, in a general quantum  
basis set $\{\lvert j\rangle\}$, obtained for the time-independent portion of the Hamiltonian, $\mathcal{H}_0$,  one can always expand the Floquet sidebands as a superposition 
$\lvert \alpha_l\rangle=\sum_{j}f_{\alpha lj}\lvert j\rangle$ (where the unknown coefficients $f_{\alpha lj}$'s are to be found by the completeness and orthonormality of the Floquet modes and the time-independent basis, with $\sum_{lj}\lvert f_{\alpha lj}\rvert^2=1$), and hence expand the Floquet modes in terms of the \emph{time-independent dressed basis}. 
This expansion is the representation of the Floquet drive-induced modes over the interaction-induced dressed basis.
It essentially provides the \emph{external dressing} while simultaneously accounting for the \emph{internal dressing} on an equal footing; consequently, a \emph{double dressing} framework is established.

Inserting Eq.~\eqref{eq:FloquetMode_FourierExpansion} in Eq.~\eqref{eq:FloquetTDSE} yields a quasienergy eigenvalue equation, similar to the time-independent Schr\"{o}dinger equation, in an extended (spatial and temporal) Hilbert space (or Sambe space)~\cite{Sambe_Steady_1973}. By solving this eigenvalue problem, one finds quasienergies and Floquet sidebands and, accordingly, can construct the Floquet modes and states~\cite{Eckardt_High-frequency_2015,Restrepo_Quantum_2018,Restrepo_Driven_2019}.

Since $\{\lvert\psi_\alpha(t)\rangle\}$, with $\alpha=0,1^\pm,\dots$ and dimension of $\dim(\mathbb{H}_{\rm dressed})$, form a complete time-dependent basis at any time, $t$, it follows from the orthonormality condition $\langle\psi_\alpha(t)\vert\psi_\beta(t)\rangle=\delta_{\alpha\beta}$, that the Floquet modes are also orthonormal $\langle\alpha(t)\vert\beta(t)\rangle=\delta_{\alpha\beta}$ [as also directly concluded from the eigenvalue problem Eq.~\eqref{eq:FloquetTDSE}].
In addition,  $\sum_\alpha\varepsilon_\alpha=\sum_j E_j$ modulus $\omega_{d}$~\cite{Shirley_Solution_1965}. 
Consequently, the total solution is written in terms of the Floquet states as 
\begin{equation}
\lvert\psi(t)\rangle=\sum_\alpha c_\alpha\lvert\psi_\alpha(t)\rangle,
\end{equation}
where $c_\alpha$'s are complex coefficients to be found via initial conditions, $c_\alpha=\langle\alpha\vert\psi(0)\rangle$ with $\lvert \alpha\rangle\equiv\lvert\alpha(0)\rangle$. However, since the quasienergies are time-independent, there exists a simpler way to transform the Floquet eigenvalue problem in Eq.~\eqref{eq:FloquetTDSE} into a totally time-independent matrix eigenvalue problem  and solve it (though with higher dimension).

If the system is prepared in a coherent superposition of several Floquet states, the time evolution will no longer be periodic in general and will instead be determined by two contributions~\cite{Eckardt_Atomic_2017}: 
(i) a contribution that stems from the periodic time dependence of the Floquet modes $\lvert\alpha(t)\rangle$ (called the micromotion contribution), and (ii) a contribution which leads to deviations from a periodic evolution, originating from the
relative dephasing of the factors $\mathrm{e}^{-\ii\varepsilon_\alpha t}$. Thus, beyond the periodic micromotion, the time evolution of a Floquet system is governed by the quasienergies $\varepsilon_\alpha$ of the Floquet states in much the same way that the time
evolution of an autonomous system (with time-independent Hamiltonian) is governed by the energies of the
stationary states~\cite{Eckardt_High-frequency_2015}.

\subsection{Floquet extended (Sambe) space formalism}
The periodicity of the Floquet modes allows us to map the eigenvalue problem of Eq.~\eqref{eq:FloquetTDSE} to a time-independent one in an extended Hilbert space, known
as Sambe space~\cite{Sambe_Steady_1973}. This is one of the major strengths of Floquet theory since all the
tools known for {\it time-independent} quantum problems can be used to obtain the solution
of a time-dependent problem.
In other words, Floquet analysis provides an alternative representation of energy states (Floquet quasienergy states) of the system, which gives an $l$ manifold of potential optical transitions 
seen in the spectrum 
by exploiting the discrete time-translational symmetry of the Hamiltonian $\mathcal{H}_{\rm DQRM}(t + nT ) = \mathcal{H}_{\rm DQRM}(t )$ for $n \in \mathbb{Z}$.
The essential part of Floquet theory is the transformation of this time-dependent problem
to a time-independent one given by an infinite matrix deduced
by Fourier expansion of $\vert\psi(t)\rangle$. 
This general result is made possible by
the Floquet theorem~\cite{Floquet_FloquetTheory_1883,Shirley_Solution_1965,Sambe_Steady_1973}.

Now we introduce the temporal Hilbert space $\mathbb{H}_{\rm temp}$ of the $T $-periodic functions, with the inner product:
\begin{equation}
    (f,g)=\frac{1}{T}\int_0^T dt\,f^*(t)g(t).
\end{equation}
Exploiting the Floquet theorem, one can 
consider the Fourier sum ($l$ sum) as an expansion of the time-functionality of the states in terms of the time-independent basis for the time-space $\mathbb{H}_{\rm temp}$, i.e., $\mathsf{B}_\mathbb{temp}=\{\lvert l)\}_{l\in\mathbb{Z}}$, such that $(t\vert l)=\mathrm{e}^{-\ii l\omega_{d}t}$, with $t$ indicating time. Hence, the extended space is spanned by the joint basis of 
\begin{equation}
    \begin{split}
       \mathsf{B}_\mathrm{ex}&=\mathsf{B}_\mathrm{temp}\otimes\mathsf{B}_\mathrm{dressed}
       \\
       &=\{\Fket{l,j}\equiv\lvert l )\otimes\lvert j \rangle;\, j=0,1^\pm,2^\pm,\dots\,\&\,l\in\mathbb{Z}\}. 
    \end{split}
\end{equation}

In addition, if an operator has a periodic time
dependence, $O(t) = O(t + T )$, its form in extended space is defined in terms of its Fourier components such that if $O(t) = \sum_{l}\mathrm{e}^{-\ii l\omega_{d}t}O_l$, then~\cite{Sambe_Steady_1973,Restrepo_Driven_2019}
\begin{equation}
\begin{split}
O_\mathrm{ex}=\sum_l F_l\otimes O_l,
\end{split}
\label{}
\end{equation} 
where $F_l$ is the (quantum) shift operator acting on the temporal space with the time-independent basis~\cite{Restrepo_Driven_2019,Restrepo_Quantum_2018}, so that 
\begin{equation}
   F_k\lvert l)=\lvert l+k).
   \label{eq:ShiftOp}
\end{equation}
Furthermore, the scalar product in the extended space is 
\begin{equation}
\begin{split}
\langle\!( O,O')\!\rangle\equiv\frac{1}{T}\int_T dt\,\langle O(t),O'(t)\rangle,
\end{split}
\label{}
\end{equation} 
for time-periodic operators.

With the extended Hilbert space and its relevant calculus, it is now possible to write the Floquet quasienergy operator $\mathcal{H}^{\rm F}_\mathrm{ex}$ (that is the representation of the Floquet Hamiltonian in the extended space basis) and the Floquet eigenvalue problem in $\mathbb{H}_{\rm ex}$, via
\begin{equation}
\begin{split}
\Fbra{l,j} \mathcal{H}^{\rm F}_\mathrm{ex}\Fket{l',j'}&=\frac{1}{T}\int_T dt\,\mathrm{e}^{\ii l\omega_dt}\langle j\vert \mathcal{H}_{\rm DQRM}(t)-\ii\partial_t\vert j'\rangle\mathrm{e}^{-\ii l'\omega_dt}
\\
&=\langle j\lvert \mathcal{H}_{l-l'}\rvert j'\rangle-\delta_{jj'}\delta_{ll'}l\omega_d,
\end{split}
\label{}
\end{equation} 
hence, we obtain:
\begin{equation}
\begin{split}
\mathcal{H}^{\rm F}_\mathrm{ex}&=\sum_{l}F_l\otimes \mathcal{H}_l-\omega_d\,\mathcal{N}_{\rm temp}\otimes\mathbf{1}_\mathrm{dressed},
\end{split}
\label{}
\end{equation} 
where $\mathcal{H}_l$'s are the Fourier components of the Hamiltonian
and $\mathcal{N}_{\rm temp}$ is a (number) operator in the temporal space which acts on the time-independent basis~\cite{Restrepo_Driven_2019} via Eq.~\eqref{eq:Ntemp}.
Accordingly, the Floquet quasienergy eigenvalue problem in the extended space reads
\begin{equation}
\begin{split}
\mathcal{H}^\mathrm{F}(t)\vert \alpha(t)\rangle&=\varepsilon_\alpha\vert \alpha(t)\rangle\,\to\,\mathcal{H}^{\rm F}_{\rm ex}\Fket{l\alpha}=\varepsilon_{l\alpha }\Fket{l\alpha}.
\end{split}
\label{eq:FloquetEigenvalueProblem_ex}
\end{equation}

Due to the dimension of the matrices in the extended space, we immediately see that this eigenvalue problem has ${\rm Dim}(\mathbb{H}_{\rm ex})$ solutions. The number of solutions seems to be redundant considering the size of the original Hilbert space $\mathbb
H_{\rm dressed}$. 
Note that these solutions are not linearly independent but only ${\rm Dim}(\mathbb{H}_{\rm dressed})$ of them.  In other words, ${\rm Dim}(\mathbb{H}_{\rm dressed})$ of the solutions are replicated ${\rm Dim}(\mathbb{H}_{\rm temp})$ times throughout the energy scale, a concept that constructs the Brilliuon zones (BZs), similar to the Bloch theory for the momentum axis, but in the energy axis. To be more precise, the mode $\lvert\alpha^{[n]}(t)\rangle\equiv \mathrm{e}^{-\ii n\omega_dt}\lvert \alpha(t)\rangle$ (i.e., translation to the $n$th BZ that connects Floquet modes in different BZs) yields a solution of \eqref{eq:TDSE} physically identical to \eqref{eq:FloquetState} but with the shifted quasienergy $\varepsilon_{n\alpha}\equiv\varepsilon_{\alpha}+n\omega_d$.
Furthermore, $\varepsilon_{\alpha}\equiv\varepsilon_{0\alpha}$ and $\lvert\alpha(t)\rangle\equiv \lvert \alpha^{[0]}(t)\rangle$. 
This property builds up the BZs, and thus, it is sufficient to only examine the set of eigenvalues $\{\varepsilon_{n\alpha}\}$ within one BZ, e.g., $-\omega_d/2\leq\varepsilon_{n\alpha}<\omega_d/2$ to construct the time-dependent SE solution. 

It is convenient for future calculations to write the Floquet modes in the extended space. With $\varepsilon_{\alpha}\equiv\varepsilon_{0\alpha}$ and $
\Fket{\alpha}\equiv\Fket{0\alpha}$, we have
\begin{equation}
\begin{split}
\Fket{\alpha}=\sum_l\vert l)\otimes\lvert\alpha_l\rangle.
\end{split}
\label{}
\end{equation}

The eigenvectors of $\mathcal{H}^{\rm F}_{\rm ex}$ in the extended (composite) space satisfy the orthonormality condition and form a complete set, 
\begin{equation}
\begin{split}
\langle\!( \alpha,\alpha')\!\rangle=\delta_{\alpha\alpha'},
\,\,\,\,\,\,\,\,\,\,\,
\sum_{\alpha}\Fket{\alpha}\Fbra{\alpha}=\mathbf{1}_{\rm ex},
\end{split}
\label{}
\end{equation} 
respectively.
In the extended space basis, we find
$\langle j\vert\alpha_l\rangle=\Fbraket{l,j}{\alpha}$.
Thus, we know how to expand $\Fket{\alpha}$ in the Floquet extended basis $\{\Fket{l, j}\}$ (or, also $\{\Fket{l, \gamma}\}$):
\begin{equation}
\begin{split}
\Fket{\alpha}
&=\sum_l\langle j\vert\alpha_l\rangle\,\Fket{l,j}
\\
&=\sum_l\Fbraket{l,j}{\alpha}
\,\Fket{l,j}.
\end{split}
\label{}
\end{equation} 

We see that the Fourier coefficients are the expansion coefficients, thus,  
an equivalent
way to derive the relation between the Fourier coefficients and the Floquet states can be
achieved by considering the inverse Fourier transform,
\begin{equation}
\begin{split}
    \langle j\vert\alpha_k\rangle&
    =\frac{1}{T}\int_0^T dt\,\mathrm{e}^{\ii k\omega_dt}\,\langle j\vert\alpha (t)\rangle
    \\&
    =\frac{1}{T}\int_0^T dt\,\langle j\vert\alpha^{[-k]} (t)\rangle
    \\&
=\Fbraket{k,j}{\alpha}.
\end{split}
\end{equation}
Furthermore, the extended space basis helps us to write the Floquet mode via
\begin{equation}
\begin{split}
\vert \alpha(t)\rangle
&=(t\Fket{\alpha}
\\
&=\sum_{jl}\mathrm{e}^{-\ii l\omega_d t}\,\Fbraket{l,j}{\alpha}\,\lvert j\rangle.
\end{split}
\label{}
\end{equation}

Thanks to the extended space formalism, many time-dependent calculations can be assisted
and made simpler. One of the most important algebras is to calculate the projection of Floquet modes at different times, i.e., 
\begin{equation}
\begin{split}
\langle\beta\vert \alpha(t)\rangle&=\langle\beta\vert\sum_{j l}\mathrm{e}^{-\ii l\omega_{d}  t}\,\langle j\vert \alpha_{l}\rangle\,\lvert j\rangle
\\
&=\sum_{l}\mathrm{e}^{-\ii l\omega_{d}  t}\,\langle\beta\vert {\alpha_l}\rangle
\\
&=\sum_{l}\mathrm{e}^{-\ii l\omega_{d}  t}\,\Fbraket{-l,\beta}{\alpha}.
\end{split}
\label{}
\end{equation}
 
The connection among the Floquet modes in different BZs and the relation between the Floquet modes and sidebands in the extended space can be made from:
\begin{equation}
\begin{split}
\Fket{n\alpha}=\sum_{l}\lvert l)\otimes\vert \alpha_l^{[n]}\rangle,
\end{split}
\label{}
\end{equation}
along with the Fourier relation between them in the $n$the BZ $\vert\alpha^{[n]}(t)\rangle=\sum_l\mathrm{e}^{\ii l\omega_dt}\,\vert \alpha^{[n]}_l\rangle$ [similar to Eq.~\eqref{eq:FloquetMode_FourierExpansion}],
facilitates the connection among the Floquet sidebands in different BZs,
\begin{equation}
\begin{split}
\vert \alpha_l^{[n]}\rangle
=\vert \alpha_{l-n}\rangle.
\end{split}
\label{FSB_differentBZs}
\end{equation}
One must also note that the selection rules for strongly driven systems are determined by symmetries of $\mathcal{H}^{\rm F}_{\rm ex}$ in the extended space $\mathbb{H}_{\rm ex}=\mathbb{H}_{\rm temp}\otimes \mathbb{H}_{\rm dressed}$, not by those
of $\mathcal{H}_{\rm DQRM}$ in $\mathbb{H}_{\rm dressed}$!

Lastly, a valuable utilization of the extended space formalism is to calculate the widely used Floquet extended space matrix elements of a system operator, given in Eq.~\eqref{eq:FelementS_ex}, is
\begin{equation}
    S_{\alpha\beta l}=\Fbra{\alpha}F_{-l}\otimes S\Fket{\beta}.
\end{equation}

\subsection{Time-evolution operator}
The time-periodic Hamiltonian satisfying the unitary time evolution generates the operator,
 \begin{equation}
 \begin{split}
U(t,t_0)
    &=\mathcal{T}\exp\lcb{-\ii\int_{t_0}^t dt'\,\mathcal{H}_{\rm DQRM}(t')}\rcb,
    \end{split}
\end{equation}
where $\mathcal{T}$ is the Wick's time-ordering operator.
To find the evolution of an arbitrary state $\lvert\psi(t)\rangle$, we use the fact that the set of solutions to Schrödinger's equation, $\lvert\psi_\alpha(t)\rangle$. forms a complete basis, so that $U(t,t_0)=\sum_\alpha\lvert\psi_\alpha(t)\rangle\langle\psi_\alpha(t_0)\rvert$. This operator satisfies the usual properties of the evolution operator. Thus we have the time-transition property $U(t, t_0)= U(t, t_1)U(t_1, t_0)$, as well as the unitary property $U^\dagger(t, t_0)= U(t_0, t)=U^{-1}(t, t_0)$.
Also, the normalization condition $U(t_0, t_0)={\bf 1}$ is satisfied at any time.
In terms of the Floquet modes, then, the time evolution operator expressed as
\begin{equation}
\begin{split}
U(t,t_0)&=\sum_\alpha \mathrm{e}^{-\ii\varepsilon_\alpha(t-t_0)}\,\lvert\alpha(t)\rangle\langle\alpha(t_0)\rvert.
\end{split}
\label{U}
\end{equation}

In Floquet theory, the system time evolution operator can be decomposed as~\cite{Eckardt_High-frequency_2015}
\begin{equation}
    U(t,t_0)=U^{\rm F}(t,t_0)\,\exp\lcb{-\ii H_{\rm F}(t-t_0)}\rcb,
\end{equation}
where the \emph{micromotion} operator $U^{\rm F}(t,t_0)=\exp\lcb{-\ii K^{\rm F}(t-t_0)}\rcb$ (with $K^{\rm F}(t)$ being the so-called the stroboscopic kick operator) is a $T$-periodic time-evolution operator for the Floquet modes within one period and satisfying $U^{\rm F}(t+T,t)={\bf 1}$. The $U^{\rm F}(t,t_0)$ operator is the time evolution operator for the Floquet modes, i.e., $\lvert\alpha(t)\rangle=U^{{\rm F}}(t,t_0)\lvert\alpha(t_0)\rangle$. 
 Such a \emph{two-point micromotion operator} can be defined by~\cite{Eckardt_High-frequency_2015}
\begin{equation}
\begin{split}
U^{\rm F}(t)&=\sum_\gamma \lvert\gamma(t)\rangle\langle\gamma\rvert
\\
&=\sum_{l} \mathrm{e}^{-\ii l\omega_dt}\,U^{\rm F}_{l},
\end{split}
\label{}
\end{equation}
where
\begin{equation}
\begin{split}
U^{\rm F}_{l}&=\sum_\gamma\lvert\gamma_l\rangle\langle\gamma\rvert
=\sum_{\gamma l'}\lvert\gamma_l\rangle\langle\gamma_{l'}\rvert
,
\end{split}
\label{}
\end{equation}
so that, by construction, it evolves the Floquet modes in time.
It is periodic in both arguments, $U^{\rm F}(t,t_0)=U^{\rm F}(t+T,t_0)=U^{\rm F}(t,t_0+T)$. Accordingly, its matrix element in the Floquet basis can be obtained via $U^{\rm F}_{\alpha\beta}(t)\equiv\langle\alpha(t)\vert U^{\rm F}(t) \lvert\beta(t)\rangle=\langle\alpha\vert U^{\rm F}(t) \lvert\beta\rangle$ as
\begin{equation}
\begin{split}
U^{\rm F}_{\alpha\beta}(t)&=
\langle\alpha\vert\beta(t)\rangle
\\
&=\sum_{l}U^{\rm F}_{\alpha\beta l}\mathrm{e}^{-\ii l\omega_dt},
\end{split}
\label{}
\end{equation}
where
\begin{equation}
\begin{split}
U^{\rm F}_{\alpha\beta l}&=\frac{1}{T}\int_0^T dt\,\langle\alpha(t)\vert\mathrm{e}^{\ii l\omega_dt} U^{\rm F}_{l}\vert\beta(t)\rangle
\\
&=\Fbra{\alpha}F_{-l}\otimes U^{\rm F}_{l}\Fket{\beta}.
\end{split}
\label{}
\end{equation}

The matrix element of the time evolution operator in the Floquet basis is of great significance. It turns out that this matrix element in the initial time and later times are closely related, i.e., 
\begin{equation}
\begin{split}
\langle\alpha\lvert U(t)\rvert\beta\rangle&=\mathrm{e}^{-\ii\varepsilon_\beta t}\langle\alpha\vert\beta(t)\rangle
\\
&=\sum_\gamma \mathrm{e}^{-\ii\varepsilon_\gamma t}\,\langle\alpha\lvert\gamma(t)\rangle\langle\gamma\rvert\beta\rangle
\\
&=\mathrm{e}^{-\ii\varepsilon_\beta t}\,
\sum_{k}\mathrm{e}^{-\ii k\omega_d t}\,\Fbraket{-k,\alpha}{\beta}
,
\end{split}
\label{}
\end{equation}
and, 
\begin{equation}
\begin{split}
U_{\alpha\beta}(t)&\equiv\langle\alpha(t)\lvert U_{\rm S}(t)\rvert\beta(t)\rangle
\\
&=\mathrm{e}^{-\ii\varepsilon_\alpha t}\langle\alpha\vert\beta(t)\rangle
\\
&=\sum_\gamma \mathrm{e}^{-\ii\varepsilon_\gamma t}\,\langle\alpha(t)\lvert\gamma(t)\rangle\langle\gamma\rvert\beta(t)\rangle
\\
&=\mathrm{e}^{-\ii\varepsilon_\alpha t}\,
\sum_{l}\mathrm{e}^{-\ii l\omega_d t}\,\Fbraket{-l,\alpha}{\beta}.
\end{split}
\label{}
\end{equation}
Hence we have the relations,  $\langle\alpha(t)\lvert U(t)\rvert\beta(t)\rangle=\mathrm{e}^{-\ii(\varepsilon_\alpha-\varepsilon_\beta)t}\langle\alpha\lvert U(t)\rvert\beta\rangle$ and $\langle\psi_\alpha(t)\lvert U(t)\rvert\psi_\beta(t)\rangle=\langle\alpha\lvert U(t)\rvert\beta\rangle$.

The matrix element of the time evolution operator in the QRM-dressed basis brings many practical utilities. Generally, in terms of the extended space, one obtains
\begin{equation}
\begin{split}
U_{\alpha\beta}(t)&=\sum_n\mathrm{e}^{-\ii k\omega_dt}\,\Fbra{-n,\alpha} \mathrm{e}^{-\ii \mathcal{H}^{\rm F}_{\rm ex} t}\Fket{0,\beta}.
\end{split}
\label{}
\end{equation}
Note that if we choose the states of initial time-independent Hamiltonian, i.e., $\alpha=j=i$ and $\beta=j'=f$, we can calculate the time-evolution from two eigenstates rather than two quasienergy states.

\subsection{Transition probability}
The transition probability from an initial quantum state of the renormalized static Hamiltonian, $\lvert j=i\rangle$,
to a final quantum state, $\lvert j=f\rangle$, is closely related to the matrix elements of the time evolution operator $U_{fi}(t,t_0)\equiv \langle f\lvert U(t,t_0)\rvert i\rangle$. In  Floquet theory, $U_{fi}(t,t_0)$ can be interpreted as the amplitude that the system initially in the Floquet joint state $\lvert i\rangle$ and zero mechanical oscillation at time $t_0$ evolves to the Floquet joint state $\lvert f\rangle$ and $k$ mechanical oscillation by time $t$ according to the time-independent Floquet Hamiltonian $\mathcal{H}_\mathrm{F}$ with some weighting factors that must be summed over all final quantum processes and states~\cite{Chu_Beyond_2004,Son_Floquet_2009,Satanin_Amplitude_2012}. The transition probability
going from the initial quantum state and a coherent mechanical oscillation state to the final
quantum state summed over all final mechanical oscillation quantum process orders reads $P_{f\leftarrow i}(t,t_0)=\lvert U_{fi}(t,t_0)\rvert^2$.

One key quantity of experimental interest, however, is the transition probability
averaged over initial times to (or equivalently averaged over the initial phases of the external drive seen by the system), keeping the elapsed time $(t-t_0)$ fixed. 
Moreover, averaging over $(t-t_0)$, one obtains the long-time average transition
probability~\cite{Chu_Beyond_2004,Son_Floquet_2009,Satanin_Amplitude_2012}, 
$\overline{P}_{f\leftarrow i}=\sum_{l,l'}\overline{P}^{(l,l')}_{f\leftarrow i}$, with
\begin{equation}
\begin{split}
\overline{P}^{(l,l')}_{f\leftarrow i}&=\sum_\alpha\lvert\langle f\vert\alpha_{l}\rangle\langle\alpha_{l'}\vert i\rangle\rvert^2.
\end{split}
\label{AveP}
\end{equation}
 Also, the order of the external photons involved in the quantum process for the higher-order processes is portrayed in the number $n=l-l'$ with the associated probability of 
 \begin{equation}
    \overline{P}^{(n)}_{f\leftarrow i}=\sum_{l-l'=n}\overline{P}^{(l,l')}_{f\leftarrow i}, 
 \end{equation}
 so that 
 \begin{equation}
     \overline{P}_{f\leftarrow i}=\sum_n\overline{P}^{(n)}_{f\leftarrow i}.
 \end{equation}

The initial-time-averaged (averaged over the initial time $t_0$ while keeping the elapsed time $(t-t_0)$ fixed) and the long-time-averaged transition probability can be also expanded in terms of the extended space~\cite{Chu_Quantum_1982,Chu_Beyond_2004,Chu_Recent_1985}, using
\begin{equation}
\begin{split}
{P}^{(n)}_{f\leftarrow i}(t-t_0)&=\lvert\Fbra{n,f}\exp\lcb{-\ii \mathcal{H}^{\rm F}_{\rm ex} (t-t_0)}\rcb\Fket{ 0,i}\rvert^2,
\end{split}
\label{}
\end{equation}
and, by performing the long time average over $(t-t_0)$, then
\begin{equation}
\begin{split}
\overline{P}^{(n)}_{f\leftarrow i}&=\sum_{l\alpha} \Fbraket{n,f}{l\alpha}\Fbraket{l\alpha}{0,i}\rvert^2.
\end{split}
\label{}
\end{equation}
We use this computation-friendly formula in the extended space for the calculation of the excitation probability.

Once an initial state $\ket{j=i}$ is set for the QRM-type Hamiltonian, the unitary Floquet probability factor   $\overline{P}_{f\leftarrow i}$ gives an ideal number for the populations of the final states $\ket{j=f}$, compared to that with dissipation given by the long-time-averaged populations $\overline{\rho}_{ff}$ obtained from, for example, an appropriate master equation.

\subsection{Floquet master equation for a closed system}
Knowing that the Floquet states construct a complete basis, it is desirable to expand the density matrix in the Floquet representation as
${\rho}(t)=\sum_{\alpha\beta}\rho_{\alpha\beta}(t)\,\lvert \alpha(t)\rangle\langle \beta(t)\rvert$, where $\rho_{\alpha\beta}(t)\equiv\langle \alpha(t)\rvert\rho(t)\lvert \beta(t)\rangle$. 
This allows us to write the von-Neumann equation for the system density matrix as~\cite{Hausinger_Dissipative_2010}
\begin{equation}
    \begin{split}
        \partial_t{\rho}_{\alpha\beta}(t)&=-\ii (\varepsilon_\alpha-\varepsilon_\beta)\rho_{\alpha\beta}(t),
    \end{split}
    \label{}
\end{equation}
for a closed system without any dissipation, where it admits the analytical solution,
\begin{equation}
    \begin{split}
        {\rho}_{\alpha\beta}(t)&=\rho_{\alpha\beta}(0)\,\exp\lcb{-\ii (\varepsilon_\alpha-\varepsilon_\beta)t}\rcb,
    \end{split}
    \label{}
\end{equation}
up to a given initial condition.
In order to configure this initial condition into the above solution, one should express the initial condition in Floquet modes
 basis. 
To do so, we obtain the Floquet matrix elements of both sides to yield:
$\rho_{\alpha\beta}(0)=\langle \alpha\vert j=0\rangle\langle j=0\vert \beta\rangle$.

\subsection{Expectation (average) value of an operator}
A general observable, $\langle O(t)\rangle\equiv\langle\psi(t)|O|\psi(t)\rangle$, is not necessarily time periodic due to the presence of off-diagonal terms ($\alpha\neq\beta$)
in the Floquet eigenbasis~\cite{Eckardt_Atomic_2017}, given by 
\begin{equation}
\langle O(t)\rangle=\sum_{\alpha\beta}c^*_\alpha c_\beta\,O_{\alpha\beta}(t)\,\mathrm{e}^{\ii(\varepsilon_\alpha-\varepsilon_\beta)t},
\end{equation}
where $O_{\alpha\beta}(t)\equiv\langle\alpha(t)\lvert O\rvert\beta(t)\rangle$.
However,
in real open systems,
 the off-diagonal terms are suppressed,
and the time evolution of observables
often becomes periodic with the same periodicity as the driving
field. 
This (prethermalization process) can be stimulated by
adding a phenomenological damping rate, $\Gamma(\omega)$ with $\Gamma(0)=0$, that yields to kill off 
the non-diagonal terms,
which are then damped out 
after a sufficiently long time~\cite{Chu_Beyond_2004,DeLiberato_Virtual_2017,Eckardt_Atomic_2017}.

Subsequently, we can derive the expectation values,
\begin{equation}
\langle O(t)\rangle=\sum_{\alpha\beta}c^*_\alpha c_\beta\,O_{\alpha\beta}(t)\,\mathrm{e}^{\ii(\varepsilon_\alpha-\varepsilon_\beta)t-\Gamma(\varepsilon_\alpha-\varepsilon_\beta) t},
\end{equation}
and obtain the 
steady-state values: $\langle O(t)\rangle_{\rm ss}\equiv\langle O(t> t_{\rm ss})\rangle=\sum_{\alpha}\lvert c_\alpha\rvert^2\,O_{\alpha\alpha}(t)$.
 We then define the 
 the mean real excitation number 
 from
 $ \overline{O}=\frac{1}{T}\int_{t_\mathrm{ss}}^{t_\mathrm{ss}+T} dt\,\langle O(t)\rangle_{\rm ss}$, where $t_{\rm ss}$ and $T$ are long enough to yield a temporal average.
 For our case, the result is exactly $T$ periodic,
 so we only have to use one period. 
Hence, the long-time-averaged observable reads

\begin{equation}
\begin{split}
    \overline{O}&=\sum_{\alpha l}\lvert c_\alpha \rvert^2\,\langle \alpha_l\lvert O\rvert \alpha_l\rangle.
\end{split}
\end{equation}
Using the extended space formalism~\cite{Restrepo_Driven_2019,Restrepo_Quantum_2018}, then
\begin{equation}
\begin{split}
    \overline{O}&=\sum_{\alpha}\lvert c_\alpha \rvert^2\,\Fbra{\alpha}\lvert F_0\otimes O\Fket{\alpha}.
\end{split}
\end{equation}

Any observable-related predictions can also be computed
from the 
master equation or the density matrix (with a finite basis).
Having obtained the time evolution of the density matrix, we can calculate the average value of observables throughout the time progression. In the Schr\"{o}dinger picture, for any arbitrary operator $O$, we have: $\langle O(t)\rangle=\mathrm{tr}\lcb O\rho(t)\rcb$. We note that the periodicity of an averaging quantity concerning an observable can be attained by the same technique of adding dissipation as mentioned in the main text.

One can find the density matrix and then calculate the expectation value in the TI basis of the DC Hamiltonian. However, knowing that the change of basis does not change the trace, we can find a simpler and more 
computationally efficient formula using the Floquet basis via
\begin{equation}
    \begin{split}
       \langle O(t)\rangle
       &=\sum_{\alpha\beta} O_{\alpha\beta}(t)\rho_{\beta\alpha}(t)
\\
&=\sum_{\alpha\beta l} O_{\alpha\beta l}\,\rho_{\beta\alpha}(0)\,\exp\lcb{\ii(\varepsilon_\alpha-\varepsilon_\beta+l\omega_d)t}\rcb,
    \end{split}
    \label{}
\end{equation}
where $O_{\alpha\beta l}=(1/T)\int_0^T dt\,\mathrm{e}^{\ii l\omega_d t}\,O_{\alpha\beta}(t)$.
Therefore, in the steady state, we obtain $\langle O(t)\rangle_{\rm ss}=\sum_{\alpha l} O_{\alpha\alpha l}\rho_{\alpha\alpha}(0)\,\mathrm{e}^{-\ii l\omega_dt}$, and the average of the expectation value reads 
\begin{equation}
    \begin{split}
\overline{O}&\equiv\frac{1}{T}\int_0^Tdt\,\langle O(t)\rangle_{\rm ss}
\\
&=\sum_{\alpha} O_{\alpha\alpha 0}\rho_{\alpha\alpha}(0).
    \end{split}
    \label{}
\end{equation}

This formula is equivalent to the formulae presented above.
Note, in general, $\overline{N}\propto \sum_{fn}\overline{P}^{(n)}_{f\leftarrow0}$. The proportionality becomes equality when the observable operator and the drive operator are the same.

\section{Derivation of Floquet-Markov generalized master equation for driven-dissipative quantum Rabi model}
\label{secS:FGME}

We wish to obtain a master equation for the case of a time-periodic driven open quantum system, with general dissipation baths. To do so, we follow the standard approach in Refs.~\onlinecite{Breuer_Theory_2002,Eckardt_High-frequency_2015,Restrepo_Driven_2019}. 

We begin by setting up the total open system Hamiltonian, 
\begin{equation}
    \begin{split}
        \mathcal{H}_\mathrm{tot}(t)=\mathcal{H}_{\rm DQRM}(t)+\mathcal{H}_{\rm B}+\mathcal{H}_{\rm SB}(t),
    \end{split}
\end{equation}
where $\mathcal{H}_{\rm DQRM}(t)=\mathcal{H}_{0}+\mathcal{H}_{d}(t)$ is the Hamiltonian of the driven system such that $\mathcal{H}_{d}(t+T)=\mathcal{H}_{d}(t)$ with $T=2\pi/\omega_d$, $\mathcal{H}_{\rm B}$ is the Hamiltonian of the environment (bath) and ${H}_{\rm SE}(t)$, with $\mathcal{H}_{\rm SE}(t)=\mathcal{H}_{\rm SB}(t+T)$, is the interaction between system and reservoir which is considered to be weak.
It is assumed that the periodic driving is applied only to the system of interest, but we allow the interaction Hamiltonian $\mathcal{H}_{\rm SE}(t)$ to depend on the time to consider applications to the Floquet engineering.
It is sometimes convenient to move to a suitable rotating frame. The interaction Hamiltonian in a rotating frame generally depends
on time, even though it is time-independent in the laboratory frame. Thus, allowing the time dependence of $\mathcal{H}_{\rm SB}(t)$ enables us to consider the problem in a rotating frame. 

We consider a generic system with $N_\Lambda$ ($\Lambda\in\{1,\dots, N_\Lambda\}$ indexes the subsystems) interacting components (subsystems) or degrees of freedom. Each $\Lambda$ component is weakly coupled to an independent bath in the environment. Thus, $\mathcal{H}_\mathrm{B}=\sum_\Lambda \mathcal{H}^\Lambda_\mathrm{B}$ with the bosonic components $\mathcal{H}^\Lambda_\mathrm{B}=\sum_k\omega^\Lambda_k b^{\Lambda\dagger}_kb^{\Lambda}_k$, or fermionic ones $\mathcal{H}^\Lambda_\mathrm{B}=\sum_k
\omega^\Lambda_k c^{\Lambda\dagger}_kc^{\Lambda}_k$.

The evolution of the total system plus reservoir is identified by the celebrated von-Neumann equation~\cite{Breuer_Theory_2002}, assuming that 
the in initial state of the dissipative system there is no correlation between the system and the environment. Furthermore, we assume the environment component fulfills the following stationary conditions~\cite{Farina_Open-quantum-system_2019}: ($i$) invariance under the action of the local Hamiltonian, $\lsb\rho_\mathrm{B}(t_0),\mathcal{H}_\mathrm{B}\rsb=0$, meaning that the bath is in equilibrium, and ($ii$) zero expectation value of any bath operators $B$, $\mathrm{tr}_\mathrm{E}\lcb\rho_\mathrm{B}(t_0)B\rcb=0$.

We can start with the ME within the Born-Markov approximation, in the interaction picture,
\begin{equation}
    \begin{split}
        \partial_t\widetilde{\rho}(t)&=-
        \int_{0}^\infty \! ds\,\mathrm{tr}_{\rm B}\lcb\lsb\widetilde{\mathcal{H}}_{\rm SB}(t),\lsb\widetilde{\mathcal{H}}_{\rm SB}(t-s),\widetilde{\rho}(t)\otimes{\rho}_{\rm B}\rsb\rsb\rcb,
    \end{split}
    \label{RedfieldEqn}
\end{equation}
known as the Redfield equation~\cite{Breuer_Theory_2002}, where
\begin{equation}
    \begin{split}
       \widetilde{\rho}(t)&=
       U^{-1}(t,t_0)U^{-1}_{\rm B}(t,t_0){\rho}(t)U_{\rm B}(t,t_0)U(t,t_0),
       \\
       \widetilde{\mathcal{H}}_{\rm SE}(t)&=
       U^{-1}(t,t_0)U_{\rm B}^{-1}(t,t_0)\mathcal{H}_{\rm SB}(t)U_{\rm B}(t,t_0)U(t,t_0),
    \end{split}
\end{equation}
with $U(t,t_0)$ being the driven cavity-QED system evolution operator and $U_{\rm B}(t,t_0)=\exp\lcb{-\ii \mathcal{H}_{\rm B}(t-t_0)}\rcb$ being the environment time-evolution operator.
The Markov approximation is justified when $\tau_{\rm B}/\tau_{\rm R}\gg1$ as well as $(t-t_0)\gg\tau_{\rm B}$. This works for a timescale $\tau_R$ known as the system relaxation time due to dissipation, where the integral disappears sufficiently fast for $s\gg \tau_\mathrm{B}$, with $\tau_\mathrm{B}$ being the environment correlation time. 

 Letting $S^\Lambda_{\rm B}$ and $B^\Lambda$ be the system and bath operators to couple in the Schrodinger picture, respectively, for each $\Lambda$ system component (subsystem), we have
\begin{equation}
    \begin{split}
        \mathcal{H}_\mathrm{SB}(t)&=\sum_\Lambda S^{\Lambda}_{\rm B}\otimes B^{\Lambda}.
    \end{split}
    \label{}
\end{equation}
Without loss of generality, one always can choose these operators to be Hermitian,  so that  $S^{\Lambda\dagger}_{\rm B}=S^\Lambda_{\rm B}$ and $B^{\Lambda\dagger}=B^\Lambda$. 
Thus, in the interaction picture,

\begin{equation}
    \widetilde{H}_\mathrm{SB}(t)=\sum_\Lambda \widetilde{S}^{\Lambda}_{\rm B}(t)\otimes \widetilde{B}^{\Lambda}(t),
\end{equation} 
where 
\begin{equation}
    \begin{split}
        \widetilde{S}^\Lambda_{\rm B}(t)&=U^{-1}(t,0)\,{S}^\Lambda_{\rm B}\,U(t,0), 
        \\
        \widetilde{B}^\Lambda(t)&=U_\mathrm{B}^{-1}(t,0)\,{B}^\Lambda\, U_\mathrm{B}(t,0).
    \end{split}
\end{equation}

Subsequently, the Redfield equation reads
\begin{equation}
    \begin{split}
        \partial_t\widetilde{\rho}(t)
        &=\sum_{\Lambda\Lambda'}
        \int_{0}^\infty ds\,\lsb\widetilde{S}^{\Lambda}_{\rm B}(t-s)\widetilde{\rho}(t)\widetilde{S}_{\rm B}^{\Lambda'\dagger}(t)\,C^{\Lambda'\Lambda}(s)
        \right.
        \\
        &\hspace{2.25cm}-\widetilde{S}^{\Lambda'\dagger}_{\rm B}(t)\widetilde{S}^{\Lambda}_{\rm B}(t-s)\widetilde{\rho}(t)\,C^{\Lambda'\Lambda}(s)
        \\
        &\hspace{2.25cm}+\widetilde{S}^{\Lambda}_{\rm B}(t)\widetilde{\rho}(t)\widetilde{S}^{\Lambda'\dagger}_{\rm B}(t-s)\,C^{\Lambda'\Lambda}(-s)
        \\
        &\hspace{2.25cm}\left.
        -\widetilde{\rho}(t)\widetilde{S}^{\Lambda'\dagger}_{\rm B}(t-s)\widetilde{S}^{\Lambda}_{\rm B}(t)\,C^{\Lambda'\Lambda}(-s)\right]
        ,
    \end{split}
    \label{RedfieldEqn}
\end{equation}
with $C^{\Lambda\Lambda'}(s)=\mathrm{tr}_\mathrm{B}\lcb\widetilde{B}^{\Lambda\dagger}(t)\widetilde{B}^{\Lambda'}(t-s)\rho_\mathrm{E}\rcb$ being the bath correlation function. Assuming that the bath is in equilibrium, the time-translational symmetry is held, $\langle\widetilde{B}^{\Lambda\dagger}(t)\widetilde{B}^{\Lambda'}(t-s)\rangle=\langle\widetilde{B}^{\Lambda\dagger}(s) {B}^{\Lambda'}\rangle$. Moreover, $\lsb C^{\Lambda\Lambda'}(s)\rsb^*=C^{\Lambda'\Lambda}(-s)$.

One can exploit the periodic time dependence in the system Hamiltonian, $\mathcal{H}_\mathrm{DQRM}(t)$, and use Eq.~\eqref{U} along with Eq.~\eqref{eq:FloquetMode_FourierExpansion} in $\widetilde{S}^\Lambda_{\rm B}(t)=U^{\dagger}(t){S}^\Lambda_{\rm B}U(t)$ to obtain~\cite{Bukov_Universal_2015,Restrepo_Driven_2019}
\begin{equation}
    \begin{split}
        \widetilde{S}^\Lambda_{\rm B}(t)
        &=\sum_{\Delta}\mathrm{e}^{-\ii \Delta t}\,{S}^\Lambda_{\rm B}(\Delta),
    \end{split}
    \label{S_Delta}
\end{equation}
where $\Delta\equiv\Delta_{\alpha\beta l}=\varepsilon_\beta-\varepsilon_\alpha+l\omega_d$ (this is the quantity to represent the net energy difference between two Floquet sidebands, thus in the extended space it can be seen as $\Delta_{\alpha\beta l}=\varepsilon_{m\alpha}-\varepsilon_{n\beta}$ with $l=m-n$) and
\begin{equation}
    \begin{split}
        {S}^\Lambda_{\rm B}(\Delta)&={S}^\Lambda_{{\rm B},\alpha\beta l}\,\lvert \alpha\rangle\langle \beta\rvert,
    \end{split}
\end{equation}
with
\begin{equation}
    \begin{split}
        {S}^\Lambda_{{\rm B},\alpha\beta l}&=\frac{1}{T}\int_0^T dt\,\mathrm{e}^{\ii l\omega_{d} t}\,\langle \alpha(t)\rvert S^\Lambda_{\rm B}\lvert \beta(t)\rangle.
    \end{split}
\end{equation}

The expression for $S_{\alpha\beta l}$ is a matrix element in Floquet basis which is defined with a time component in the extended space. In order to calculate it, we first calculate the matrix element in the Floquet basis ${S}^\Lambda_{{\rm B},\alpha\beta}(t)\equiv\langle \alpha(t)\rvert S^\Lambda_{\rm B}\lvert \beta(t)\rangle$, and due to its periodicity, it can be expanded in Fourier series as ${S}^\Lambda_{{\rm B},\alpha\beta}(t)=\sum_l\mathrm{e}^{-\ii l\omega_dt}\,S^\Lambda_{{\rm B},\alpha\beta l}$. Hence, using the extended space calculus~\cite{Restrepo_Driven_2019,Restrepo_Quantum_2018},
\begin{equation}
    \begin{split}
        {S}^\Lambda_{{\rm B},\alpha\beta l}
        &=\langle\langle\alpha\vert F_{-l}\otimes S^\Lambda_{\rm B}\vert\beta\rangle\rangle.
    \end{split}
\end{equation}
With the spectral decomposition of the system operators, one obtains the system-bath interaction Hamiltonian in the interaction picture as
\begin{equation}
    \begin{split}
        \widetilde{H}_\mathrm{SB}(t)&=\sum_{\Lambda}\sum_\Delta \mathrm{e}^{-\ii\Delta t}\,{S}^{\Lambda}_{\rm B}(\Delta)\otimes \widetilde{B}^\Lambda(t),
    \end{split}
    \label{}
\end{equation}
and due to the stationary condition $\langle\widetilde{B}^\Lambda(t)\rangle_\mathrm{B}=\mathrm{tr}\lcb\widetilde{B}^\Lambda(t)\rho_\mathrm{B}\rcb=0$.

Inserting these operators into the Redfield equation, we obtain
$\partial_t\widetilde{\rho}(t)=\widetilde{\mathcal{L}}_\mathrm{diss}(t)\widetilde{\rho}(t)$,
where the dissipative Liouvillian term in the interaction picture reads:
\vspace{0.3cm}
\begin{widetext}
    \begin{equation}
    \begin{split}
      \widetilde{\mathcal{L}}_\mathrm{diss}(t)\widetilde{\rho}(t)
      &=\sum_{\Lambda\Lambda'}\sum_{\Delta\Delta'}\mathrm{e}^{-\ii(\Delta-\Delta') t}\,\left[\Gamma_{\Lambda'\Lambda}(\Delta)\,\lcb S_{\rm B}^{\Lambda}(\Delta) \widetilde{\rho}(t)S_{\rm B}^{\Lambda'\dagger}(\Delta')
        -S_{\rm B}^{\Lambda'\dagger}(\Delta')S_{\rm B}^\Lambda(\Delta)\widetilde{\rho}(t)
        \rcb\right.
        \\
        &\hspace{2.0cm}\left. +\Gamma_{\Lambda\Lambda'}^*(\Delta')\lcb S_{\rm B}^\Lambda(\Delta)\widetilde{\rho}(t)S_{\rm B}^{\Lambda'\dagger}(\Delta')
- \widetilde{\rho}(t)S_{\rm B}^{\Lambda'\dagger}(\Delta')S_{\rm B}^\Lambda(\Delta)\rcb\right],
    \end{split}
    \label{}
\end{equation} 
\end{widetext}
with $\Delta'\equiv\Delta_{\alpha'\beta' l'}=\varepsilon_{\beta'}-\varepsilon_{\alpha'}+l'\omega_d$, and
\begin{equation}
    \begin{split}
\Gamma_{\Lambda\Lambda'}(\Delta)&=\int_0^\infty ds\,\mathrm{e}^{\ii\Delta s}\,C_{\Lambda\Lambda'}(s),
    \end{split}
    \label{}
\end{equation}
or equivalently, $\Gamma_{\Lambda\Lambda'}(\Delta)=2\pi\lvert g_{\Lambda\Lambda'}(\Delta)\rvert^2 {D}_{\Lambda\Lambda'}(\Delta)$, where $g$ is the system-bath coupling  and ${D}$ is the bath density of states (DOS).
Note here that, in the F-GME, the \emph{full secular approximation} means $\Delta=\Delta'$ and the \emph{partial secular approximation} means only $l=l'$.

We now take the GME to the Schrodinger picture without any secularization.
Applying the canonical time transformation to the system operators and knowing that~\cite{Restrepo_Driven_2019}
$U(t)[\sum_{\Delta}\mathrm{e}^{-\ii\Delta t}\,{S}_{\rm B}^{\Lambda}(\Delta)] U^{-1}(t)=S_{\rm B}^{\Lambda}$
and 
\begin{equation}
    \begin{split}
        &U(t)\lcb\sum_{\Delta}\mathrm{e}^{-\ii\Delta t}\,\Gamma_{\Lambda'\Lambda}(\Delta){S}_{\rm B}^{\Lambda}(\Delta)\rcb U^{-1}(t)
\\
&\hspace{2.5cm}=\sum_{\alpha\beta l}\mathrm{e}^{-\ii l\omega_d t}\,\Gamma_{\Lambda'\Lambda}(\Delta){S}_{\rm B}^{\Lambda}(\Delta;t),
    \end{split}
    \label{}
    \newline
\end{equation}
 \vspace{0.3cm}
with 
\begin{equation}
    \begin{split}
        {S}_{\rm B}^{\Lambda}(\Delta;t)
        &= {S}^{\Lambda}_{{\rm B},\alpha\beta l}\,\vert \alpha(t)\rangle\langle \beta(t)\vert
        .
    \end{split}
    \label{S_Delta_t}
\end{equation}

This yields the \emph{Floquet-Markov generalized master equation} 
\begin{equation}
\partial_t{\rho}(t)=\mathcal{L}(t)\rho(t),
\end{equation}
where 
\begin{equation}\mathcal{L}(t)\rho(t)=\mathcal{L}_{\rm S}(t)\rho(t)+\mathcal{L}_{\rm diss}(t)\rho(t),
\end{equation} 
with
\begin{equation}
    \begin{split}
        \mathcal{L}_{\rm S}(t)\rho(t)&=-\ii\lsb \mathcal{H}_\mathrm{DQRM}(t),\rho(t)\rsb,
            \end{split}
    \label{}
\end{equation}
and the dissipator is obtained from
\begin{widetext}
\begin{equation}
    \begin{split}
      {\mathcal{L}}_\mathrm{diss}(t){\rho}(t)  
&=\sum_{\Lambda\Lambda'}\sum_{\Delta\Delta'}\left\{\mathrm{e}^{-\ii l\omega_d t}\,\Gamma_{\Lambda'\Lambda}(\Delta)\,\lsb S_{\rm B}^{\Lambda}(\Delta;t) {\rho}(t)S_{\rm B}^{\Lambda'\dagger}(\Delta')
        -S_{\rm B}^{\Lambda'\dagger}(\Delta')S_{\rm B}^\Lambda(\Delta;t){\rho}(t)
        \rsb\right.
        \\
        &\hspace{2.0cm}\left. +\mathrm{e}^{\ii l'\omega_d t}\,\Gamma_{\Lambda\Lambda'}^*(\Delta')\lsb S_{\rm B}^\Lambda(\Delta){\rho}(t)S_{\rm B}^{\Lambda'\dagger}(\Delta';t)
- {\rho}(t)S_{\rm B}^{\Lambda'\dagger}(\Delta';t)S_{\rm B}^\Lambda(\Delta)\rsb\right\}.
    \end{split}
    \label{DissipativeLindblad}
\end{equation}
\end{widetext}

We highlight that ${S}_{\rm B}^{\Lambda}(\Delta;t)$ is the time-evolved version of the operator ${S}_{\rm B}^{\Lambda}(\Delta)$ with only the system model (within one period) time-evolution operator, not by the total system plus bath time-evolution operator.
Also, the spectral decomposition of the (time-independent) system operator in the Schrodinger picture reads $S^\Lambda_{\rm B}=\widetilde{S}^\Lambda_{\rm B}(t=0)=\sum_{\Delta}S^\Lambda_{\rm B}(\Delta)$.

It is useful to identify several symmetries  and discuss them  in more detail. By comparing Eqs.~\eqref{S_Delta} and \eqref{S_Delta_t} with their Hermitian conjugate, we
find $S_{\rm B}^{\Lambda\dagger}(\Delta) = S^\Lambda_{\rm B}(-\Delta)$. Furthermore, by summing $S_{{\rm B}, l}(\omega)\equiv S_{\rm B}(\Delta)$ over all $\omega\equiv\varepsilon_\alpha-\varepsilon_\beta$ (quasienergy levels difference), we have $\sum_\omega S^\Lambda_{{\rm B},l}(\omega)=S^\Lambda_{{\rm B},l}$ with the property $S^{\Lambda\dagger}_{{\rm B},l}= S^\Lambda_{{\rm B},-l}$, and by summing $S^\Lambda_{{\rm B},l}(\omega)$ over all $l$'s (sidebands), we have $\sum_l S^\Lambda_{{\rm B},l}(\omega)=S^\Lambda_{{\rm B}}(\omega)$ with the property $S^{\Lambda\dagger}_{{\rm B}}(\omega)= S^\Lambda_{{\rm B}}(-\omega)$.
For $\Delta>0$, $S^\Lambda_{{\rm B}}(\Delta)$ is a net positive-frequency operator that
takes the system from a state (in Floquet extended basis) with lower energy to one with higher energy. Conversely, for $\Delta<0$, $S^\Lambda_{{\rm B}}(\Delta)$ is a net negative-frequency operator which produces a transition to a lower-energy state. In the following, to emphasize these properties, we introduce the notation~\cite{Settineri_Dissipation_2018}
\begin{equation}
    \begin{split}
        S^{\Lambda+}_{{\rm B}}(\Delta)&=S^\Lambda_{{\rm B}}(\Delta),\,\,\,\,\,\Delta>0,\\
        S^{\Lambda-}_{{\rm B}}(\Delta)&=S^\Lambda_{{\rm B}}(-\Delta),\,\,\,\,\,\Delta>0,\\
        S^{\Lambda(0)}_{{\rm B}}(\Delta)&=S^\Lambda_{{\rm B}}(\Delta),\,\,\,\,\,\Delta=0. \\
        &\phantom{gg}
    \end{split}
    \label{}
\end{equation}
We  see that, simply, 
$S^\Lambda_{{\rm B}}(-\Delta)=S^{\Lambda\dagger}_{{\rm B}}(\Delta)$ and $S^{\Lambda-}_{{\rm B}}(\Delta)=[S^{\Lambda+}_{{\rm B}}(\Delta)]^\dagger$.

Expanding Eq.~\eqref{DissipativeLindblad}, we obtain terms oscillating at frequencies $\pm(\Delta\pm\Delta')$ arising from products of ${S}_{{\rm B}}^{\Lambda\pm}(\Delta;t)$ and ${S}_{{\rm B}}^{\Lambda'\mp}(\Delta')$. We also obtain terms oscillating at frequencies $\pm\Delta$, $\mp\Delta'$ arising from products of $S^{\Lambda\pm}(\Delta;t)$ or $S_{{\rm B}}^{\Lambda'\pm}(\Delta)$ with $S_{{\rm B}}^{(0)}$, and non-oscillating terms arising from products between zero-frequency operators $S^{\Lambda(0)}_{{\rm B}}$ and $S^{\Lambda'(0)}_{{\rm B}}$~\cite{Settineri_Dissipation_2018}. Moreover, considering a system with well-separated energy levels ($\Delta\gg\Gamma$), the terms oscillating at $\pm(\Delta+\Delta')$, $\pm\Delta$ and $\pm\Delta'$ can be considered as rapidly oscillating and can be neglected. Moreover, the spectral function $\Gamma^{\Lambda\Lambda'}(\Delta)$ is only nonzero for positive arguments. 
This means we consider no upward rates (no absorption), such as a $T\approx 0$ bath, and also, each downward transition frequency 
$\Delta>0$, the corresponding spectral rate is positive (the environment only causes emission, never excitation).
Therefore, it picks up the system operators only with $\Delta>0$. 

Including only those terms providing non-negligible contributions to the dynamics, the Liouvillian in Eq.~\eqref{DissipativeLindblad} can then be written as
\vspace{0.2cm}
\begin{widetext}
\begin{equation}
    \begin{split}
      {\mathcal{L}}_\mathrm{diss}(t){\rho}(t)          &=\sum_{\Lambda\Lambda'}\sum_{\Delta,\Delta'>0}\left\{\mathrm{e}^{-\ii l\omega_d t}\,\Gamma_{\Lambda'\Lambda}(\Delta)\,\lsb S_{\rm B}^{\Lambda+}(\Delta;t) {\rho}(t)S_{\rm B}^{\Lambda'-}(\Delta')
        -S_{\rm B}^{\Lambda'-}(\Delta')S_{\rm B}^{\Lambda+}(\Delta;t){\rho}(t)
        \rsb\right.
        \\
        &\hspace{2.0cm}\left. +\mathrm{e}^{\ii l'\omega_d t}\,\Gamma_{\Lambda\Lambda'}^*(\Delta')\lsb S_{\rm B}^{\Lambda+}(\Delta){\rho}(t)S_{\rm B}^{\Lambda'-}(\Delta';t)
- {\rho}(t)S_{\rm B}^{\Lambda'-}(\Delta';t)S_{\rm B}^{\Lambda+}(\Delta)\rsb\right\}.
    \end{split}
    \label{Dissipator_FGME}
\end{equation}
\end{widetext}

Assuming that the bath correlation functions are for the independent baths, they are diagonal in the bath index so that: $C_{\Lambda\Lambda'}(s)=\delta_{\Lambda\Lambda'}C_\Lambda(s)$ and $\Gamma_{\Lambda\Lambda'}(\omega)=\delta_{\Lambda\Lambda'}\Gamma_\Lambda(\omega)$, which yields precisely the F-GME used in the main text.

We note that as assumed $S^{\Lambda}_{\rm B}=S^{\Lambda\dagger}_{\rm B}$, we see that $S^\Lambda_{{\rm B},\alpha\beta l}=S_{{\rm B},\beta\alpha,-l}^{\Lambda*}$. Accordingly, $\Delta_{\alpha\beta l}=\varepsilon_\beta-\varepsilon_\alpha+l\omega_d=-\Delta_{\beta\alpha,-l}$. Hence, $\Delta\to-\Delta$ means $S_{\rm B}(\Delta)\to S_{\rm B}(-\Delta)=S^\dagger_{\rm B}(\Delta)$ and $S^\Lambda_{\rm B}(\Delta;t)\to S^\Lambda_{\rm B}(-\Delta;t)=S^{\Lambda\dagger}_{\rm B}(\Delta;t)$.

\subsection{Periodicity of the Liouville superoperator and numerical implementation of the F-GME}
Assuming that the system dynamics is periodic, we can write
\begin{equation}
    \begin{split}
        \mathcal{L}_{\rm S}(t)\rho(t)&=-\ii\lsb \mathcal{H}_\mathrm{DQRM}(t),\rho(t)\rsb
        \\
        &=\sum_m\lcb-\ii\lsb \mathcal{H}_m,\rho(t)\rsb\rcb\,\mathrm{e}^{-\ii m\omega_d t}.
            \end{split}
    \label{}
\end{equation}
Hence, $\mathcal{L}_{\rm S}(t)\rho(t)=\sum_m\mathcal{L}_{{\rm S},m}(t)\rho(t)$, with
\begin{equation}
    \begin{split}
        \mathcal{L}_{{\rm S},m}(t)\rho(t)
        &=-\ii\lsb \mathcal{H}_m,\rho(t)\rsb.
            \end{split}
    \label{}
\end{equation}

The  super-operator (dissipator) is $T$-periodic so that it can be expanded as ${\mathcal{L}}_\mathrm{diss}(t)\rho(t)=\sum_n\mathrm{e}^{-\ii n\omega_dt}{\mathcal{L}}_{\mathrm{diss},n}\rho(t)$. To find the dissipator's Fourier component ${\mathcal{L}}_{\mathrm{diss},n}$, we first expand the time-dependent system-bath coupling operator in the Fourier series as $S^\Lambda_{{\rm B}}(\Delta;t)=\sum_n\mathrm{e}^{-\ii n \omega_dt}\,S^\Lambda_{{\rm B},n}(\Delta)$, with
\begin{equation}
    \begin{split}
      S^\Lambda_{{\rm B},n}(\Delta) 
        &=\sum_{l'}S^\Lambda_{{\rm B},\alpha\beta l}\,\lvert\alpha_{l'}\rangle\langle\beta_{l'-n}\rvert=\sum_{l'}S^\Lambda_{{\rm B},\alpha\beta l}\lvert\alpha_{l'}\rangle\langle\beta_{l'}^{[n]}\rvert.
    \end{split}
    \label{}
\end{equation}

Then, we can use this in Eq.~\eqref{Dissipator_FGME} to obtain
\begin{equation}
    \begin{split}
          {\mathcal{L}}_{\mathrm{diss}}(t){\rho}(t) &=\sum_{\Lambda\Lambda'}\sum_{\Delta\Delta'}\sum_{kk'}\lcb\mathrm{e}^{-\ii(l+k)\omega_dt}\,\Gamma_{\Lambda'\Lambda}(\Delta)\right.
          \\
          &\hspace{-1cm}\times\lsb{S}_{{\rm B},k}^{\Lambda}(\Delta) {\rho}(t){S}_{\rm B}^{\Lambda'\dagger}(\Delta')
          -{S}_{\rm B}^{\Lambda'\dagger}(\Delta'){S}_{{\rm B},k}^{\Lambda}(\Delta){\rho}(t)\rsb 
          \\
          &\hspace{2cm}+\mathrm{e}^{\ii(l+k')\omega_dt}\,\Gamma^*_{\Lambda\Lambda'}(\Delta')
          \\
          &\hspace{-1cm}\left.\times\lsb{S}_{{\rm B}}^{\Lambda}(\Delta) {\rho}(t){S}_{{\rm B},k'}^{\Lambda'\dagger}(\Delta')
          -{\rho}(t){S}_{{\rm B},k'}^{\Lambda'\dagger}(\Delta'){S}_{{\rm B}}^{\Lambda}(\Delta)\rsb \rcb.
    \end{split}
    \label{Dissipator_FGME_FourierSeries}
\end{equation}

By setting  $l+k=n$ in the $k$-sum, and $l'+k'=-n$ in the $k'$-sum, and using  
\begin{equation}
    \begin{split}
        S^\Lambda_{{\rm B},{n-l}}(\Delta) 
        &=\sum_{l'}S^\Lambda_{{\rm B},\alpha\beta l}\,\lvert\alpha_{l'}\rangle\langle\beta_{l+l'-n}\rvert=\sum_{l'}S^\Lambda_{{\rm B},\alpha\beta l}\lvert\alpha_{l'}\rangle\langle\beta_{l'}^{[n-l]}\rvert
        \\
        &=\sum_{l'}S^\Lambda_{{\rm B},\alpha\beta l}\,\lvert\alpha_{l'-l}\rangle\langle\beta_{l'-n}\rvert=\sum_{l'}S^\Lambda_{{\rm B},\alpha\beta l}\lvert\alpha_{l'}^{[l]}\rangle\langle\beta_{l'}^{[n]}\rvert,
    \end{split}
    \label{}
\end{equation}
according to Eq.~\eqref{FSB_differentBZs},we obtain the final form of the dissipator's Fourier components. We next apply the operators' positive and negative decomposition, to write
\begin{widetext}
\begin{equation}
    \begin{split}
      {\mathcal{L}}_{\mathrm{diss},n}\rho(t)
        &=\sum_{\Lambda\Lambda'}\sum_{\Delta\Delta'}
        \lcb\Gamma_{\Lambda'\Lambda}(\Delta)\lsb{S}_{{\rm B},n-l}^{\Lambda+}(\Delta){\rho}(t){S}^{\Lambda'-}_{{\rm B}}(\Delta')-{S}^{\Lambda'-}_{\rm B}(\Delta'){S}_{{\rm B},n-l}^{\Lambda+}(\Delta){\rho}(t)\rsb\right.
        \\
        &\hspace{5cm}\left.+\Gamma^*_{\Lambda\Lambda'}(\Delta')\lsb{S}^{\Lambda+}_{\rm B}(\Delta){\rho}(t){S}_{{\rm B},-l'-n}^{\Lambda'-}(\Delta')-{\rho}(t){S}_{{\rm B},-l'-n}^{\Lambda'-}(\Delta'){S}^{\Lambda+}_{\rm B}(\Delta)\rsb\rcb,
    \end{split}
    \label{Dissipator_FGME_n}
\end{equation}
\end{widetext}
where
\begin{equation}
    \begin{split}
        S^{\Lambda+}_{\mathrm{B},{n-l}}(\Delta) 
        &=\sum_{k}S^\Lambda_{{\rm B},\alpha\beta l}\,\lvert\alpha_{k}\rangle\langle\beta_{l+k-n}\rvert
        \\
        &=\sum_{k}S^\Lambda_{{\rm B},\alpha\beta l}\,\lvert\alpha_{n-l+k}\rangle\langle\beta_{k}\rvert,\,\,\,\,\,\Delta>0,
    \end{split}
    \label{}
\end{equation}
and 
\begin{equation}
    \begin{split}
        S^{\Lambda+}_{{\rm B},{-l-n}}(\Delta) 
        &=\sum_{k}\lp S^\Lambda_{{\rm B},\alpha\beta l}\rp^*\,\lvert\beta_{k+n+l}\rangle\langle\alpha_{k}\rvert,\,\,\,\,\,\Delta>0.
    \end{split}
    \label{}
\end{equation}

With this, we finalize the expression for the F-GME in the original time-independent eigenstates of the $\mathcal{H}_0$ Hamiltonian. This form is convenient for numerical implementation, particularly, in QuTiP~\cite{Johansson_Qutip_2012,Johansson_Qutip_2013,Lambert_Qutip5_2026}. However, one can also express the GME in the Floquet basis.

\subsection{Floquet GME in the Floquet basis}
The fact that the Floquet states $\lvert\psi_\alpha(t)\rangle$ solve the Schrödinger equation allows for a formal simplification of the above master equation. We can represent the density operator in this basis, i.e.,
\begin{equation}
    \begin{split}
    \rho(t)&=\langle\psi_\alpha(t)\vert\rho(t)\vert\psi_\beta(t)\rangle\vert\psi_\alpha(t)\rangle\langle\psi_\beta(t)\vert\\
&=\rho_{\alpha\beta}(t)\,\vert \alpha(t)\rangle\langle \beta(t)\vert
,
\end{split}
    \label{}
\end{equation}
with $\rho_{\alpha\beta}(t)=\langle \alpha(t)\vert\rho(t)\vert \beta(t)\rangle$. Then, the GME reads~\cite{Hausinger_Dissipative_2010}
\begin{equation}
    \begin{split}
        \partial_t\rho_{\alpha\beta}(t)
        &=-\ii (\varepsilon_\alpha-\varepsilon_\beta)\rho_{\alpha\beta}(t)+\langle \alpha(t)\vert \mathcal{L}_{\rm diss}(t)\rho(t)\vert \beta(t)\rangle,
        \end{split}
    \label{}
\end{equation}
where
\begin{widetext}
\begin{equation}
    \begin{split}
    \langle\lambda(t)\vert \mathcal{L}_{\rm diss}(t)\rho(t)\vert\lambda'(t)\rangle
&=\sum_{\Lambda\Lambda}\sum_{\alpha\beta l}\sum_{\alpha'\beta'l}\sum_\sigma\Big\{
\\
&\kern-1em\mathrm{e}^{-\ii l\omega_{d}t}\Gamma^{\Lambda'\Lambda*}(\Delta)\lsb U^{{\rm F}}_{\lambda\alpha}(t){S}^{\Lambda+}_{\alpha\beta l} U^{{\rm F}\dagger}_{\beta\sigma}(t) \rho_{\sigma\alpha'}(t) {S}^{\Lambda'-}_{\alpha'\beta' l'}\delta_{\beta'\lambda'}-\delta_{\lambda\alpha'}{S}^{\Lambda'-}_{\alpha'\beta' l'}U^{{\rm F}}_{\beta'\alpha}(t){S}_{\alpha\beta l}^{\Lambda+}U^{{\rm F}\dagger}_{\beta\sigma}(t){\rho}_{\sigma\lambda'}(t) \rsb
\\
&\kern-1em+\mathrm{e}^{\ii l'\omega_{d}t}
\Gamma^{\Lambda\Lambda'}(\Delta')\lsb \delta_{\lambda\alpha}{S}^{\Lambda+}_{\alpha\beta l}{\rho}_{\beta\sigma}(t)U^{{\rm F}\dagger}_{\sigma\alpha'}(t){S}^{\Lambda'-}_{\alpha'\beta' l'}U^{{\rm F}}_{\beta'\lambda'}(t)- \rho_{\lambda\sigma}(t) U^{{\rm F}\dagger}_{\sigma\alpha'}(t){S}^{\Lambda'-}_{\alpha'\beta' l'}U^{{\rm F}}_{\beta'\alpha}(t) {S}^{\Lambda+}_{\alpha\beta l} \delta_{\beta\lambda'}\rsb\Big\}.
    \end{split}
    \label{}
\end{equation}
\end{widetext}

Consequently, the dissipator can be written in the form,  
\begin{equation}
\mathcal{L}_{\rm diss}(t)\rho(t)
=\sum_{\lambda\lambda'\nu\nu'}\mathcal{L}_{\lambda\lambda',\sigma\sigma'}(t)\rho_{\sigma\sigma'}(t)\,\lvert\lambda(t)\rangle\langle\lambda'(t)\rvert .
\end{equation}

The master equation now takes the general form of 
\begin{equation}
\partial_t\rho_{\lambda\lambda'}(t)=\lsb-\ii (\varepsilon_\lambda-\varepsilon_{\lambda'})\delta_{\lambda\sigma}\delta_{\lambda'\sigma'}+\mathcal{D}_{\lambda\lambda',\sigma\sigma'}(t)\rsb\rho_{\sigma\sigma'}(t),
\end{equation}
where
\begin{widetext}
\begin{equation}
    \begin{split}
       \hspace{-0.2cm} \mathcal{D}_{\lambda\lambda',\nu\nu'}(t)
&=\sum_{\Lambda\Lambda}\sum_{\alpha\beta l}\sum_{\alpha'\beta'l}\sum_{\sigma\sigma'}\Big\{
\mathrm{e}^{-\ii l\omega_{d}t}\Gamma^{\Lambda'\Lambda*}(\Delta)\lsb U^{{\rm F}}_{\lambda\alpha}(t){S}^{\Lambda+}_{\alpha\beta l} U^{{\rm F}\dagger}_{\beta\sigma}(t) \delta_{\sigma\nu}\delta_{\nu'\alpha'} {S}^{\Lambda'-}_{\alpha'\beta' l'}\delta_{\beta'\lambda'}-\delta_{\lambda\alpha'}{S}^{\Lambda'-}_{\alpha'\beta' l'}U^{{\rm F}}_{\beta'\alpha}(t){S}_{\alpha\beta l}^{\Lambda+}U^{{\rm F}\dagger}_{\beta\sigma}(t)\delta_{\sigma\nu}\delta_{\nu'\lambda'}\rsb
\\
&\hspace{0.5cm}+\mathrm{e}^{\ii l'\omega_{d}t}
\Gamma^{\Lambda\Lambda'}(\Delta')\lsb \delta_{\lambda\alpha}{S}^{\Lambda+}_{\alpha\beta l}\delta_{\beta\nu}\delta_{\nu'\sigma}
U^{{\rm F}\dagger}_{\sigma\alpha'}(t){S}^{\Lambda'-}_{\alpha'\beta' l'}U^{{\rm F}}_{\beta'\lambda'}(t)- \delta_{\lambda\nu}\delta_{\nu'\sigma} U^{{\rm F}\dagger}_{\sigma\alpha'}(t){S}^{\Lambda'-}_{\alpha'\beta' l'}U^{{\rm F}}_{\beta'\alpha}(t) {S}^{\Lambda+}_{\alpha\beta l} \delta_{\beta\lambda'}\rsb\Big\}.
    \end{split}
    \label{LindbladElements_F}
\end{equation}
\end{widetext}

Conveniently, the matrix form of the Floquet-Markov GME,  in the Floquet basis, reads 
\begin{equation}
    \begin{split}
\dot{{\varrho}}(t)
&=-\ii[\mathcal{H}_{\rm F},{\varrho}(t)]
\\
&\hspace{0.25cm}
+\sum_{\Lambda\Lambda'}
\{\mathcal{S}^{\Lambda'-}_{U}(t)\,\varrho(t)\,\mathcal{S}^{\Lambda\Lambda'+}_{\rm B}(t)-\varrho(t)\,\mathcal{S}^{\Lambda\Lambda'+}_{\rm B}(t)\,\mathcal{S}^{\Lambda'-}_{U}(t)
\\
&\hspace{0.45cm}
+\mathcal{S}^{\Lambda\Lambda'-}_{\rm B}(t)\,\varrho(t)\,\mathcal{S}^{\Lambda+}_{U}(t)-\mathcal{S}^{\Lambda+}_{U}(t)\,\mathcal{S}^{\Lambda\Lambda'-}_{\rm B}(t)\,\varrho(t)\},
    \end{split}
    \label{eq:FGME_FloquetBasisRep}
\end{equation}
where ${\varrho}(t)$ is a matrix with the matrix elements of $\rho(t)$ in the Floquet basis, i.e., 
\begin{equation}
    {\varrho}(t)\equiv[\rho_{\alpha\beta}(t)].
\end{equation}

Similarly, 
\begin{equation}
    \begin{split}
       &\mathcal{H}_\mathrm{F}\equiv\mathrm{diag}\{\varepsilon_\alpha\}\quad \text{(only in the first BZ)}, 
       \\
       &\mathcal{S}^{\Lambda\Lambda'+}_{\rm B}(t)\equiv[\sum_{l}\mathrm{e}^{\ii l\omega_{d}t}
\Gamma^{\Lambda'\Lambda*}(\Delta)S^{\Lambda}_{{\rm B},\alpha\beta l}],
\quad \Delta>0, 
\\
&\mathcal{S}^{\Lambda\Lambda'-}_{\rm B}(t)=[\mathcal{S}^{\Lambda'\Lambda+}_{\rm B}(t)]^\dagger,
\quad \Delta>0,
\\
&\mathcal{S}^{\Lambda+}_{U}(t)\equiv [U^{{\rm F}\dagger}_{\alpha\beta}(t)][\sum_{l}S^\Lambda_{{\rm B},\alpha\beta l}][U^{\rm F}_{\alpha\beta}(t)],
\quad \Delta>0, 
    \end{split}
\end{equation}
and $U^{\rm F}(t)$ being the one-period time-evolution operator of the Floquet modes within the closed system, i.e., 
\begin{equation}
   \lvert\alpha(t)\rangle=U^{\rm F}(t)\lvert\alpha\rangle, 
\end{equation}
is the canonically transformed system operator in Floquet picture, and $\mathcal{S}^{\Lambda-}_{U}(t)=[\mathcal{S}^{\Lambda+}_{U}(t)]^\dagger$.

Finally, we also note that the operator, $S^{\Lambda\pm}_U(t)$, is periodic so that it can be written as 
\begin{equation}
\mathcal{S}^{\Lambda\pm}_U(t)=\sum_l\mathrm{e}^{\ii l\omega_dt}\,\mathcal{S}^{\Lambda\pm}_{U,l}, 
\end{equation}
where
$\mathcal{S}^{\Lambda\pm}_{U,l}
=({1}/{T})\int_0^T dt\,\mathrm{e}^{-\ii l\omega_dt}\,\mathcal{S}^{\Lambda\pm}_U(t)$. 
Alternatively, the Floquet matrix elements of the canonically transformed system operator in the expended space can be obtained via
\begin{equation}
    \begin{split}
\mathcal{S}^{\Lambda\pm}_{U,\alpha\beta l}&=\frac{1}{T}\int_0^T dt\,\mathrm{e}^{-\ii l\omega_dt}\,U^{{\rm F}\dagger}(t)[\sum_{n}S^{\Lambda\pm}_{{\rm B},\alpha\beta n}]U^{\rm F}(t)
\\
&=\frac{1}{T}\int_0^T dt\,\mathrm{e}^{-\ii l\omega_dt}\,\langle\alpha(t)\vert\lp\sum_{\alpha'\beta'l'}S^{\Lambda\pm}_{{\rm B},\alpha'\beta' l'}\vert\alpha'\rangle\langle\beta'\vert\rp\vert\beta(t)\rangle
\\
&=\Fbra{\alpha} F_{-l}\otimes\lp\sum_{\alpha'\beta'l'}S^{\Lambda\pm}_{{\rm B},\alpha'\beta' l'}\vert\alpha'\rangle\langle\beta'\vert\rp\Fket{\beta}.
    \end{split}
    \label{}
\end{equation}

With this relation, we can also write the whole dissipation term in a Fourier series, using the Floquet basis.

\subsection{Steady-state formalism}
We have seen that, in the Schrodinger picture, 
\begin{equation*}
  \rho(t)=\sum_{\alpha\beta} \mathrm{e}^{-\ii(\varepsilon_\alpha-\varepsilon_\beta)t}\,\rho_{\alpha\beta}(t)\,\lvert\alpha(t)\rangle\langle\beta(t)\rvert.  
\end{equation*}
 In general, for harmonic driving the density matrix can be written in the form of $\rho(t)=\sum_n\rho_n(t)\,\mathrm{e}^{\ii n\omega_dt}$; however, in the steady state the density matrix is diagonal in the
interaction picture~\cite{Restrepo_Driven_2019}, so that it is  pure periodic in the Schrodinger picture, so  $\rho_n(t)\to\rho_n$, and
\begin{equation}
\begin{split}
    \rho(t>t_{\rm ss})&=\sum_\alpha \rho_{\rm ss}\,\lvert\alpha(t)\rangle\langle\alpha(t)\rvert
    \\
    &=\sum_n\rho_n\,\mathrm{e}^{\ii\omega_dt}.
\end{split}
\end{equation}

Consequently, the F-GME at late times reads
\begin{equation}
    \begin{split}
        \ii n\omega_d\rho_n&=\sum_{n'}\mathcal{L}_{n'}\rho_{n-n'},
    \end{split}
    \label{GME_SteadyState}
\end{equation}
where $\rho_n$ and $\mathcal{L}_n$ are Fourier components of the density matrix in the long-time limit and the Liouvillian, respectively.

In the extended space, Eq.~\ref{GME_SteadyState} has the form
\begin{equation}
    \begin{split}
        \lsb\sum_n F_n\otimes\mathcal{L}_n-\ii\omega_d F_z\rsb\,\vec{\rho}=0,
    \end{split}
    \label{GME_SteadyState_ex}
\end{equation}
with ${\vec{\rho}}=[\rho_n]^T$ denoting a vector containing all Fourier components of $\rho(t)$ in the steady state. By going to the
extended space we have made the task of finding the long-time limit state of the F-GME,
time-independent, similar to the Floquet's eigenvalue problem to find the Floquet's quasienergies and modes, which is  discussed in the next section.

\section{Floquet--Liouville formalism for periodically driven quantum Rabi model}
\label{secS:FL}

We next summarize the Floquet--Liouville (FL) formalism 
[see Refs.~\onlinecite{Chu_Beyond_2004} and
\onlinecite{Akbari_Quasienergy-Resolved_2026}] used to compute periodic steady states, two-time correlation functions, spectra, and mode-resolved spectral decompositions. The FL modal decomposition also reveals the hybridization of the Floquet channels into FL modes. Starting from the nonsecular F-GME,
\begin{equation}
\partial_t \rho(t)=\mathcal{L}(t)\rho(t),
\qquad
\mathcal{L}(t+T)=\mathcal{L}(t),
\qquad
T=\frac{2\pi}{\omega_d},
\label{eq:FGME_periodic_app}
\end{equation}
we vectorize the density operator in Liouville space,
\begin{equation}
\rho(t)\rightarrow \Lket{\rho(t)}\in\mathbb{H}_{\rm L},
\qquad
\Lbraket{A}{\rho}=\Tr{A^\dagger\rho}.
\end{equation}
Here, $\mathbb{H}_{\rm L}$ is the Liouville (vector) space which takes the operators/matrices (linear transformations) in the physical space (here, the physical space is the dressed space $\mathbb{H}_{\rm dressed}$) as the vectors and the superoperators (supermatrices) in the physical Hilbert space, as the operators (matrices). 

The Liouville space is then spanned by $\mathsf{B}_{\rm L}=\lcb\Lket{jk}\equiv\ket{j}\bra{k}\rcb$, with the dimension $\dim(\mathbb{H}_{\rm L})=\dim(\mathbb{H}_{\rm dressed})\times\dim(\mathbb{H}_{\rm dressed})$.
The master equation then becomes
\begin{equation}
\partial_t \Lket{\rho(t)}
=
\mathscr{L}(t)\Lket{\rho(t)},
\qquad
\mathscr{L}(t)
=
\sum_{n\in\mathbb{Z}}
e^{-in\omega_d t}\mathscr{L}_n ,
\label{eq:Liouville_Fourier_app}
\end{equation}
where $\mathscr{L}(t)$ is the Liouville-space matrix representation of the time-periodic superoperator $\mathcal{L}(t)$.

\subsection{Floquet--Liouville eigenvalue problem}

Equation~\eqref{eq:Liouville_Fourier_app} is a linear differential equation with periodic coefficients, thus Floquet's theorem can be applied directly in Liouville space. Accordingly, the right FL modes are written as
\begin{equation}
\Lket{\rho_\mu(t)}
=
e^{\lambda_\mu t}\Lket{R_\mu(t)},
\qquad
\Lket{R_\mu(t+T)}=\Lket{R_\mu(t)},
\label{eq:FL_ansatz_app}
\end{equation}
which gives the time-domain eigenvalue problem,
\begin{equation}
\left[\mathscr{L}(t)-\partial_t\right]\Lket{R_\mu(t)}
=
\lambda_\mu \Lket{R_\mu(t)}.
\label{eq:FL_time_eig_app}
\end{equation}
Similarly, the corresponding left modes satisfy
\begin{equation}
\Lbra{L_\mu(t)}
\left[\mathscr{L}(t)-\partial_t\right]
=
\lambda_\mu \Lbra{L_\mu(t)}.
\label{eq:FL_left_time_app}
\end{equation}
Since the Liouvillian is generally non-Hermitian, left and right modes are chosen biorthonormal over one period, and satisfy
\begin{equation}
\frac{1}{T}\int_0^T dt\,
\Lbraket{L_\mu(t)}{R_\nu(t)}
=
\delta_{\mu\nu}.
\label{eq:FL_biorth_time_app}
\end{equation}

To obtain a time-independent eigenvalue problem, we introduce the temporal space spanned by $\{|n)\}$, with
\begin{equation}
(t|n)=\exp\lcb{-\ii n\omega_d t}\rcb.
\end{equation}
Similar to the conventional Floquet extended space, the FL extended space is
\begin{equation}
\mathbb{H}_{\rm ex}^{\rm FL}
=
\mathbb{H}_{\rm temp}\otimes\mathbb{H}_{\rm L}.
\end{equation}
This space is spanned by the basis set 
\begin{equation}
    \mathsf{B}_{\rm FL}=\lcb\FLket{l,jk}\equiv\tket{l}\otimes\ket{j}\bra{k}\rcb,
\end{equation} 
and thus, all other time-periodic Liouville-space vectors can be identified accordingly in the FL extended space.
The periodic modes are expanded as
\begin{equation}
\Lket{R_\mu(t)}
=
\sum_n e^{-\ii n\omega_d t}\Lket{R_{\mu,n}},
\ \
\Lbra{L_\mu(t)}
=
\sum_n e^{+\ii n\omega_d t}\Lbra{L_{\mu,n}},
\end{equation}
and represented in the extended space as
\begin{equation}
\FLket{R_\mu}
=
\sum_n |n)\otimes \Lket{R_{\mu,n}},
\qquad
\FLbra{L_\mu}
=
\sum_n (n|\otimes \Lbra{L_{\mu,n}}.
\end{equation}

Using the temporal shift and number operators, given in Eqs.~\eqref{eq:ShiftOp} and \eqref{eq:Ntemp}, respectively, the extended-space FL 
{\it supermatrix} is
\begin{equation}
{
\mathscr{L}_{\rm ex}^{\rm F}
=
\sum_{n\in\mathbb{Z}}
F_n\otimes \mathscr{L}_n
+
\ii\omega_d \mathcal{N}_{\rm temp}\otimes \mathbf{1}_{\rm L},
}
\label{eq:LFL_def_app}
\end{equation}
where $\mathbf{1}_{\rm L}$ is the identity (unit operator) in the Liouville space.
The right and left eigenvalue problems are, then,
\begin{equation}
\mathscr{L}_{\rm ex}^{\rm F}\FLket{R_\mu}
=
\lambda_\mu\FLket{R_\mu},
\qquad
\FLbra{L_\mu}\mathscr{L}_{\rm ex}^{\rm F}
=
\lambda_\mu\FLbra{L_\mu},
\label{eq:FL_eig_app}
\end{equation}
and the eigenvectors satisfy:
\begin{equation}
\FLbraket{L_\mu}{R_\nu}=\delta_{\mu\nu},
\qquad
\sum_\mu \FLket{R_\mu}\FLbra{L_\mu}
=
\mathbf{1}_{\rm FL},
\label{eq:FL_complete_app}
\end{equation}
where $\mathbf{1}_{\rm FL}$ is the indentity (unit operator) in the FL extended space.

As in ordinary Floquet theory, the spectrum is replicated in vertical strips:
\begin{equation}
\lambda_\mu^{[k]}=\lambda_\mu+\ii k\omega_d,
\qquad
k\in\mathbb{Z}.
\label{eq:FL_replicas_app}
\end{equation}
One can therefore fold the FL extended space eigenvalues and eigenvectors into FL Brillouin strips.
Yet, the FL extended space is complex and the modal decomposition in a particular computational setting might not organize the modes as one wishes, and certainly, is not as handy as it was in the Floquet-Hamiltonian extended space case. Hence, a more robust approach to identify BZs is that to, first, sort the FL eigenvalues, and then pick the middle sector of $\dim(\mathbb{H}_{\rm L})$ eigenvalues and their corresponding eigenvectors. Then, build a ladder of $\pm\ii l\omega_d$ for the consecutive BZs to find the other BZs' eigenvalues and then match their corresponding eigenvectors. 

Nevertheless, one can predict all the observables without worrying about the order of the BZs, but working in the full extended space. We note that the structure of the FL modes in the primary BZ is such that the coherence mode is the mode with $\lambda_\mu=0$ and the population mode is with $\mathrm{Re}[\lambda_\mu]=0$ ($\alpha=\beta$), and the other modes represent the transitions ($\alpha\neq\beta$). This trend is repeated by $\mathrm{Im}[\lambda_\mu]\pm\ii l\omega_d$ in the other BZs throughout the full extended space. 

We now write
\begin{equation}
\lambda_\mu=-\gamma_\mu-\ii\Delta_\mu,
\qquad
\gamma_\mu\ge 0,
\label{eq:lambda_gamma_omega_app}
\end{equation}
so that $\Delta_\mu$ gives the oscillation frequency of the FL mode, and $\gamma_\mu$ gives its decay rate or spectral linewidth.

\subsection{Periodic steady state and averaged observables}

\subsubsection{Floquet--Liouville modal decomposition of \(N(t)\)}
We let the density matrix in the FL extended space be expanded as
\begin{equation}
\FLket{\rho(t)}
=
\sum_{\mu}
c_{\mu}\,
e^{\lambda_{\mu}t}
\FLket{R_{\mu}},
\end{equation}
where
$c_{\mu}
=\FLbraket{L_{\mu}}{\rho(0)}$.
For an observable \(N\), one can define the corresponding bra,
\begin{equation}
\langle\!\langle\!\langle N|_{0}
=
(0|
\otimes
\langle\!\langle \mathbf{1}|\,N_{\rm L},
\end{equation}
where \((0|\) selects the zero Fourier component and $N_{\rm L}=\mathbf{1}_{\rm dressed}\otimes N$ if Liouville space representation of the dressed basis number operator. Then
\begin{equation}
N(t)
=
\FLbra{N}_{0}
\FLket{\rho(t)}
=
\sum_{\mu}
\mathcal{W}_{\mu}^{N}
e^{\lambda_{\mu}t},
\end{equation}
with modal weights
\begin{equation}
\mathcal{W}_{\mu}^{N}
=
\FLbra{N}_{0}
\FLket{R_{\mu}}
\,
\FLbraket{L_{\mu}}{\rho(0)}.
\end{equation}

\subsubsection{Long-time periodic steady state}
In the long-time limit, all modes with
\(\mathrm{Re}[\lambda_{\mu}]<0\) decay. The remaining neutral Floquet family satisfies
\begin{equation}
\lambda_{\mu_l}
\simeq
\ii l\omega_d ,
\qquad
l\in\mathbb{Z}.
\end{equation}
Thus, the periodic steady-state occupation is
\begin{equation}
N_{\rm ss}(t)
=
\sum_{l}
\mathcal{W}_{\mu_l}^{N}
e^{-\ii l\omega_d t}.
\end{equation}
Equivalently, the steady-state waveform is built from the neutral FL modes.

\subsubsection{Cycle-averaged occupation}
The long-time period-averaged occupation is 
\begin{equation*}
\overline{N}
=
\frac{1}{T}
\int_{0}^{T}
N_{\rm ss}(t)\,dt ,
\qquad
T=\frac{2\pi}{\omega_d}.
\end{equation*}
Substituting the neutral-family expansion gives
\begin{equation}
\overline{N}
=
\frac{1}{T}
\sum_l
\mathcal{W}_{\mu_l}^{N}
\int_T
e^{-\ii l\omega_d t}\,dt
=
\mathcal{W}_{\mu_0}^{N},
\end{equation}
where we used
\begin{equation*}
\frac{1}{T}
\int_T
e^{-\ii l\omega_d t}\,dt
=
\delta_{l0}.
\end{equation*}
Thus, the scalar quantity \(\overline{N}\) is determined only by the zero FL mode,
\begin{equation}
\lambda_{\mu_0}\simeq0 .
\end{equation}

Similar calculations for  the average number of excitations can be done directly in the extended space,  from a slightly different perspective. For a trace-preserving master equation, the identity defines the left trace mode,
\begin{equation}
\FLbra{\mathbf{1}_{\rm FL}}
=
(0|\otimes \Lbra{\mathbf{1}},
\qquad
\FLbra{\mathbf{1}_{\rm FL}}\mathscr{L}_{\rm ex}^{\rm F}=0.
\label{eq:trace_bra_app}
\end{equation}
The periodic steady state corresponds to the right zero mode,
\begin{equation}
\mathscr{L}_{\rm ex}^{\rm F}\FLket{R_{\rm ss}}=0,
\qquad
\FLbraket{\mathbf{1}_{\rm FL}}{R_{\rm ss}}=1,
\label{eq:ss_mode_app}
\end{equation}
and the physical steady state is reconstructed as
\begin{equation}
\Lket{\rho_{\rm ss}(t)}
=
\sum_n e^{-\ii n\omega_d t}\Lket{R_{{\rm ss},n}}.
\label{eq:ss_reconstruct_app}
\end{equation}

For a one-time observable $N$, the period-averaged expectation value is
\begin{equation}
\overline{N}
=
\frac{1}{T}\int_0^T dt\, \Tr{N\rho_{\rm ss}(t)}.
\label{eq:Nbar_def_app}
\end{equation}
Using Eq.~\eqref{eq:ss_reconstruct_app}, only the zeroth Fourier component contributes,
\begin{equation}
{
\overline{N}
=
\Lbraket{N}{R_{{\rm ss},0}}
=
\FLbraket{N_{\rm FL}}{R_{\rm ss}},
\qquad
\FLbra{N_{\rm FL}}=(0|\otimes\Lbra{N}.
}
\label{eq:Nbar_FL_app}
\end{equation}

Thus, averaged populations probe the zero-harmonic component of the periodic steady state. The steady-state mode $\FLket{R_{\rm ss}}$ in the FL extended space is the mode with the corresponding $\lambda_\mu=0$ eigenvalue (repeated throughout the extended space by the replica of $\ii l\omega_d$).
Whereas the population modes in each extended space are the modes with $\mathrm{Re}[\lambda_\mu]=0$ with the smae replicas of the imaginary parts, thus, the population modes, in general, have the form $\lambda_\mu^{[l]}=\mathrm{Im}[\lambda_\mu]+\ii l\omega_d$.
Note, for simplicity in the computation, wake a unified index $\mu$ for all the FL extended-space index $l\mu$.

\subsubsection{Mode-channel decomposition of the average number of excitation.}
Each FL right eigenmode can further be resolved into quasienergy-transition components as
\begin{equation}
\FLket{R_{\mu}}
=
\sum_{\alpha\beta l}
R_{\mu}^{\alpha\beta l}
|l)
\otimes
|\alpha\rangle\langle \beta| .
\label{eq:R_expand_ab_l_app}
\end{equation}
The normalized component weight is
\begin{equation}
r_{\mu}^{\alpha\beta l}
=
\frac{
|R_{\mu}^{\alpha\beta l}|^2
}{
\sum_{a'b'l'}
|R_{\mu}^{\alpha'\beta' l'}|^2
},
\qquad
\sum_{\alpha\beta l}r^{\alpha\beta l}_{\mu}=1.
\label{eq:w_ab_l_app}
\end{equation}
For \(\overline{N}\), the relevant mode is the zero mode \(\mu_0\). Since \(N\) is a population-like observable, its dominant zero-mode components are typically diagonal population channels,
\begin{equation}
(\alpha,\alpha,l),
\end{equation}
rather than transition coherences \((\alpha,\beta,l)\) with \(\alpha\neq \beta\). Hence, the decomposition of \(\FLket{R_{\mu_0}}\) reveals which Floquet-dressed populations build the final cycle-averaged occupation.

\subsection{Two-time correlations and emitted spectra}

Let $s^-$ and $s^+$ be system operators (for convenience we drop the superscript $\Lambda$ and subscript ${\rm B}$). The steady-state two-time correlator is evaluated using the quantum regression theorem,
\begin{equation}
G_{s^-s^+}^{\rm ss}(t,\tau)
=
\langle s^-(t+\tau)s^+(t)\rangle_{\rm ss}
=
\Lbra{s^-}\mathscr{U}(t+\tau,t)s^+\Lket{\rho_{\rm ss}(t)},
\label{eq:GAB_QRT_app}
\end{equation}
where $\mathscr{U}(t+\tau,t)$ is the (Liouville-space) propagator generated by $\mathscr{L}(t)$,
\begin{equation}
\partial_\tau \mathscr{U}(t+\tau,t)
=
\mathscr{L}(t+\tau)\mathscr{U}(t+\tau,t),
\qquad
\mathscr{U}(t,t)=\mathbf{1}_{\rm L}.
\end{equation}

The relevant stationary correlator is the period average:
\begin{equation}
\overline{G}_{s^-s^+}(\tau)
=
\frac{1}{T}\int_0^T dt\,
G_{s^-s^+}^{\rm ss}(t,\tau).
\label{eq:Gbar_app}
\end{equation}
Using the FL spectral decomposition of the propagator,
\begin{equation}
\mathscr{U}(t+\tau,t)
=
\sum_\mu
e^{\lambda_\mu\tau}
\Lket{R_\mu(t+\tau)}
\Lbra{L_\mu(t)},
\label{eq:U_FL_app}
\end{equation}
that is expressed in the Liouville space,
one obtains
\begin{equation}
\overline{G}_{s^-s^+}(\tau)
=
\sum_\mu e^{\lambda_\mu\tau}\mathcal{W}_{\mu},
\label{eq:Gbar_modes_app}
\end{equation}
with the residues
\begin{equation}
\mathcal{W}_{\mu}
=
\frac{1}{T}\int_T dt\,
\Lbraket{s^-}{R_\mu(t)}
\Lbraket{L_\mu(t)}{s^+\rho_{\rm ss}(t)}.
\label{eq:W_time_app}
\end{equation}
Equivalently, in the FL extended space,
\begin{equation}
{
\mathcal{W}_{\mu}
=
\FLbraket{s^-_{\rm FL}}{R_\mu}
\FLbraket{L_\mu}{s^+_{\rm FL}R_{\rm ss}},
\qquad
\FLbra{s^-_{\rm FL}}=(0|\otimes\Lbra{s^-},
}
\label{eq:W_FL_app}
\end{equation}
where $s^+_{\rm FL}$ denotes the lifted superoperator acting on the Liouville component of the extended vector.

The (stationary) emission spectrum is defined from
\begin{equation}
\mathsf{S}_{s^-s^+}(\omega)
\propto
\mathrm{Re}\int_0^\infty d\tau\,
e^{i\omega\tau}\overline{G}_{s^-s^+}(\tau),
\label{eq:S_def_app}
\end{equation}
and using Eq.~\eqref{eq:Gbar_modes_app}, we obtain
\begin{equation}
{
\mathsf{S}_{s^-s^+}(\omega)
\propto
\mathrm{Re}
\sum_\mu
\frac{\mathcal{W}_{\mu}}
{-\lambda_\mu-\ii\omega}.
}
\label{eq:S_FL_app}
\end{equation}

Moreover, using Eq.~\eqref{eq:lambda_gamma_omega_app}, this becomes
\begin{equation}
\mathsf{S}_{s^-s^+}(\omega)
\propto
\mathrm{Re}
\sum_\mu
\frac{\mathcal{W}_{\mu}}
{\gamma_\mu-\ii(\omega-\Delta_\mu)}.
\label{eq:S_res_app}
\end{equation}
Each FL mode therefore contributes a resonance centered at $\Delta_\mu$, with linewidth $\gamma_\mu$, and complex residue $\mathcal{W}_{\mu}$.

Writing
\begin{equation}
\mathcal{W}_{\mu}
=
\mathcal{W}^{({\rm r})}_{\mu}
+\ii\mathcal{W}^{({\rm i})}_{\mu},
\end{equation}
one may express the contributions explicitly as
\begin{equation}
\mathsf{S}_{s^-s^+}(\omega)
\propto
\sum_\mu
\frac{
\mathcal{W}^{({\rm r})}_{\mu}\gamma_\mu
+
\mathcal{W}^{({\rm i})}_{\mu}(\Delta_\mu-\omega)
}{
(\omega-\Delta_\mu)^2+\gamma_\mu^2
}.
\label{eq:S_explicit_app}
\end{equation}
The real part of the residue gives the symmetric Lorentzian contribution, while the imaginary part gives a dispersive correction. For an isolated resonance, the approximate peak-prominence metric used in the main text is
\begin{equation}
\mathcal{R}_\mu
\equiv
\frac{\mathcal{W}^{({\rm r})}_{\mu}}{\gamma_\mu},
\label{eq:Rmetric_app}
\end{equation}
where $\omega=\Delta_\mu$ is set in Eq.~\eqref{eq:S_explicit_app}.
When several modes overlap, however, the full sum in Eq.~\eqref{eq:S_explicit_app} must be retained and no unique peak height can be assigned to a single mode.

\subsection{Incoherent spectra and population modes}

The incoherent spectrum is obtained from fluctuation operators,
\begin{equation}
\delta s^-(t)=s^--\langle s^-(t)\rangle_{\rm ss},
\qquad
\delta s^+(t)=s^+-\langle s^+(t)\rangle_{\rm ss}.
\end{equation}

This yields
\begin{equation}
\overline{G}^{\,{\rm inc}}_{AB}(\tau)
=
\overline{G}_{s^-s^+}(\tau)
-
\frac{1}{T}\int_0^T dt\,
\langle s^-(t+\tau)\rangle_{\rm ss}
\langle s^+(t)\rangle_{\rm ss},
\label{eq:Ginc_sub_app}
\end{equation}
where subtraction removes the coherent elastic contribution associated with the periodic steady-state motion (of a CW drive).

In the FL spectrum, these coherent contributions appear as population-like modes, dominated by diagonal Floquet components $(\alpha=\beta)$ and harmonically locked frequencies $\Delta_\mu\simeq m\omega_d$. Numerically, such modes may be identified by small decay rates and harmonic frequency locking,
\begin{equation}
|\mathrm{Re}[\lambda_\mu]|<\epsilon_\gamma,
\qquad
\min_{m\in\mathbb{Z}}
|\mathrm{Im}[\lambda_\mu+\ii m\omega_d]|<\epsilon_\omega .
\label{eq:pop_mode_criteria_app}
\end{equation}

Denoting the set of such modes by $\mathcal{P}$, the incoherent spectrum can be constructed from
\begin{equation}
{
\mathsf{S}^{\rm inc}_{s^-s^+}(\omega)
\propto
\mathrm{Re}
\sum_{\mu\notin\mathcal{P}}
\frac{\mathcal{W}_{\mu}}
{-\lambda_\mu-\ii\omega}.
}
\label{eq:Sinc_app}
\end{equation}
This procedure removes spurious delta-like or nearly elastic harmonics and isolates the fluorescence signal produced by decaying coherence modes.

\subsection{Connection to quasienergy-resolved channels}
\subsubsection*{Mode-channel decomposition of the correlators and spectra}
The FL eigenmodes can be projected onto the natural operator basis generated by the Floquet states and temporal harmonics, $|l)\otimes|\alpha\rangle\langle\beta|$, according to Eq.~\eqref{eq:R_expand_ab_l_app}.
In the absence of dissipation, the basis component
$|l)\otimes|\alpha\rangle\langle\beta|$ corresponds to the bare quasienergy transition channel
$\Delta_{\alpha\beta l}
=
\varepsilon_\beta-\varepsilon_\alpha+l\omega_d$.
With dissipation included, the nonsecular FL Liouvillian generally couples different channels, so a mode $\mu$ need not correspond to a single triplet $(\alpha,\beta,l)$.

A useful normalized measure of the contribution of channel $(\alpha,\beta,l)$ to mode $\mu$ is, therfore, from Eq.~\eqref{eq:w_ab_l_app}.
Sorting these weights identifies the dominant quasienergy channels composing each FL mode. Later, in the Additional Results, we tabularize the dominant modes with their most contributing Floquet components. To do so, in the tables, we retain components satisfying:
\begin{equation}
r^{\alpha\beta l}_\mu
\ge \eta_{\rm cut},
\qquad
\eta_{\rm cut}=0.1\times\mathrm{max}[r^{\alpha'\beta' l'}_\mu],
\label{eq:w_cut_app}
\end{equation}
so that only contributions carrying at least $10\%$ of the modal norm are shown.

If a single component dominates, the FL frequency is approximately identified with the corresponding quasienergy gap,
\begin{equation}
\Delta_\mu \approx \Delta_{\alpha\beta l}.
\end{equation}
If several components have comparable weights, the mode represents a hybridized dissipative resonance formed from multiple Floquet channels. This hybridization is produced by nonsecular Liouvillian couplings and is enhanced when different quasienergy gaps are nearly resonant. Therefore, a spectral peak should generally be interpreted as an FL eigenmode rather than as an isolated quasienergy transition.

The diagonal components $\alpha=\beta$ correspond to population harmonics and coherent elastic response, while off-diagonal components $\alpha\neq\beta$ correspond to Floquet coherences that generate inelastic fluorescence sidebands. Thus, the projection weights $R^{\alpha\beta l}_{\mu}$ provide a practical diagnostic for identifying whether a given mode contributes mainly to averaged populations, coherent scattering, or incoherent emission.

\subsection{Practical details for numerical implementations}

The numerical procedure used in the main text follows these general guidelines:
\begin{enumerate}
\item Construction of the Fourier components $\mathscr{L}_n$ of the periodic Liouvillian;
\item Building the truncated FL matrix $\mathscr{L}_{\rm ex}^{\rm F}$ from Eq.~\eqref{eq:LFL_def_app};
\item Diagonalizing the non-Hermitian eigenvalue problem to obtain $\lambda_\mu$, $\FLket{R_\mu}$, and $\FLbra{L_\mu}$;
\item 
Biorthonormalize the left and right eigenvectors;
\item Identifying and normalizing the zero mode $\FLket{R_{\rm ss}}$;
\item Computing the averaged observables from Eq.~\eqref{eq:Nbar_FL_app};
\item Computing the residues from Eq.~\eqref{eq:W_FL_app} and spectra from Eq.~\eqref{eq:S_FL_app} for the total spectra or Eq.~\eqref{eq:Sinc_app} for the incoherent spectra; 
\item Projecting the FL modes onto $(\alpha,\beta,l)$ channels using Eq.~\eqref{eq:w_ab_l_app}, and grant that;
\item The same procedure can be applied to the TI-GME, with the only modification that there is only a zero-component in the Fourier expansion of the dissipative Liouvillian, $\mathcal{L}_{{\rm diss},0}=\mathcal{L}_{\rm diss}$ given in Eq.~\eqref{eq:Ldiss_TIGME}.
\end{enumerate}

This compact FL formulation provides the basis for the mode-resolved interpretation of the spectra in the main text: $\Delta_\mu$ gives the spectral position, $\gamma_\mu$ gives the effective linewidth, $\mathcal{W}_{\mu}$ gives the observable residue, and $R^{\alpha\beta l}_\mu$, with the effective weight $r^{\alpha\beta l}_\mu$, identifies the quasienergy-resolved channels forming each dissipative resonance.

\section{Additional Results and Insights}
\label{secS:AdditionalResults}
In this section, we provide some extra numerical results and calculations that expand and complement the  discussions and examples in the main text.

To perform calculations either in the main text or here, we use the following numerical specifications: $m_{\rm max}=1$ for the optical drive and $m_{\rm max}=20$ for the mechanical drive. To grant the numerical convergence, we use a minimum basis size of  
\begin{equation*}
\begin{split}
    &\dim(\mathbb{H}_{\rm dressed})=8\quad (j,\alpha=0,1^\pm,2^\pm,3^\pm,4^-),
    \\
    &l_{\rm max}=30\quad (l=-30,\dots,0,\dots,30),
\end{split}
\end{equation*} 
so that 
\begin{equation*}
    \begin{split}
        &\dim(\mathbb{H}_{\rm temp})=2l_{\rm max}+1=61,
        \\
        &\dim(\mathbb{H}_{\rm ex})=\dim(\mathbb{H}_{\rm temp})\times \dim(\mathbb{H}_{\rm dressed})=488,
        \\
        &\dim(\mathbb{H}_{\rm L})=64,
        \\
        &\dim(\mathbb{H}_{\rm ex}^{\rm FL})=3904\quad (\mu=0,\dots,3903).
    \end{split}
\end{equation*}

\subsection{Optical drive}
We first note that the
energy state labels based on their orders in Fig.~\ref{fig:EnergyBasis_Optical}(a), from bottom to top at $\eta=0.5$ (vertical dotted black line), read: 
$j=0,1^{-},2^{-},1^{+},3^{-},2^{+},4^{-},3^+$.
Likewise, the quasienergy state labels from their orders in the range $[-\omega_d/2,\omega_d/2)$ from Fig.~\ref{fig:EnergyBasis_Optical}(b), from bottom to top at $\eta_d=0.3$ (vertical dotted black line) are:
$\alpha=2^{-},1^{+},3^{-},3^{+},1^{-},2^{+},4^{-},0$, listed on the bottom-right corner of panel (c).

\subsubsection{Variation of the Floquet coherent pump process versus\\ the frequency of the optical drive $\omega_d$}
\begin{figure*}[htbp]
\includegraphics[width=.92\linewidth]{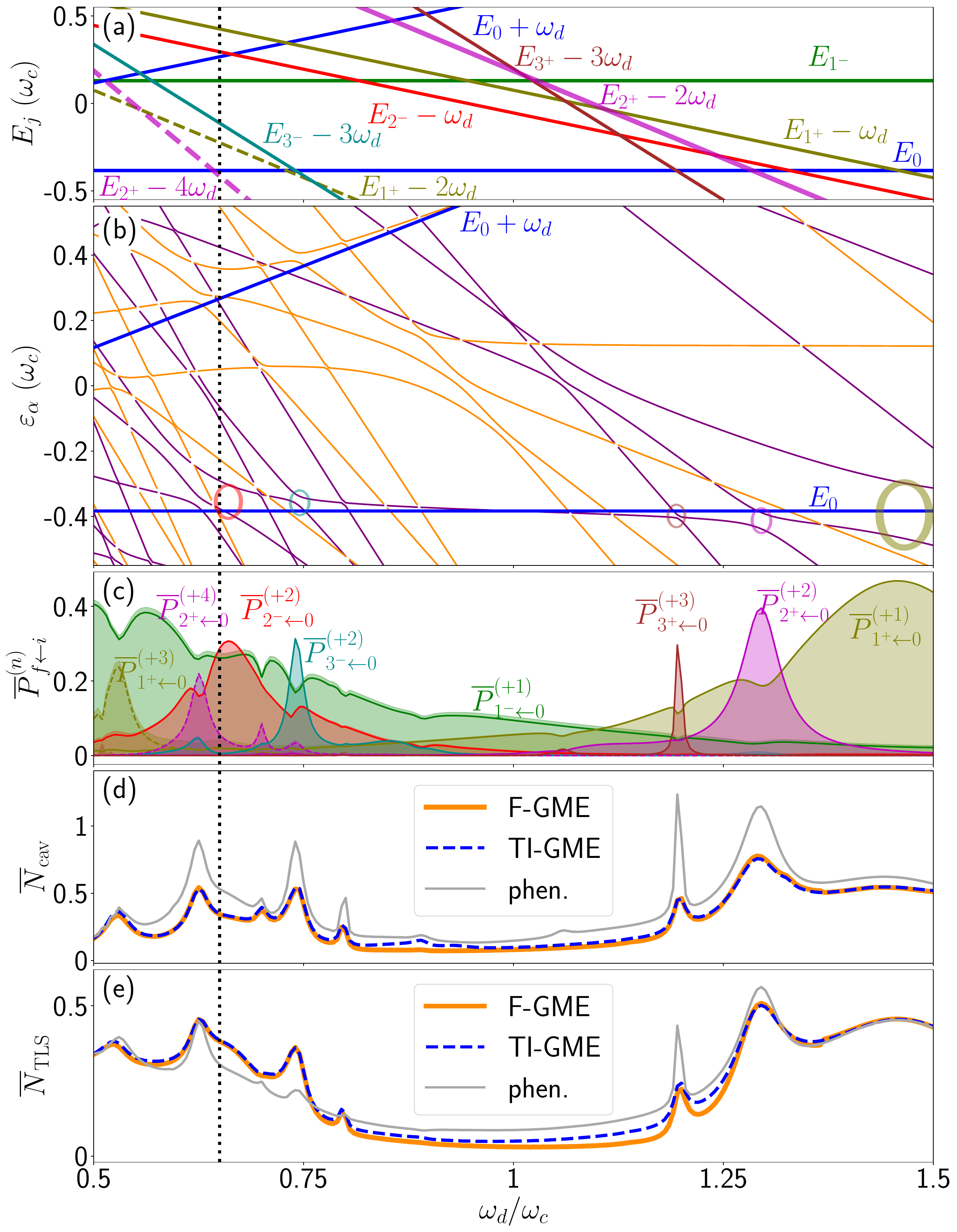}
\caption[]{\textbf{
Optically driven cavity-QED examples: Eigenenergies, transition probabilities and populations as a function of $\omega_d$.} (a) Shifted eigenenergies to show the initial crossings/anticrossings of the energy states that seed the quasienergy anticrossings. (b) Floquet quasienergies in the primary BZ; circles mark anticrossings associated with dominant drive-assisted excitation transitions. (c) Floquet transition probabilities for the most dominant upward externally-assisted (order-resolved) transitions corresponding to the anticrossings in panel (b); the shaded areas show the sum of all multi-photon transitions. (d,e) Long-time averaged cavity and TLS excitation numbers, respectively, obtained from the phenomenological Floquet treatment (thin gray), TI-GME (dashed blue), and F-GME (orange), for flat baths with small loss rates $\kappa=\gamma=0.01\omega_a$.
The other simulation parameters are $\omega_a=\omega_c$, $\eta=0.5$ (dipole-cavity coupling), and $\eta_d=0.3$ (optical drive strength).
}
    \label{figS:Fig_wd_Optical}
\end{figure*}
Figure~\ref{figS:Fig_wd_Optical} provides additional Floquet-resolved information for the optically driven cavity-QED system discussed in the main text. We consider the resonant USC case, $\omega_a=\omega_c$, with $\eta=0.5$ and optical drive strength $\eta_d=0.3$, and analyze the response as a function of the optical drive frequency $\omega_d$.

Panel (a) of Fig.~\ref{figS:Fig_wd_Optical} shows the relevant shifted dressed-state energies of the QRM, which identify the initial crossings and avoided crossings that seed the corresponding quasienergy anticrossings. Panel (b) shows the Floquet quasienergies in the primary Brillouin zone, with the highlighted anticrossings associated with the dominant drive-assisted (upward) transitions. In the zero-drive limit, the quasienergies reduce to the folded dressed energies,
$\varepsilon_{l\alpha}\rightarrow E_{j=\alpha}+l\omega_d$,
up to the usual quasienergy-zone redundancy. Turning on the optical drive hybridizes the corresponding Floquet replicas and produces avoided crossings, whose separation and curvature encode the strength of the associated multiphoton processes.

Panel (c) of Fig.~\ref{figS:Fig_wd_Optical} shows the long-time averaged transition probabilities $\overline{P}^{(n)}_{f\leftarrow i}$ for the dominant upward drive-assisted processes, where $n$ denotes the number of drive quanta exchanged with the system. The shaded same-color area represents the total transition probability for a corresponding transition, $\overline{P}_{f\leftarrow i}=\sum_n\overline{P}^{(n)}_{f\leftarrow i}$. 
The unitary Floquet probability diagnoses which dressed states are most strongly accessed by the coherent drive, independently of the choice of dissipative master equation. Since the coherent Hamiltonian is the same in the TI-GME and F-GME, these excitation pathways provide a common reference for both descriptions. However, $\overline P_{f\leftarrow i}^{(n)}$ does not determine the final steady-state population by itself; the actual open-system population is set by the competition between coherent excitation and the dissipative redistribution/depletion encoded in the corresponding GME.
These probabilities provide a microscopic interpretation of the peak and valley structures observed in the time-averaged cavity and TLS populations in panels (d) and (e). 
In general, higher-order processes are narrower and more localized in $\omega_d$ than the lowest-order resonances, but they become important when the drive-induced anticrossings are sufficiently strong or when several nearby channels overlap.

It is important to note that the transition probabilities shown in panel (c) of Fig.~\ref{figS:Fig_wd_Optical} are obtained from the unitary Floquet theory of the corresponding closed system and therefore quantify only the coherent drive-induced hybridization between Floquet manifolds. They do not by themselves include dissipative broadening or lifetime effects. In the open system, the actual visibility and influence of a given resonance are determined by the competition between its coherent transition strength and its effective linewidth relative to the bath-induced decay rates. 

When the dominant coherent process is spectrally narrower than the relevant dissipation scale, its contribution can be strongly suppressed, and a broader neighboring resonance with slightly smaller unitary transition probability may become the most influential channel in the steady-state populations or spectra. For this reason, in the present figure we deliberately choose a reduced dissipation rate, $\gamma=0.01\omega_a$ and $\kappa=0.01\omega_c$, so that coherent Floquet resonances remain well resolved and the predictions of both GMEs approach the phenomenological Floquet benchmark. This serves as an additional consistency check that the generalized master-equation frameworks correctly recover the weak-damping limit where coherent Floquet physics is expected to dominate.

The comparison between panels (c)--(e) of Fig.~\ref{figS:Fig_wd_Optical} shows that the population response is not determined by a single resonance, but by the competition and interference among several Floquet-assisted excitation pathways. The phenomenological Floquet result tracks the coherent transition structure, while the GME calculations show how these drive-assisted resonances are modified by dissipation. For the weak flat-bath losses used here, the TI-GME and F-GME give nearly identical averaged populations, consistent with the flat-bath optical-drive behavior discussed in the main text. The same channel structure, however, remains essential for interpreting the frequency-resolved spectra in Fig.~\ref{fig:Spectracav_Optical_wd065}.

\subsubsection{Floquet–Liouville  modal decomposition and quasienergy-resolved processes for the most influential spectral peaks for the optical drive}
The modal decompositions reported in Tabs.~\ref{tab:FLmodes_cavity_kappa01Flat}--\ref{tab:FLmodes_cavity_kappa02LorOhmic1} provide a microscopic interpretation of the cavity spectra shown in the main text for the optical-drive case, respectively, in Fig.~\ref{fig:Spectracav_Optical_wd065}(a)-(d). Each spectral peak is not viewed merely as a transition line, but as a FL relaxation mode $\mu$, characterized simultaneously by a center frequency $\Delta_\mu$, linewidth $\gamma_\mu$, and residue weight $\mathcal{W}_\mu$. By projecting the FL eigenmodes onto the quasienergy transition basis $(\alpha,\beta,l)$, one can identify which drive-assisted processes [depicted by downward arrows in Fig.~\ref{fig:EnergyBasis_Optical}(c)], are responsible for each observed resonance and whether the corresponding spectral feature is generated by a nearly pure quasienergy channel or by a hybridized superposition of several channels.
Note we keep the components with $R^{\alpha\beta l}_\mu\geq 0.1\times\mathrm{max}[R_\mu^{\alpha'\beta'l'}]$, meaning we keep the components that are no less than $10\%$ of the most dominant component and discard the lower $\alpha\beta l$-contributions in a $\mu$-mode.

\begin{table*}[htbp]
\centering
\caption{Dominant Floquet--Liouville modes contributing to the cavity spectrum for the case of optical coherent pumping the TLS. The table is corresponding to the Fig.~\ref{fig:Spectracav_Optical_wd065}(a) in the main text, with $(\eta,\eta_d,\omega_d)=(0.5,0.3,0.65\omega_c)$ for a USC resonant ($\omega_a=\omega_c$) cavity-QED.
Dissipation is characterized via $\kappa=0.1\omega_c$ for the flat cavity bath and $\gamma=0.01\omega_a$ for the flat TLS bath. The Floquet--Liouville modes are ranked by $\mathrm{Re}[\mathcal{W}_\mu]/\gamma_\mu$, for which, the contributing Floquet channels are listed with no less than $10\%$ contribution.}
\label{tab:FLmodes_cavity_kappa01Flat}

\renewcommand{\arraystretch}{1.32}
\setlength{\tabcolsep}{5pt}

\footnotesize
\begin{tabular*}{\textwidth}{@{\extracolsep{\fill}} c c c c l c c c}
\hline\hline
$\mu$ &
$\Delta_\mu/\omega_c$ &
$\gamma_\mu/\omega_c$ &
$\mathrm{Re}[\mathcal{W}_\mu]/\gamma_\mu$ &
Dominant component $(\alpha,\beta,l)$ &
$\Delta_{\alpha\beta l}/\omega_c$ &
$r_\mu^{\alpha\beta l}$ &
$|R_\mu^{\alpha\beta l}|$ \\
\hline

3800 & 0.874 & 0.035 & 1.173 & $(1^{-},0,1)$     & 0.873 & 0.880 & 1.201 \\
     &       &       &       & $(1^{-},2^{-},2)$ & 0.958 & 0.089 & 0.383 \\
\hline
3677 & 0.650 & 0.039 & 0.505 & $(2^{-},0,0)$     & 0.565 & 0.407 & 0.489 \\
     &       &       &       & $(0,2^{-},2)$     & 0.735 & 0.407 & 0.489 \\
\hline
3779 & 0.426 & 0.035 & 0.303 & $(0,1^{-},1)$     & 0.427 & 0.880 & 1.201 \\
     &       &       &       & $(2^{-},1^{-},0)$ & 0.342 & 0.089 & 0.383 \\
\hline
3799 & 1.726 & 0.035 & 0.140 & $(0,1^{-},3)$     & 1.727 & 0.880 & 1.201 \\
     &       &       &       & $(2^{-},1^{-},2)$ & 1.642 & 0.089 & 0.383 \\
\hline
3505 & 0.728 & 0.056 & 0.124 & $(0,2^{-},2)$     & 0.735 & 0.896 & 1.324 \\
\hline
3590 & 0.953 & 0.042 & 0.096 & $(1^{-},2^{-},2)$ & 0.958 & 0.968 & 1.278 \\
\hline
3617 & 1.647 & 0.042 & 0.047 & $(2^{-},1^{-},2)$ & 1.642 & 0.968 & 1.278 \\
\hline
3295 & 0.927 & 0.068 & 0.043 & $(1^{+},1^{-},1)$ & 0.927 & 0.990 & 1.155 \\
\hline
3270 & 1.673 & 0.068 & 0.034 & $(1^{-},1^{+},3)$ & 1.673 & 0.990 & 1.155 \\
\hline
3035 & 1.153 & 0.090 & 0.034 & $(1^{+},0,1)$     & 1.149 & 0.873 & 1.296 \\
\hline
3167 & 1.370 & 0.088 & 0.033 & $(2^{-},1^{+},2)$ & 1.365 & 0.908 & 1.191 \\
\hline
3613 & 0.347 & 0.042 & 0.029 & $(2^{-},1^{-},0)$ & 0.342 & 0.968 & 1.278 \\
\hline
3809 & 2.174 & 0.035 & 0.028 & $(1^{-},0,3)$     & 2.173 & 0.880 & 1.201 \\
     &       &       &       & $(1^{-},2^{-},4)$ & 2.258 & 0.089 & 0.383 \\
\hline
3400 & 0.650 & 0.059 & 0.026 & $(2^{+},0,1)$     & 0.725 & 0.236 & 0.183 \\
     &       &       &       & $(0,2^{+},1)$     & 0.575 & 0.236 & 0.183 \\
     &       &       &       & $(2^{-},0,0)$     & 0.565 & 0.206 & 0.171 \\
     &       &       &       & $(0,2^{-},2)$     & 0.735 & 0.206 & 0.171 \\
     &       &       &       & $(1^{+},1^{-},1)$ & 0.927 & 0.028 & 0.063 \\
     &       &       &       & $(1^{-},1^{+},1)$ & 0.373 & 0.028 & 0.063 \\
\hline
3051 & 1.447 & 0.090 & 0.025 & $(0,1^{+},3)$     & 1.451 & 0.873 & 1.296 \\
\hline\hline

\end{tabular*}
\end{table*}

\begin{table*}[htbp]
\centering
\caption{Dominant Floquet--Liouville modes contributing to the cavity spectrum for the case of optical coherent pumping the TLS. The table is corresponding to the Fig.~\ref{fig:Spectracav_Optical_wd065}(b) in the main text, with $(\eta,\eta_d,\omega_d)=(0.5,0.3,0.65\omega_c)$ for a USC resonant ($\omega_a=\omega_c$) cavity-QED.
Dissipation is characterized via $\kappa=0.1\omega_c$ for the Lorentzian--Ohmic cavity bath centered at
$\omega_0=\omega_c$ and $\gamma=0.01\omega_a$ for the flat TLS bath. The Floquet--Liouville modes are ranked by $\mathrm{Re}[\mathcal{W}_\mu]/\gamma_\mu$, for which, the contributing Floquet channels are listed with no less than $10\%$ contribution.}
\label{tab:FLmodes_cavity_kappa01LorOhmic1}

\renewcommand{\arraystretch}{1.32}
\setlength{\tabcolsep}{5pt}

\footnotesize
\begin{tabular*}{\textwidth}{@{\extracolsep{\fill}} c c c c l c c c}
\hline\hline
$\mu$ &
$\Delta_\mu/\omega_c$ &
$\gamma_\mu/\omega_c$ &
$\mathrm{Re}[\mathcal{W}_\mu]/\gamma_\mu$ &
Dominant component $(\alpha,\beta,l)$ &
$\Delta_{\alpha\beta l}/\omega_c$ &
$r_\mu^{\alpha\beta l}$ &
$|R_\mu^{\alpha\beta l}|$ \\
\hline

3679 & 0.874 & 0.010 & 7.240 & $(1^{-},0,1)$     & 0.873 & 0.991 & 1.218 \\
\hline

3500 & 1.450 & 0.013 & 1.549 & $(0,1^{+},3)$     & 1.451 & 0.993 & 1.214 \\
\hline

3477 & 1.150 & 0.013 & 1.089 & $(1^{+},0,1)$     & 1.149 & 0.993 & 1.214 \\
\hline

3661 & 0.426 & 0.010 & 1.022 & $(0,1^{-},1)$     & 0.427 & 0.991 & 1.218 \\
\hline

3753 & 1.673 & 0.008 & 0.731 & $(1^{-},1^{+},3)$ & 1.673 & 0.998 & 1.076 \\
\hline

3061 & 1.366 & 0.023 & 0.721 & $(2^{-},1^{+},2)$ & 1.365 & 0.943 & 1.030 \\
\hline

3583 & 0.650 & 0.011 & 0.640 & $(0,2^{-},2)$     & 0.735 & 0.287 & 0.051 \\
     &       &       &       & $(2^{-},0,0)$     & 0.565 & 0.287 & 0.051 \\
     &       &       &       & $(0,2^{+},1)$     & 0.725 & 0.108 & 0.031 \\
     &       &       &       & $(2^{+},0,1)$     & 0.575 & 0.108 & 0.031 \\
     &       &       &       & $(2^{+},1^{-},1)$ & 0.502 & 0.047 & 0.021 \\
     &       &       &       & $(1^{-},2^{+},1)$ & 0.798 & 0.047 & 0.021 \\
\hline

3681 & 1.726 & 0.010 & 0.504 & $(0,1^{-},3)$     & 1.727 & 0.991 & 1.218 \\
\hline

3309 & 0.957 & 0.021 & 0.400 & $(1^{-},2^{-},2)$ & 0.958 & 0.949 & 1.059 \\
\hline

3695 & 2.174 & 0.010 & 0.291 & $(1^{-},0,3)$     & 2.173 & 0.991 & 1.218 \\
\hline

2954 & 0.650 & 0.033 & 0.237 & $(0,2^{-},2)$     & 0.735 & 0.429 & 0.202 \\
     &       &       &       & $(2^{-},0,0)$     & 0.565 & 0.429 & 0.202 \\
\hline

3762 & 0.927 & 0.008 & 0.234 & $(1^{+},1^{-},1)$ & 0.927 & 0.998 & 1.076 \\
\hline

3157 & 0.735 & 0.023 & 0.191 & $(0,2^{-},2)$     & 0.735 & 0.988 & 1.095 \\
\hline

3303 & 1.643 & 0.021 & 0.187 & $(2^{-},1^{-},2)$ & 1.642 & 0.949 & 1.059 \\
\hline

3183 & 0.565 & 0.023 & 0.104 & $(2^{-},0,0)$     & 0.565 & 0.988 & 1.095 \\
\hline\hline

\end{tabular*}
\end{table*}

\begin{table*}[htbp]
\centering
\caption{Dominant Floquet--Liouville modes contributing to the cavity spectrum for the case of optical coherent pumping the TLS. The table is corresponding to the Fig.~\ref{fig:Spectracav_Optical_wd065}(c) in the main text, with $(\eta,\eta_d,\omega_d)=(0.5,0.3,0.65\omega_c)$ for a USC resonant ($\omega_a=\omega_c$) cavity-QED.
Dissipation is characterized via $\kappa=0.2\omega_c$ for the flat cavity bath and $\gamma=0.01\omega_a$ for the flat TLS bath. The Floquet--Liouville modes are ranked by $\mathrm{Re}[\mathcal{W}_\mu]/\gamma_\mu$, for which, the contributing Floquet channels are listed with no less than $10\%$ contribution.}
\label{tab:FLmodes_cavity_kappa02Flat}

\renewcommand{\arraystretch}{1.32}
\setlength{\tabcolsep}{5pt}

\footnotesize
\begin{tabular*}{\textwidth}{@{\extracolsep{\fill}} c c c c l c c c}
\hline\hline
$\mu$ &
$\Delta_\mu/\omega_c$ &
$\gamma_\mu/\omega_c$ &
$\mathrm{Re}[\mathcal{W}_\mu]/\gamma_\mu$ &
Dominant component $(\alpha,\beta,l)$ &
$\Delta_{\alpha\beta l}/\omega_c$ &
$r_\mu^{\alpha\beta l}$ &
$|R_\mu^{\alpha\beta l}|$ \\
\hline

3725 & 0.885 & 0.061 & 0.700 & $(1^{-},0,1)$     & 0.873 & 0.638 & 1.313 \\
     &       &       &       & $(1^{-},2^{-},2)$ & 0.958 & 0.323 & 0.935 \\
\hline

3785 & 0.650 & 0.058 & 0.589 & $(2^{-},0,0)$     & 0.565 & 0.405 & 0.921 \\
     &       &       &       & $(0,2^{-},2)$     & 0.735 & 0.405 & 0.921 \\
     &       &       &       & $(1^{-},0,1)$     & 0.873 & 0.043 & 0.302 \\
     &       &       &       & $(0,1^{-},1)$     & 0.427 & 0.043 & 0.302 \\
\hline

3741 & 0.415 & 0.061 & 0.231 & $(0,1^{-},1)$     & 0.427 & 0.638 & 1.313 \\
     &       &       &       & $(2^{-},1^{-},0)$ & 0.342 & 0.323 & 0.935 \\
\hline

3719 & 1.715 & 0.061 & 0.084 & $(0,1^{-},3)$     & 1.727 & 0.638 & 1.313 \\
     &       &       &       & $(2^{-},1^{-},2)$ & 1.642 & 0.323 & 0.935 \\
\hline

3405 & 0.650 & 0.115 & 0.055 & $(2^{-},0,0)$     & 0.565 & 0.350 & 0.695 \\
     &       &       &       & $(0,2^{-},2)$     & 0.735 & 0.350 & 0.695 \\
     &       &       &       & $(0,2^{+},1)$     & 0.575 & 0.059 & 0.285 \\
     &       &       &       & $(2^{+},0,1)$     & 0.725 & 0.059 & 0.285 \\
\hline

3290 & 0.928 & 0.130 & 0.015 & $(1^{+},1^{-},1)$ & 0.927 & 0.971 & 1.341 \\
\hline

3177 & 1.389 & 0.166 & 0.013 & $(2^{-},1^{+},2)$ & 1.365 & 0.656 & 1.706 \\
     &       &       &       & $(0,1^{+},3)$     & 1.451 & 0.270 & 1.094 \\
\hline

3308 & 1.672 & 0.130 & 0.012 & $(1^{-},1^{+},3)$ & 1.673 & 0.971 & 1.341 \\
\hline

3708 & 2.185 & 0.061 & 0.010 & $(1^{-},0,3)$     & 2.173 & 0.638 & 1.313 \\
     &       &       &       & $(1^{-},2^{-},4)$ & 2.258 & 0.323 & 0.935 \\
\hline

3788 & 1.950 & 0.058 & 0.008 & $(0,2^{-},4)$     & 2.035 & 0.405 & 0.921 \\
     &       &       &       & $(2^{-},0,2)$     & 1.865 & 0.405 & 0.921 \\
     &       &       &       & $(0,1^{-},3)$     & 1.727 & 0.043 & 0.302 \\
     &       &       &       & $(1^{-},0,3)$     & 2.173 & 0.043 & 0.302 \\
\hline

3306 & -0.372 & 0.130 & 0.003 & $(1^{+},1^{-},-1)$ & -0.373 & 0.971 & 1.341 \\
\hline

3154 & 1.211 & 0.166 & 0.003 & $(1^{+},2^{-},2)$ & 1.235 & 0.656 & 1.706 \\
     &       &       &       & $(1^{+},0,1)$     & 1.149 & 0.270 & 1.094 \\
\hline

3074 & 1.173 & 0.178 & 0.003 & $(1^{+},0,1)$     & 1.149 & 0.609 & 1.781 \\
     &       &       &       & $(1^{+},2^{-},2)$ & 1.235 & 0.275 & 1.198 \\
\hline

3400 & 1.950 & 0.115 & 0.002 & $(0,2^{-},4)$     & 2.035 & 0.350 & 0.695 \\
     &       &       &       & $(2^{-},0,2)$     & 1.865 & 0.350 & 0.695 \\
     &       &       &       & $(0,2^{+},3)$     & 1.875 & 0.059 & 0.285 \\
     &       &       &       & $(2^{+},0,3)$     & 2.025 & 0.059 & 0.285 \\
\hline

3742 & -0.885 & 0.061 & 0.002 & $(0,1^{-},-1)$     & -0.873 & 0.638 & 1.313 \\
     &       &       &       & $(2^{-},1^{-},-2)$ & -0.958 & 0.323 & 0.935 \\
\hline\hline

\end{tabular*}
\label{tab:kappa02flat_optical}
\end{table*}

\begin{table*}[htbp]
\centering
\caption{Dominant Floquet--Liouville modes contributing to the cavity spectrum for the case of optical coherent pumping the TLS. The table is corresponding to the Fig.~\ref{fig:Spectracav_Optical_wd065}(d) in the main text, with $(\eta,\eta_d,\omega_d)=(0.5,0.3,0.65\omega_c)$ for a USC resonant ($\omega_a=\omega_c$) cavity-QED.
Dissipation is characterized via $\kappa=0.2\omega_c$ for the Lorentzian--Ohmic cavity bath centered at
$\omega_0=\omega_c$ and $\gamma=0.01\omega_a$ for the flat TLS bath. The Floquet--Liouville modes are ranked by $\mathrm{Re}[\mathcal{W}_\mu]/\gamma_\mu$, for which, the contributing Floquet channels are listed with no less than $10\%$ contribution.}
\label{tab:FLmodes_cavity_kappa02LorOhmic1}

\renewcommand{\arraystretch}{1.32}
\setlength{\tabcolsep}{5pt}

\footnotesize
\begin{tabular*}{\textwidth}{@{\extracolsep{\fill}} c c c c l c c c}
\hline\hline
$\mu$ &
$\Delta_\mu/\omega_c$ &
$\gamma_\mu/\omega_c$ &
$\mathrm{Re}[W_\mu]/\gamma_\mu$ &
Dominant component $(\alpha,\beta,l)$ &
$\Delta_{\alpha\beta l}/\omega_c$ &
$r_\mu^{\alpha\beta l}$ &
$|R_\mu^{\alpha\beta l}|$ \\
\hline

3656 & 0.888 & 0.031 & 1.020 & $(1^{-},0,1)$     & 0.873 & 0.872 & 1.440 \\
     &       &       &       & $(1^{-},2^{-},2)$ & 0.958 & 0.105 & 0.501 \\
\hline

3670 & 0.412 & 0.031 & 0.572 & $(0,1^{-},1)$     & 0.427 & 0.872 & 1.440 \\
     &       &       &       & $(2^{-},1^{-},0)$ & 0.342 & 0.105 & 0.501 \\
\hline

3776 & 1.674 & 0.024 & 0.154 & $(1^{-},1^{+},3)$ & 1.673 & 0.993 & 1.228 \\
\hline

3666 & 1.712 & 0.031 & 0.133 & $(0,1^{-},3)$     & 1.727 & 0.872 & 1.440 \\
     &       &       &       & $(2^{-},1^{-},2)$ & 1.642 & 0.105 & 0.501 \\
\hline

3472 & 1.435 & 0.048 & 0.115 & $(0,1^{+},3)$     & 1.451 & 0.879 & 1.522 \\
     &       &       &       & $(2^{-},1^{+},2)$ & 1.365 & 0.095 & 0.501 \\
\hline

3569 & 0.650 & 0.039 & 0.108 & $(0,2^{-},2)$     & 0.735 & 0.438 & 0.588 \\
     &       &       &       & $(2^{-},0,0)$     & 0.565 & 0.438 & 0.588 \\
\hline

3171 & 0.706 & 0.073 & 0.053 & $(0,2^{-},2)$     & 0.735 & 0.870 & 1.306 \\
\hline

3273 & 0.650 & 0.064 & 0.053 & $(2^{-},0,0)$     & 0.565 & 0.334 & 0.346 \\
     &       &       &       & $(0,2^{-},2)$     & 0.735 & 0.334 & 0.346 \\
     &       &       &       & $(0,2^{+},1)$     & 0.725 & 0.098 & 0.188 \\
     &       &       &       & $(2^{+},0,1)$     & 0.575 & 0.098 & 0.188 \\
\hline

3480 & 1.165 & 0.048 & 0.039 & $(1^{+},0,1)$     & 1.149 & 0.879 & 1.522 \\
     &       &       &       & $(1^{+},2^{-},2)$ & 1.235 & 0.095 & 0.501 \\
\hline

2957 & 0.650 & 0.104 & 0.037 & $(0,2^{-},2)$     & 0.735 & 0.419 & 0.830 \\
     &       &       &       & $(2^{-},0,0)$     & 0.565 & 0.419 & 0.830 \\
\hline

3357 & 1.659 & 0.058 & 0.035 & $(2^{-},1^{-},2)$ & 1.642 & 0.617 & 1.256 \\
     &       &       &       & $(0,1^{-},3)$     & 1.727 & 0.353 & 0.949 \\
\hline

3646 & 2.188 & 0.031 & 0.023 & $(1^{-},0,3)$     & 2.173 & 0.872 & 1.440 \\
     &       &       &       & $(1^{-},2^{-},4)$ & 2.258 & 0.105 & 0.501 \\
\hline

3062 & 1.380 & 0.074 & 0.022 & $(2^{-},1^{+},2)$ & 1.365 & 0.596 & 1.272 \\
     &       &       &       & $(0,1^{+},3)$     & 1.451 & 0.354 & 0.980 \\
\hline

3764 & 0.926 & 0.024 & 0.017 & $(1^{+},1^{-},1)$ & 0.927 & 0.993 & 1.228 \\
\hline

3770 & -0.374 & 0.024 & 0.012 & $(1^{+},1^{-},-1)$ & -0.373 & 0.993 & 1.228 \\
\hline\hline

\end{tabular*}
\end{table*}

A robust feature across all four tables is that the most influential positive-frequency cavity peak is associated with the process $(1^{-},0,1)$, yielding $\Delta_{\alpha\beta l}\simeq0.873\omega_c$. This channel remains the leading contribution under both flat and structured baths and for both damping values $\kappa=0.1\omega_c$ and $0.2\omega_c$. Physically, it corresponds to a first-order drive-assisted emission pathway linking the Floquet replica of the ground-state manifold to the $1^{-}$ polaritonic branch. Its persistence demonstrates that this resonance is primarily determined by coherent Floquet hybridization and is therefore structurally stable against moderate changes in the reservoir model.

The companion lower-frequency peak near $\Delta_\mu\simeq0.42\omega_c$ is consistently dominated by the reverse channel $(0,1^{-},1)$, with a small admixture from $(2^{-},1^{-},0)$. This reveals that the lower sideband and the main $\sim0.87\omega_c$ line form a correlated pair originating from the same dressed manifold. In the spectrum, these two peaks therefore should not be interpreted as unrelated resonances, but rather as complementary Floquet sidebands generated by the same underlying hybridized polaritonic sector.

Another important recurring structure is the feature around $\Delta_\mu\simeq0.65\omega_c=\omega_d$. Unlike the peaks above, this resonance is not tied to a single dominant process. Instead, it repeatedly appears as an almost symmetric mixture of $(2^{-},0,0)$ and $(0,2^{-},2)$, with additional smaller admixtures depending on bath type and damping strength. This is a direct signature of quasienergy folding: two distinct physical transitions, separated in the undriven picture, are mapped into the same observation frequency through the absorption or emission of two drive quanta. Consequently, the spectral line near $0.65\omega_c$ is intrinsically composite and cannot be assigned to a unique bare transition.

Higher-frequency peaks near $\Delta_\mu\simeq1.72\omega_c$ and $2.17\omega_c$ are likewise robust throughout the tables and are mainly governed by $(0,1^{-},3)$ and $(1^{-},0,3)$, respectively. These correspond to third-order drive-assisted replicas of the lower polariton branch. Their clear presence confirms that even for the moderately strong amplitude $\eta_d=0.3$, higher-order Floquet processes already participate substantially in the resonance fluorescence spectrum. This explains why time-independent approaches that neglect explicit sideband structure become inaccurate in the frequency-resolved response, even when they may still approximate integrated populations.

Processes involving the $1^{+}$ branch, such as $(1^{+},0,1)$, $(0,1^{+},3)$, and $(1^{-},1^{+},3)$, generate peaks around $1.15\omega_c$, $1.45\omega_c$, and $1.67\omega_c$, respectively. Although these lines are weaker than the dominant $1^{-}$-sector resonances, they are systematically present in all bath configurations. Their appearance shows that the optical drive redistributes spectral weight into both parity sectors of the dressed QRM manifold, allowing transitions that would remain much less visible in an undriven or weakly driven setting.

The dependence on the cavity bath structure is especially illuminating. For the Lorentzian--Ohmic reservoir centered near $\omega_0=\omega_c$, the mode near $\Delta_\mu\simeq0.874\omega_c$ becomes dramatically enhanced, reaching $\mathrm{Re}(\mathcal{W}_\mu)/\gamma_\mu=7.240$ for $\kappa=0.1\omega_c$, far exceeding the corresponding flat-bath value. This occurs because the bath spectral density selectively amplifies dissipation channels whose frequencies lie close to the reservoir maximum while simultaneously narrowing their linewidths. Hence, the structured environment does not merely broaden lines; it reshapes modal hierarchy by favoring certain quasienergy transitions over others.

Increasing the cavity damping from $\kappa=0.1\omega_c$ to $0.2\omega_c$ naturally increases $\gamma_\mu$, lowers modal rankings, and enhances channel mixing. For example, the leading mode at $\Delta_\mu\approx0.88\omega_c$ remains dominant, but its purity decreases as the admixture of $(1^{-},2^{-},2)$ becomes more pronounced. Similarly, several peaks that were nearly single-channel at weak damping become visibly hybridized at stronger damping. This demonstrates that dissipation acts not only as spectral broadening, but also as a mechanism that mixes nearby quasienergy decay pathways into collective Liouvillian modes.

Negative-frequency entries such as $\Delta_\mu=-0.372\omega_c$ and $-0.885\omega_c$, appearing for larger $\kappa$, correspond to the mirrored absorption sector of the full susceptibility. Their presence confirms that the FL decomposition naturally captures both emission and absorption poles within one unified framework. Although these modes carry smaller positive weights in the cavity emission spectrum, they are important for preserving the complete causal analytic structure of the response function.

Overall, the tables establish that the cavity spectrum in the optical-drive regime is organized by a small number of dominant Floquet--Liouville poles, each of which can be traced to specific drive-assisted quasienergy transitions. Some peaks are nearly pure single-channel processes, while others are genuinely hybrid resonances formed from several folded transitions. This channel-resolved viewpoint explains the origin of the spectral peaks in the main text, clarifies the influence of bath engineering, and demonstrates why a quasienergy-resolved dissipative treatment is necessary for quantitatively reliable spectroscopy in driven ultrastrong-coupling cavity QED systems.

\newpage
\subsubsection{Cavity bright modes}
\begin{figure*}[htbp]
\centering
\includegraphics[width=0.98\linewidth]
{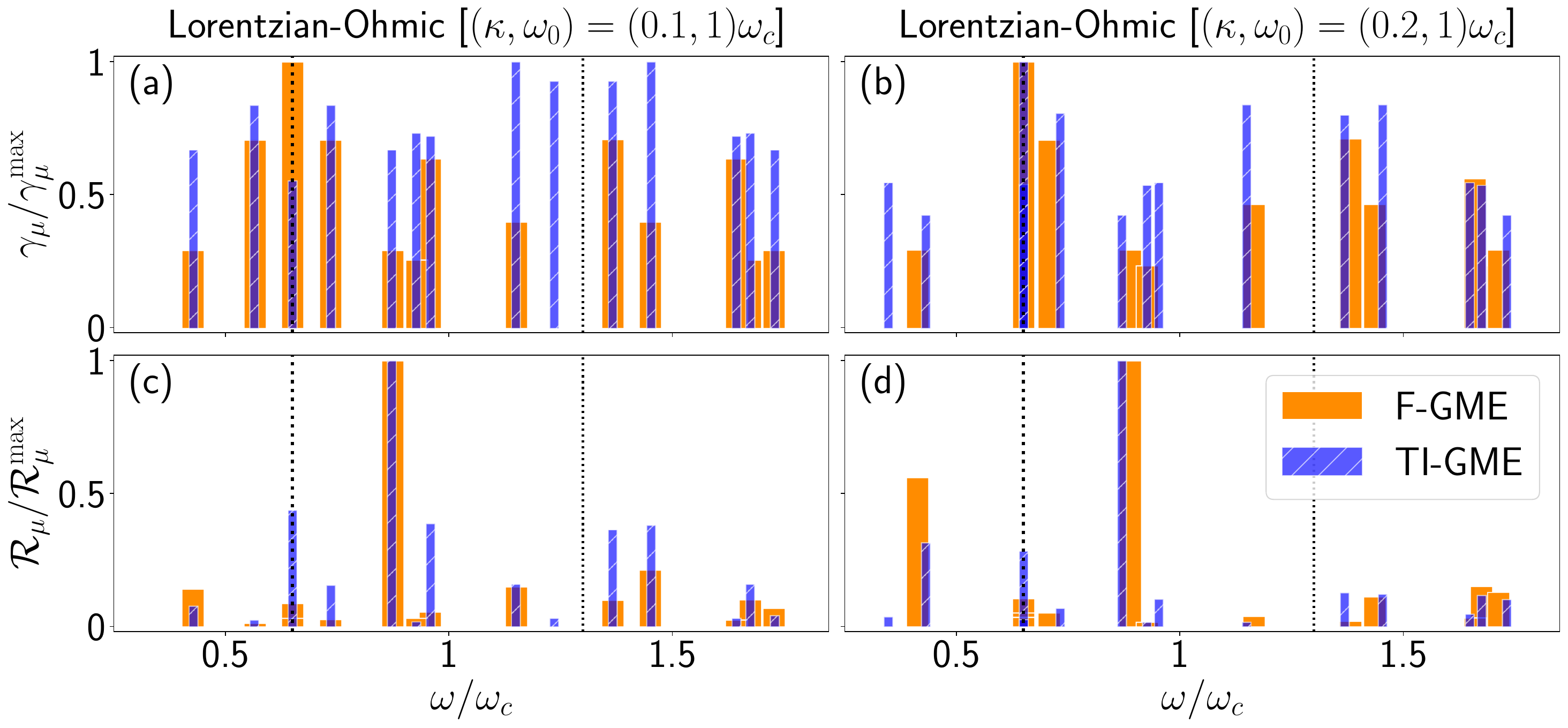} 
\caption[]{\textbf{Optically driven open cavity-QED in USC: 
cavity Bright modes.} Shown in panels (a) and (c) are the spectral modal linewidth and peak prominence of Fig.~\ref{fig:Spectracav_Optical_wd065}(b), respectively.
Panels (b) and (d) show the spectral modal linewidth and peak prominence of Fig.~\ref{fig:Spectracav_Optical_wd065}(d), respectively. A comparison between the TI-GME results (dashed blue) and F-GME results (orange) is made.
}
\label{figS:CavityBrightChannels_Optical}
\end{figure*}
Figure~\ref{figS:CavityBrightChannels_Optical} presents the dominant bright channels contributing to the cavity emission spectrum for the optically driven case at $\omega_d=0.65\omega_c$, obtained from the Floquet--Liouville modal decomposition. Here, each visible spectral peak is resolved into the principal FL poles and their associated quasienergy transitions $(\alpha,\beta,l)$, thereby providing a microscopic channel-by-channel interpretation of the cavity fluorescence discussed in the main text.

The term \emph{bright channel} refers to a dissipative Floquet transition whose associated residue weight is sufficiently large to generate an observable contribution to the spectrum. In the present framework, brightness is controlled not only by the transition matrix element, but also by the modal linewidth $\gamma_\mu$, spectral interference with neighboring poles, and the overlap of the FL eigenmode with the cavity emission operator. Consequently, the most visible peak is not always generated by the channel with the largest unitary transition probability alone.

A central result of Fig.~ \ref{figS:CavityBrightChannels_Optical}
is that the strongest emission line near $\omega \simeq 0.87\omega_c$ is dominated by the channel $(1^{-},0,1)$, consistent with the tabulated FL analysis. This process corresponds to a first-order drive-assisted transition linking the ground Floquet manifold to the lower polaritonic branch and remains the primary bright channel across the different bath models. Its prominence explains the robust main peak observed in the cavity spectrum.

The lower-frequency feature near $\omega \simeq 0.43\omega_c$ is mainly generated by the companion channel $(0,1^{-},1)$, while the resonance around $\omega \simeq 0.65\omega_c$ emerges from a hybrid combination of $(2^{-},0,0)$ and $(0,2^{-},2)$. The latter is especially important because it illustrates how two distinct physical transitions, folded by the drive into the same quasienergy window, combine into a single observable spectral structure. Thus, some peaks are intrinsically collective Floquet resonances rather than isolated lines.

Higher-frequency bright channels near $\omega \simeq 1.73\omega_c$ and $2.17\omega_c$ are associated with $(0,1^{-},3)$ and $(1^{-},0,3)$, respectively. These are higher-order sideband processes involving three exchanged drive quanta. Their visible contribution demonstrates that the moderate optical drive amplitude $\eta_d=0.3$ already activates substantial nonlinear Floquet scattering pathways, which are inaccessible within a purely time-independent picture.

 Figure~\ref{figS:CavityBrightChannels_Optical} also compares the channel decompositions obtained from the Floquet generalized master equation (F-GME) and the time-independent generalized master equation (TI-GME), shown here for the Lorentzian--Ohmic bath in order to directly expose the origin of their discrepancies. The F-GME organizes dissipation in the quasienergy basis and therefore assigns spectral weight to the physically correct drive-dressed channels. By contrast, the TI-GME distributes decay through undriven dressed transitions and then projects this dynamics into the driven spectrum. This can misallocate brightness among neighboring peaks, distort linewidths, and either suppress or exaggerate specific sidebands.

Most noticeably, when several channels are nearly resonant or strongly hybridized, the TI-GME may predict the correct approximate peak location while assigning an incorrect modal hierarchy. In other words, the disagreement is often not simply a frequency shift, but a redistribution of spectral intensity among competing bright channels. This directly supports the main claim of the paper that quasienergy-resolved dissipation is essential for quantitatively reliable spectroscopy in driven open USC systems.

It is also important to note that the transition probabilities underlying these channels originate from unitary Floquet theory of the closed system. In the actual open system, observability depends on whether the corresponding coherent channel survives dissipative broadening. If the nominally strongest process is too narrow compared with the relevant decay scale, it can be effectively washed out, allowing the next broader channel to become the dominant bright contributor. Hence, spectral brightness is determined by the competition between coherent transition strength and linewidth.

Overall, Fig.~\ref{figS:CavityBrightChannels_Optical} converts the cavity spectrum from a set of peaks into a resolved map of underlying quasienergy decay pathways. It shows which transitions are bright, which are hybridized, how bath structure reshapes their hierarchy, and why only a Floquet-resolved dissipative formalism can faithfully capture the observed emission landscape.

\subsubsection{Long-time averaged number of emitter
(TLS) excitation}
\begin{figure*}[htbp]
\centering
\includegraphics[width=0.98\linewidth]
{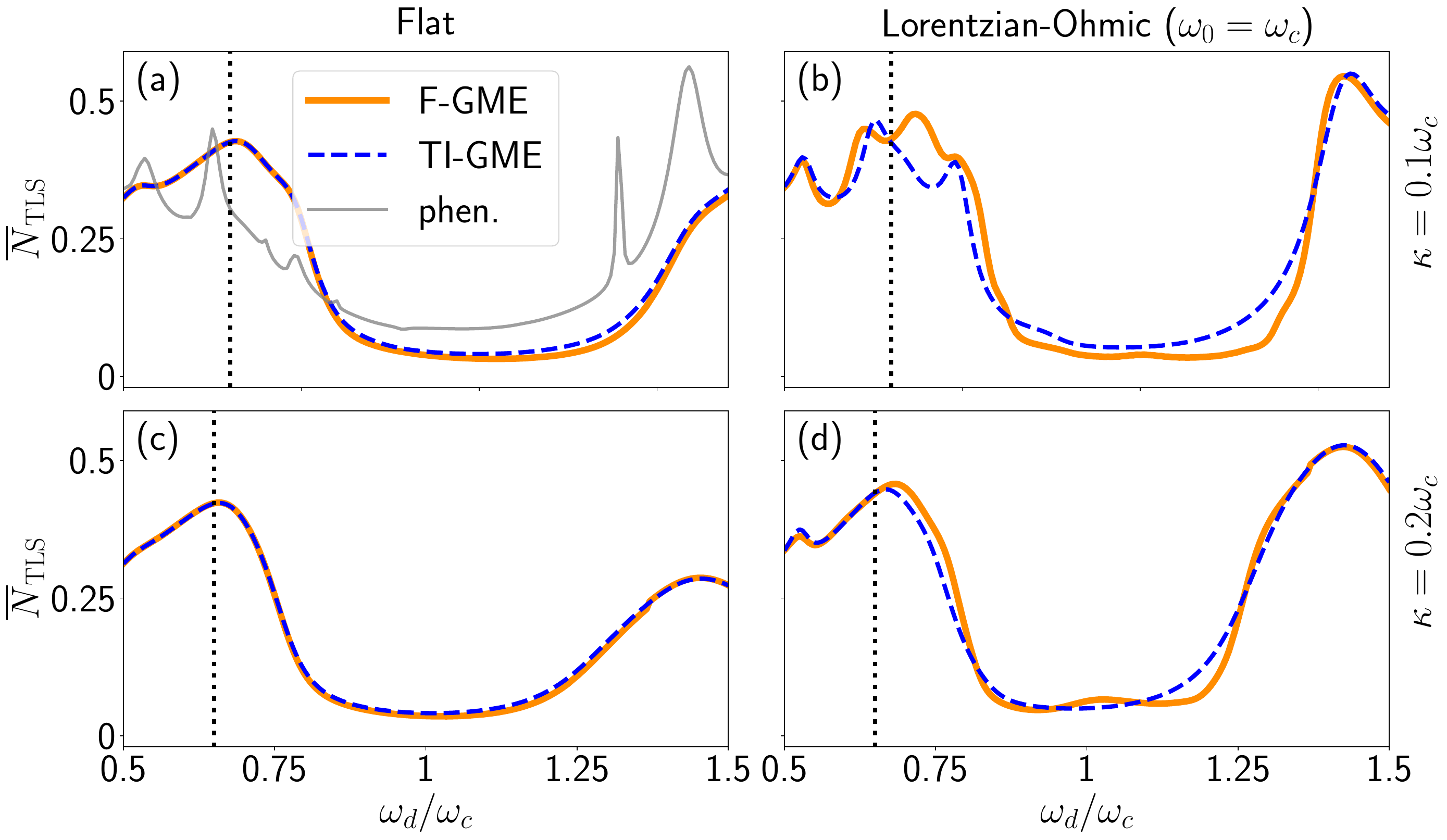} 
\caption[]{\textbf{Optically driven open cavity-QED in USC: 
TLS average number of excitations versus the optical drive frequency $\omega_d$.} Same as Fig.~\ref{fig:AveNcav_wd_Optical} but for the TLS excitations instead of cavity excitation.
}
\label{figS:AveNTLS_wd_Optical}
\end{figure*}
Figure~\ref{figS:AveNTLS_wd_Optical} shows the long-time averaged excitation number of the TLS as a function of the optical drive frequency, complementing the cavity population results presented in the main text. Since the external coherent pump acts directly on the TLS sector, this observable provides a more direct probe of how efficiently the drive injects energy into the hybrid light--matter system before that excitation is redistributed through the cavity coupling and dissipation channels.

The resonance peaks occur at frequencies closely aligned with those identified in the cavity excitation spectrum, confirming that both observables originate from the same Floquet quasienergy anticrossings and drive-assisted transition pathways. In particular, the dominant structures near the principal one-photon resonance and the lower-frequency multiphoton resonances appear in both cavity and TLS observables. However, their relative amplitudes differ substantially because the underlying dressed Floquet states possess different atomic and photonic compositions.

This distinction is a hallmark of the ultrastrong-coupling regime. The relevant eigenstates are polaritonic superpositions rather than separable atomic and cavity excitations, so the cavity and TLS populations correspond to different projections of the same nonequilibrium steady state. Resonances associated with matter-like Floquet replicas are enhanced in the TLS excitation number, whereas those with stronger photonic character are more pronounced in the cavity population of Fig.~\ref{fig:AveNcav_wd_Optical}.

The comparison among the phenomenological approach, the Floquet generalized master equation (F-GME), and the time-independent generalized master equation (TI-GME) follows the same qualitative trends discussed for the cavity observable. Away from strongly hybridized resonances, all three approaches may agree reasonably well. However, near quasienergy anticrossings or in frequency windows where multiple sidebands compete, the TI-GME can noticeably overestimate or underestimate the TLS excitation number. This deviation arises because a time-independent dissipator cannot faithfully capture how the periodic drive redistributes populations among Floquet replicas carrying different atomic weight.

In contrast, the F-GME treats relaxation directly in the quasienergy basis and therefore tracks the drive-dressed excitation pathways more accurately. This leads to substantially improved agreement with the phenomenological benchmark, especially in the vicinity of sharp resonances where channel selectivity is important.

It is also useful to emphasize that the relatively small loss rates used here serve as a consistency test of the open-system theories. By deliberately reducing $\kappa$ and $\gamma$, the predictions of both generalized master-equation approaches move closer to the phenomenological results. This demonstrates that the remaining discrepancies at stronger damping originate from the treatment of driven dissipation rather than from numerical instability or uncontrolled behavior. In this sense, the weak-loss regime provides a numerical check for the validity of the GME frameworks.

Finally, although the resonance positions are rooted in the closed-system Floquet transition structure, the observed TLS peaks are determined by the competition between coherent excitation strength and dissipative broadening. A nominally dominant unitary transition can become practically unimportant if its linewidth is too narrow compared with the relevant decay scale, allowing a broader neighboring channel to dominate the measured response. Hence, the open-system population spectrum reflects not only where transitions occur, but also which of them remain spectrally resolvable under dissipation.

Figure~\ref{figS:AveNTLS_wd_Optical} confirms that the resonance landscape induced by optical driving is a genuine Floquet many-channel phenomenon visible in both matter and light observables, while also reinforcing the necessity of quasienergy-resolved dissipation for quantitatively reliable predictions of driven ultrastrong-coupling cavity-QED dynamics.

\begin{figure*}[htbp]
\centering
\includegraphics[width=.98\linewidth]
{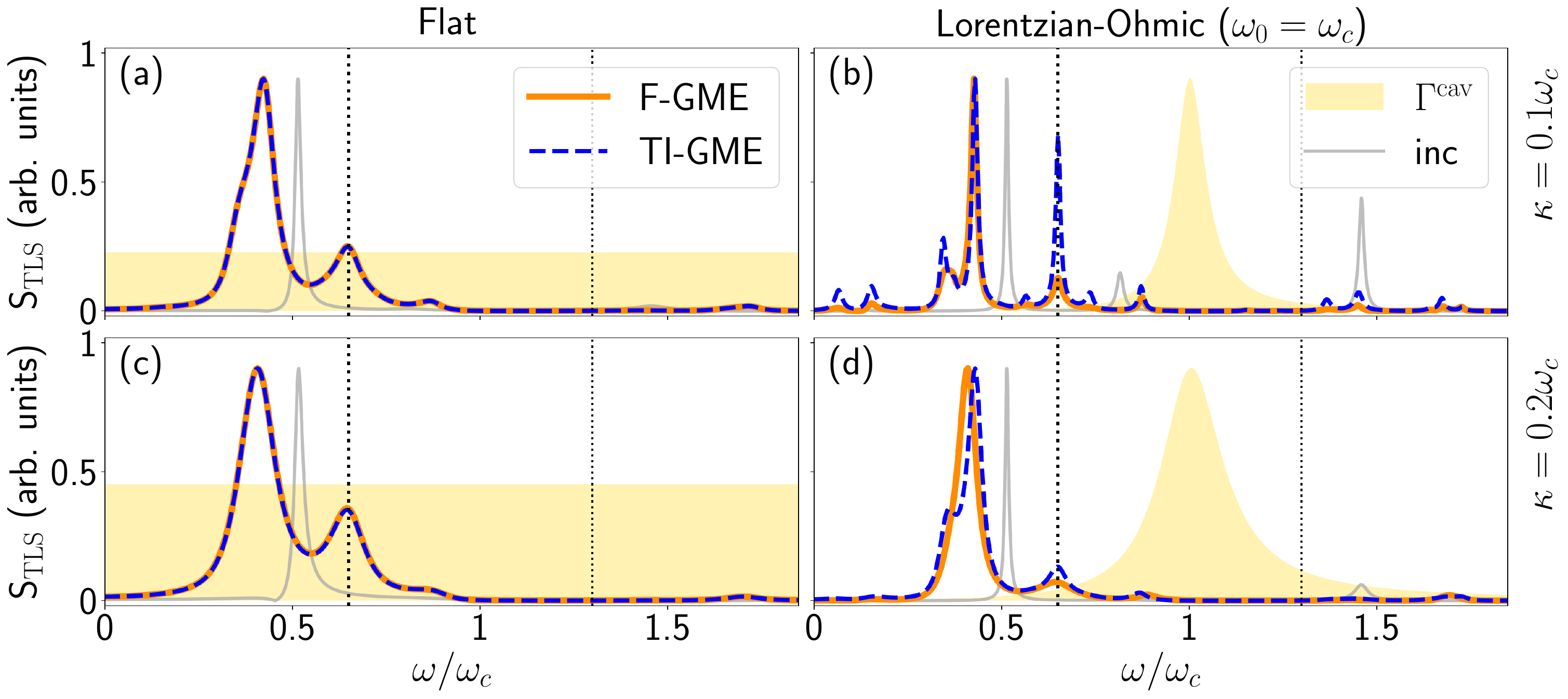} 
\caption[]{\textbf{Optically driven cavity-QED in USC: TLS (incoherent) spectra.} Same as Fig.~\ref{fig:Spectracav_Optical_wd065} but for the TLS spectra instead of cavity spectra.
}
\label{figS:SpectraTLS_Optical_wd065}
\end{figure*}
Figure~\ref{figS:SpectraTLS_Optical_wd065} shows the incoherent TLS fluorescence spectrum for the optically driven system at $\omega_d=0.65\omega_c$, complementing the cavity spectrum discussed in the main text. Since the TLS operator directly probes the matter component of the polaritonic excitations, this spectrum reveals how the same driven steady state manifests when observed through the atomic sector rather than through the cavity field.

The dominant resonance frequencies are broadly consistent with those of the cavity spectrum, confirming that both observables are governed by the same underlying Floquet quasienergy transitions. Peaks associated with the principal lower-polariton channel, its sidebands, and higher-order replicas remain visible in both spectra. However, their relative intensities differ substantially because each dressed Floquet state contains a different admixture of atomic and photonic character. Consequently, transitions that are bright in the cavity channel need not be equally bright in the TLS channel, and vice versa.

This contrast is especially pronounced in the ultrastrong-coupling regime, where the emission operators are strongly dressed and no longer correspond to simple bare annihilation or spin-lowering transitions. The TLS detection operator carries significant overlap with several dressed transitions and Floquet harmonics simultaneously. Therefore, the TLS incoherent spectrum is generally more sensitive to interference among nearby channels and to the redistribution of spectral weight across Floquet sidebands.

A particularly important observation of
TLS excitations is that, unlike the cavity case, visible discrepancies between the TI-GME and F-GME can persist even under flat-bath conditions. This does not contradict the comparatively good agreement sometimes seen in long-time averaged TLS populations. Rather, it reflects the fundamental difference between scalar steady-state observables and frequency-resolved two-time correlation functions. Integrated populations can average over micromotion details, whereas spectra remain directly sensitive to the full time-periodic structure of the regression dynamics.

Within the F-GME, the quantum regression is performed using the periodic Liouvillian and quasienergy-resolved jump operators, so the TLS correlator retains cross-harmonic interference and micromotion-induced sideband structure. By contrast, the TI-GME uses a stationary regression based on time-independent dressed operators. Even when the bath rates are frequency independent, this procedure effectively compresses the harmonic content into a static representation and can therefore miss or misweight sideband contributions. Hence, flat baths remove spectral selectivity of the reservoir, but they do not remove the intrinsic Floquet time structure of the driven operator dynamics.

The transition from flat to Lorentzian--Ohmic baths modifies the TLS spectra in a manner analogous to the cavity case: peaks whose frequencies lie near the bath maximum become sharper and brighter, while others are relatively suppressed. However, because the TLS operator samples a different projection of the same dressed states, the quantitative reshaping of the spectrum can differ significantly from the cavity observable.

Figure~\ref{figS:SpectraTLS_Optical_wd065} demonstrates that agreement in time-averaged populations does not guarantee agreement in spectroscopy. The TLS fluorescence remains strongly sensitive to Floquet micromotion and channel-resolved dissipation, reinforcing the need for a genuinely quasienergy-resolved open-system treatment.


\subsubsection{Emitter (TLS) bright modes}
\begin{figure*}[htbp]
\centering
\includegraphics[width=0.98\linewidth]
{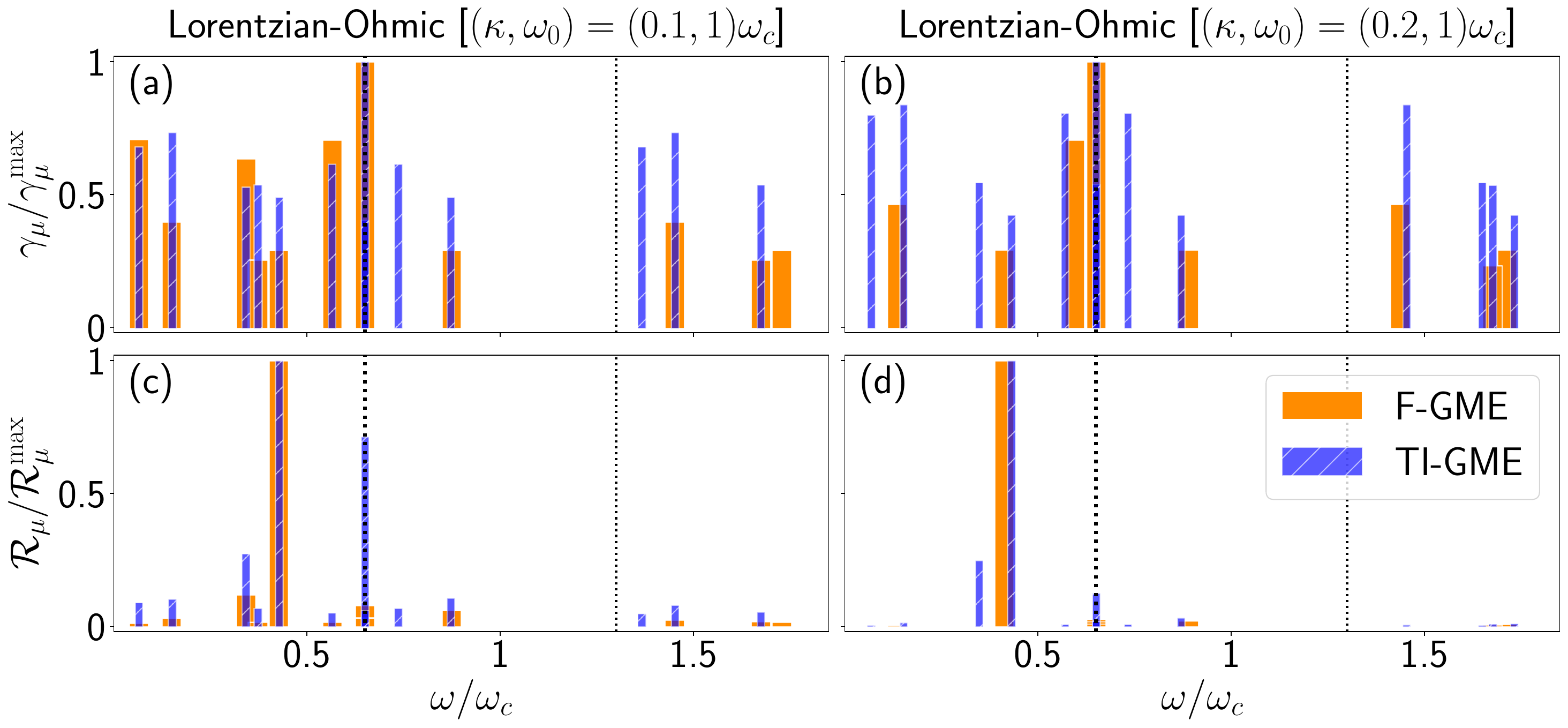} 
\caption[]{\textbf{Optically driven open cavity-QED in USC: 
TLS Bright modes.} 
Shown in panels (a) and (c) are the spectral modal linewidth and peak prominence of Fig.~\ref{figS:SpectraTLS_Optical_wd065}(b), respectively.
(b) and (d) show the spectral modal linewidth and peak prominence of Fig.~\ref{figS:SpectraTLS_Optical_wd065}(d), respectively. Comparison between the TI-GME results (dashed blue) and F-GME results (orange) is made.
}
\label{figS:TLSBrightChannels_Optical}
\end{figure*}
Figure~\ref{figS:TLSBrightChannels_Optical} resolves the TLS fluorescence spectrum into its dominant bright channels using the Floquet--Liouville modal decomposition. As in the cavity analysis, each visible spectral structure is decomposed into the principal FL poles and their associated quasienergy transitions $(\alpha,\beta,l)$. This reveals which drive-assisted decay pathways are responsible for the atomic emission peaks and how their hierarchy differs from the cavity observable.

Many of the dominant channels coincide with those found in the cavity spectrum, since both detection scenarios probe the same driven steady state. In particular, the main lower-polariton resonance and its sidebands remain central contributors. However, the brightness ordering changes because the TLS operator overlaps more strongly with matter-like dressed states. As a result, channels involving stronger atomic participation are enhanced here, while more photonic channels can become comparatively weaker than in the cavity spectrum.

 Figure~\ref{figS:TLSBrightChannels_Optical} therefore provides a direct visualization of polaritonic state composition. A given Floquet transition may be visible in both spectra, yet with different residue weights depending on whether the emitted signal is measured through the cavity field or through the TLS dipole. This confirms that cavity and TLS spectroscopy offer complementary windows into the same nonequilibrium Floquet polariton manifold.

The comparison between F-GME and TI-GME bright-channel decompositions is particularly revealing. For the TLS observable, discrepancies in modal hierarchy and linewidth assignment can be more pronounced than for the cavity field. The reason is that the TLS emission operator carries broader harmonic content in the dressed basis, so errors caused by collapsing the time-periodic dynamics into a stationary regression become more visible. The TI-GME may approximately reproduce some peak locations while still assigning incorrect channel strengths or suppressing interference-enhanced resonances.

It is again important to emphasize that the brightest observed channel is determined not only by coherent transition probability, but also by dissipation. A nominally strong unitary channel can be rendered ineffective if it is too narrow relative to the open-system linewidth scale, while a broader neighboring process may dominate the actual measured spectrum. Thus, brightness reflects the interplay of transition matrix elements, quasienergy hybridization, and modal damping.

Overall, Fig.~\ref{figS:TLSBrightChannels_Optical} confirms that the TLS spectrum is generated by the same Floquet transition network as the cavity spectrum, but filtered through a different operator sensitivity. This further illustrates that spectroscopy in driven ultrastrong-coupling systems is operator dependent, and that only a Floquet-resolved dissipative formalism can consistently capture both matter and photonic emission channels.

\subsection{Periodic mechanical drive}
For the (periodic) mechanical drive case, the
energy state labels based on their orders from Fig.~\ref{fig:EnergyBasis_Mechanical}(a) is the same as that of the optical drive case, as the static QRM eigenenergies remain the same. However, the quasienergy state labels from their orders in the range $[-\omega_d/2,\omega_d/2)$ from Fig.~\ref{fig:EnergyBasis_Mechanical}(b), from bottom to top at $\eta_M=0.3$ (vertical dotted black line), are:
$\alpha=2^{-},1^{+},3^{-},3^{+},1^{-},4^{-},2^{+},0$, listed on the bottom-right corner of panel (c).

\subsubsection{Variation of the Floquet engineering process versus\\ the amplitude of the dynamical coupling parameter $\eta_M$}
 

\begin{figure*}[htbp]
\includegraphics[width=.92\linewidth]{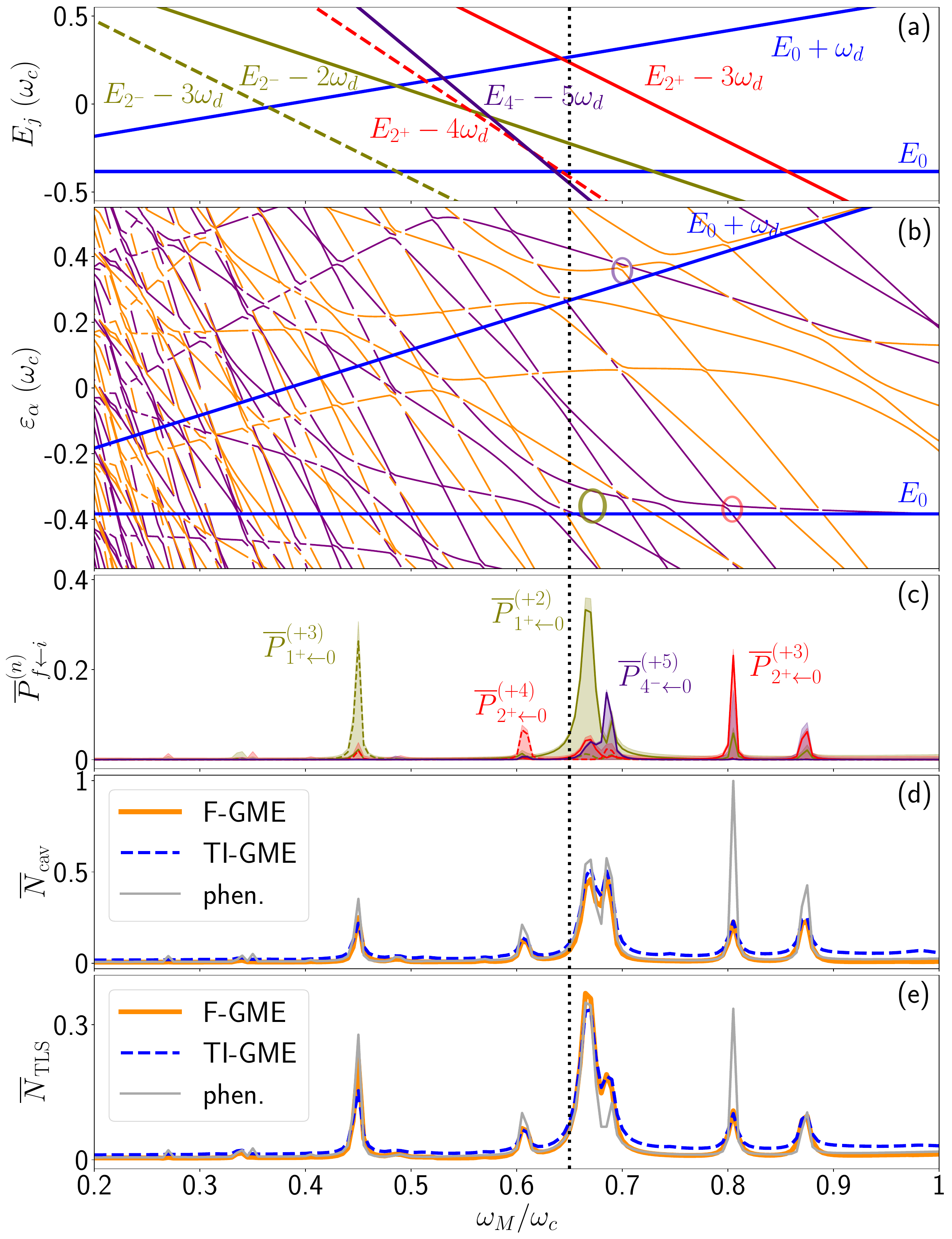}
\caption[]{\textbf{Mechanically driven cavity-QED: Eigenenergies, transition probabilities and populations as a function of the drive frequency $\omega_M$.} Same as Fig.~\ref{figS:Fig_wd_Optical}, but for the Floquet-engineered (mechanically driven) QRM, with $(\eta_0,\eta_M)=(0.5,0.3)$.
}
    \label{figS:Fig_wd_Mechanical}
\end{figure*}
In Fig.~\ref{figS:Fig_wd_Mechanical}, we analyze the spectral and population features as a function of the external mechanical modulation frequency $\omega_M$ (in units of $\omega_c$) for fixed values of $(\eta_0,\eta_M)=(0.5,0.3)$ and $\omega_a=\omega_c$. In contrast to the optical-drive case, the periodic perturbation now acts by modulating the light--matter interaction strength itself. Consequently, the drive does not directly inject excitations into either subsystem, but instead Floquet-engineers the hybridization structure of the USC Hamiltonian.

Since the switching process already occurs in the USC regime, with a sizable static coupling $\eta_0=0.5$, the time-dependent modulation can efficiently convert virtual dressing processes into real nonequilibrium excitations. As the modulation frequency increases, the injected energy generally promotes the production of real cavity and TLS excitations, except in frequency windows where destructive interference, nonlinear competition between channels, or anticrossing-induced redistribution suppresses the response~\cite{Akbari_Floquet_2025}.

As in the coherent optical drive case, the resonance peaks can be directly connected to the quasienergy spectrum. Multi-phonon (multi-mechanical-quantum) resonances occur when the modulation bridges quasienergy gaps between dressed polaritonic states, generating characteristic peak sequences in the long-time averaged observables. These resonances become narrower at higher order and typically appear at lower modulation frequencies, reflecting the larger number of exchanged drive quanta required to satisfy the transition condition.

The time-independent eigenenergy spectra remain independent of $\omega_M$, since the bare static Hamiltonian contains no dependence on the modulation frequency. More precisely, the standard QRM eigenenergies depend only on the static coupling strength, while the renormalized time-averaged Hamiltonian depends on the static and dynamical amplitudes $(\eta_0,\eta_M)$, but still not on $\omega_M$~\cite{Akbari_Floquet_2025}. Therefore, any frequency-dependent structures in the populations or spectra must originate from genuine Floquet effects rather than from static spectral motion.

Since the dynamical modulation is finite, $\eta_M\neq0$, the renormalized static spectrum differs from the unmodulated QRM spectrum through the modification of the effective DC component of the coupling. This shift is analogous to the ground-state and virtual-excitation renormalization effects discussed previously, and it provides the static backdrop from which the driven quasienergy anticrossings emerge~\cite{Akbari_Floquet_2025}.


Figure~\ref{figS:Fig_wd_Mechanical} provides a microscopic interpretation of the mechanical-drive response. The shifted eigenenergies in panel (a) indicate the crossings and avoided crossings that seed the quasienergy anticrossings shown in panel (b). Once the modulation is turned on, these quasienergy anticrossings signal efficient exchange of mechanical drive quanta with the dressed polaritonic system and therefore mark the locations of enhanced excitation production.

The transition probabilities shown in panel (c) 
of Fig.~\ref{figS:Fig_wd_Mechanical} quantify the dominant upward Floquet-assisted channels from the ground-state sector. Their summed contribution explains the peak structure in the cavity and TLS long-time averaged excitation numbers displayed in panels (d) and (e). As in the optical case, these probabilities are derived from the unitary Floquet theory of the closed system. In the open system, however, the actually observed peak strengths depend additionally on modal linewidths and dissipation. If a nominally strongest coherent channel is too narrow compared with the relevant damping scale, a broader neighboring process may become the practically dominant observable resonance.

The comparison between the phenomenological treatment, the F-GME, and the TI-GME again highlights the importance of quasienergy-resolved dissipation. Near isolated weak resonances, all approaches may show reasonable agreement. However, near strong quasienergy hybridization or overlapping sideband regions, the TI-GME can misestimate both the amplitude and detailed structure of the excitation peaks because it lacks an explicit treatment of the periodic dissipative channels generated by the modulation of the coupling itself.

In contrast, the F-GME incorporates relaxation directly in the driven quasienergy basis and therefore reproduces the phenomenological benchmark more faithfully. In particular, this agreement supports the interpretation that the observed structures are genuine Floquet-engineered USC phenomena rather than artifacts of a specific approximation scheme.

Broadly, Fig.~\ref{figS:Fig_wd_Mechanical} demonstrates that periodic modulation of the interaction strength provides a powerful route to engineer nonequilibrium ultrastrong-coupling physics. The resonance landscape is governed by mechanically assisted quasienergy transitions, interference among competing sidebands, and dissipation-induced selection of observable channels.

\subsubsection{Floquet--Liouville modal decomposition and quasienergy-resolved processes for the most influential spectral peaks for the mechanical drive}
\begin{table*}[htbp]
\centering
\caption{Dominant Floquet--Liouville modes contributing to the cavity spectrum for the case of optical coherent pumping the TLS. The table is corresponding to the Fig.~\ref{fig:Spectracav_Mechanical_wM065}(a) in the main text, with $(\eta,\eta_M,\omega_M)=(0.5,0.3,0.65\omega_c)$ for a USC resonant ($\omega_a=\omega_c$) cavity-QED.
Dissipation is characterized via $\kappa=0.1\omega_c$ for the flat cavity bath and $\gamma=0.01\omega_a$ for the flat TLS bath. The Floquet--Liouville modes are ranked by $\mathrm{Re}[\mathcal{W}_\mu]/\gamma_\mu$, for which, the contributing Floquet channels are listed with no less than $10\%$ contribution.}
\label{tab:FLmodes_cavity_kappa01flat_Mechanical}

\renewcommand{\arraystretch}{1.32}
\setlength{\tabcolsep}{5pt}

\footnotesize
\begin{tabular*}{\textwidth}{@{\extracolsep{\fill}} c c c c l c c c}
\hline\hline
$\mu$ &
$\Delta_\mu/\omega_c$ &
$\gamma_\mu/\omega_c$ &
$\mathrm{Re}[\mathcal{W}_\mu]/\gamma_\mu$ &
Dominant component $(\alpha,\beta,l)$ &
$\Delta_{\alpha\beta l}/\omega_c$ &
$r_\mu^{\alpha\beta l}$ &
$|R_\mu^{\alpha\beta l}|$ \\
\hline

3766 & 0.786 & 0.015 & 0.848 & $(2^{+},4^{-},1)$ & 0.787 & 0.972 & 1.040 \\
\hline
3793 & 1.164 & 0.015 & 0.455 & $(4^{-},2^{+},2)$ & 1.163 & 0.972 & 1.040 \\
\hline
3755 & 0.514 & 0.015 & 0.255 & $(4^{-},2^{+},1)$ & 0.513 & 0.972 & 1.040 \\
\hline
3047 & 0.941 & 0.102 & 0.044 & $(3^{+},4^{-},1)$ & 0.939 & 0.951 & 1.223 \\
\hline
3764 & -0.514 & 0.015 & 0.043 & $(2^{+},4^{-},-1)$ & -0.513 & 0.972 & 1.040 \\
\hline
3482 & -0.751 & 0.061 & 0.032 & $(2^{-},4^{-},-2)$ & -0.753 & 0.961 & 1.136 \\
\hline
3351 & 0.830 & 0.068 & 0.030 & $(2^{+},0,1)$ & 0.830 & 0.922 & 1.246 \\
\hline
3777 & 1.436 & 0.015 & 0.021 & $(2^{+},4^{-},2)$ & 1.437 & 0.972 & 1.040 \\
\hline
3473 & -2.051 & 0.061 & 0.019 & $(2^{-},4^{-},-4)$ & -2.053 & 0.961 & 1.136 \\
\hline
3770 & 1.814 & 0.015 & 0.018 & $(4^{-},2^{+},3)$ & 1.813 & 0.972 & 1.040 \\
\hline
3767 & 0.136 & 0.015 & 0.014 & $(2^{+},4^{-},0)$ & 0.137 & 0.972 & 1.040 \\
\hline
3468 & 1.401 & 0.061 & 0.008 & $(4^{-},2^{-},3)$ & 1.403 & 0.961 & 1.136 \\
\hline
3483 & 1.199 & 0.061 & 0.008 & $(2^{-},4^{-},1)$ & 1.197 & 0.961 & 1.136 \\
\hline
3489 & -0.101 & 0.061 & 0.007 & $(2^{-},4^{-},-1)$ & -0.103 & 0.961 & 1.136 \\
\hline
3341 & 1.120 & 0.068 & 0.006 & $(0,2^{+},2)$ & 1.120 & 0.922 & 1.246 \\
\hline
1850 & 0.982 & 0.153 & 0.006 & $(1^{+},2^{+},1)$ & 0.990 & 0.277 & 1.173 \\
     &       &       &       & $(3^{+},4^{-},1)$ & 0.939 & 0.244 & 1.100 \\
     &       &       &       & $(3^{+},0,1)$     & 0.981 & 0.105 & 0.723 \\
     &       &       &       & $(0,3^{+},2)$     & 0.969 & 0.102 & 0.712 \\
     &       &       &       & $(4^{-},3^{+},2)$ & 1.011 & 0.084 & 0.645 \\
     &       &       &       & $(2^{+},1^{+},2)$ & 0.960 & 0.076 & 0.615 \\
     &       &       &       & $(2^{+},0,1)$     & 0.830 & 0.047 & 0.484 \\
     &       &       &       & $(0,2^{+},2)$     & 1.120 & 0.046 & 0.476 \\
\hline
3035 & -1.009 & 0.102 & 0.004 & $(3^{+},4^{-},-2)$ & -1.011 & 0.951 & 1.223 \\
\hline
3055 & -2.959 & 0.102 & 0.004 & $(3^{+},4^{-},-5)$ & -2.961 & 0.951 & 1.223 \\
\hline
3031 & -1.659 & 0.102 & 0.004 & $(3^{+},4^{-},-3)$ & -1.661 & 0.951 & 1.223 \\
\hline
3039 & -0.359 & 0.102 & 0.004 & $(3^{+},4^{-},-1)$ & -0.361 & 0.951 & 1.223 \\
\hline\hline

\end{tabular*}
\end{table*}

\begin{table*}[htbp]
\centering
\caption{Dominant Floquet--Liouville modes contributing to the cavity spectrum for the case of optical coherent pumping the TLS. The table is corresponding to the Fig.~\ref{fig:Spectracav_Mechanical_wM065}(b) in the main text, with $(\eta,\eta_M,\omega_M)=(0.5,0.3,0.65\omega_c)$ for a USC resonant ($\omega_a=\omega_c$) cavity-QED.
Dissipation is characterized via $\kappa=0.1\omega_c$ for the Lorentzian--Ohmic cavity bath centered at
$\omega_0=\omega_c$ and $\gamma=0.01\omega_a$ for the flat TLS bath. The Floquet--Liouville modes are ranked by $\mathrm{Re}[\mathcal{W}_\mu]/\gamma_\mu$, for which, the contributing Floquet channels are listed with no less than $10\%$ contribution.}
\label{tab:FLmodes_cavity_kappa01LorOhmic1_Mechanical}

\renewcommand{\arraystretch}{1.32}
\setlength{\tabcolsep}{5pt}

\footnotesize
\begin{tabular*}{\textwidth}{@{\extracolsep{\fill}} c c c c l c c c}
\hline\hline
$\mu$ &
$\Delta_\mu/\omega_c$ &
$\gamma_\mu/\omega_c$ &
$\mathrm{Re}(\mathcal{W}_\mu)/\gamma_\mu$ &
Dominant component $(\alpha,\beta,l)$ &
$\Delta_{\alpha\beta l}/\omega_c$ &
$r_\mu^{\alpha\beta l}$ &
$|R_\mu^{\alpha\beta l}|$ \\
\hline

3767 & 0.787 & 0.005 & 5.793 & $(2^{+},4^{-},1)$ & 0.787 & 0.997 & 1.017 \\
\hline
3791 & 1.163 & 0.005 & 1.675 & $(4^{-},2^{+},2)$ & 1.163 & 0.997 & 1.017 \\
\hline
3789 & 0.513 & 0.005 & 0.844 & $(4^{-},2^{+},1)$ & 0.513 & 0.997 & 1.017 \\
\hline
3121 & 0.828 & 0.026 & 0.469 & $(2^{+},0,1)$     & 0.830 & 0.881 & 1.105 \\
     &       &       &       & $(2^{+},1^{+},2)$ & 0.960 & 0.103 & 0.377 \\
\hline
3676 & -0.754 & 0.007 & 0.203 & $(2^{-},4^{-},-2)$ & -0.753 & 0.994 & 1.096 \\
\hline
3769 & 1.437 & 0.005 & 0.176 & $(2^{+},4^{-},2)$ & 1.437 & 0.997 & 1.017 \\
\hline
3668 & -2.054 & 0.007 & 0.168 & $(2^{-},4^{-},-4)$ & -2.053 & 0.994 & 1.096 \\
\hline
3787 & -0.513 & 0.005 & 0.122 & $(2^{+},4^{-},-1)$ & -0.513 & 0.997 & 1.017 \\
\hline
3696 & 1.196 & 0.007 & 0.108 & $(2^{-},4^{-},1)$ & 1.197 & 0.994 & 1.096 \\
\hline
2423 & 0.975 & 0.082 & 0.087 & $(0,3^{+},2)$     & 0.969 & 0.165 & 0.521 \\
     &       &       &       & $(3^{+},0,1)$     & 0.981 & 0.165 & 0.521 \\
     &       &       &       & $(2^{+},1^{+},2)$ & 0.960 & 0.147 & 0.492 \\
     &       &       &       & $(1^{+},2^{+},1)$ & 0.990 & 0.147 & 0.492 \\
     &       &       &       & $(2^{+},0,1)$     & 0.830 & 0.086 & 0.375 \\
     &       &       &       & $(0,2^{+},2)$     & 1.120 & 0.086 & 0.375 \\
     &       &       &       & $(3^{+},4^{-},1)$ & 0.939 & 0.067 & 0.332 \\
     &       &       &       & $(4^{-},3^{+},2)$ & 1.011 & 0.067 & 0.332 \\
     &       &       &       & $(1^{+},3^{+},1)$ & 0.839 & 0.021 & 0.188 \\
     &       &       &       & $(3^{+},1^{+},2)$ & 1.111 & 0.021 & 0.188 \\
\hline
3675 & -0.104 & 0.007 & 0.074 & $(2^{-},4^{-},-1)$ & -0.103 & 0.994 & 1.096 \\
\hline
3778 & 1.813 & 0.005 & 0.073 & $(4^{-},2^{+},3)$ & 1.813 & 0.997 & 1.017 \\
\hline
2744 & 0.958 & 0.053 & 0.032 & $(2^{+},1^{+},2)$ & 0.960 & 0.916 & 1.124 \\
\hline
3768 & 0.137 & 0.005 & 0.026 & $(2^{+},4^{-},0)$ & 0.137 & 0.997 & 1.017 \\
\hline
3796 & 2.087 & 0.005 & 0.024 & $(2^{+},4^{-},3)$ & 2.087 & 0.997 & 1.017 \\
\hline
2719 & 0.992 & 0.053 & 0.022 & $(1^{+},2^{+},1)$ & 0.990 & 0.916 & 1.124 \\
\hline
2127 & 0.942 & 0.122 & 0.021 & $(3^{+},4^{-},1)$ & 0.939 & 0.963 & 1.038 \\
\hline
3109 & 1.122 & 0.026 & 0.021 & $(0,2^{+},2)$     & 1.120 & 0.881 & 1.105 \\
     &       &       &       & $(1^{+},2^{+},1)$ & 0.990 & 0.103 & 0.377 \\
\hline
3673 & 0.546 & 0.007 & 0.019 & $(2^{-},4^{-},0)$ & 0.547 & 0.994 & 1.096 \\
\hline
3116 & 1.478 & 0.026 & 0.018 & $(2^{+},0,2)$     & 1.480 & 0.881 & 1.105 \\
     &       &       &       & $(2^{+},1^{+},3)$ & 1.610 & 0.103 & 0.377 \\
\hline\hline

\end{tabular*}
\end{table*}

Tables~\ref{tab:FLmodes_cavity_kappa01flat_Mechanical} and \ref{tab:FLmodes_cavity_kappa01LorOhmic1_Mechanical} provide the FL modal decomposition of the cavity emission spectrum for the mechanically driven case, thereby revealing the microscopic quasienergy-resolved channels responsible for the spectral peaks discussed in the main text. In contrast to the optical-drive configuration, where the dominant channels were largely connected to transitions involving the low-lying states $(0,1^\pm,2^-)$, the mechanical modulation activates a qualitatively different family of processes dominated by transitions among already hybridized excited polaritonic states, most notably the $(2^{+},4^{-})$ manifold. This directly reflects the fact that the drive modulates the light--matter coupling itself and therefore couples dressed sectors through the interaction operator rather than directly pumping the TLS.

The most influential resonance in both baths is the mode centered near
$\Delta_\mu \simeq 0.787$, whose dominant component is
$(2^{+},4^{-},1)$. Its nearly pure weight
$(R_\mu^{(\alpha\beta l)}\approx 0.97$ for the flat bath and (0.997) for the Lorentzian--Ohmic bath) demonstrates that this spectral line originates from a highly resolved single Floquet sideband transition between two excited quasienergy states mediated by one mechanical quantum. This channel is therefore the mechanical analogue of a bright polaritonic sideband and explains the strongest cavity peak in the mechanically driven spectrum.

The next strongest resonances at
$\Delta_\mu \simeq 1.163$ and $0.513$ arise from the reciprocal channels
$(4^{-},2^{+},2)$ and $(4^{-},2^{+},1)$, respectively. Together with the dominant $(2^{+},4^{-},1)$ line, these peaks reveal a structured ladder generated by repeated sideband exchange between the same two quasienergy states. Their simultaneous visibility indicates that once the $(2^{+},4^{-})$ manifold becomes bright, multiple harmonics $l=1,2,3,\dots$ can contribute observably to the cavity emission. This is a clear signature of genuine Floquet engineering rather than a simple static dressed-state transition.

Another important family of channels involves the ground-state-connected process
$(2^{+},0,1)$ near $\Delta_\mu \simeq 0.83$. This mode appears more strongly in the Lorentzian--Ohmic bath than in the flat bath, indicating that spectral filtering by the structured reservoir enhances transitions whose emitted frequency lies closer to the bath center $\omega_0=\omega_c$. In physical terms, the reservoir selectively amplifies channels near resonance while suppressing off-resonant ones, thereby reordering the modal hierarchy without changing the underlying Floquet eigenstructure.

The tables also reveal several negative-frequency modes such as
$(2^{-},4^{-},-2)$, $(2^{+},4^{-},-1)$, and
$(2^{-},4^{-},-4)$. These correspond to the Hermitian-conjugate or reverse-frequency sectors of the two-time correlator and encode the full Floquet regression structure. While they are less dominant in the positive-frequency cavity spectrum, their presence confirms that the Liouvillian eigenmodes naturally organize into forward and backward sideband partners.

A particularly informative feature is the appearance of mixed modes, especially around
$\Delta_\mu \approx 0.97$ in the Lorentzian--Ohmic case (mode $\mu=2423)$ and similarly in the flat bath (mode $\mu=1850$). These modes contain comparable contributions from several channels:
$(0,3^{+},2)$,
$(3^{+},0,1)$,
$(2^{+},1^{+},2)$,
$(1^{+},2^{+},1)$,
$(3^{+},4^{-},1)$, etc. Such states are not associated with a single bare transition, but instead represent genuinely nonsecular Floquet--Liouville hybrid modes produced by dissipative mixing of nearby quasienergy channels. They become especially relevant when multiple transitions cluster spectrally and their linewidths overlap. This provides direct evidence for the importance of retaining nonsecular couplings in the master equation.

Comparing the two baths clarifies the role of linewidth engineering. Under the flat bath, the leading mode at $\Delta_\mu\simeq0.786$ has
$\gamma_\mu=0.015$, whereas in the Lorentzian--Ohmic bath the corresponding linewidth narrows dramatically to $\gamma_\mu=0.005$, causing its ranking metric
$\mathrm{Re}(\mathcal{W}_\mu)/\gamma_\mu$ to jump from $0.848$ to $5.793$. Similar sharpening occurs for the other dominant $(2^{+},4^{-})$ sidebands. Hence, the structured bath does not merely rescale the spectrum; it selectively increases coherence times of favored channels and thereby enhances their visibility.

Overall, the results in the tables show that the mechanically driven cavity spectrum is governed by quasienergy transitions between excited polaritonic sectors, with the $(2^{+},4^{-})$ ladder forming the principal bright manifold. Structured dissipation then acts as a spectral selector, narrowing and amplifying specific sidebands. This behavior differs qualitatively from the optical-drive case, where transitions involving the ground-state sector dominated more strongly. The mechanical modulation therefore realizes a distinct Floquet-engineering regime in which the periodic control of the interaction strength promotes higher polaritonic manifolds into the main radiative channels.


\subsubsection{Cavity bright modes with mechanical driving}
\begin{figure*}[htbp]
\centering
\includegraphics[width=0.95\linewidth]
{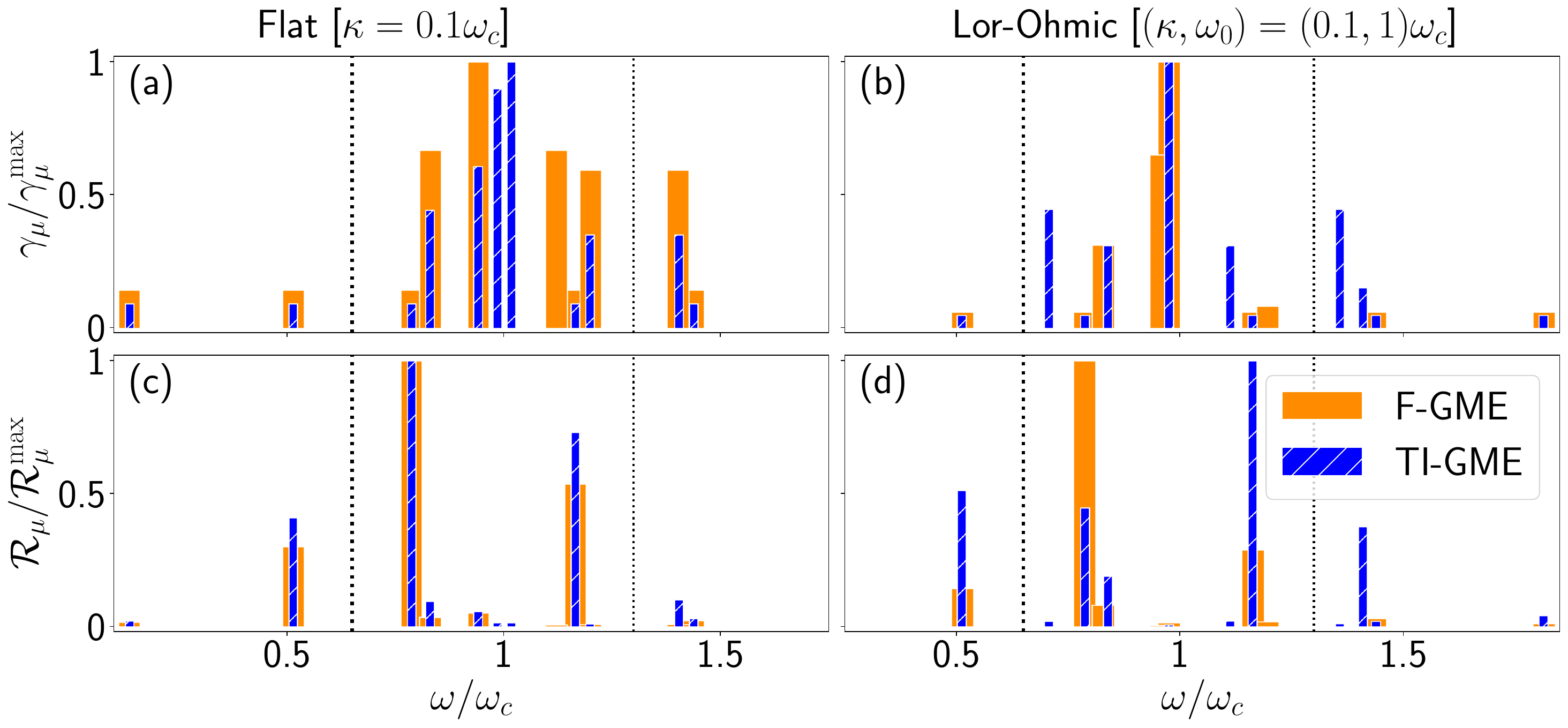} 
\caption[]{\textbf{Mechanically driven open cavity-QED in USC: 
cavity bright modes.} 
Shown in panels (a) and (c) are the spectral modal linewidth and peak prominence of Fig.~\ref{fig:Spectracav_Mechanical_wM065}(a), respectively.
(p) and (d) show the spectral modal linewidth and peak prominence of Fig.~\ref{fig:Spectracav_Mechanical_wM065}(b), respectively. We show a comparison between the TI-GME results (dashed blue) and F-GME results (orange).
}
\label{figS:CavityBrightChannels_Mechanical}
\end{figure*}

Figure~\ref{figS:CavityBrightChannels_Mechanical} presents the bright-channel decomposition of the cavity spectrum for the mechanically driven system, complementing the spectral results in the main text and the modal analysis of Tabs.~\ref{tab:FLmodes_cavity_kappa01flat_Mechanical} and \ref{tab:FLmodes_cavity_kappa01LorOhmic1_Mechanical}. The figure resolves the total cavity emission into its dominant Floquet transition channels, thereby identifying which quasienergy processes are responsible for each visible spectral feature.

The most prominent contribution originates from the channel
$(2^{+},4^{-},1)$, which generates the strong line near
$\omega \simeq 0.79,\omega_c$. As shown already by the tables, this transition is nearly pure and possesses the largest radiative weight. Its dominance confirms that the mechanical modulation preferentially activates transitions between excited polaritonic manifolds rather than only transitions connected directly to the ground state. In this regard, the drive reshuffles the hierarchy of radiative channels and promotes higher dressed sectors into bright emission pathways.

A second family of visible peaks is associated with the reciprocal sideband ladder built from the same pair of quasienergy states, notably
$(4^{-},2^{+},1)$,
$(4^{-},2^{+},2)$, and
$(4^{-},2^{+},3)$, appearing near
$\omega \simeq 0.51$, $1.16$, and $1.81$, respectively. The simultaneous presence of these harmonically related lines demonstrates that the periodic modulation of the coupling strength induces multiple Floquet replicas of the same underlying polaritonic transition. This is one of the clearest signatures of genuine Floquet engineering in the mechanically driven USC regime.

Additional bright channels involving lower states, such as
$(2^{+},0,1)$ and $(0,2^{+},2)$, contribute around
$\omega \simeq 0.83$ and $1.12$. These channels are weaker than the dominant $(2^{+},4^{-})$ ladder but remain spectroscopically relevant because they connect excited polaritons with the vacuum sector and therefore carry appreciable cavity dipole weight. Their coexistence with the higher-manifold transitions illustrates that the emitted spectrum is built from several competing quasienergy families rather than from a single resonance.

%
\begin{figure*}[htbp]
\centering
\includegraphics[width=.95\linewidth]
{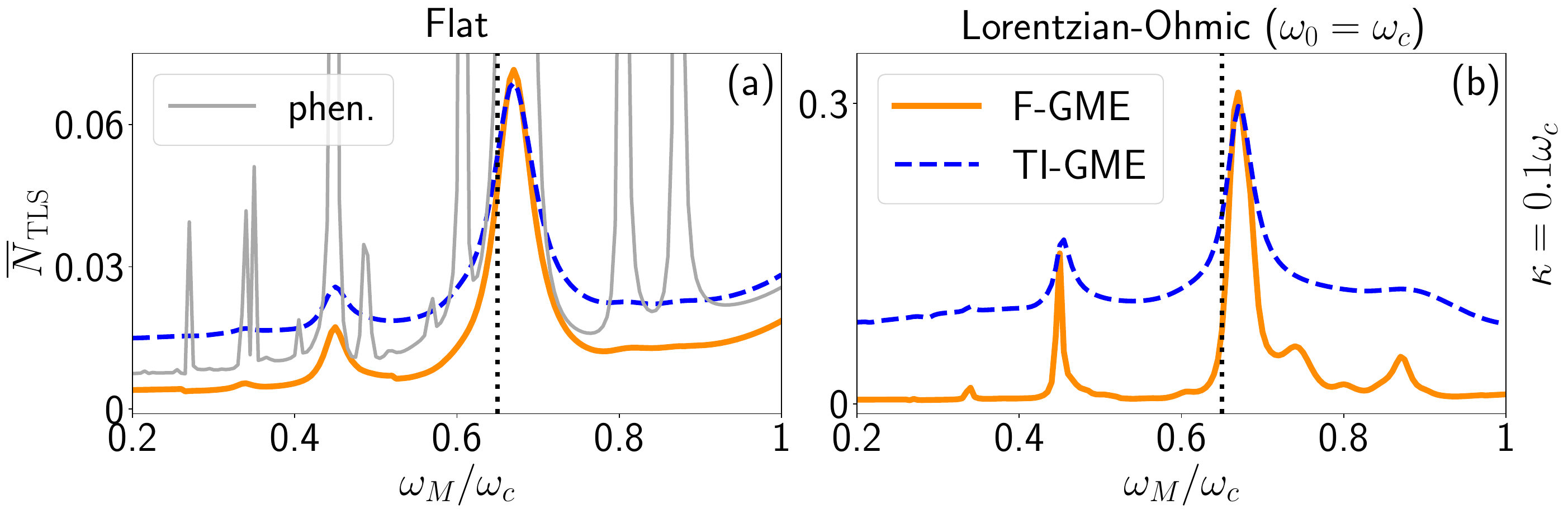} 
\caption[]{\textbf{Mechanically driven open cavity-QED in USC (Floquet-engineering the dissipative QRM in USC): TLS average number of excitations vs. mechanical drive's frequency $\omega_M$.} Same as Fig.~\ref{fig:AveNcav_wd_Mechanical}, but for the Floquet-engineered (mechanically driven) QRM, with $(\eta_0,\eta_M)=(0.5,0.3)$.
}
\label{figS:AveNTLS_wd_Mechanical}
\end{figure*}

Figure \ref{figS:CavityBrightChannels_Mechanical} also helps explain the difference between the flat-bath and Lorentzian--Ohmic results. Under the structured bath, channels whose frequencies lie closer to the bath center $\omega_0=\omega_c$ are selectively enhanced and narrowed, while off-resonant channels are relatively suppressed. Consequently, peaks near
$\omega\sim0.8 - 1.2$  become sharper and more dominant, whereas more distant harmonics receive less weight. This bath filtering is consistent with the linewidth reductions and ranking changes observed in the tables.

Comparisons between the F-GME and TI-GME decompositions (shown for the Lorentzian--Ohmic case) further clarify the origin of discrepancies between the two approaches. The F-GME preserves the full time-periodic jump structure and therefore resolves distinct Floquet sidebands separately, assigning spectral weight to individual channels such as
$(2^{+},4^{-},1)$ and
$(4^{-},2^{+},2)$. By contrast, the TI-GME projects the dynamics onto a static dressed basis and tends to merge or redistribute these sideband contributions. This can shift relative peak heights and obscure channels that are intrinsically Floquet in origin.

Overall, Fig.~\ref{figS:CavityBrightChannels_Mechanical} shows that the mechanically driven cavity spectrum is controlled by a bright polaritonic ladder centered on the $(2^{+},4^{-})$ manifold, supplemented by weaker vacuum-connected channels. The decomposition demonstrates explicitly that modulation of the light--matter interaction creates radiative pathways absent in the static problem, while the environment determines which of these engineered channels become experimentally visible.

\subsubsection{Average number of emitter (TLS) excitation with mechnical driving}

Figure~\ref{figS:AveNTLS_wd_Mechanical} shows the long-time averaged TLS excitation number for the mechanically driven (Floquet-engineered) cavity-QED system in the USC regime, complementing the cavity population results presented in the main text (Fig.~\ref{fig:AveNcav_wd_Mechanical}). While the cavity observable primarily tracks the photonic content of the driven polaritonic states, the TLS excitation number probes how the periodic modulation redistributes population into Floquet states with stronger matter character. Overall, the two observables provide a more complete picture of the Floquet-engineered nonequilibrium steady state.

As in the cavity case, the TLS population exhibits a sequence of resonant peaks and valleys when swiping the modulation frequency $\omega_M$. These features arise when the drive becomes resonant with quasienergy gaps and efficiently activates mechanically assisted transitions between dressed sectors (cf. Fig.~\ref{figS:Fig_wd_Mechanical}). Since the modulation acts on the light--matter coupling itself, the relevant resonances are not simple bare atomic transitions; instead, they involve hybrid polaritonic states whose cavity and TLS composition varies strongly with frequency. Consequently, a peak in the TLS population need not coincide exactly in magnitude with the corresponding cavity peak, even when both originate from the same quasienergy anticrossing, or the same-order quantum processes in the transition probability display.

The lower-frequency resonances typically correspond to first-order sideband processes and therefore generate broader and stronger responses, while higher-order multiphonon processes appear as narrower features at smaller or intermediate values of $\omega_M$. This hierarchy mirrors the Floquet transition probabilities discussed previously: lower-order channels generally dominate unless they are weakened by destructive interference, parity constraints, or competition from nearby resonances. When several channels overlap, the TLS population may display asymmetric line shapes or local suppressions rather than isolated monotonic peaks.

A notable aspect of the TLS observable is its enhanced sensitivity to the matter fraction of the participating Floquet states. Some mechanically induced resonances may produce sizable cavity excitation while yielding only modest TLS population if the populated states are predominantly photonic. Conversely, channels that weakly affect the cavity can still produce visible TLS peaks when they access states with stronger atomic polarization. This explains why the cavity and TLS population curves, although correlated, are not identical.

The comparison among the phenomenological theory, TI-GME, and F-GME again provides a useful consistency check. When the dissipation rates are deliberately reduced, the GME results move closer to the phenomenological curves, especially around well-isolated resonances. This indicates that once broadening becomes sufficiently weak, the quasienergy transition structure dominates over dissipative corrections, and all approaches recover the same principal resonant landscape. Residual discrepancies occur mainly near dense clusters of crossings or strongly mixed Floquet regions, where time-dependent dissipation channels and nonsecular effects become more important.

Compared with the optical-drive case, the mechanically driven TLS population more directly reflects genuine Floquet engineering of the interaction Hamiltonian. Here the modulation does not merely inject excitations into the TLS; it reshapes the dressed spectrum itself and thereby alters which polaritonic states are most accessible in the steady state. The resulting TLS population therefore serves as a sensitive probe of interaction-driven quasienergy restructuring.

Altogether, Fig.~\ref{figS:AveNTLS_wd_Mechanical} confirms that periodic modulation of the coupling strength can control not only cavity photons but also atomic excitation content in the ultrastrong-coupling regime. The frequency dependence of the TLS population provides complementary evidence for the resonance network, sideband activation, and nonequilibrium state engineering discussed throughout the main text.

\subsubsection{Emitter (TLS) incoherent spectra}

\begin{figure*}[htbp]
\centering
\includegraphics[width=.95\linewidth]
{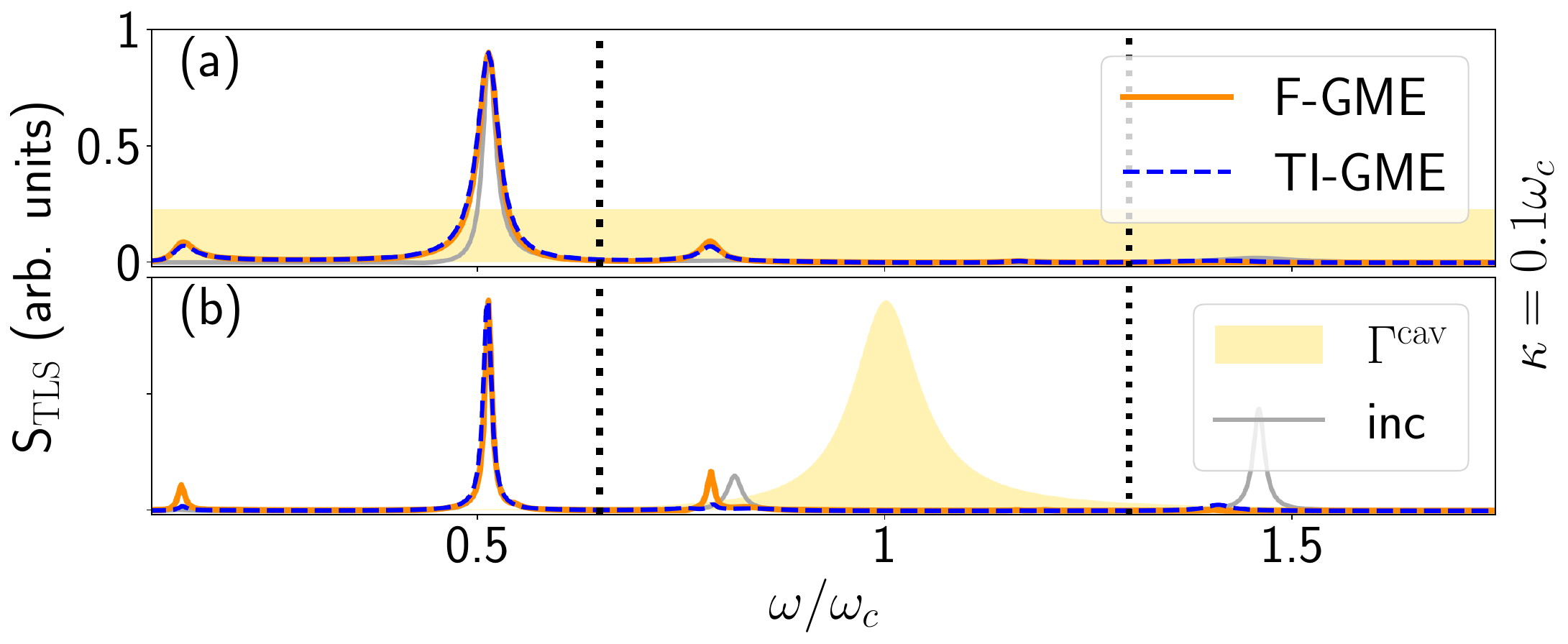} 
\caption[]{\textbf{
Mechanically driven open cavity-QED in USC (Floquet-engineering the dissipative QRM in USC): TLS (incoherent) spectra.} Same as Fig.~\ref{fig:Spectracav_Mechanical_wM065}, but for the Floquet-engineered (mechanically driven) QRM, with $(\eta_0,\eta_M)=(0.5,0.3)$.
}
\label{figS:SpectraTLS_Mechanical_wM065}
\end{figure*}

%
\subsubsection{Emitter (TLS) bright modes with mechanical driving}
\begin{figure*}[htbp]
\centering
\includegraphics[width=0.95\linewidth]
{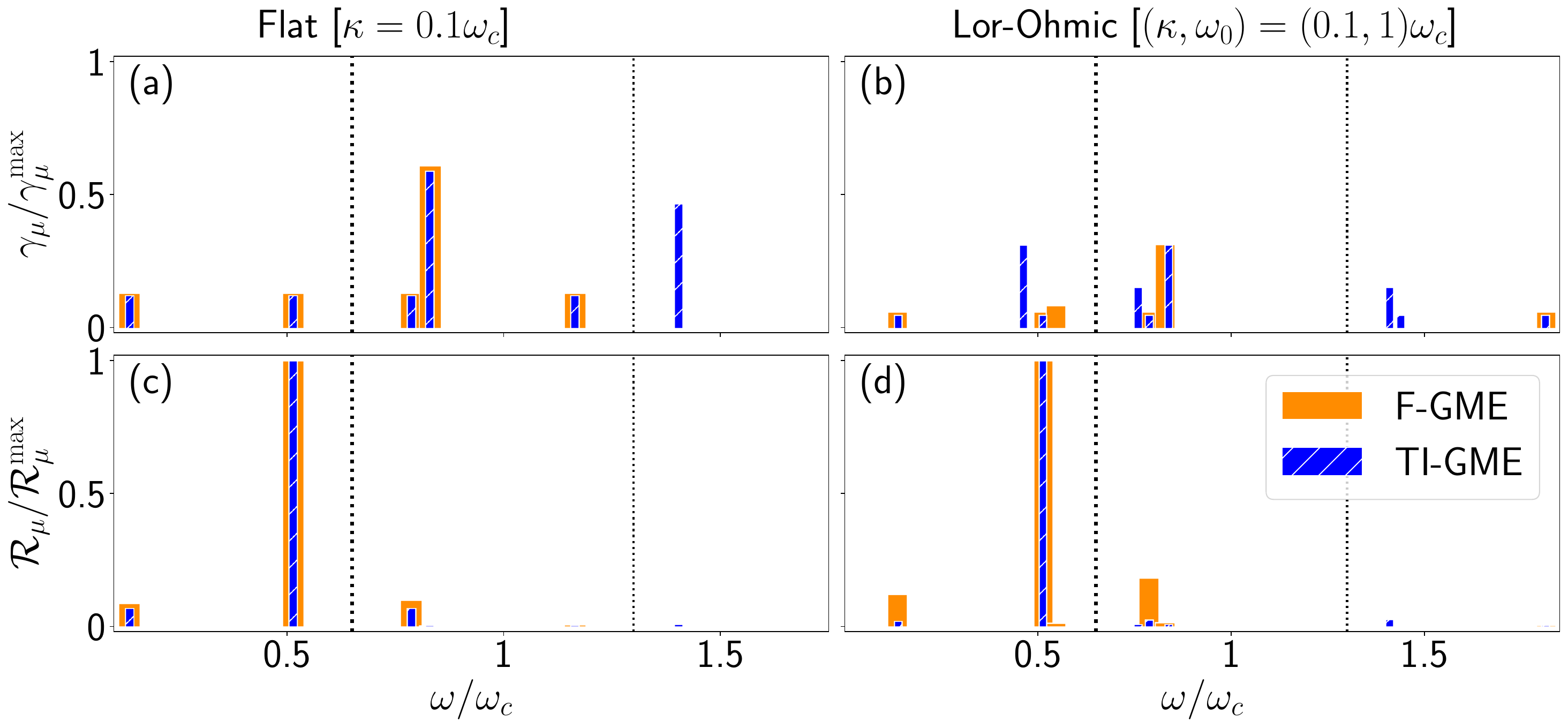} 
\caption[]{\textbf{Mechanically driven open cavity-QED in USC: 
TLS bright modes.} 
Shown in panels (a) and (c) are the spectral modal linewidth and peak prominence of Fig.~\ref{figS:SpectraTLS_Mechanical_wM065}(a), respectively.
(p) and (d) show the spectral modal linewidth and peak prominence of Fig.~\ref{figS:SpectraTLS_Mechanical_wM065}(b), respectively. Comparison between the TI-GME results (dashed blue) and F-GME results (orange) is made.
}
\label{figS:TLSBrightChannels_Mechanical}
\end{figure*}

Figure~\ref{figS:SpectraTLS_Mechanical_wM065} displays the incoherent TLS emission spectrum for the mechanically driven cavity-QED system in the USC regime, to provide the matter-sector counterpart of the cavity spectra discussed in the main text (Fig.~\ref{fig:Spectracav_Mechanical_wM065}). While the cavity spectrum emphasizes radiative channels weighted by photonic (cavity-field) matrix elements, the TLS spectrum probes the same Floquet steady state through the atomic (spin-like) operator and therefore highlights transitions with stronger matter participation. The comparison between the two spectra is especially valuable in the USC regime, where light and matter are strongly hybridized but not identically represented in different observables.

The dominant spectral lines remain associated with the same quasienergy transition network identified in the cavity spectrum, particularly the mechanically activated sideband ladder involving the $(2^{+},4^{-})$ manifold. Accordingly, peaks appear near frequencies corresponding to channels such as
$(2^{+},4^{-},1)$,
$(4^{-},2^{+},1)$,
$(4^{-},2^{+},2)$, and higher harmonics. 
However, their relative amplitudes differ from the cavity case because the TLS detection operator weights the atomic component of the participating Floquet states rather than their cavity content. As a result, transitions that are strongly bright in the cavity spectrum need not be equally dominant in the TLS spectrum, and vice versa.

This distinction is particularly important in the USC regime, where dressed states can vary substantially in their photonic and matter composition. Some mechanically engineered resonances primarily populate polaritons with larger cavity fraction and therefore appear stronger in the cavity output channel, whereas others carry stronger TLS polarization and become comparatively enhanced here. The TLS spectrum thus provides an independent spectroscopic map of the same Floquet manifold structure.

As in the optically driven cavity-QED, the TLS spectrum is generally more sensitive than time-averaged populations to whether the dissipative treatment retains the full periodic structure of the dynamics. The F-GME preserves time-dependent jump operators and the full harmonic content of the quantum regression problem, thereby resolving micromotion-induced sidebands and interference among nearby channels. In contrast, the TI-GME projects the dissipative dynamics onto a static dressed basis and can redistribute spectral weight among neighboring lines. Consequently, even when integrated observables agree reasonably well, visible differences may remain in peak heights, widths, or local line-shape asymmetries.

The structured Lorentzian-Ohmic bath again acts as a frequency filter. Peaks whose emission frequencies lie near the bath center are sharpened and relatively enhanced, while more detuned sidebands are suppressed. Because several mechanically generated TLS lines cluster around intermediate frequencies, this filtering can significantly reorder the apparent prominence of spectral features. Hence, the bath does not merely broaden the TLS spectrum; it selects which Floquet-engineered matter-sector channels become experimentally most visible.

Another notable feature is the persistence of higher-order sidebands. Since the coupling modulation directly reshapes the interaction Hamiltonian, multiple Floquet harmonics can acquire measurable weight. In the TLS spectrum these channels may appear as weaker satellite peaks or shoulders around the principal resonances, signaling that the matter subsystem retains clear memory of the periodic interaction engineering.

Indeed, Fig.~\ref{figS:SpectraTLS_Mechanical_wM065} confirms that mechanical modulation in the USC regime engineers not only the photonic emission landscape but also the atomic response. The TLS incoherent spectrum exposes complementary bright channels, highlights the matter content of the Floquet polaritons, and further demonstrates the necessity of a fully Floquet-resolved dissipative treatment for quantitatively reliable frequency-domain predictions.

Figure~\ref{figS:TLSBrightChannels_Mechanical} presents the bright-mode decomposition of the TLS incoherent spectrum for the mechanically driven cavity-QED system in the USC regime. This figure is the matter-analogue of Fig.~\ref{figS:CavityBrightChannels_Mechanical} and complements Fig.~\ref{figS:SpectraTLS_Mechanical_wM065} by explicitly resolving the incoherent TLS emission into the dominant FL modes that can possibly hybridize Floquet transition channels. It therefore identifies which quasienergy processes are most visible when the system is monitored through the TLS operator rather than the cavity output field.

As in the cavity spectrum, the principal bright channels are organized around the mechanically induced $(2^{+},4^{-})$ transition ladder. Channels such as
$(2^{+},4^{-},1)$,
$(4^{-},2^{+},1)$,
$(4^{-},2^{+},2)$, and higher harmonics generate the strongest TLS lines, confirming that the periodic modulation of the coupling activates repeated Floquet sidebands between the same pair of dressed manifolds. Their presence in both cavity and TLS decompositions demonstrates that these are genuine system resonances rather than detector-specific artifacts.

However, the relative brightness of the channels differs from the cavity case because the TLS observable probes the matter component of each Floquet polariton. Consequently, transitions involving states with larger atomic polarization are comparatively enhanced here, while channels dominated by photonic weight can be reduced even if they remain strong in the cavity output spectrum. This distinction is a hallmark of the USC regime, where light and matter excitations are inseparably hybridized but retain observable-dependent signatures.

The decomposition also highlights transitions connected to lower manifolds, such as channels involving the vacuum or first excited strata. These processes may be secondary in the cavity spectrum yet become more visible in the TLS signal when the participating states possess stronger spin-like character. Thus, the TLS bright-channel map provides complementary information about the composition of the Floquet steady state and about which engineered polaritons carry matter excitation.

For the Lorentzian-Ohmic bath, comparisons between  F-GME and TI-GME channel decompositions again clarify the origin of discrepancies between the two dissipative treatments. The F-GME retains the explicit time-periodic jump operators and therefore assigns spectral weight to distinct Floquet harmonics, separately. In contrast, the TI-GME collapses these processes into an effective static dressed-state picture, which can merge nearby sidebands or misallocate their intensities. The resulting differences are particularly noticeable in the TLS channel map because the matter operator is dense in the dressed basis and couples to multiple harmonics simultaneously 
(note the bare-state representation of the matter system operator $\sigma_x$ in comparison with the light system operator $\Pi$).

The features in Fig.~\ref{figS:TLSBrightChannels_Mechanical} further show that structured dissipation acts as a selective visibility filter. Channels whose emission frequencies lie near the bath maximum are sharpened and amplified, whereas more detuned transitions are weakened. This reweighting can change which TLS sidebands appear dominant experimentally, even though the underlying quasienergy transition network remains the same.

The cavity and TLS bright-mode figures, together, i.e., Figs~\ref{figS:CavityBrightChannels_Mechanical} and \ref{figS:TLSBrightChannels_Mechanical}, demonstrate that mechanical Floquet engineering in the USC regime creates a common set of quasienergy radiative pathways whose observed prominence depends on the measurement operator or the observable of choice, whether it is a light-like observable or a matter-like observable. The cavity output emphasizes photonic content, while the TLS emission emphasizes matter content. Their combined analysis therefore provides a fuller tomography of the Floquet-engineered polaritons and further supports the need for a fully Floquet-resolved open-system treatment.


\vspace{0.1cm}

\bibliography{main}

\end{document}